\documentclass[12pt]{book}
\usepackage{mathtext}
\usepackage{amsfonts}
\usepackage{amssymb}
\usepackage{amsmath}
\newcommand{\nnn}{\bigskip}
\newcommand{\nn}{\medskip} 
\newcommand{\n}{\smallskip}

\renewcommand{\d}{\delta}
\renewcommand{\b}{\beta}
\newcommand{\g}{\gamma}

\newcommand{\Rev}{{\rm Rev}}
\newcommand{\grad}{{\rm grad }}
\newcommand{\card}{{\rm card}}
\newcommand{\ora}{{\rm or}}

\newcommand{\QFT}{{\rm QFT}}
\newcommand{\D}{\Delta}

\newcommand{\e}{\epsilon}
\newcommand{\FT}{{\rm FT}}

\newcommand{\Qu}{{\rm Qu}}

\newcommand{\ar}{\longrightarrow}
\newcommand{\w}{\omega}
\newcommand{\s}{\sigma}
\newcommand{\la}{\lambda}
\renewcommand{\a}{\alpha}
\begin{document}
\title{{ {\bf {\LARGE Constructive physics}}}}
\author{Y.I.Ozhigov}
\date{}
\maketitle
\newpage
\nnn

Supported by NIX Computer Company, grant $\# $F793/8-05.
\newpage

\nnn

\ \ \ \ \ \ \ \ \ \ \ \ \ \ \ \ \ {\it I dedicate this book to my teachers}
\newpage

\tableofcontents
\newpage

{\bf Annotation}
\nnn

At first, I intended to call the book ''Algorithmic physics''. Then I replaced algorithmic by constructive in the title. The point is that the constructivism is the direction in mathematics, which contains algorithms but it is irreducible to them only. The book is devoted to the analysis of the mathematical apparatus of quantum theory, and it contains arguments for the necessity and desirability of its replacement by the constructive mathematics. I take the main examples just from quantum mechanics, but factually, it is a question of the mathematical apparatus of all physics.
The row of principal phenomena of the collective type belongs to the area where become valuable such elements of reality that are not contained in the standard mathematical abstractions. It is the evidence of the serious drawbacks of the standard mathematical apparatus, used in physics, which rests on the standard mathematical analysis, algebra and classical logic. 
The most known example on which these drawbacks become evident is the famous quantum computer, because the traditional methods are not applicable to the investigation of its scalability. 

The problem of quantum computer scalability represents factually the old question of the description of the measurements and decoherence in quantum theory. This obstacle is very serious. We could obtain the possibility to investigate properly the complex systems, which belong to chemistry and biology only if we overcome this obstacle. I am sure that just the switch of physics to the constructive way makes it possible. In this book, I explain how to make quantum physics constructive. The uncommonness of the aim I set in writing of the text requires the more detailed explanation of steps than it is customary in the physical and mathematical literature. This book is rather the manifest, but not as the final answer to the question, how constructive physics looks. The real building of this science requires the big efforts, including the serious and captivating work of programmers. 

Only the rebuilding of the gigantic construction of the modern physics in the constructive manner can open doors to the understanding of the complex processes in the sense of exact sciences. The matter concerns the new science - quantum physics of complex systems. Only constructivism makes possible to build it. I hope that the reading of this book will inspire an inquisitive reader to the practical participation in this important and exciting work. 

\section{Algorithms and the future of physics}

Physics, which was the unconditional leader in the science of twentieth century now gives its visible positions to such disciplines as biology. This change is natural, and it does not mean that the subject of our science loses actuality or the community already takes no interest how the world is constructed. On the contrary, just now the interesting process goes which concerns some reconsideration of foundations. Such ideas in general never disappear (at least in talks) but just now, they obtain the degree of ripeness, which makes possible to give them the form of real turn. This turn has not yet happened. Nevertheless, we can in general predict this turn from the existing situation. I attempt to do it in this book. 

The modern situation in physics looks like a crisis, and the genealogy of this crisis is the same as for the crisis in mathematics in the first third of 20 century: this is the crisis of the axiomatic method. It concerns the integrity of natural sciences and the crisis in mathematics, which was not completely resolved (see below). The axiomatic method that typically serves as the standard of reliability is not the same in reality, and the address to the experiment for the checking of appropriateness of one or another mathematical apparatus becomes not only legal, but also unavoidable. 

The mathematical community realized the situation after the failure of Hilbert program that we discuss later. The most radical method proposed for the overcoming this difficulty is important for us, because theoretical physics would have to follow this way. This method as applied to mathematical logic sounds as follows: we must detruncate the possibilities of formalism by the explicit introduction of effective procedures to it. This reduction is called the constructive mathematics. Today we possess the more exact kit of instruments of the constructive mathematics, and we can formulate the requirement of constructivism more definitely: {\bf algorithms must replace formulas}. Of course, it does not mean that we refuse from formulas, but we must treat a formula as some algorithm of its application.\footnote{ In the midst of pure mathematicians constructivism is not generally accepted though it is already well elaborated. Physicists are factually more familiar with it, because the relation between formulas and experimental data presumes some a priori procedural character.} Hence, conditions of the physical constructivism are more auspicious, and certainly, its ideas here will obtain the further development.

Why the dawning of the constructive age in physics is unavoidable? Because the simple problems with one or two particles (that are solvable by the functions permitting the expansion of Schmidt type, e.g., factually, by one particle functions) are already solved, and the formulas do not work for more complex problems. Here I must explain what the complex problems are. Let us consider the gas consisting of $10^{25}$ the small particles of the form of sphere (their size is much less than the distance between them), and with the simple law of collision. For this system we can ask what will be the mean energy of particles in the unit of volume, what will be the temperature, the pressure, etc. There are the questions about the statistical values and it is the subject of statistical physics. In this case, the rare events are out of the consideration, for example will ever these particles obtain the shape of human body, etc. 

We can set the other problem: what is the comparative probability that some three fixed particles once appear at one line in comparison with the other three particles, provided the initial states of all particles are determined with the known accuracy. The variant of this problem is the acceptance of the form of amino acid or nucleotide. This is the question about rare events, but for its solution we need the enormous computational recourses because we have to search among the total number of variants of the order of exponential of $10^{25}$. One could treat such questions as the empty exercises, which have no attitude to the reality, but it is wrong. Here is the version of such inaccessibly difficult problem which solution touches all of us. Can we guarantee that in course of the life of one generation the Earth orbit will not radically change that makes our life impossible? The problem of tree bodies in classical mechanics have no analytical solution, and the problem of more complex ensembles have no even the reliable numerical solution, even for the fairly small time segments. The movement of the Earth in the solar gravitational field with the influence of the Moon, Jupiter and Saturn belongs to this type of problems. 

Despite the unconditional importance of the cosmological problems, I think (and many others share this opinion) that we have more chances for the success in the areas where the human practice can help us more effectively, as in chemistry and biology. It is not the discrimination of cosmology. Merely the construction of big bodies and their evolution is more complex than for atoms and molecules despite their evolution is macroscopic, and hence in the cosmology the role of observation is large. It is more difficult to develop the detailed theory in cosmology than to solve the questions about the rare events in the gas\footnote{Here is the example for the detailed theory in cosmology: are there planets in our Galaxy where a human can live without environmental suit? The solution of it is beyond our possibilities. We do not know how to develop cosmology for that because the human practice cannot help us here. In contrast, in chemistry this practice plays the key role. We can expect that if we learn how to solve the problems in chemistry and biology, e.g., how to control processes that we meet every day and where we have the possibility for the various experiments, it would essentially help us in the cosmological problems as well. At least I do not see any other way.}. 

The explicit demand of practice is then the solution the problems belonging to chemistry and biology. The example is the creation of the model of a living cell that possesses the ability to predict its behavior. The cell can abruptly change its behavior due to a single photon of the certain frequency that hits to its membrane and starts the cascade of chemical reactions leading to the change of movement of the cell. The robust model must predict all these reactions, and in what follows, I will consider the possible approach to the creation of such models.\footnote{But not simple animations in sense of moving pictures which cannot have the predicting force.} We can expect that the modern level permits the creation of such models, e.g. the full account of the quantum character of all elementary particles composing a living cell gives such a model. For this, we must be able to apply quantum physics to the systems consisting of millions of particles. The attempts to build the model of living things without the usage of the whole arsenal of quantum physics mean the waste of time. It is also senselessly to reduce the quantum nature of reality to some method of the computation of the elastic forces acting between the atoms in molecules. Quantum physics contains the principal nonlocal constituent which sense is that an ensemble is irreducible to the simple collection of its members. It means that the role of quantum theory is not in the so-called quantum effects, which disappear at the big distance or for the big velocities. Its role is the main in the investigation of the complex, potentially – macroscopic sized systems at the level of exactness making possible to build their models with the predicting force. 

The main intrigue is thus that the physics must deal with the complex processes involving millions of particles, with the processes, which now belongs to the formal sphere of chemistry and biology. The real take-over of these processes requires the creation of their models by means of quantum physics. However, the modern quantum physics is not ready to this role. Its subject of investigation consists of simple systems, of one or two particles typically, or systems reducible to one or two particle. Nevertheless, quantum theory has discovered such properties of ensembles that are not explainable in terms of the separate particles in these ensembles. This is the existence of entangled quantum states. These principally multi particle phenomena lie in the basement of chemical reactions, and in the basement of the process of life. The real understanding of these processes lies along the line of development of physics, and I will try, with reasonable limits of my possibility, to convince the reader of this theses. However, for this the physics must acquire the constructive form. 

We can treat the physical constructivism in a simple style. The formed situation says that the computers will play the key role in physics. In general, nobody contests its auxiliary role in all scientific areas. The peculiarity is that just in physics they will play not auxiliary, but organizing role. This opinion could seem strange because in the fundamental physics this role always belongs to the mathematical apparatus. I will try to substantiate the idea that just the mathematical apparatus of physics must change by means of computerization. This physics can be called algorithmic, because algorithms will play here the role of formulas. But more right to call it constructive physics, because the main here is the continuation of the great traditions of constructivism, laid in mathematics after the realization of the crisis of its foundations following to the failure of Hilbert program of axiomatization of the natural sciences. 

This turn in the modern physics is difficult but unavoidable. One of the reasons of my confidence is that the mathematical apparatus for this is already ready. This is the constructive mathematics, including constructive mathematical logic, constructive mathematical analysis and the elements of constructive algebra, practically completed (excluding a few open problems) algorithm theory, and the developed technique of the industrial programming with the sufficiently powerful computers. I will try to prove that this basement is quite sound and the modern physics can be transferred to constructive language without serious problems. The constructivism then rests on the reliable foundation enough elaborated for the beginning of its regular implementation in physics. I will consider quantum theory but it does not mean that this is limits the expansion of constructivism; its area is the whole physics and even more. 

I think that the reader is aware that the complete transfer of physics to the constructive (factually – to the algorithmic language) represent the problem feasible to the big party of investigators. In this book, I try to convince him that this transfer is realizable in the framework of routine work, and that this transfer is extremely desirable for those who decided to connect their life with the science and for those who expect some real output from the science. The most important argument for the constructive physics is that it gives us the visual picture of the reality without the loss in the accuracy. A book can give the perception of clearness for static images only, for the dynamics – it is the destination of video film. The strong side of algorithmic approach just in that it gives the visual representation of the dynamical scenarios of the complex systems, and it is perhaps the most cogent argument for this approach. The form of a book does not give this possibility to the author. My aim here is slightly different.
I will bring the formal arguments for algorithmic approach by the proof of its full legality in the structure of physical and mathematical disciplines. This aim is important because the constructive mathematics is not contained in the typical courses for physics, and there is some fear in the using of algorithmic language in theory. Such a fear is groundless: algorithms are even more reliable than formulas. We will see that the algorithmic form of physics preserves the customary tools of the work of theorists practically without changes. At the same time, algorithmic approach makes possible to establish order where now the known disorder reigns. This role is peculiar just to the mathematical apparatus, but not to the technical service. 

One of the incentives to the writing of this book is the recent appearance of the possibility of the direct comparison of the traditional apparatus with algorithmic approach, and this possibility resulted from the known project of a scalable quantum computer. Such a possibility of the direct comparison has never appeared earlier. For me it is the evidence that this book will meet the rational interpretation, and the reader will be able to assess the value of the brought arguments for the physical constructivism independently.

\section{Preface}

The modern development of natural sciences and technique compels us to extend the exact methods of theoretical physics to chemistry and biology. The descriptive character of our representation about the design and functionality of the complex systems in Nature does not permit us to advance in the control over them in such degree that the society expects from the science. The interdisciplinary approach or the synthesis of sciences usually proposed for this aim is the right general idea, but to obtain the concrete results we must extend the exact methods of physics to the complex systems. Any other approach ignoring the exact methods in theory is doomed to failure. When speaking about the known reconsideration of the basements of quantum theory we will keep in mind the modification of the mathematical apparatus, but not the refusal from it. The modification we have to realize is sufficiently serious. One of the aims of this book is to show that this modification is the single real possibility to include the complex systems in the area of physics. Only this aim will constitute justification of the whole work. Naturally, there arises the question of succession, namely the guarantees that the planned modification will not throw overboard anything already reached by our science. This question is the most difficult and one cannot answer to it as to the scientific problem with exact formulation, because it is impossible to build the mathematical apparatus like one of physical theories. \footnote{Using analogy from programming we can compare the mathematical apparatus with the operation system.}

Factually, we have no choice at all, because the all possibilities of the constructing of new mathematics are already realized. This new mathematics is the constructive mathematics. The crisis I speak about is not new. The first its manifestation in the history is the failure of Hilbert program of axiomatization of the natural sciences. Constructivism has appeared as the possible outlet from this crisis in the area of formal logic; and the following development of constructive methods in physics represents just this new mathematical apparatus, which we have to accept. 

\subsection{Constructivism and its role in quantum theory}  

The idea of introduction of the constructive methods in physics has its own history. In the work \cite{NB} J. von Neumann and G. Birkgoff proposed to replace the classical logic in quantum mechanics by the constructive mathematical logic. \footnote{I met this ideas for the first time at the seminar of A.G.Dragalin in 70-th, which was devoted to the possible application of intuitionism in quantum theory, and from that time realized how determinant role plays what is called the foundations of mathematics in theoretical physics. This role is so that it would be more correct to speak about the common basements of physics and mathematics.} We know today more than in that times. The whole areas have been arisen: the algorithm theory, the constructive mathematical analysis and the constructive (or algorithmic) algebra, based on the notion of algorithm. 

It is well known that the main obstacle in the extension of exact methods to the complex systems is the computational complexity, which arises on this way. The development of the science about the computations must thus play the fundamental role in this process, and the computations themselves must be not the technical service, but the integral part of the new physics – the physics of complex systems. The constructivism gives the necessary theoretical substantiation for that in sense that it determines the general form of mathematical instruments, which must replace the traditional language of formulas and classical proofs. However, this is not sufficient for the obtaining of the exact results forming the aim of physics.  

Evidently, in complex systems, we cannot expect to obtain the exact results like the form of atomic spectrums, and we must reformulate what the exactness means for these systems. The exactness means the fidelity of the reproduction of the dynamical scenario. If we obtain the method of creation of the right scenario, we would be able to obtain the exact values in the traditional sense of this term. One must keep in mind that the constructivism contains the computational part of mathematical formalism, which we can always attract for the obtaining of the numerical evaluations in traditional sense. \footnote{It would be naive to compete with the efficiency, for example, of analytical methods for the electron in the Coulomb field of motionless charge, or standard tasks of this type. All the analytic technique contains in the constructivism and we can use it, with some precautions, which we discuss in what follows. Our aim is not to compete with analytical technique, but to ensure the mating of it with the world of many particles where this technique means nothing.} Just the obtaining of the right scenario for the complex systems represents the problem, and we need constructivism for this aim. This is our understanding of the exactness for the complex system. 

Here the peculiarity of constructive physics becomes apparent. In the standard quantum theory, it is possible to obtain the results on the tip of a pencil, in sense that the theory immediately gives the prediction of the result of experiment. In the constructive physics, there is the serious mediator between the theories and experiments: the computer. Its role in the traditional physics is technical and subservient; it realizes there the so-called mathematical or computational methods. We will see that it is not appropriate for complex systems. Constructivism is not the perfection of the computational methods of quantum theory, but its new form.\footnote{The standard program packages used in quantum theory do not presume the direct participation of programmers. The necessity to attract programmers is the practical criterion determining the border with constructivism, here the participation of programmers is necessary, and it bears the conceptual burden.}
Here the creating of computer programs and the work with them practically belongs to the theory as the technique of the solution of differential equations belongs to the arsenal of a traditional physicist theorist. 

Constructive mathematics does not give us immediately the computer programs that we can realize using the existing program packages. It does not give even the algorithms, which we must build using the so-called heuristics. The programming required here is thus the conceptual but not the technical because the algorithms themselves will become more exact. The advance here requires the industrial form of programming designed for the work of the teams of programmers at the same project. The tools for such a work appeared comparatively recently. Just the existence of such tools gives me the confidence that it is possible to divide the problem of the description of complex evolutions to the certain and transparent tasks, without which the physical constructivism would be the empty talk. \footnote{The reader can find many examples of such talks in the history of physics; I do not intend to discuss it. One must clearly distinguish the talks from the concrete flow-blocks of algorithms which I name the constructive heuristics and which serves as the foundation for the needed programming.}
We will not consider the questions connected with the programming here, but instead take up the constructive heuristics, and determine how must they look for the complex quantum systems. 

We can speak about the traditional form of the theory only when we substitute heuristics, in their applications to simple systems to which quantum theory is applicable and where the complete accordance with the known experimental data is required. For the complex systems where the traditional theory is inapplicable only algorithms based on our heuristics will be at our disposal with some possibility to revise the heuristics themselves. 

\subsection{My vision of the history of question}

The starting point in the writing of this book was my own work in quantum informatics and quantum computations. Just it leaded me to the clear understanding of the real nature of the problems of quantum informatics as the area of many particle quantum phenomena, of the stunning possibilities which the development of computers opens for the natural sciences, and of what a scanty part of these possibilities is mastered by the existing mathematical methods. 

The immediate reason for the address to the constructivism is connected with the problem of quantum computations and quantum computers. Quantum computer was proposed by Feynman (\cite{Fe2}), and by other authors (see for example \cite{Be}) as the necessary instrument for the simulation of quantum many particle physics, because it is impossible to overcome the exponential computational barrier arising in the complex systems with many particles on a classical computer. The evolution of my understanding of this problem repeats its history. It goes from the initial idea about the necessity of a quantum computer, formalization of this notion, euphoria of 90-th to the more deep view to the problem in the light of the difficulties of hardware-based realization of a quantum computer, and finally – to the necessity of the revision of the basement of quantum theory by the algorithmic approach.

Quantum computations first arise as the mathematical theory based on the traditional Hilbert formalism for quantum systems. However, the most valuable here became the so-called problem of decoherence, e.g., the deviation of the real quantum evolutions of complex systems from the ideal unitary dynamics which forms the basement not of only quantum computing but of the analytical part of quantum theory as a whole. The physics of quantum computers became much more fundamental subject than quantum computations themselves. One of the first who attracted my attention to it was K.A.Valiev, who first in Russia started to study the physical side of quantum computing, e.g., the problem of hardware realization of quantum gates.

\subsubsection{About the history of constructivism}

 Ripen serious turn in the development of quantum theory factually touches all natural sciences and has the deep roots. It continues Hilbert program aimed to the conquest of all natural sciences beginning with physics by mathematics. This idea lies in the basement of his famous program, formulated in the beginning of 20-th century. Hilbert program of the transformation of all sciences failed because it based on the axiomatic approach. Goedel has shown that the axiomatic method is inappropriate even for the substantiation of mathematics itself, let alone the extension of its methods to physics and chemistry. 

Nevertheless, this failure did not decrease the value of Hilbert idea. His program led to the surprising: appearance of the new mathematics, which is accepted to call constructive mathematics. The idea of constructivism historically appeared slightly earlier than the algorithm theory; it appeared as the answer to Hilbert program of axiomatization. This idea conquered its place first in the heated debates between Hilbert and Kronecker, then in the appearance of Brouwer intuitionism, then in the appearance of the exact definition of the constructive procedure - algorithm, which authors were Turing, and also Post and Markov young. After that already the constructive mathematical analysis and constructive algebra have appeared that has finalized the creation of surprisingly harmonious and powerful construction of the constructive mathematics.
Constructivism puts on the top of the list procedures of the building of needed object but not the logical substantiation of their existence. This approach to the formal apparatus is much more physical than the traditional. The consideration of the different parts of physics through the constructivism eliminates formal collisions and inaccuracies that unavoidably arise in classical formal apparatus. One of examples is the technique of Feynman path integrals. Nowadays we can certainly assert that the constructive mathematics represents potentially more modern and more convenient apparatus for physics than the traditional (classic) mathematics. 

Constructivism has its own history, which began before the formal definition of algorithm as Turing machine. This history is connected with the basement of constructive mathematics - constructive mathematical logic. Logical constructivism recognizes as true only the facts, which can be established by the formal procedures with the exact determined details. For example, in constructivism only such proof of existence is valid, which gives us the procedure of obtaining of the target object. Constructivism does not recognize the proofs of the pure existence. In the mathematical logic the constructivism has the form of intuitionism\footnote{Founders of intuitionism were Brouwer (see \cite{Br}) and Kronecker. It is known the heat polemics of the past with Hilbert about the legality of the proofs of pure existence, which are the proud of the axiomatic approach defended by Hilbert.}. 
For us it is necessary the constructive approach to mathematical analysis that I briefly describe in Chapter 2. 

The educational system due its unavoidable inertia yet partially keeps constructivism in the background, despite it represents the mathematical apparatus in the form convenient for physicists-theorists: the formalism free of the "mathematical lyrics" \footnote{Words of L.D.Landau.}. Meanwhile, there are strong reasons to suppose that just systematic application of this formalism can give us the exact knowledge about the nature of complex systems, which are yet investigated exclusively by the experimental means. We trace such few corollaries from the constructive approach that can be done by hand; for the further advance, the programming is needed and it oversteps the limits of book. We will see that the constructivism makes possible to reduce the big part of investigation to the creation of the determined requirements specifications for programmers. Here it is only possible to bring the general arguments for that such tasks will lead us to the aim, and I will try to do it everywhere. 

The real triumph of constructive is ahead. It is connected with the new possibilities in the area of industrial programming which were lacking even ten years ago, and which permit to join the natural sciences. There are the row of evidences that such joining of the natural sciences on the basis of programming technology will happen sufficiently soon and it will have no direct analog in the past.

\subsection{Brief review of contents}

The feature of the physical constructivism is that we make the demands of universality and scalability for it. In the ideal case, it must embrace all known scales and types of interactions learned in the natural sciences. One can thus treat our attempt as the electronic systematization of science that differs from the electronic archive only in that in some cases, (as I hope) it would help the authors to write these articles. We relay on the traditional hierarchy of the natural knowledge where the physics of elementary particles forms the zero level, then follows the physics of atoms and molecules, chemistry, biochemistry and at last biology. This hierarchy dictates the organization of our material and the fact that almost all relates to quantum physics. 

The most fundamentality of physics does not mean that the other science have the less value. This hierarchy has ordering character and results from that atoms consists of nuclei and electrons, molecules – from atoms, complex compounds from the simple molecules, living cells – from polymers and great number of more simple compounds, etc. This organization of natural knowledge follows from the atomistic hypothesis stated in antiquity by Democritus, which sense is that all the substances consists of elementary indivisible objects called atoms. Physics of 19-th and 20-th centuries completely confirmed atomistic hypothesis and made it the main scientific paradigm. In the framework of this paradigm, the fundamentality of the main micro objects means not their physical indivisibility but the border of applicability of the corresponding physical laws. It presumes that the fundamental objects can have their components but these components obey the completely different laws. This understanding of the atomistic hypothesis is very fruitful, as we will see, for the treatment of quantum theory. The atomistic hypothesis will be guiding line for us in the building of algorithms simulating the behavior of complex systems. The main notion in our approach will be the notion of particle, e.g., the point wise object that possesses, beyond the coordinates, some additional characteristics, as mass, electric charge, spin, and the set of its components, for example, the indication that the atom consists of certain nucleus and electrons, etc. We will see that the atomistic approach in this form is very convenient for the description of physical systems independently of how we treat them: as classical or as quantum. 

The main way of the realization of algorithmic approach will be the creation of computer Meta program serves for the building of the dynamical model of the considered systems. Without such Meta program, it is impossible to obtain essentially new results. All our considerations will be the substantiation of the appropriateness of the proposed algorithmic methods. Algorithmic approach to physics this represents the big project aimed to the creation of the instrument of developer of the simulation programs. 

It is substantial that the subject of the simulation will be the dynamical scenario of the evolution of the considered systems. The computation of the stationary states parameters, for example, electron eigen states, bound energies, stationary configurations of molecules etc., we treat not as the final product of simulation but as the tool of tuning the models of dynamical scenario, which will be our main aim. 

It makes the visualization the key step in the simulation of scenario. The necessity of visualization follows from the nature of algorithmic approach to physics. The point is that there is no universal method to learn the result of the work of algorithm but to launch it and to observe its work gradually. The visualization of the result of simulation is thus the single method for the determining of adequacy of the algorithmic models of dynamics. Meanwhile, the visualization is necessary not only to verification of the final product of the model. It is important in the intermediate steps as well. For example, the visualization is necessary for the decision of what object to treat as particles. Visual representation is the universal interface of the relation between the user and the program, and it causes the division of the model to two parts: user and administrative. We treat as the user part of the model all the dynamical scenario accessible for the user observation. The administrative part of the model contains the realization of computational algorithms building this scenario. With this division, we can acquire the physical sense to the user part only. 

The constructive form of physics gives us the framework of the organization of physical knowledge, which allows its expansion to all natural knowledge. This way is feasible, and we will certain that. 

The main attention in the book I devote to algorithmic physics as the basement for the further advancement of the algorithmic approach to chemistry and biology. In the first chapter, we go into the general principles of the dynamical scenarios including the genetic method. The brief description of the classical algorithm theory is done in the second chapter. In the third chapter we recall the simulation of processes based on the classical physics, the main important of which are the movement of many bodies interacting with each other through the field they create, and the movement of media consisting of huge number of particles with the different laws of interaction leading to the equations of diffusion and oscillations. 

In the forth chapter we look into the simulation of simple quantum processes for that we recall some material from quantum mechanics. The text is not the introduction to quantum theory, is designed for the aims of constructivism. In particular, we will omit the analytical computations but go into the important for us methods, at first into Feynman method of path integrals. In the fifth chapter, we learn quantum computers and quantum computations. We investigate what perspectives the orthogonal quantum theory gives to the simulation of processes in chemistry and biology. We devote the special attention to the attempts to treat decoherence in the framework of Hilbert formalism of quantum theory, and will discuss the principal drawbacks of these treatment in the light of the experimental results in quantum computing. 

In the sixth chapter, we consider the algorithmic approach to quantum theory, which aspires to replace Hilbert formalism of quantum mechanics in the area of many particles. This chapter is the key for algorithmic physics. Here we show the method of the contraction of standard Hilbert formalism for quantum theory of many particles which permits to immerse quantum theory in the common programming container with the other natural sciences which are formulated on the language of particles, not by wave functions. 

In the seventh chapter, we describe the structure of the program container for natural science (PCS) and the questions of agreement of the languages for the different languages of it. Such a container represents the programming case built on the basement of atomistic hypothesis into which one can immerse the separate modules on physics and chemistry that makes possible to create the working models of systems containing up to $10^{27}$ atoms, e.g., to the systems of macroscopic size.  

\chapter{Simulation of dynamic scenarios}

The used type of mathematics determines the general format in which are formulated the criteria of the success of physical theories. In the traditional mathematics, this format is a real number. Correspondingly, the theory is considered as successful if it gives the value of one or another physical magnitude with the high precision and with the little cost in the computational resources: time and space. The simplest way to obtain the single number is its representation by a formula; hence, the analytical technique is in the foreground in the traditional physics. The framework of the applications of this physics are strictly limited by the narrow range of problems traditional for physics. The more reliable the result obtained in this physics is, the more severely these limitations act. It would be well even not to dream about the inclusion to physics such a close area as the chemical reactions. For this expansion of physics its constructive version is required, e.g., the transfer to the constructive mathematics. We must pay a cost for it: refuse from the numerical criteria. 

Instead of precision of the numerical evaluations, we have to accept the different criterion: reliability of the dynamic scenario. It does not mean the complete refusal from the old criterion. The numerical evaluations remains but will play the auxiliary role in the debugging of the dynamical picture. 

{\bf The main criterion in the constructive physics: does the theory give immediately the right video film of the considered process.}

If we turn to the classical mechanics, we see that it is possible basing on its formulas by the simple technical tricks to build the realistic video film of the flight of artificial sputnik of Earth controlled by reactive force of its jet engines, and it determines the success of cosmonautics. However, if we try using the program technology of the same type to build the model of association of two hydrogen atoms to the molecule, even with quantum mechanics, we fail. It is easier to control over the cosmic flights than to control over chemical reactions. This is the practical application of the criterion of dynamical scenario. The second task is much more complex than the first one, because for its solution is not sufficient to use formulas.\footnote{The success of chemistry results not from good formulas but from the possibility to launch the huge number of the repetitions of chemical reactions and to observe their results. If only the fulfillment of such association of the molecules is such difficult as the cosmic flight, the successes of chemistry would be much more modest.} This chapter is devoted to the general questions of the simulation of scenario.

\section{What does the simulation of processes mean}

The simulation is applied so widely that the sense of this word becomes blurred. We clarify what mean the simulation in this book. By the simulation, we mean the creation of algorithm giving the objective picture of the evolution of the considered system in the numerical or in the visual form. Here the objectivity means that our algorithm rests upon the limited tricks of the definite form and has the universality, e.g., one can apply it to the wide range of systems and it will always give the right picture of the real evolution. 

In the ideal case, the area of application of such algorithm must contain all the systems investigated in the natural sciences. My firm conviction of the existence of such algorithm and the possibility to build it practically represents the strongest incentive for the writing of this book. Why this possibility is so important? The point is that it means that we can completely transfer the natural knowledge to algorithmic language and fulfill the main point in Hilbert program \footnote{In times of Hilbert himself the idea of algorithms was in embryo, and he did not mean this apparatus speaking about the mathematization of the natural sciences. He assigned primary importance to the axiomatic method that caused the failure of this mathematization in the narrow sense of word.}. 

Algorithms are the most natural formal tool of the joining of natural sciences that exists nowadays. We sequentially describe the beginning of this process that touches quantum mechanics as we represent it today. Here the very important preliminary step will be the fixation of the exact notions for each area for which is the simulation algorithm designed. This work on the formalization of physics has not yet complete. The existing formalization corresponds to the standard apparatus of axiomatic building of theoretical physics on the basement of analysis and algebra, and it is not sufficient for the aims of algorithmic approach. The required formalization consists in some additional constriction of the formal possibilities of the mathematical apparatus that makes possible to speak about its modernization or about the replacement of this apparatus by the more advanced. 

Here the first, preliminary step is the representation of traditional apparatus in the so-called qubit form. This form appeared in quantum informatics but it plays the universal role. The qubit form of quantum theory is the version of standard Hilbert formalism arranged for the needs of computer simulation. The qubit form is not equivalent to the constructivism. The passage to the qubit representation is only the first and the easiest step of constructivism, though even this step has not yet fully accomplished. 

The constriction of the possibilities of mathematical apparatus required by the constructivism corresponds to what is happening in the passage from the mathematics of formulas to the mathematics of computer programs. The main lost notion will be the notion of actual infinity, which serves the obvious basement of mathematical analysis. All the objects we consider will be finite. In particular, the notion of the limit by Cauchy will be transformed radically. Factually, these losses are negligible from the viewpoint of theoretical physics. Indeed, it is possible to formulate the mathematical analysis in the form of the so-called nonstandard analysis, where the infinitely large and infinitely small values will became simply the special terms or numbers added to the ordinary numbers \footnote{I will not consider this way, because it will not in use further. The reader can take a closer look at it by the book \cite{Us}.}. Nonstandard analysis overcomes the methodical difficulties of the introduction of actual infinity but is the equivalent language for the description of the standard analysis. 

There is the other, radical approach to the treatment of analysis, which is really its algorithmic constriction. This is the so-called constructive mathematical analysis. It considers only such numbers or functions, which are the limits of sequences of numbers or functions, generated by some algorithms (see the chapter 2). There is no infinities at all in the world of constructive analysis, and all constructive functions are continuous that radically differs the constructive analysis from the standard analysis and have a good agreement with physics \footnote{It is well known what difficulties appear due to the existence of actual infinity in physics. There are the convergence of the rows in quantum electrodynamics, the impossibility to normalized the function of the plane wave type, the infinite values of the own energy, etc.}
 The constructive reduction of mathematical analysis completely preserves its computational part, which the sole represents the real value for physics. It completely preserves the instruments of integration and differentiation, differential equations and the practical methods of their solution, including purely analytical part and the numerical methods as well. The constructive mathematical analysis completely preserves also the algebraic apparatus of computations, in particular, the symmetric properties of equations, permitting to use the group theory, etc. It makes possible to state the full succession in the passage from the standard analysis to the constructive analysis. Physicists-theorists accustomed to their apparatus have not then the reasons to feel any principal discomfort. All the factual they have created remain true in the constructive physics. 

However, the algorithmic constriction is not the mere formality. It removes the possibility to operate with arbitrarily large and arbitrarily small numbers. How important this possibility is? One could find that it narrows the scope and blocks some important ways of development. 
However, this is the illusion. Manipulations with actual (accessible) infinities were good in times of the formation of analytical methods in 19-th century. Nowadays these manipulations are empty but not fully harmless exercises. We will see that they lead to the experimentally unobserved phenomena. It is the evidence of the serious defect in the traditional mathematical apparatus of physics. 

The next negative sign is that they block the development of the algorithmic methods in simulation. The seducing, at first glance, opportunities of analytics and algebraic apparatus represent the serious illusion, and the deliverance of this illusion would be quite auspicious for the future of the physics theory. There is no reason for regret for anything serious supposedly lost in the passage to algorithms. In this passage, we loose only illusions, as the project of perpetual mobile. One must not think that the algorithmic formalism yields to the traditional apparatus in the strictness or in aesthetics. Algorithmic formalism is the language of the formal texts of program instructions, and it is the same aesthetically beautiful, and permits the same strict check than the standard language of formulas. 

In this chapter, we begin to consider the main theses of algorithmic physics, with the general principles of the simulation of the dynamics of systems consisting of many particles. The simplest object of the simulation is the scenario, or the simple model of dynamics. We call a scenario a sequence  $\bar S$
\begin{equation}
(\bar S):\ \ S_1, S_2,\ldots,S_t
\label{scenario}
\end{equation}
of the states of the considered system. It is supposed that the state $S_j$ corresponds to the time instant $t_j$ such that $t_j=t_{j-1}+\Delta t,\ j=1,2,\ldots,t$. Following the accepted algorithmic treatment we will encode the states $S_j$ in the form of sequences of ones and zeroes and consider any procedure determining states $S_j$ as some classical algorithm, which finds the codes of these states. We take the segment $\Delta t$ sufficiently small for that the scenario \ref{scenario} gives the right representation about the real dynamics of the considered system after its reasonable visualization. We fix the value of this segment. We say that the scenario well describes the real behavior of the system if the state $S_j$ sufficiently easy to find given the state $S_{j-1}$ in the preceding time instant. The scenario well describes the behavior of classical system without singular points, and if $S_j$ denote the wave functions of the simulation system, the behavior o non relativistic quantum particle in its unitary evolution. 

In the relativistic case we have to suppose that $\Delta t =\infty$, e.g., the passage from $S_{j-1}$ to $S_j$ is one of the channels of the scattering process. Since there can be many different channels of the process, and each of which has its own probability, in the relativistic case one scenario will not already give the good description of the real dynamics. In the non relativistic case with decoherence we meet the analogous difficulty because the wave function of state in the following time instant is not determined with certainty but with some probability only, and we have to consider the mixed states.

However, even in case of classical physics a scenario is not applicable everywhere. For example, in case of singular points in the potential of the force field the significant role belongs to the value of time interval $\Delta t$ and to the accuracy of the determining of the state $S_j$, e.g., the length of the sequence of ones and zeroes encoding this state. For example, in case of movement of the ball in gravity field on a curve surface with local tops we must know the speed of the ball, generally speaking, with the infinite accuracy, because in the opposite case we could not determine is it able to overpass one or other top or roll back. The rigorists usually say that in such special cases quantum effects become significant despite the initial dynamical problem formulation contains the macroscopic size objects. This is correct because the speed of the ball in the singular point of its trajectory becomes zero and hence its action on some interval $[t_j,t_{j-1}]$ is less than Planck constant that just presumes the necessity of the application of quantum theory to the ball instead of classical (see the chapter 4).

The simple models of the dynamics has the analogous drawbacks also for the systems consisting of very large objects, for example, planetary systems. Correspondingly, we complicate the simple model of dynamics supplementing it with some possibility to choose scenario. 

We call a model of dynamics a sequence of scenario
\begin{equation}
(M):\ \ \bar S_1,\bar S_2,\ldots,\bar S_\w
\label{model}
\end{equation}
constructed by certain rules. The model of dynamics must give us the full description of the considered system. We illustrate it on the following simple example (see the picture).

\begin{figure}
\centering
\caption{Finding of the curvature of the surface}
\vspace{150mm}
\makebox[380mm][l]{\includegraphics{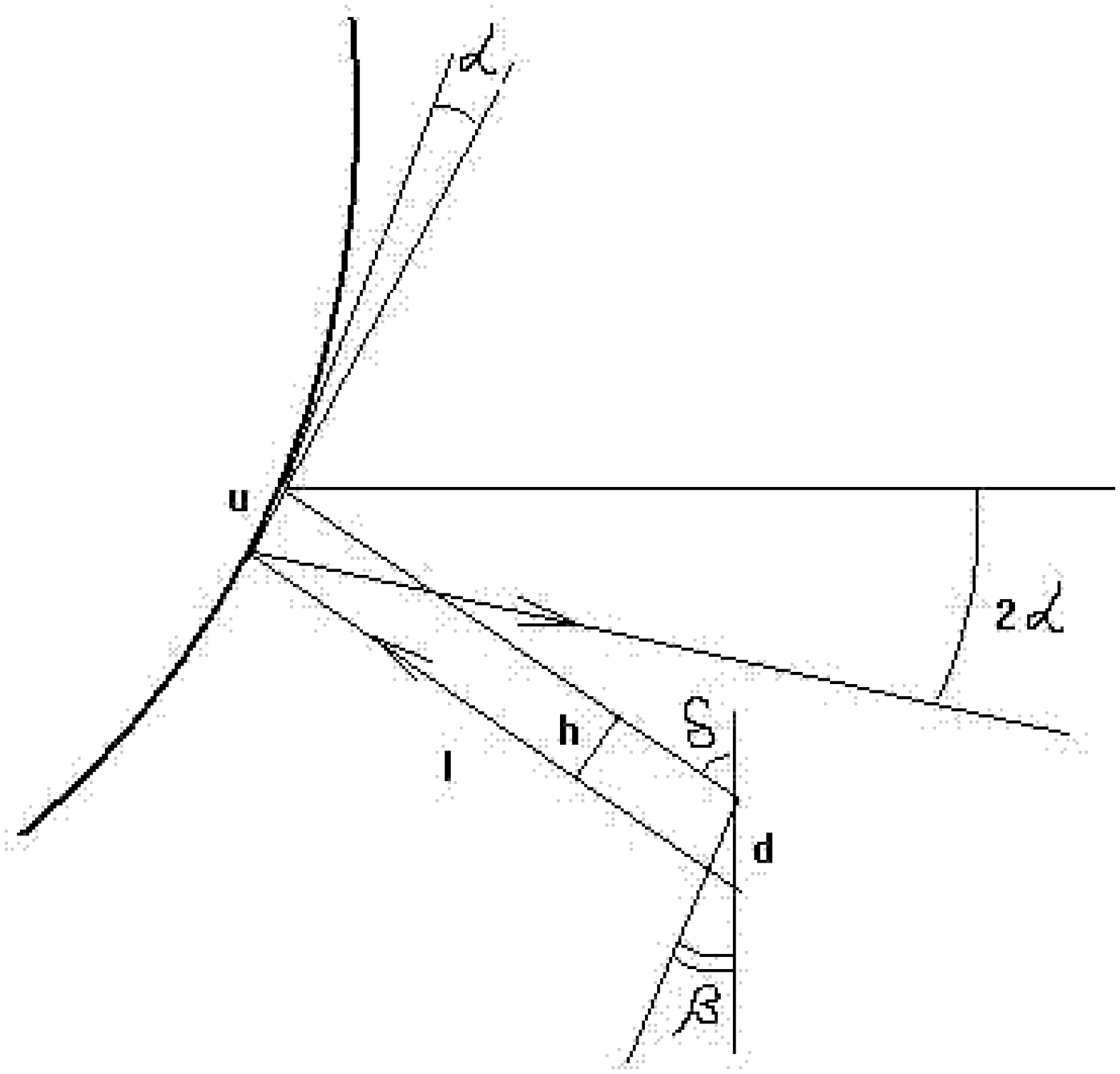}}%
\end{figure}

Let us consider the recognition of the form of remote object by means of the directed separate particles able to reflect from its surface. We can direct the particles to the investigated object, choosing their directions of flight and detect them as they reflect from the surface. The task is to find the curvature $C=\frac{1}{R}$ of this surface where $R$ is its radius. We have:
\begin{equation}
\frac{R}{u}\approx\sin\ 2\a,\ u=\frac{h}{\sin (\d+\b )},\ h=d\ \sin\ \d, R\approx\frac{d\ \sin\ \d\ sin\ 2\a}{sin\ (\d+\b)}
\label{curve}
\end{equation}
With this purpose we launch two particles which reach $P$, and after the reflection return back. Let the coordinate axes be disposed as shown at the picture. Knowing the angles $\a$ and $\b$, we can by the last formula \ref{curve} find the target radius of curvature. Here we, of course, suppose that the angle $\a$ is small.

Now we put attention to one important detail. If we have to restore by this method the form of the entire surface $P$, we could consider the plane $\pi_y$, orthogonal to the axis $y$ and split it to the squares with the side $d$. And then to launch the detecting particles from each vertex of the division of $\pi_y$ and, sequentially applying the formulas \ref{curve}, obtain the curvature $C(X,Y)$ of the surface $P$, using as the coordinates $X,Y$ on this surface the coordinates of the starting point of the particle on the plane $\pi_y$. The trajectory of separate particle here plays the role of scenario, and the sequence of launches of particles needed for the restoration of the shape of $P$, plays the role of the model. 

This simple example is factually very indicative since instead of particles one could mean the wave spreading in the elastic media, phonons in the solid state or photons in vacuum, at last merely atoms or molecules. Furthermore, the measurement of some quantum state in some basis depending on the initial point $X,Y$ can play the role of the separate scenario in this experiment. This process will then describe the procedure of quantum tomography of the unknown quantum state. At last, one could treat the separate scenario as the chemical reaction, where the point $X,Y$ determines the states of reagents. This scheme will then describe the search of the required states of the reagents for obtaining the needed product. 

It is important to mention that the separate scenario in the model of the dynamics $(\bar M)$ are disposed sequentially. We perform on the considered system the sequential experiments $\bar S_1,\bar S_2,\ldots$ and at each step store the information the current experiment in the memory of special computer. Here the state of the memory can, in general case, influence to initial conditions of the next experiment. For example, if we find that the angle $\g$ becomes too large for the application of the approximate equation \ref{curve}, we should decrease the value of step $d$ in order to have possibility to use the formulas \ref{curve} for finding of the curvature.  

The following requirements to the structure of the model follow from this example. 
\begin{itemize}
\item The model consists of the sequential scenario.
\item The result of each scenario is kept in the memory of computer and is in use for the determining of the initial conditions for the new scenario.
\end{itemize}
These two conditions we call the genetic approach to the simulation. The scenario play the role of genes here, and the model plays the role of evolution process. In the simplest case, which represents for example by the quantum tomography (see the chapter 6) the evolution is reducible to the accumulation of information about the resulting states of the considered system scenario in the memory of controlling computer. This information is in use in the finding if the results of the simulation but does not influence to the next experiment. In the more complex cases this information influence to the arrangement of the next experiment, e.g., to the choice of the next scenario. 

\section{Visualization and the role of user}

We now consider how we can process the information about scenarios. The most general way to represent the scenario $\bar S_j$ is the classical algorithm which generates the states of our system corresponding to this scenario step by step. Accordingly to the famous result of Turing (see the chapter 2), there is no way to learn how the algorithm will work but to launch it and to observe its work directly. This means that in general case the processing of the scenario at hand cannot have the form of computer program. The processing of the scenarios in the general case is the prerogative of human, as the user of the model. 
We thus establish the necessity to visualize scenarios produced by our model for their representation in the form convenient to the user. 

The visualization is the separate and integral element of the algorithmic approach and it distinguishes this approach principally from the standard form of the organization of physics based on formulas. 
In the solution of every physical problem, the formulas compose only the part solution. The principal question is how exactly these formulas must be applied, what mean the signs occurred in them, why we should use one formula or another, etc. All these things are typically fixed by the ordinary text. It is generally accepted that such texts – instructions cannot be formalized because here we meet the thing, which is customary to name the physical intuition, or something of this sort. In the constructive physics the non formalized part of physical texts can be represented in the form of expert estimation of the user of algorithmic model that rest ultimately on the visual perception of the picture. The rest of physical texts can be transferred to algorithms. The exact meaning of these dim thesis’s is described in the chapter 2 where we take a closer look at the computations with oracles. 

The non-algorithmic part of the work of physicists is thus reducible to the expert estimation of the visual images. This picture is ideal and thus very schematic. However, just this ideal situation reflects what we mean by the constructive physics. It means practically the strategy of the theoretical investigation based on algorithmic formal apparatus. There is no analog of this strategy in the standard theoretical physics there is no formal procedure determining the application of analytic and algebraic apparatus for the computation of physical values and the interpretation of these values. 

We will take up the formalization of this remaining a mystery procedure. Here our ideal aim is the creation of the algorithm with the perfect user interface, which requires only one fro the work: the expert user estimation of the video film generated by this algorithm. The further steps of this algorithm depend on this estimation. The algorithm determines the form of this expert estimation as well. In the ideal case this estimation is reducible to the choice between ‘’ yes’’ and ‘’no’’. Of course, in practices this simplest form is not convenient at the same degree as the usage of Turing machines for the building of algorithms. 

For the work of this model the presence of user is necessary, who takes up the observation on its behavior.\footnote{It looks like the variations of the many world interpretation of quantum mechanics (see, for example, \cite{Me}). The principal difference of our situation is that we do not operate with the metaphysical notions connected with the demonstration of the "free will". In the algorithmic approach, there is the expert estimation of visual images and this notion is completely constructive.}. We can divide the model to two parts: administrative and user. The first part consists of algorithms creating the visual image, the second one – this image itself. The formalized part of physical intuition belongs to the administrative part, non-formalized – to the second. 

This division in its turn requires the fairly larger degree of formalization of the initial notions than in standard formalism. These notions must be reduced to the objects, which algorithms can operate with. Such elementary ideal objects have no correspondence in the real world. They have no physical meaning and their role consists in their work in the simulation algorithm. The administrative part as a whole has no physical meaning; this part of the model is the formal apparatus of physics. Only the things that user observes have the physical meaning. The border between the administrative and the user parts of the model is thus the border of applicability of the physical meaning of objects. 

We illustrate the role of the user and administrative parts of the model on the example of simulation of the detection of photon pair in entangled state (we consider quantum states in more details in the chapter 4).
Let the photon pair be given generated in the nonlinear optical crystal after its irradiation by the laser. By means of the mirrors we can achieve that two photons flight in the opposite directions such that we can detect them by two detectors disposed for the distance $D$ that as a few kilometers, so that the crystal emitting photons lies exactly in the middle between the detectors, at the distance $D/2$ from each of them. Each detector can be tuned in certain direction of polarization, which means that it clicks only if the photon hit in it has this polarization. An entangled photon pair differs from the ordinary in that independently on their orientation when the tuning of two detectors coincides, they click simultaneously, or they both keep silence. In the work of detectors, the concrete outcome: the click or silence is the pure randomness which probability results from the tuning of the detectors. The instant of the photon emission is the pure randomness as well which distribution results from the uncertainty relation of the form ‘’ time – energy’’ (see the chapter 4). 

We now suppose that there are two generators of random numbers associated with each detector. These generators chooses the tuning of the corresponding detector in time interval $\D t < D/2c$, where $c$ is the speed of life. The information about the concrete positions of detectors cannot thus reach the point of photon emission because the speed of the transfer of the locally created information cannot exceed the speed of light. In these conditions, we make the row of independent experiments and after that, we choose those experiments in which the orientations of detectors were the same. In each thus selected experiment, the both detectors behave strictly the same: either the both keep silence, or the both click. If we consider the other experiments, where the detectors have the different states, they will behave randomly: situation when one detector clicks and the other does not will have the same frequency as the situation when they click or keep silence simultaneously.

\begin{figure}
\centering
\caption{Detectors}
\vspace{250mm}
\makebox[180mm][l]{\includegraphics{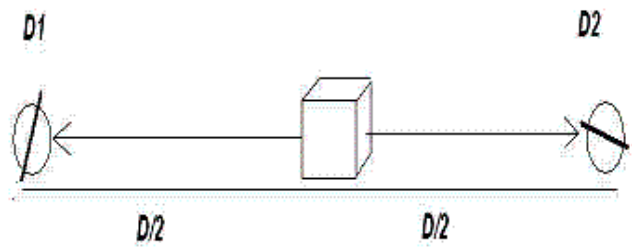}}%
\end{figure}

Without going into the description of this experiment from the quantum viewpoint, we try to understand how computers can simulate it. We presume that the simulation must be full, e.g., it must include not only detectors and photons, but the environment with all distances, it must account the relativistic limitation of the speed of spreading of the information transfer mentioned above. In addition, we must not have at our disposal anything but the simulating computers and the wires connecting them, e.g., we cannot use the entangled photon pairs themselves. If our model choose the independent states of the both detectors we must determine whether the clicks of these detectors coincide or not. To do this we must compare these states. If our model is exactly realistic our computers must be disposed at the $D$, and be have to spend the time $D/2c$ to compare the states of detectors. This time exceeds the time of choosing the states therefore, the simulation in the real time is impossible. It means the following. 
Every realistic simulation must give us the video film, which a user observes. This user cannot interfere in the process of demonstration and change something in the course of it. Every change has the following form: we formulate the new conditions; the model then prepares the new film, which we then view in the same regime. 

This is the work of the realistic model divided to the user and administrative parts. We bear in mind this scheme in the description of our models. 

The visualization represents the interface of interaction with a user who obtains the status of expert. It requires the more careful elaboration of thee details of visualization in comparison with the traditional approach where these details are of secondary importance.\footnote{"We can understand what we cannot imagine" – this phrase of L.D.Landau expresses the spirit of Copenhagen quantum theory.} 

The visualization can rest only on the atomic representation of the particles as point wise objects, or the collections of the point wise objects. Just so classical and quantum particles look in the method of collective behavior (see the chapter 6). Hilbert formalism for quantum theory of many bodies based on the tensor products of spaces of states does not permit the visualization in principle, because it contains such objects as the unlimitedly scalable quantum computer (see the chapter 5). Taking the path of visualization of the physical processes, we assume the following serious obligations. We must point such a contraction of Hilbert formalism, which at first permits to represent the quantum ensembles of many particles in the form of visual images, and in second permits to transfer all the reliable physics to the algorithmic way\footnote{By the reliable we mean the part of physics reliably verified in experiments.}, keeping safe all its results. 

The confidence in the practicability of this plan rests on two things:
\begin{itemize}
\item – all what we call the physical intuition rest on the visual images,
\item – no one of computational methods of the modern physics lie beyond the framework of the effective classical algorithms (see the chapter 2).
\end{itemize}
We consider the different types of physical objects and the corresponding ways of their visual representation.

{\bf 1. Point wise particles with nonzero mass.} Their visual representation makes no problem, for example, there are no difficulties with the change of the size with the distance like in perspective, etc.

{\bf 2. Classical lengthy particles with nonzero mass.} Visualized by the templates corresponding to the types of these particles. The type must contain the full information of the component parts of these particles and their interactions. 

{\bf 3. Quantum particles with nonzero mass.} Represented in the form of the ensembles of point wise samples of this particle connected by the bonds. The density of points must equal the probability density of this particle in the corresponding point of space (see the work \cite{OAO} for the details). In the chapter 5 it is described more detailed.

{\bf 4. Photons.} Visual images of photons are their samples connected by the special threads directed along the photon impulse vector. The length of thread inversely proportional to the mean square of the photon impulse dispersion (the more exactly we determine the impulse, the longer the thread is). The direction of the photon impulse determines the direction of the movement of this thread. The samples move with the thread and simultaneously oscillate in the direction of photon polarization that is transverse to the impulse. Here the impulse of concrete samples transfers along the thread goes with very high speed (Coulomb field spreads instantly – see the chapter 5), and the transverse oscillations spread slowly – with the speed of light. We assume that the longitudinal impulse determines Coulomb field and the transversal oscillations – the vector potential. This thread represents the electromagnetic field, and its division to photons results from the existence of the charged particles; at last, just the necessity of the good visualization of the quantum electro dynamical processes determines the division to photons. This division does not thus follow from any logic considerations but the needs of visualization. To represent Maxwell equations \ref{maxwell}, taking into account that $rot\ A$ is the vortex of the field $A$, it is reasonable to represent each photon sample in the form of two hooked gear wheels with some fixed point in each of them. One from these wheels determines the intensity of the electric field $E$, the other – the intensity of the magnetic field $B$; the fixed points connect one sample with the others. The heuristics of electrodynamics rests on such representations, which Maxwell used for derivation of equations \ref{maxwell}. In the constructivism, the heuristics has the different sense, we must consider it separately and accurately; it is outside the scope of our considerations.\footnote{D.Hilbert said: ‘’ When the house complete the scaffolding should be removed.’’ In the constructivism the scaffolding is the equation \ref{maxwell}, the house – is the heuristics.}

We mention that the consideration of particles as classical or quantum depends on the conditions of problem. The relation between the value of action $S[\g]=\int\limits_{t_0}^{t_1}L(x,x^{'}_t,t)dt$ of the simulated system along the considered elementary trajectory $\g$, and Plank constant $h$, where  $L$ is Lagrangian of the system, $x$ - the set of its coordinates. Since the property to be elementary depends on the necessity of its visual representation only, we can conclude that the ascription of particles to quantum or classical depends ultimately on the requirement of the visual representation of the dynamics. 

\section{Evolutionary principle in the simulation of dynamics}

The particular role of the visualization in the algorithmic physics is connected with the other its peculiarity, which we by convention name the evolutionary principle. It consists in that the sequential scenarios form the model of the dynamics already, so that each of which represents some specification of the previous. Such a changing of scenarios we call an evolution. 

There is one important circumstance here connected with the space and time. The point is that each scenario $S_j$, generally speaking, contains its own space and time which we denote as usual: $x,t$. This space-time of one fixed scenario has the physical sense only in the case if the user sees the film devoted just to this scenario. If this scenario is in use in the evolutionary process only in the preparation of some other more advanced scenario then the time $t$ must be considered as administrative and we can ascribe no physical sense to it. The space-time of one scenario can be called local \footnote{Some authors (for example, \cite{KMB}) call this time the internal time of quantum system.} in the sense that the initial point $O_j$ for this space-time is determined by the administrative part of the model for just this scenario. Here the states of all elements of the real world, which do not touch this scenario are accounted only in the determining of $O_j$. 

The passage from one scenario to the next one goes also in a time, but this is the other time. We denote this time by $\tau$. This time is global. It is the internal characteristics of our model. In the consideration of on scenario we cannot endow $\tau$ with any physical sense. It determines the sequential change of scenarios. However, if the user sees a long film obtained by the gluing together many scenarios, just the time $\tau$ is the user time and has the physical sense, whereas the time $t$ is the user, or administrative time.   

We leave beyond the bounds of consideration the question about the possibility to ascribe the physical sense to the administrative time in all cases. \footnote{This question is not senseless, especially for the joining of quantum mechanics with the general relativity theory. It also lies out of our attention.} Instead of all this, we will follow the given rule for the determining what time has the physical sense, and what time has the administrative sense. 

Somebody could express the perplexity: do we have the rights to decide what has the physical sense, and what has not? Especially, it seems strange if the matter concerns such a fundamental value as the time. I mention the following. In the reality, the mathematical apparatus is the integral part of physics. It merely does not exist without this apparatus. Hence, the prerogative to confer the physical sense belongs to this apparatus. The reader can think over this thesis that will be important for the further advance. If we use the traditional mathematical apparatus, all the work goes in its framework that we merely do not note. If we need to replace the mathematical apparatus, we have to submit to look at the things more abstractly. We will touch this theme in the section devoted to the mathematical logics, which deals with such situations. The reader must not have the perception about the presence of some suspicious places here. All what is needed for the further reading is the division of the model to the user and administrative parts, and the thesis’s from the above. I hope that the further material will give confidence to the reader that this way is right. 

We treat the sequence of scenarios composing the model of dynamics as the sequence of steps in the preparation of the film, which a user of the model sees. In some cases, (see the chapter 5) the regular scenario can be obtained from a few predecessors if we consider this sequence as the repetitions of previous scenarios with the more fine spatial and time resolution. In the other cases, the consequent scenarios are obtained from the preceding by the addition of the new particles, for example, the account of photons in the reactions of association and dissociation of molecules. 
At last, it is possible to specify the model by the addition of the new states of the same particles. Sometimes this specification of the model may be minor. For example, if we take into account the nuclear spins in the chemical reactions it makes the negligible change in the scenario. But the account of electron spins makes much more significant correction because just the electron spins determine the states of many electron systems, which create the chemical bonds. 

\subsection{Cauchy problem}

Cauchy problem is the standard form of the description of physical systems evolution in the classical and the simple (without QED) quantum mechanics. It consists in the finding of solution of the equation 
\begin{equation}
F(\bar x(t),t)=0
\label{cau}
\end{equation}

with the initial and border conditions of the form $\bar x(0)=\bar x_0$ and $\forall \bar r\in \partial D \bar x(\bar r,t)=\bar g (\bar r,t)$, where $F$ - is the time dependent unknown function of the form $D\times \Delta\ar H_2$, where $D\subseteq H_1$, $\Delta$ - time segment, $H_1, H_2$ - are Hilbert spaces, $\partial D$ - is the border of the area $D$, $\bar x_0(\bar r)$, $\bar g (\bar r,t)$ - are the given functions. 

Cauchy problem is to find the unknown evolution of the physical system provided we know the evolutionary law $F$, its initial state $\bar x_0$ and the regime on the border $\bar g (\bar r,t)$. The space $H_1$ is the generalized configuration space of this system. It means that all possible coordinates of the components of this system span it. The generalized configuration space differs from the ordinary in that it contains not only coordinates of objects, but their derivatives, for example, $x'_t, x''_{tt}$, the derivatives on coordinates of the solution, etc. The evolutionary law $F$ in standard mathematical analysis results by the passage to limit from some scheme of finite elements $S(\delta r)$, depending on the grain of spatial resolution in the generalized configuration space $\delta r$, when $\delta r\ar 0$. Hence, the functions $F$ and $\bar g$, in the general case contain the partial derivatives of the function $\bar x$ taken along all the coordinates of configuration space and time. 

The principal property of quantum mechanics is that the function $F$ is linear on $\bar x$. It means that if some $\bar x_1$ and $\bar x_2$ are the solutions of equation \ref{cau}, then for any complex numbers $\la,\ \mu$ the function $\la\bar x_1+\mu\bar x_2$ is its solution as well. This property is called the principle of superposition. We assume this principle as the fundamental thesis concerning not only quantum theory but also every model that we can consider. Further we treat the more general models (see the chapter 6), but they completely rest on quantum theory and in this sense the principle of superposition is applicable to them as well. 

All the algebra of quantum physics rests on the principle of superposition, and it lies in the basement of quantum interference effects. We can assert that the effectiveness of the algebraic apparatus in quantum physics rests just on the principle of superposition. It does not contradict to using of non linear equations; it only means that the basic equations must be always linear. Non linear equations appear when we are short of the possibilities of the algebraic language in the description of real processes. It is just the case when we really need the language of algorithms instead the traditional formulas. 

\subsection{When the evolutionary simulation is required}

Creation of the dynamical scenario model for Cauchy problem requires not only the exact knowledge of the evolutionary law $F$, but also the initial conditions and boundary regime. 
It is impossible to determine exactly any of these three types of data for the complex systems. The form of Cauchy problem is thus applicable to the narrow circle of the simple systems, which we call the model systems. Cauchy problem is correct, if a small change of $F$, $\bar x_0$ and $\bar g$ results in the small change of the solution $\bar x$\footnote{I do not concretize here what the small changes mean.}. In the opposite case the problem is non correct. If the notion of "small changes" does not contain the possibility to change the law $F$, then the correct Cauchy problem is the problem which solutions are stable in sense of Liapunov. In the reality the law cannot be known with the absolute accuracy, hence the notion of correctness is more practical. 

The law $F$ cannot be known exactly because it is the result of the limit passage in the scheme of finite elements (in particular, in the scheme of differences). In this passage the mechanism of action of the law on the distances to many orders less than the grain $\delta r$ of spatial resolution becomes actual. The experience of quantum physics itself evidences about the illegality of such extrapolations. It is impossible to determine the boundary regime exactly because the boundaries of real system are usually unclear, e.g., we cannot exactly determine what particles this system contains. This uncertainty is the more evident due to the existence of the so-called entangled quantum states where the state of pair of particles is irreducible to the pair of their separate states. At last, nobody can determine the initial condition exactly due to the fundamental reasons. The point is that in quantum electrodynamics the administrative and real times are the different notions, and we cannot treat the state $\bar x_0(\bar r)$ as the initial state of the system. 

One could object in the sense that the effects we speak about are very small and thus make no influence to the result of the simulation. I underline once more: this objection is true for the narrow set of systems, which we join under the common term the model systems. The model systems can contain the large number of particles, for example, the ideal crystal lattice, conducting the heat or oscillations. The notion of small changes in the model system can have the wide sense as well. For example, we can know the non-stable trajectories and the types of instability, limits of the admissible changes of $F$ (like the small coefficient at the major derivative), etc. The feature of the model systems is that the evolution appears in one-step: as the solution of the equation with the known boundary and initial conditions. 

Therefore, almost all problems we meet in practice are non-correct. We can describe some of them approximately by the model systems. We can extend the notion of the model systems including to them those systems for which the evolutionary model appears from the averaging of scenarios with some dispersion of the initial and boundary conditions, and of the evolutionary law in some fixed limits. Though such models do not immediately result from the solution of Cauchy problem, they are reducible to it by the simple averaging. The development of the idea of averaging leads to some methods of smoothing of the curves \footnote{The known method of the solution of the non-correct problems given in \cite{Ti} belongs to this type. The regularization is the kind of smoothing of the trajectories of the non-correct problem. Like the non linear equation, it is the answer to the drawbacks of algebraic methods by means of classical mathematics. The application of these means makes possible to reduce the non-correct problems to the proper algebraic form. Nevertheless, these methods are not possible even in principle to open any new phenomena. The methods of the type of regularization just aimed to cut off such new phenomena from the consideration.}.

Factually, such averaging methods represent the most primitive version of evolutionary methods. A method of averaging is applicable only if we know completely the mechanism of evolution $F$, and if the problem is reducible to the solution of equation \ref{cau} and the investigation of the non stable trajectories. However, the point is that for the complex systems we do not know $F$. Due to the know non locality of quantum physics and the principal character of relativity in quantum electrodynamics (see below) we cannot even theoretically not only separate the evolutionary law from the initial conditions, but determine where the initial conditions are. \footnote{I do not touch the question of the insurmountable difficulties in the determining of the initial conditions even in non relativistic case, especially for the large quantity of particles in the system.}.

The methods of averaging is not thus appropriate to advance in knowing of $F$. Moreover, it even worsens the deal. To illustrate this, we consider the simple example of two bodies tied by the flexible rope. We assume that only one of the bodies is observable when the other is invisible. We observe some movement of the single body that looks like chaotic and can apply the averaging to the movement that shows how the center of masses of the real system moves. Since we do not know about the invisible body, we have to make some random factor responsible for the chaotic movement, or even to ascribe the fundamental character to this factor. We then turn to the investigation of the system of $n$ bodies tied sequentially by the flexible ropes where only $m$ are observable. This system obeys the same laws as the previous system, but we are not able to apply immediately the experience with the previous system where only one body was observable. Now we have to consider the certain influence of one body to another. To make the example more obvious we assume that the observable particles have the large mass, and the invisible have the less mass. The averaging methods treat the influence of the invisible bodies as the random noise. In any case to find the law of elastic force of Guk and to understand the behavior of the system with $n$ bodies we need to include to our model the particles which we cannot directly observe. 
   
The situation in quantum mechanics is the same. The naive model of quantum system where the particles move under the influence of the potential created by the other particles (but not by themselves, because the self-action erect the irresolvable difficulties) is not appropriate for the investigation of the complex systems because it is not completely scalable. Really, to find the field acting on each of $n$ particles from the side of the others we need about $O(n^2)$ elementary steps that is too prodigally for the large systems where the complexity must not exceed $O(n)$. If we begin to consider such a system in more details, including to it photons creating the field, entangled states, etc., we loose the possibility to apply the formulation of Cauchy problem! We see that Cauchy problem is inappropriate for the analysis of the complex systems for which we should use the approach with the evolutionary simulation. 

The method of density functional represents the attempt to bypass the difficulty we mentioned. We can assume that the density of particle creating the field is the same in all points of its support and equals its density in the point where the probe particle is located. It simplify the computations but gives the large error in case when the difference of density is big (for example, for the electron density in atoms the method of density functional works bad, for the conducting zones in metals it works much better). The simplification of computations is often more important than the accuracy, and the method of density functional is often applied in the solution of the model problems with many bodies on supercomputers.

The evolutionary simulation with expertise of a user gives the more general type of models. Evolutionary models better correspond to what we call the constructive physics, because these models factually allows the different mechanisms for the different sizes of grain of spatial and time resolution, that agree with the spirit of the constructive mathematical analysis. We will certain of it in the next sections. Here we mention only that the evolutionary principle allows the including of new elements to the model, the checking of new hypothesis, e.g., to vary the strict limits of Cauchy problem.

\section{Summary of the simulation of dynamical scenarios}

The evolutionary principle of the simulation of dynamical scenarios makes possible to investigate the complex systems of which the form of Cauchy problem is inapplicable. Such simulation explicitly includes the time as the administrative parameter that allows the investigation of such systems which complexity makes the averaging methods non-efficient. The evolutionary method presumes the accumulation of the information about the scenarios for its further usage in the building of new scenarios. This method is typical for quantum electrodynamics; it is universal in comparison with the language of scenarios without evolution. The simulation of scenarios explicitly requires the presence of a user who plays the role of expert of the found scenarios.

\chapter{Constructive mathematics}
\nnn

\ \ \ \ \ \ \ \ \ \ \ \ \ \ {\it "we require too much from a formula: it has the good view but leads to

\ \ \ \ \ \ \ \ \ \ \  misunderstanding, because it forces us to think that we know something 

\ \ \ \ \ \ \ \ \ \ that we do not know."
\nn

\ \ \ \ \ \ \ \ \ \ \ \ \ \ \ \ \ \ \ \ \ \ \ \ \ \ \ \ \ \ \ \ \ \ \ \ \ \ W. Huckel}
\nnn

Constructive physics is based on the constructive mathematics. This trend in mathematics is well known but due to the inertia of traditions is not yet widely represented in the education. It is important for us to take a closer look at the basement of the constructive mathematics, because it represents the necessary foundation for the constructive methods in physics we develop.

\subsection{Review of mathematical constructivism}

The constructive mathematics consists of the constructive mathematical logics, the algorithm theory, the constructive mathematical analysis and the constructive algebra. Learning of the constructivism one must keep in mind that physics and mathematics represent the same science. \footnote{This truth is not generally accepted in the both camps.} The already mentioned crisis of the program of axiomatization after Gedel theorem clearly pointed to the constructivism as to the way out of this deadlock. Constructivism rests on the constriction of the formal apparatus of the proof theory. This constriction permits to avoid the consideration of such notion as the actual infinity by the replacement of it by the concrete algorithm, which step by step gives the values converging to the infinity in sense of ordinary mathematical analysis. The passage to constructivism means that we replace the abstract operations with infinities (for example, the limits of functions by Cauchy - Weierstrass) by the concrete manipulations with algorithms giving its approximations.

The meaning of this turn is that the deductive apparatus that had complete domination in the classical mathematics unavoidable looses its position in favor of the application to physical experiments. It becomes apparent in that the behavior of algorithm in contrast to formula is unpredictable. We can substitute $t=+\infty$ in the formula which expresses the value of some function $f(t)$ with argument $t$, and see what results. However, it is impossible to fulfill this trick with an algorithm. If some algorithm $A$ gives step by step the sequential values of this function $f$, we, generally speaking, principally cannot prove this given the text of commands of $A$. Correspondingly, we are unable to predict the asymptotic behavior of this algorithm. We can only observe the work of algorithm as a physicist observes the evolution of a real system. 

 The limitation of the procedure of mathematical applications to the reality by the formal proofs leads to the deadlock due to the clear reason: the impossibility to verify the results. The verification or the checking of mathematical proofs is necessary condition in mathematics itself. It is possible when its complexity is substantially less than the complexity of the proof. In the modern mathematics the lower bound on the quantity of pages for a proof is a few hundreds.\footnote{Of course, this evaluation is rough, it depends on the concrete situation and I show its approximate value.} However, there are the examples of works exceeding two thousands pages. To verify (and to build) such proofs the using of computer program is necessary. This causes the discussions among the mathematicians.\footnote{The famous problem of the four colors serves the example.} 

Moreover, the verification makes sense only in the case when sufficiently many specialists can participate in it.\footnote{The reader can imagine the situation when only the author and his nearest friend understand the proof.} Here, as in the statistical physics (and also in quantum mechanics) to speak about the reliability we have to be sure that the key procedure can be repeated many times, at least in theory. The complex proofs of modern mathematics, probably, have reached the limits of this possibility. 

However, there is the other important circumstance. A proof itself appears in mathematics as the most forcible argument in the scientific discussions, and its special place is still connected with this. Here the set of axioms dictated by the practice is often more important than the logical consistency. A theory can be inconsistent but in spite of it successful. However, the length (complexity) of the deduction of the contradiction plays the main role here. Of course, nobody wants to use a theory which asserts that $2\times 2=3.9999$. However, if a theory has the simple axioms, gives the easy explanation and can predict the results of many independent experiments, whereas the deduction of contradiction in it requires about five thousands pages, then this theory would gain many supporters who interest not in the complex proofs, but in these experiments. All the successful physical theories are of this type. As for the complex systems in our sense then the visual demonstration, e.g., the checking of its axioms has the more important role than the formal proofs at all. This is why the natural sciences investigating the complex systems: chemistry and biology do not use the formal proofs. We can certainly assert that just visual demonstration of the complex system behavior plays the role of mathematical proofs for these systems. The passage to the regular application of visual images will be the gigantic step ahead in comparison with the descriptive character of these sciences now. 

Constructivism just points us the way to the building of these systems of visual demonstrations, which preserve the level of strictness accepted in mathematics. It serves the comprehensive substantiation of the legal place of constructivism in the modern physics and in the system of natural sciences at all. 

Constructive mathematics rests on the notion of algorithm, not on the notion of sets as the classical mathematics. Constructivism admits only those constructions that can be described by the certain instructions for some computer, at least in principle. Therefore there in not the notion of actual infinity in the constructive mathematics, but there is the notion of the potential reach ability, which means, for example, that a type of Turing machine is infinite because we can always, if it is necessary, add to this (finite) type the additional cells, one by one from the right. 

Algorithmic approach to physics rests on the mathematical constructivism in general. Constructive mathematical analysis is the most valuable part of constructivism for the interpretation of physics.  
The notion of algorithm is the working tool of constructivism. Algorithms were introduced by Turing, Post and Markov the young. It puts the constructive mathematics into complete shape and makes it ready to play the role of the mathematical apparatus in physics. I will not describe the history of constructivism, which is full of dramatic fight with the traditional (classical) mathematics for the right for existence. I mention only about the discussion between Hilbert and Kronecker when the last one defended the ideas of constructivism long before the appearance of exact notion of computability. From those times, the mathematical constructivism obtained civic rights in mathematics and became its integral part. 
We go into it with the certain aim: to make quantum theory constructive. Nowadays the classical mathematics reigns in quantum theory that leads to the various collisions and forces to sweep difficulties under the rug. The main is that the classical mathematics is not able to open doors to the understanding of complex phenomena.

Mathematical constructivism is not only a substitute of the standard formal apparatus. It contains something principally new in comparison with standard mathematics, and it shows already at the level of constructive mathematical logics. This new is the fundamental idea of pluralism. The plurality of true evaluations is the feature of the intuitionistic logics. The idea of the logical plurality is similar to the so-called many world interpretation of quantum theory. This parallelism has the far-reaching consequences yet not estimated by physics. 

We now return to the genesis of the constructive mathematics and to the severe crisis caused by the collapse of the famous Hilbert program that gave birth to the appearance of mathematical constructivism. Mathematicians began to recognize this crisis simultaneously with the understanding of the sense of Goedel theorem about the proofs of consistency. The reliability of mathematic itself was called in question. Kronecker and Brouwer proposed the radical remedy for this crisis - intuitionism, or constructive mathematical logics. The main conclusion that we should derive from the past discussions of the founder of constructivism looks simple. The role of single arbiter determining the right mathematical apparatus belongs to the experiments, which means that physics is the factual arbiter. 

The modern development of physics enters the zone of conflict with its mathematical apparatus. It concludes in the unavoidable choice: either we must revise the notion of effective classical algorithm, or we have to modify the mathematical apparatus of physics.  The most evident reflection of this choice we see in the project of a scalable quantum computer. The proved absence of the direct ban of its existence (see \cite{VK}) means that either such a computer appears, or the change of mathematical formalism for quantum theory becomes the first point of the agenda. The fundamentality of quantum theory makes possible to speak about the revision of its mathematical apparatus if the experiments in quantum computing give no cogent arguments for the first hypothesis. 

On the other the mathematical constructivism including constructive versions of logics, mathematical analysis and algebra, is not complete. It lacks for the main link: constructive physics. The corresponding rebuilding of quantum theory does not follow immediately from the formal apparatus of constructivism we are going to learn. This apparatus merely sets the more general notions for the required rebuilding: algorithms, constructive functions, and constructive algebraic systems. At the same time, a physical theory must give the exact predictions. To obtain the physical results it is necessary to develop the applied apparatus of constructivism. It is necessary to build the concrete algorithms, which work results in the numerical results. I mentioned above that it is impossible to do it in a book. This is a feature of constructivism: in contrast to the standard physics where a reader could be able to check computations and to compare them with the experimental data, in constructivism it is impossible. I do not put this problem: it is the aim of the special works. My aim is to proof the full legality of constructivism and the reality of its realization. I ask the reader to assume this treatment of the approaches to the physical constructivism we will consider, including the so-called heuristics of the constructive physics. 

\section{Algorithms and computable functions}

We give the brief outlook of the basic notions and results in the theory of classical algorithms required for the understanding of the following. 

Algorithms represent the main tool of constructivism. Their physical sense is that 

{\bf Every real evolution of a physical system in time is determined by some algorithm.}

The constructivism of physics consists in the development of this thesis. 

Algorithms were in use in mathematics from the ancient times, without the definition. The need in the exact definition of algorithm appeared only to prove that some algorithm does not exist. The formal definition of algorithm acquired the new sense when the needs aroused to have a convenient language, describing algorithm for their creation and analysis. Programmers who create the programs on the different programming languages typically do such a work. For us the formal description of algorithms is important also because of the other reason. Such a description contains the important peculiarities making possible to give them the universal properties, and to apply the same algorithm or the same subroutine in it for the simulation of the different phenomena. There are the factual pointers to the mechanisms of physical laws. These mechanisms contains in the administration part of the model and do not visible to a user.\footnote{The explicit description of these mechanisms is called the constructive heuristics.} Nevertheless, their role is very important because just they define the concrete way of realization of the general theses of this book, and thus the quality of the dynamical models.

There are several formalizations of algorithms, from which we consider only three: Turing machines, cellular automata and Markov normal algorithms.

{\bf Turing machines}

Turing machine consists of the input tape, head and the set of commands. The input tape (we can treat it as the limited, for example, from the left, without loss of generality) consists of cells each of which contains one the letters of the special input alphabet $a_0,a_1,\ldots,a_n$, the first from these letters means the blank. The tape is potentially infinite, which means that it is possible to add to it the additional cells with blanks if necessary, from the right. The head can observe exactly one cell of the tape and must be in one of states of the head: $q_0,q_1,\ldots,q_k$, the first of which is the initial, the last is the final. A state of Turing machine is the contents of its tape and the state and position of its head.
\footnote{There can be many heads, the tape can be unlimited from the both sides, etc.; all these details do not play the significant role. We specify them only for the distinctness.}. We will consider only the states where all the tape beginning from some place is filled with blanks. Hence, any state of Turing machine can be encode by the finite cortege from ones and zeroes.  

Given a word $x$ in the input alphabet without blanks, the initial state $S_0(x)$ of the machine is the state in which the word $x$ is written on the tape beginning from the second cell. In the first cell is the blank, and only blanks follow to the word $x$, whereas the head observes the first cell of the tape and is in the initial state $q_0$. If all these properties are fulfilled but the head is in state $q_k$, we denote this state by $S_{fin}(x)$ and call it the final state. 

The commands of Turing machine are the strings of the form
\begin{equation}
a_j,q_i\ \ar\ a_J,q_I, (R,L,\emptyset ),
\end{equation}
where $j,J=0,1,2,\ldots,n;\ i,I=0,1,2,\ldots,k$, and the last stands one of symbols $R,L,\emptyset$ of the set of special symbols, means sequentially: the shift of head to right, the shift of head to left, and the conservation of head at the same place. This command means the sequence of operations: to write in this cell the symbol  $a_J$, to transfer the head into the state $q_I$, and to shift it accordingly to the special symbol. The set of commands is called the program corresponding to this Turing machine. We assume that the program also contains in Turing machine. We require that every combination of numbers $j,i$, included in the pointed ranges occurs in one of the commands from the program. If any pair $j,i$ uniquely determines a command from the program, this Turing machine is called deterministic, otherwise it is non deterministic machine. 

A computation on the Turing machine is such a sequence of its states of the form
$$
S_0,S_1,\ldots,S_T
$$
where $S_0=S_0(x) $ and $S_T=S_{fin}(y)$ for some words $x,y$ in alphabet $a_1,a_2,\ldots, a_n$, there is not the final state among the states of head in $S_0,S_1,\ldots,S_{T-1}$, and each passage $S_j\ar S_{j+1}$ in this sequence corresponds to some command from the program of this machine. If for all such words $x,y$ the equality $y=f(x)$ is true, where $f$ is some (in general case not everywhere defined) word function, we say that this Turing machine computes the function $f$. If $T$ denotes this Turing machine, then $f$ we denote by $f_T$. Here if $f$ is the characteristic function of some set of words, we say that tis Turing machine computes this set. To compute a set means to have an algorithm, which for a given word determine does it belong to this set or not. 

A function computable by some Turing machine is called computable.  

We say that Turing machine $T$ enumerate some set of words $A$, is $A$ is the image of the set of all words in the mapping $f_T$. In the other words, an enumerable set is a set of members in the sequence $f_T(W_1),f_T(W_2),\ldots$ for some Turing machine $T$, where $W_1,W_2,\ldots$ is the set of all words in the input alphabet without blanks taken in lexicographic order. 

It follows from the definitions that each enumerable set is computable. The reverse is not true. We consider, for example, the set of all pairs of the form (code of Turing machine $T$ , word $x$), such that $T$ is applicable to $x$, e.g., $x\in Dom\ f_T$. This set is enumerable, because all such pairs can be enumerated applying $T$ to $x$ step by step. Nevertheless, this set is ton computable. Really, let us suppose that it is computable. By $[T]$ we denote the finite binary cortege encoding Turing machine $T$. Since the total number of its commands is finite, this code always exists. There exists then such machine $T'$, that determines, is $T$ applicable to $x$ or not given a pair $[T]$, $x$. We then can build Turing machine $T_0$, which is applicable to a number $[T]$ if and only if $T$ is not applicable to $[T]$. The building of such a machine $T_0$ is the routine procedure that uses our supposition about the existence of machine $T'$. Let us now check, is the machine $T_0$ applicable to its own code $[T_0]$. It follows from its definition that every choice here leads to the contradiction. We thus have proven that the problem of applicability is non computable: there is no Turing machine determining the applicability of any other machine to a given word.\footnote{This method is called the diagonal method of Cantor. The main two methods for the proof of non-existence of some Turing machine and for the establishing of the lower bounds for the complexity of computations are the diagonal Cantor method and the usage of Dirichlet principle.}

\section{Church-Turing-Markov thesis}

Turing machine formalizes the notion of algorithm. After some training one could detect that each function computable in the naive sense of this word, e.g., for which some exact method exists how obtain the value of function given a value of argument, is computable in sense of Turing machines. 
 (see \cite{AHU}, \cite{AHU}, \cite{Ma}, \cite{Us2}). However, the programming on Turin machines is not the convenient way for the creation of programs. The reason is in that the design of the real computers is far from the abstract scheme with the tape and head. There are many other ways to formalize the notion of algorithm. We consider two of them - the normal algorithm of Markov and cellular automata. 

{\bf Markov normal algorithms}

Markov normal algorithms differ from Turing machines only in that instead of the tape we have a word $A$ in the input alphabet without blanks. There is no head and commands have the form of substitutions $x\ar x'$, where $x,x'$ are words in the input alphabet without blanks. On each step of the work we fulfill such substitution that the first occurrence of the word $x$ in the current word $A$ stands the first from all substitutions, and we apply this substitution to this first occurrence. The definition of computation looks like for Turing machines provided we fix some special letter in the input alphabet, which plays the signaling role for the end of computation. It follows straightforwardly from the definitions that the computability in sense of Markov algorithms is equivalent to the computability in sense of Turing machines, because we can easily encode the position of head in Turing machine in terms of special signaling letters. Of course, we should narrow (or extend) the input alphabet because the work with words requires the special symbols, which do not occur in the words $x$ и $y$. 

{\bf Cellular automata}

Cellular automata give the different formalisation of algorithms. A cellular automaton, as Turing machine has the tape in which cells letters from the input alphabet stand. There is no head. The content of every $j$-th cell in each time instant $t$ is the function of its content and the content of its neighbors in the previous time instant $t-1$: $a_j(t) = F(a_{j-1}(t-1), a_j(t-1), a_{j+1}(t-1))$. The time evolution of the cellular automaton is thus defined. The end of computation is determined as in the case of Markov normal algorithms, by the appearance of the signal letter. Extending the input alphabet, we can encode positions of the imaginary head of Turing machine in terms of cellular automaton states. Hence, a cellular automaton can simulate the work of Turing machine in real time, e.g., step by step. The reverse simulation is possible as well, but with quadratic slowdown: the head of Turing machine must go through all the tape to fulfill all the work a cellular automaton makes on one-step. 

In contrast with normal algorithms which mutually simulate Turing machines with only linear slowdown, cellular automata cause the quadratic slowdown in its simulation on Tiring machines: if $T$ is the time of work of a cellular automaton, then the time of the work of Turing machine simulating it will be $O(T^2)$.
We can regard cellular automata on any discrete manifold in which the notion of neighborhood of cells is properly defined. 

In each case, a function computable in one model of algorithms will be computable in any other model. This fact is true for all known models of algorithms: Post machines, Kolmogorov-Uspenski algorithms, quantum computers, etc. If we formulate this thesis for all models of algorithms including those, which may be built in future, we obtain the Mehta mathematical principle called Turing - Church-Markov thesis: 
"The notion of algorithm and computable function is unified and does not depend on the method of formalization". 

However, the different models of algorithms have their peculiarities that touch the computational complexity. In what follows we call the concrete formalization of algorithms the computational machine, or the computer, because we interest in the physical side of its work. We thus consider the formal mathematical sides of the computations from the physical viewpoint. 

Let we are given an input word $x$ and let $S_0(x),S_1,\ldots,S_T$ be a computation in some algorithm model. The number $T$ is then called the complexity of the work of this algorithm $M$ on the word $x$, and we denote it by $s_M(x)$. If we consider all the words of the length not exceeding $n$, Then the maximal complexity of the work on such words is treated as the complexity of this machine: $s_M(n)=max_{x:\ |x|<=n}s_M(x)$. The complexity of a given algorithm is thus the function of the natural argument: the length of input word. Correspondingly, we have the linear, quadratic, polynomial, exponential, etc. complexities, and also of the more fine scales.\footnote{For every reasonable scale it can be proved that this scale is exact: there are the function with just this lower bound of complexity (\cite{Ko}).} For any function on words, its computational complexity in a given class of computational machines is the least possible complexity of the machine computing this function. For any function $f$ of the natural argument with the sufficiently fast growth it is possible to find the function on words with the complexity approximately equal $f$. 

However, the classification of problems on their complexity is not complete. The definition of complexity itself by the fastest algorithm solving the problem is not constructive, because it requires in general case the search through all algorithms and the choice of optimal among them. Just this makes our definition declarative, let alone the impossibility to make the required decision even for one algorithm.\footnote{The still open problem of fast factoring integers gives the good example.} In several cases, only we can define the complexity exactly. (The complexity of the computation of a set of words is the computational complexity of its characteristic function.) For example, the problem of comparison of two given words has the quadratic complexity on Turing machines with one head and the linear complexity on Turing machines with two heads and on cellular automata. This fact results in that the simulation of cellular automata cannot be sped up on Turing machines with one head, in comparison with the ordinary method giving the quadratic slowdown. \footnote{The result about the lower bound of the recognition of the coincidence of words on one head Turing machines belongs to Tseitin, it is proved by the reduction to Dirichlet principle by the account of the head states at the intersection of the border between tested words.}.

We can define the relation on the class of computable functions called the reduction. A set of words $A$ is polynomial reducible to a set $B$, if there exists the function of words $f$ of the polynomial complexity such that $x\in A$ is equivalent to $f(x)\in B$. If the set $B$ has the complexity $F$, the complexity of $A$ then does not exceed the polynomial of $F$. In particular, if $B$ is computable with polynomial complexity, the same will be the set $A$. 

A model of computations is called deterministic if the computation is defined uniquely given the initial state and non-deterministic in the other case. For example, quantum computations are the non-deterministic computational model. It is known that every deterministic model is reducible (with polynomial complexity) to any other. In particular, for Turing machines and cellular automata we established this fact earlier. We can thus separate the minimal complexity class of computable functions independent of the model of algorithm, namely, the class of functions computable with the polynomial complexity on the deterministic models. We denote it by ${\cal P}$. 

We now turn to the non-deterministic computations. There can be many non-deterministic branches of a non-deterministic computation beginning with the same initial state. We treat the set of all these branches as the computation on the non-deterministic machine. The complexity of non-deterministic computation is defined as the length of the shortest branch beginning with the initial state. Since the computability in non-deterministic models is equivalent to the computability on deterministic models, the set of computable functions will be the same in these two classes of computations. However, if we intend to simulate a non-deterministic computation on a deterministic machine we have to use, in general case, the time as exponential of the time of non-deterministic computation. It is true for non-deterministic Turing machines, cellular automata and for quantum computers as well.  

The class of sets computable on non deterministic Turing machines in polynomial time is denoted by ${\cal NP}$. The equivalent definition of the class ${\cal NP}$ is the following. A set $A$ of words belongs to class ${\cal NP}$, if there exists the set $B$ consisting of pair of words and belonging to class ${\cal P}$, such that for some polynomial $p$ the following assertion is true. For any word $x\ $ $x\in A$ if and only if there exists the word $y$ of the length not exceeding $p(|x|)$ (polynomial of the length of $x$), such that $(x,y)\in B$. The proof follows easily from the definitions. This second definition factually says that the class ${\cal NP}$ is the class of search problems: to determine does a word $x$ belongs to the set of this class it is required to search all $y$ of the sufficiently small length. Here each step of the search is checked quickly, because $B$ has the polynomial complexity. The enclosure 
$$
{\cal P}\subseteq {\cal NP}
$$
follows from the definition. A set $A$ belonging to ${\cal NP}$ is called ${\cal NP}$-complete if any ${\cal NP}$ - set is deducible to $A$ with the polynomial complexity. It is known about a few hundreds ${\cal NP}$-complete problems. To solve of them in polynomial time would suffice to solve ${\cal P}= {\cal NP}$(?) problem. 
However, the problem of coincidence ${\cal P}= {\cal NP}$(?) is still open. The best-known algorithm solving search problems is the bruit force. It clarifies why the search problems are equated to the search of the password to black box. We cannot use the knowledge of its internal design for the finding of password. This argument is metaphysical by we will show one argument for it, which does not rest on the open problems in mathematics.

\subsection{Computations with oracle}

There is no chance to solve ${\cal NP}$-complete problems without the bruit force. The direct search algorithm examining step by step all possibilities represents the fundamental algorithm plainly connected with the evolutionary method of the dynamical models building. The search through all evolutionary scenarios lies in the basis of algorithmic physics. In each step of this search, a user fulfills the expert estimation of the current scenario. The estimation influences to the further building of scenarios. 

 There is the convenient form of the formalization of this process, the simulation with user - expert. It is the notion of computations with oracle.

Let us consider an arbitrary set of words $O$, called an oracle. We redefine the notion of computation and computable function in order to obtain the more general notion of the computations relatively to this oracle. We add to the possible elementary steps in a computation (there are the applications of commands for Turing machine) the new action called the query to oracle. The query to oracle means the question: does a given word $z$ belong to the set $O$ or not. After this question, the computation is postponed to the instant when the oracle answers. The time of delay is the administrative time that means that it has no physical sense. After the reception of answer, the computation recommences in the usual mode. Further, the computation can generate the new word $z'$ for which the next query will be sent to the oracle, etc. The exact instant of query is lasso generated in the course of computation. We thus can suppose that there are the special registers (groups of cells) on the input tape where the word for query $z$ stands, and the special register for the determining of the moment for this query. All the cells in this registers participate in the work of algorithm, in particular, its content can change during the computation. An oracle can be treated as an external device towards the machine at hand. 

The complexity of computation with oracle is the number of queries in the course of computation. This definition resulted from the consideration of the complexity as the administrative time of the work of algorithm provided the time spent to the processing of queries is much greater than the time of the fulfillment of the ordinary command of the computational machine. On the other hand, we factually suppose that the oracle is much more complex object than the computational machine itself. This supposition agrees with the destination of a user to the role of oracle, that we made. 

We thus obtain the notion of computation relatively to a given oracle. The passage to the computations with oracle is called the relativistic version of computations. We can consider classes analogous to the early introduced classes ${\cal P},\ {\cal NP}$, but defined for the computations with some oracle $O$. Then the problem ${\cal P}= {\cal NP} (?)$ turns relativized towards this oracle. It turns that there are the oracles for which the relativization of this problem has the positive answer, and there are the other oracles, for which this relativization has the negative answer. In the other words, The relativization can convert the equation ${\cal P}= {\cal NP}$ into the true equation, and into the false equation as well.\footnote{It is known the result of Gil and Solovay accordingly to which if we choose an oracle randomly then with the probability 1 it gives the negative relativization of the problem ${\cal P}= {\cal NP}(?)$.}. This fact is meaningful. It means that every logical reasoning in which the relative computation can replace the ordinary computation without loss of strictness, cannot be applied for the solution of problem ${\cal P}= {\cal NP}(?)$ neither in the positive, nor in the negative sense. All known logical constructions operating with the notion of computation as is, allow such a substitution. Hence, this construction is useless in the solution of this problem. There are the evidences for that the statement ${\cal P}= {\cal NP}$ can be independent of the arithmetic axioms at all (it is not proved).\footnote{We see the difficulties which the axiomatic approach applied in the algorithm theory leads to. These difficulties reveal not only in the algorithm theory, but also in all mathematics as well. The more detailed discussion a reader can find in the monographs devoted to the fate of Hilbert program in the light of Goedel theorem (see, for example, \cite{She}).}.

The class of possible oracles has the continuum cardinality. At the same time, the class of computable (without oracle) sets is enumerable. Hence, the bulk of oracles represent the non-computable sets. Therefore, the computability relative to these oracles is the more wide notion than the simple computability.

The physical side of algorithms contains the explicit indication that we must treat an oracle as so complex device that we refuse to analyze it and consider it as a black box. It exactly corresponds to our division of the model to the user and administrative parts. A user participates in the building of the final model of the dynamics with the computer. The user participation as the expert shows has the form of answers to queries he receives from the computer. We see that the algorithm theory gives the formal apparatus for the building of the physical model, and the role of this apparatus consists in the ordering of considerations (as for any mathematical formalism). In the description of our algorithms, we thus should aim to the formalization of expert evaluations, and this formalization will be the universal instrument of algorithmic approach. Factually, the methods we describe below are aimed to this formalization. 

\section{Constructive mathematical logic and quantum theory}

We consider the so-called constructive mathematical logics, which plays the same role in constructive mathematics, as the ordinary mathematical logic plays in the classical mathematics. 

Historical basement of the mathematical constructivism is the field of mathematical logic called intuitionism.
The essence of intuitionism is that there is no excluded middle law in it. The logical formula $A$ or $not\ A$ is not axiom in intuitionism. To prove this formula we must prove either $A$ or $not\ A$. The models of intuitionistic theories are known as Kripke models. These models are organized as the sequences of ordinary models but in the three-valued logic, where besides the ordinary true and false exists the third logic value: uncertainty. The uncertainty may take the certain value: true or false in the next members of the sequence. It bears resemblance to quantum mechanics, where we do not know the result of measurement of the system in a quantum state.\footnote{This analogy prompted J. von Neumann and G.Birkhoff (see \cite{NB}) take up such a subject as quantum logic attracting for this aim the formal apparatus of intuitionism. To tell the truth, the axiomatic method, which buried the initial Hilbert program, here turned fruitless as well except for the substantiation of the principal idea of pluralism.}.

We also find in Kripke models the direct analogy with the dynamical models consisting of scenarios, and with the genetic method of the building of these scenarios. I think that the formal definition of dynamical models by means pf the formal apparatus of the models of intuitionistic theories is possible. It is the interesting theme belonging rather to mathematical logics and we do not stop on it\footnote{There are several works about this subject that can be found in \cite{Int}).}.

\subsection{Standard mathematical logic}

The basement of mathematical logic is the propositional logic. Its central notion is the logical formula. The definition of logical formula is inductive and looks as follows.

{\bf Basis.} Any Boolean variable from the set $\{ \a_1,\a_2,\ldots\}$ is a formula.

{\bf Step.} If $A$ and $B$ are formulas, the expressions $(A\& B)$, $A\ \ora\ B$, $not\ A$ are formulas as well. 

The logical value of a formula when its logical variables are fixed, is determined by the induction from the standard rules for elementary Boolean functions "and", "or" and "not". The following important notion of logical deduction is defined in the framework of propositional calculus. This calculus is more convenient to formulate not in the language of formulas, but by means of the so-called logical sequences. In this paragraph we mean by sequence a logical sequence. A sequence is a string of the form 
$$
A_1,A_2,\ldots, A_n\ \ar\ B_1,B_2,\ldots,B_k
$$
where $A_1,A_2,\ldots, A_n$ is the list of formulas called antecedent, $B_1,B_2,\ldots,B_k$ is a list of formulas called succeedent. The intuitive sense of the sequence is: is all the formulas from antecedent are true, then at least one of the formulas of succeedent is true. In the other words, the formulas in antecedent are joint by the logical connective "and", the formulas in succeedent - by the logical connective "or". We call the sequences of the form $A\ar A$ the axiom of propositional calculus. We say that a sequence is deducible, or is a theorem, if we can deduce it from axioms step-by-step, accordingly to the rules of deduction. We say that a formula $A$ is deducible, if the sequence $\ar A$ is deducible. The rules for deduction make possible to deduce the new sequences from the already deduced. These rules are the following: 

1) The transfer of formula: from a sequence $\ldots,A,\ldots\ar\ldots$ or $\ldots\ar\ldots,A,\ldots$ we pass to $\ldots,\ldots\ar\ not\ A,\ldots$ or $\ldots,\ not\ A\ldots\ar\ldots$ correspondingly, where dots denote unchanged strings. 

2) The introduction of conjunction or disjunction: from a sequence $A_1,A_2,\ldots\ar\ldots$ or $\ldots\ar\ldots,B_1,B_2$ we pass to $(A_1\& A_2),\ldots\ar\ldots$ or $\ldots\ar\ldots,(B_1\ \ora\ B_2)$ correspondingly. 

3) The mixing: the formulas inside of succeedent (or antecedent) can change the order; we can cancel the repeating formulas and introduce the repeating formulas separately in antecedent and succeedent. 

4) The cut: from $\bar D, A\ar B$ and $D_1\ar A,E$ we can deduce $\bar D,D_1\ar B,E$. 

The theorem of elimination of cuts is known, accordingly which if a formula is deduced with cuts, it can be deduced without cuts. However, cuts play the valuable role in the predicate calculus, which we consider further. 

We introduce to the language the following functional symbols $f_1,f_2,\ldots, f_n$ and the following predicate $g_1,g_2,\ldots,g_k$. We call the pair of these two sets a signature of logical theory. The intuitive sense of a functional symbol is a function on some abstract set of individuals, the intuitive sense of a predicate symbol is a characteristic function of some set of individuals. We mean that a predicate separate this set from all individuals. We define the notion of a term (complex function) by the following induction.

{\bf Basis}. Any variable $x_1,x_2,\ldots$ is a term.

{\bf Step}. If $t_1,t_2,\ldots,t_m$ are terms and $f_i$ is $m$ placed functional symbol, then $f_i(t_1,t_2,\ldots,t_m)$ is a term.

We now consider the propositional calculus in which the formulas of the form $(t_1=t_2)$ and $g(t_1,t_2,\ldots,t_m)$ play the role of logical variables $\a_j$ where $=$ is the special sign means "equality, $g$ is $m$ placed predicate symbol. 

We introduce as the new axioms the following sequences:

$$
\begin{array}{ll}
&(t_1=t'_1),\ldots, (t_m=t'_m)\ar (f(t_1,t_2,\ldots,t_m)=(f(t'_1,t'_2,\ldots,t'_m)),\\
&(t_1=t'_1),\ldots, (t_m=t'_m),g(t_1,t_2,\ldots,t_m)\ar g(t'_1,t'_2,\ldots,t'_m),\\
&(t_1=t_2),\ (t_2=t_3)\ar (t_1=t_3),\\
(t_1=t_2)\ar (t_2=t_1).
\end{array}
$$

At last we introduce the special signs $\forall , \ \exists$, called quantifiers of the universality and existence, and generalize the notion of formula by the introduction of the new possibilities to the inductive step in the definition of formulas. 

If $A(x)$ is a formula with the free variable $x$, then the expression $\exists x\ A(x)$ and $\forall x\ A(x)$ are the formulas in which the variable $x$ is bound (not free). The rules for the substitution of terms instead of free variables in a formula are defined by the natural way. We denote by $A(t)$ the result of the substitution of term $t$ instead of the free variable $x$ in formula $A$.

We now introduce the new rule for deduction concerning quantifiers. 

1) Is $t$ is a term, $x$ is a variable not occurring beforehand, then we can from $\bar D, A(t)\ar \bar C$ derive the sequence $\bar D, \forall x\ A(x)\ar \bar C$, from $\bar D\ar A(t),\bar C$ derive the sequence $\bar D\ar\exists x\ A(x),\bar C$.

2) If a variable $x$ does not occurs in the sequence but the pointed place, then form the sequence $\bar D,\ A(x)\ar \bar C$ we can derive $\bar D,\ \exists x\ A(x)\ar \bar C$, from $\bar D\ar A(x),\bar C$ we can derive $\bar D\ar \forall x\ A(x),\bar C$. 

Thinking a little we can understand that the introduced deductive rules exactly correspond to the natural sense of the symbols $\forall x\ A(x)$ and $\exists x\ A(x)$, if we treat the first one as the conjunction of all possibilities: $A(x_1)\& A(x_2\&\ldots$, and the second one as the disjunction of all possibilities: $A(x_1) \ or\ A(x_2)\ or\ \ldots$. 

The defined logical system is called the predicate calculus. Here the theorem of the elimination of cuts does not take place. Nevertheless, its analog is true, which says that we can derive any deducible sequence applying a cut only to the formulas $A$ of the form $(t_1=t_2)$ where $t_1,\ t_2$ are terms. 

The deductive rules give the possibility to obtain the theorems from the already proved theorems, beginning with axioms. We say that a formula $A$ is deducible if the sequence $\ar A$ is deducible. The following fact takes place: the set of deducible formulas coincides with th eset of formulas which obtain logical value truth for any logical values of its variables. For the propositional calculus this fact has the form of exact theorem. For the predicate calculus this fact is true as well, but with the appropriate meaning of words "logical variables" (there are no logical variables there, only individuals and functional and predicate symbols). We call a formula true if it is true for all models, e.g. for any possible model and all possible concretization of functions and predicates in this model. In this form, the following Goedel theorem of the completeness of the predicate calculus takes place:

A formula is true if and only if it is derivable in the predicate calculus.
I propose to a reader to find the derivation of the formula $(\a_1\ \ora\ not\ \a_1)$ in the propositional calculus as the simple excercise. 

\subsection{The problems of the consistency of logical theories}

In the classical mathematical logic, the main question traditionally is the question about the consistency of a logical theory. We call a theory consistent if there are not derivable formulas in its signature. In the opposite case, we call a theory inconsistent. If for some formula $A$ it is possible to prove it and in addition to prove its negation $not\ A$, this theory is inconsistent, e.g., in it all formulas are derivable, that can be proved by the simple application of the rules of transfer and cuts. The most interesting are the theories which signature makes possible to write formulas corresponding to the statements of arithmetic. For example, to encode the strings of symbols, state the simple facts about their structure, etc. In particular, we assume that in the theories we consider it is possible to write the statement like "this string of symbols represents the text of proof of this formula in this theory". We can build such formulas in the standard arithmetic with the addition and multiplication, it is the routine though the laborious job. I address a reader to any sufficiently detailed book on the mathematical logic in which it is fulfilled in details (for example, \cite{She}).  

It is then possible to write in the signature of such a theory the following statement:
\nnn

"This theory is consistent"
\nnn

This formula has the form: 
$$
\exists D (proof\ (D)\&\ fin(D,\ (2\cdot 2=5)),
$$
where $proof\ (D)$ is the formula claiming that a text $D$ is the proof, and $fin(D,\ C)$ means that a text $D$ ends with the logical formula $C$. The statement of the consistency of a given theory ${\cal T}$ we denote by $Consis_{\cal T}$. Of course, this formula contains the codes of axioms of ${\cal T}$, etc. 

{\bf Goedel theorem about the proofs of consistency} claims that if the formula $Consis_{\cal T}$ is derivable in a theory ${\cal T}$ then this theory is inconsistent. 

E.g., if the tools of some logical theory are sufficient to prove its own consistency, then any statement at all is derivable by these tools, and this theory is absolutely useless thing. 

If we introduce to a theory ${\cal T}$ the statement about its consistency, e.g., if we add $\ar Consis_{\cal T}$ to its axioms, we obtain the different theory ${\cal T}_1$, and the statement about its consistency will be the different: $Consis_{{\cal T}_1}$, etc. It means that the question about the consistency of logical theories has no solution in the framework of these theories. This fact immediately makes the axiomatic method unreliable. The recognition of this fact led to the crisis in mathematics we mentioned above, and it stimulated the appearance of constructivism.

In full measure, it concerns physical theories because they rest on mathematical formalism. The axiomatic method suffers with the unavoidable defect and it is not applicable beyond the connection with the so-called common sense or the intuition. If the axiomatic method is not supported by the intuition, it can lead the investigation to the dead end. At the same time, the intuition is especially weak in quantum theory. Many statements of quantum theory explicitly contradict to intuition; hence, the role of formalism here is very important. In the framework of the classical mathematics, it is impossible to separate the necessary part of formalism that the physicists always aim to, often unconsciously. The notion of actual infinity lying in the basement of mathematical analysis requires just the axiomatic approach because this notion does not allow any effective procedures in principle. 

Only constructivism based on the effective procedures is free from this drawback.

\subsection{Constructive mathematical logic}

The standard mathematical logic is not completely appropriate for the basement of constructive mathematics. For this aim, there exists the special constructive mathematical logic, which is called {\bf intuitionism}. The intuitionism does not operate directly with algorithms, but it formulates the logical rules for the work with the logical formulas of constructive mathematics. We devote a few pages to intuitionism. 

We consider one of the main experiments lying in the basement of quantum theory, the interference of one quantum particle in its passing through two slits in the screen. It is well known, that the interference picture for this experiment cannot be obtained by the simple summing of the pictures for the passing trough one and through the other slit separately. Let the logical formula $A$ mean that the particle has passed through the first slit, and the formula $B$ mean that it has passed through the second slit. The passing of the particle through the both slits can be then expressed logically by the formula $A\ \ora\ B$. This is just the treatment of the experiment with the passing through two slits from the classical physics viewpoint, where the particle means the pont wise particle. In the other words if we detect the particle behind the screen we conclude that it has passed through the first or through the second slit, e.g., the formula $A\ \ora\ B$ takes place. 

In the quantum physics, the situation is different. The appearance of particle behind the screen does not mean that it has passed through the first slit or through the second slit. It is the special event, which concerns the slits by some specific way. Here the passing of the particle as point wise through one single slit is aslo expressed by the formula $A\ \ora\ B$, but to prove it we must answer to the question, which slit exactly our particle passed through ! To prove the formula $A\ \ora\ B$ in quantum physics we must detect the particle in the instant of passing through the screen that unavoidably results in the observation of it in the first or in the second slit, and, of course, ruins the interference picture.

In the constructive mathematical logics, all is like in quantum theory. There are two ways to prove the formula $A\ \ora\ B$: to prove either $A$, or $B$. Constructivism does not admit any other way to prove a disjunction. We can do the deep conclusion from this: to prove the formula $A\ \ora \ not\ A$ in the intuitionism, we have to prove either $A$, or $not\ A$. It means that there is no excluded middle law here. At the first glance it is the shortage of intuitionism. However we remember that it is not always good to have possibility to prove much. The deductive capability in quantum mechanics directly contradicts to the Nature as in experiment with two slits, where the acceptance of $A\ \ora\ B$ excludes the real interference picture. Just constructive mathematical logic shows the best correlation to quantum physics. It was recognized already by von Neumann who tried to build quantum theory on the basement of intuitionism. This effort had no serious success despite of its right idea: to prove that the intuitionism is the real logic of quantum theory. The pioneer work \cite{NB} had no serious immediate continuation due to the difficulties in the conversion of quantum physics to the constructive way that requires also the replacement of standard mathematical analysis by the constructive analysis. The deal cannot be resolved by the simple change of the logic because simultaneously we must change the interpretation of all the computational methods used in quantum theory. 

We formulate the necessary changes in the language of propositional calculus for the conversion from the classical calculus to the constructive calculus. This is only one and very simple change. Logical sequences must now have the form 
\begin{equation}
A_1,A_2,\ldots,A_n\ar (B).
\label{intu}
\end{equation}
E.g., no more that one formula can stand in the succeedent. This simple syntactical condition: the ban of succeedent splitting gibes us the intuitionistic logical calculus instead of the classical. All the rest rules and axioms remain unchanged. In this calculus to prove the law of excluded middle is already impossible, just due to the definition of sequences by \ref{intu}. 

Intuitionistic logic has thus the same arsenal of deductive rules as the classical logic, and the restriction that intuitionism imposes to logic must no touch the bulk of the basic things. This guess finds the complete confirmation by the exact theorems. Namely, there is the so-called interpretation of the classical mathematical logic in terms of the constructive logic, which preserves the logical deduction. It means that there exists the translation of formulas of the form 
$$
Trans\ :\ \ A\ar [A]
$$
where $A$ is a formula of classical logic, $[A]$ is the corresponding formula of constructive logic called its constructive interpretation, such that $A$ is deducible in classical logic if and only if $[A]$ is deducible in the intuitionism. This result (see \cite{No}, \cite{KV}) means that the classical and the intuitionistic logic are equally consistent. If a contradiction contains in one of them, the corresponding contradiction would be in the other. 

As in the classical logic, we can add to intuitionism the axioms of arithmetic or of the set theory, which gives the constructive versions of these logical theories. In the constructive logic, it is possible to prove the theorem analogous to Goedel theorem about the completeness, but for the constructive logic the definition of model will be the different, it will be Kripke models. A model of constructive logic contains the special parameter which we can call the time $t$. The division of logical formulas to the true and the false depends on the value of $t$. In each time instant the formulas will be divided to the true, the false and the uncertain. If for the first two types the true values are determined to the current time instant, then for the uncertain formulas we assume that their true values to the instant $t$ are not yet determined. In the next time instants $t_1>t$ the values already determined remain unchanged, and for uncertain formulas their true values can obtain certainty. The set of formulas with the certain true values thus will not decrease, and the set of formulas with the uncertain true values will not increase. We will not show here the exact definition of the intuitionistic Kripke models. We limit the consideration by the mentioning that such models correspond to quantum theory in a better degree than the standard models. 

In the constructive mathematics, there are more requirements to the true formulas than in the classical. It follows from the more strict requirements to the logic deduction that is the characteristic feature of constructivism. Here we must show the explicit procedures of the building of the required objects, not simply prove that their absence leads to the contradiction, as in classical mathematics. \footnote{We speaks about those proofs of the pure existence that so annoy to physicists and which represent the subject of proud of the classical mathematics. The constructive mathematics is free of these excesses.   } We see that this peculiarity of constructivism completely corresponds to the situation in quantum theory. In quantum mechanics, many magnitudes have no the certain values but can obtain these values further, in the result of the observations fulfilled on the considered system. Moreover, the main physical magnitudes: the charge and the mass of the particles can have slightly different value dependently of the scenario in which we consider them, or factually, their personal vicinity (renormalization of charge and mass). It confirms the mind that the constructivism inheres to quantum theory. This is why the formulation of quantum physics in the constructive terms represents the deeply substantiated step. 

\subsection{Idea of pluralism and its importance for physics}

The constructive logic contains something of the same importance for physics than the rules for logical deduction. This is the fundamental idea of the logical pluralism. I dwell on it because of its big value for our aims.  

As we know from the previous paragraph, the question about the consistency of a logical theory has no solution in the framework of this theory. The question about the conditional consistency then takes on special significance. If we believe that some theory is consistent, then from this fact we sometimes can make the certain conclusion about the consistency of the other theory. We look from this viewpoint to the classical and intuitionistic logics. Since the intuitionistic logic is the contraction of the classical logic, the consistency of the classical logic guarantees the consistency of the intuitionistic. Is the opposite right? 

As we know, the answer is positive. There exists the translation of texts of the form 
\begin{equation}
D\ar \tilde D,
\label{trans}
\end{equation}
such that if a text $D$ is the proof in the classical logic, then the text $\tilde D$ is the proof in the constructive logic. The conditional consistency of the classical logic follows from this fact. It is consistent if the intuitionistic logic is consistent. This way of substantiation of classical mathematics Brouwer and the other supporters of intuitionism kept in mind. The constructive mathematical logic has the more intuitive cogency than the classical logic, it is reflected in its name. The intuitionism requires from a proof the concrete procedures of the building of target object, not only verification that this object cannot be lacking. The intuitionism does not accept the standard scheme of classical mathematics: we suppose that there is no such object and obtain a contradiction... It means that the constructive logic is more reliable just from the physical viewpoint, especially in the light of Gedel theorem about the proofs of consistency. The theorem about the translations of proofs claims that this high level of reliability of the constructivism in some sense can be extended to the classical logic! \footnote{However, there is the same problem of the proofs of consistency than in classical case. The intuitionism cannot then be, despite of von Neumann dreams, the single reliable basis of the constructive physics.}

To realize this extension in reality we need the translation of the texts of proofs of the form \ref{trans}, but is not uniquely determined. It is one of the manifestations of the pluralism of true evaluations typical for constructivism.  The other manifestation of this pluralism is the uncertainty of the true value of formulas to some time instant, as in Kripke models. The constructivism thus possesses the inbuilt property of pluralism of the true values: these values can depend on the time and on the method of interpretation of classical logic in it. It is significant that this property of intuitionism is unavoidable. We should treat it as the necessary cost for the conveniences, which intuitionism gives, namely, for the effectiveness of all procedures of finding objects and the arising from it more confidence in the reliability of logical deduction in comparison with the classical case.\footnote{This confidence is still not absolute due to the existence of Goedel construction. The single completely correct way is the division of the model to the user and administrative part and the explicit introduction of a user as an oracle.}

The fundamental property of pluralism of the constructive mathematical logic is important for physics. The pluralism establishes the limits for the application of physical theories, and especially, of the main physical concepts. First of all the matter concerns the concept of indistinguishability, or identity of the elementary particles of the same type that makes possible to apply the methods of matrix algebra. The formalism of occupation numbers substantially uses the identity of elementary particles that permits to write the physical operators in the compact form. Of course, our work in the physical constructivism needs the computational methods of this type. This is why we do not intend to refuse from the representation about the identity of elementary particles of the same type. However, the status of this representation in the constructive physics is limited by the fundamental property of pluralism, e.g., it is lower than the fundamental law. It means that the representation of identity does not belong to the new formalism, but is simply the convenient supposition for the creation of simulating algorithms in the known cases. This supposition concerns the convenience and we can refuse from it if it becomes necessary. We cannot treat as the principal step the possibility to refuse from the representation of identity, if we choose the constructivism. Hence, we cannot consider as the concept the supposition of identity in constructivism\footnote{The connection of the types of elementary particles with the irreducible representations of the Poincare group does not involve the fundamentality of the identity of particles. We always can suppose that the particles inside the same type have individuality connected with their different environment. This supposition in constructive physics is the question of effectiveness of the simulating algorithm, but not the axiom.}. 

The representation about the identity of elementary particles of the same type (for example, electrons) is the result of the simplicity of the simulating algorithms in simple cases, e.g., if we ignore the entangled quantum states of particles. Traditionally important place of this concept in quantum theory follows from this effectiveness. The situation in which the division of particles to smaller groups gives profit would break this tradition. Also the inverse situation is possible, when the joining of particles to bigger groups gives convenience.\footnote{The example of the last situation is the introduction of isospin in the nuclear physics.} The method of collective behavior gives the possible form for this, we treat it in the next chapter. In this method, we represent one particle as an ensemble consisting of its samples. It gives, in principle, the possibility to acquire samples with the individual properties, e.g., to realize just that sub-structure we keep in mind in the refusal from the concept of identity. The refusal from this representation can give the serious profit in case of the dynamical scenarios of complex systems, and we return to this question below. Now we merely state that we preserve the representation about the identity of particles for the sake of the convenience of the existing simulating algorithms.

\section{Constructive mathematical analysis}

Algorithmic approach to physics is successive towards the traditional. It results from that algorithms are applied to the standard apparatus of physicists theorists: mathematical analysis. We need such algorithmic modification of mathematical analysis that preserves its computational part, what is only important for physics. There is the abstract part in mathematical analysis of Cauchy - Weiershtrass, containing such objects as theoretical problems in the set theory, the different choices of axioms, and the things like these, which have no applications in physics. We can and must sacrifice these things to the success of algorithmic modification of mathematical analysis. This apparatus is already created, it is the constructive mathematical analysis. This section is the short introduction to this necessary for us area.\footnote{A reader who is interested can take a closer look at this subject by the book \cite{Ku}.}

\subsection{Constructive real numbers}

We remember that the real numbers can be (in the nesting order): natural, integer, rational, algebraic and transcendental. The constructivism introduces the new class constructive real numbers in addition to two last classes.

Briefly, a constructive real number is such a real number, which is the limit of a sequence of rational numbers $r_1, r_2, \ldots$, generated by some algorithm. It is clear that any algebraic real number is constructive, because solutions of equations with the integer coefficients have the approximation by the sequence of rational numbers where each member generates the next member by means of some constructive procedure. There are the examples of transcendental numbers which are constructive, for example, $e$,$\pi$, and any algebraic combinations of these numbers, Euler constant about which is not known is it rational or not, etc. 

All real numbers used in the practical computations are constructive that makes possible to state that this class of numbers is sufficient for the building of mathematical analysis. Nevertheless, the set of constructive numbers is enumerable, and the bulk of real numbers is thus non-constructive. It indicates that the constructive mathematical analysis can differ radically from the standard (and really differs) at least if we use for its description the same logical theory \footnote{It is known that every consistent theory with the enumerable signature has an enumerable model. The signature of mathematical analysis is enumerable; therefore, the enumerable model exists for the standard mathematical analysis, for example, in the formalization (axiom system) of Zermelo - Fraenkel. The question arises what then means the non-denumerability of the set of real numbers. It merely means the absence of the function establishing one-to-one correspondence with the natural numbers and the set playing the role of real numbers in this model. This function exists in the wider model but it does not play any rule because it is lacking in this model. By the way, this example demonstrates one of the features of axiomatic approach, which is unexpected from the viewpoint of the common sense: there is no common sense here.}.

It turns that it is the case.

\subsection{Constructive functions of constructive real variable}

Since the constructive mathematical analysis rests on algorithms, it must also relate to the numerical functions. 

Rational numbers have the form of pairs of integers with the sign, and they can thus serve as the domain of a function determined by algorithm. Given a function $f$, determined by an algorithm, which maps the arbitrary corteges of rational numbers to rational numbers. We assume that this function determines the map of constructive real numbers to constructive real numbers provided for any converging sequence of rational numbers generated by algorithm $x_1, x_2,\ldots$ the sequence $f(x_1), f(x_1,x_2),f(x_1,x_2,x_3),\ldots$ is converging as well, and the limit of the second sequence does not depend on the choice of the first sequence. Let $x$ and $y$ be the limits of the first and the second sequences correspondingly. These numbers are the constructive real numbers and we say that the algorithm $f$ determines the constructive real function of a constructive real variable, which we denote by the same letter: $y=f(x)$. 
The notion of constructive real function is the analogue of the function of the real variable in the constructive mathematical analysis. 

A function $y=f(x)$ of a constructive (real) variable $x$ is then a constructive if there exists an algorithm $T$, which for any approximation $r$ for $x$ gives the approximation $T(r)$ of the number $y$, and if $r\ar x$, then $T(r)\ar y$. 

There is the theorem of Markov the young and Tseitin (see \cite{Mar}, \cite{Ts}), which claims that {\bf any constructive function is continuous}. We will not show its complete proof here, and limit ourselves to the sketch of the proof. We show that the Heaviside function:

$$
\theta(x)=\left\{
\begin{array}{lll}
& 0,\ if\ & x<=0,\\
& 1,\ if\ & x>0
\end{array}
\right.
$$
is not constructive. Really, we suppose that it is constructive. Then there is Turing machine, which gives sequential approximations of $0$ or $1$ depending on to which limit the initial sequence $x_1,x_2,\ldots$ converges: to zero, or to some positive number. We take the sequence of numbers coincident with $(-1)^n/n$ on step $n$, provided some Turing machine $M$ has not finished its work on some word $a$. Since the function $f$ gives its values resting the final set $x_1,x_2,\ldots,x_n$ only, it must give the good applications of its value on the final step already, that is impoissible because in this case we would solve the problem of applicability for the machine $M$, which is in general case not computable.  

In the other words, only continuous functions exists in the constructive mathematical analysis\footnote{Of course, not all continuous functions are constructive. The total number of constructive function is enumerable.}. We can explain it as follows. A constructive function of constructive argument must give the approximations of its values knowing only the approximations of the argument. This requirement is very practical, because, in contrast to the standard mathematical analysis we know the values of all variables occurring in the statement of one or another concrete problem, only within some accuracy. Here we even do not know beforehand with what accuracy we know them. E.g., the error may be unknown for us! Nevertheless, a constructive function must give us the approximations of its values (which accuracy will therefore depend on the accuracy of the initial data). This way of determining of real functions is much more physical than its axiomatic definition in standard analysis.

In the constructive mathematical analysis, we naturally define the limit of functional sequence, where some algorithm must generate this sequence. All properties of limits from the standard analysis are preserved. Literally analogously, we introduce the notion of derivative, indefinite and definite Riemann integral, here all the computational properties, in particular, the methods of solutions of differential equations, theorems about stability, etc. remain valid, with some variations accounting Markov-Tseitin theorem. These variations do not touch the analytical computational methods. 

The notion of constructive number and constructive functions can be naturally generalized to the complex numbers and functions of complex variables. The differential properties of these functions remain valid including main theorems of the theory of functions of complex variables concerning Cauchy integral and the technique of its computation, e.g., all contents of the standard analytical apparatus important for physics remain unchanged.

We illustrate the importance of the notion of constructive functions on the simple examples.  

{\bf Example 1.} It is known that the singularity of Coulomb field of point wise charge causes many difficulties: for example, it leads to the infinite own energy of electron if we treat it as consisting from more elementary component parts. The various difficulties arise in the numerical simulation of the probe charged particle dynamics (for example, electron) in the field of Coulomb centers. For example, this dynamics becomes dependent of the exact position of charged particle. In the consideration of separate field created by all particles there arises the self action problem (see \cite{FH}). If we change a bit the position of any charged particle (or, for example, fulfill the measurement of this position with some nonzero error that always takes place) then the own Coulomb field gives so big impulse to the particle that brings to naught any attempts to find the satisfactory dynamical picture. We can go round the problem by accounting nuclear forces between the nucleus and the electron.\footnote{This is not the real solution, we show it for illustration only. In the chapter 5 we show more complete solution based on the method of collective behavior.} In this case, the potential acquires the form represented on the picture (we do not observe the relative sizes). Here in the top we find Coulomb potential of attraction of electron to the point wise proton. Lower is the potential of interaction between the nucleus and electron with the account of nuclear forces. The repulsive nuclear potential acts at the short distance. The local minimum in the center is conditional, it expresses the fact that if the electron flights to the proton with the exact aiming, it results in the appearance of the neutron. The exact aiming corresponds to the semi classical consideration of electron. It happens when it approach to the proton with the very high speed. 

\begin{figure}
\centering
\caption{Potential of electron in hydrogen atom: Coulomb and real}
\vspace{250mm}
\makebox[180mm][l]{\includegraphics{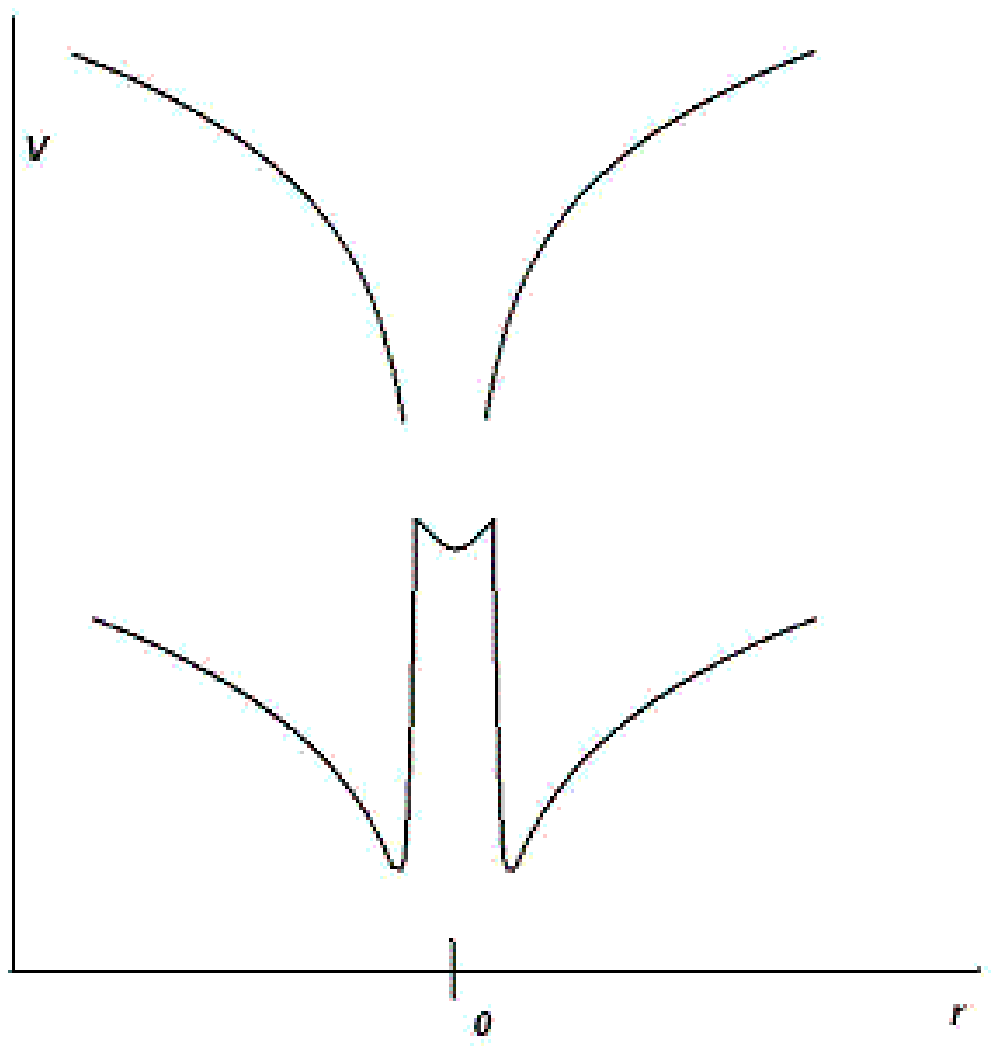}}%
\end{figure}

We can look at this example otherwise. What a force creates the potential near the nucleus? If it is the nuclear forces, the electron must rebound from the nucleus as the ball from the wall.\footnote{Of course, the analogy does not account the size: if the electron is represented by the ball, the nucleus then looks as the small weight, which is smaller than the ball but about 2000 times heaver.} It has no influence on the conclusion based on Shroedinger equation. However, these hits result from the consideration of electron as some hydro dynamical media that can create the pressure to the nucleus. We can make clear the known effect of the bigger stability of a nucleus in neutral atom than in the positive ion. If we fulfill the ionization, moving off several electrons in the heavy isotope which is stable in the neutral atom, the nuclear decay begins (\cite{Du}). It evidences about the close connections between the electromagnetism and the nuclear physics that is beyond the consideration in the standard quantum theory. Further we try to consider this phenomenon from the constructive viewpoint.

{\bf Example 2.}
We consider the charged particle in the two equivalent holes potential. Its ground state is such that the probabilities to find the particle in the both holes are the same. We now begin slowly heighten the potential barrier between the holes. Practically we could do it, for example, using the molecular ion of hydrogen with two protons and one electron. The heightening of the barrier is equivalent to the increasing of the distance between protons that we can fulfill accelerating ion in the electric field and colliding it with the chemical molecules. We can do more simply, considering two quantum dots and one electron. The heightening of barrier results from the increasing of the voltage on the controlling gates. 

At last, we obtain two holes and the probe particle finds itself in one of them. This process in the standard quantum theory is called the wave function collapse, which happens instantly. In the constructivism, this explanation is not acceptable because it does not allow the description in terms of the wave function that means the necessity to go out of the limits of standard quantum theory. 

These two examples illustrate the importance of the application of constructive functions because the real picture must not contain singularities. However, they differs one from another. The first example means that one algorithm can join the different physical laws. In the second example, the requirement of constructivism immediately leads to the going out of the frameworks of standard quantum theory. 

The most important advantage of the constructive mathematical analysis over the classical is that the constructive analysis requires the explicit models of processes scalable with the grain of spatial resolution. Really, we consider the process described by the function of real variable $f(x)$. If this is the ordinary real function, the accuracy of the definition of $x$ would not mean the accuracy of the determining $f(x)$ as it takes place in the discontinuous functions. Even if $f(x)$ is continuous but not constructive, we could not determine effectively the factual accuracy of the determining of $f(x)$ given the accuracy of knowing $x$. This possibility to give the approximation of $f(x)$ from the given approximation of $x$ is important. Let us imagine that in the reality the values of $x$ are grained, e.g., they have the form $n\e$ for the natural $n$ and very small fixed $\e$, which is so small that lies out of the immediate observation. The constructive analysis obliges us to point to the explicit mechanism of the break of accuracy in the approximation of $f(x)$ for any approximation of $x$. It means that for any small value of the grain $\e$ our algorithm determining the function $f$ must work with this grain.  We simply break the process of sequential approximations of $f(x)$ when the approximation of $x$ reaches the accuracy of the grain $\e$. The advantage here concludes in that we get rid of the necessity to consider the pathological functions $f$, existing in the standard mathematical analysis. 
It makes strict the consideration of Feynman path integrals where on the small distances the paths are replaced by the straight lines. In the opposite case using standard mathematical analysis we have to apply the tricks lacking in the mathematical correctness\footnote{In the book \cite{FH} it is indicated directly.}, which is not formally right and risky because if we use the apparatus we should satisfy its requirements.

\section{Constructive algebra for quantum mechanics}

This paragraph is devoted to the algebraic apparatus of quantum theory. At first, we describe it for the standard quantum mechanics, then we remember the basic notions from the theory of algebraic systems, and at last we define the notion of constructive versions of algebraic objects used in quantum theory. We use the term "algebraic systems", accepted in mathematics for these objects, despite only Hilbert and Euclidean spaces and groups of operators in them are really in use. 

\subsection{Algebraic apparatus of quantum theory}

We can conditionally divide the algebraic apparatus of quantum theory to the following big categories.
\begin{itemize}
\item Algebraic systems of the configuration space of physical ensemble. 
\item Algebraic systems of the space of states of considered ensemble.
\end{itemize}

Quantum theory uses the richest algebraic apparatus; for example, the gravity theory uses only the first type of algebraic systems. It results from the basic principle of the superposition of quantum states. This principle factually makes the matrix technique for the description of evolution the main technique for quantum systems; it presumes explicit introduction of the space of states of the considered system and the auxiliary algebraic systems for it. 

At first we consider the algebraic systems of the configuration space. Let $K$ denote the configuration space of the system at hand. If the system consists of $n$ particles without spins this is $R^{3n}$, if particles have spins directed along or conversely some separated axe, it is $R^{4n}$, if there is the time common for all particles, it is $R^{4n+1}$, if any particle has its own time, it is $R^{5n}$, etc. We always accept that such a space is local, that is the fixed coordinates are applicable in the small vicinity of some point $O$ only, which we treat as the reference point for this vicinity. Elements of the set $K$ we denote by $|j\rangle$. Globally, the configuration space represents the manifold $M$ of the corresponding dimensionality: $3n,\ 4n,\ldots$. We will not discuss the structure of the manifold $M$ here; for one particle it is described by the matrix of metric tensor $g_{i,j}$, satisfying the equation of the general relativity theory (see the book \cite{Di}). The question about the structure of manifold $M$ for many particles is open. This is the question about the agreement between quantum theory and general relativity theory, and its pendency nowise tell on our problems. The transformation of the passage to the other orthogonal coordinates for any particle must preserve the value of the differential form 
\begin{equation}
ds^2=dx^2+dy^2+dz^2-c^2dt^2
\label{rim}
\end{equation}
where $x,y,z,t$ are the coordinates of this particle (the first three are spatial, the fourth corresponds to the time), $c$ is the speed of light. It means that for two observers if each of them uses his own rectangular coordinates, the values of $ds^2$ computed by them must be the same. These transformation form the group called Poincare group. We can consider the derivative algebraic constructions on this group, for example, its representations, characters, etc. 

The second type of algebraic systems is connected with the space of states ${\cal H}$ of the physical ensemble at hand. The space ${\cal H}$ consists of the linear compinations of the form 
\begin{equation}
|\Psi\rangle=\sum\limits_{j\in K}\la_j|j\rangle ,
\end{equation}
in which the different states $|j\rangle$ are treated as orthogonal to each other and have the unit norm, $\la_j$ are complex numbers.

The main operators in this space are the deterministic evolutionary operators of the system that have the form

\begin{equation}
S:\ {\cal H}\ar {\cal H}
\label{evolution}
\end{equation}
 
and the random operators of measurement. 

At first we take up the deterministic operator \ref{evolution}.
This operator forms the elementary step of the evolution of the system in the real physical time. The time $t$, taking part in this operator as one of the coordinates of the configuration space $K$ of our system ( or the times, if any particle has its own time) is the administrative time of quantum system\footnote{Some authors, for example, \cite{KMB}) call it the hidden time of quantum system. In the many particle system it would be more correct to speak about the internal times; they coincide for the interacting particles only (see below).}.

The representation of evolutionary operator in the form \ref{evolution} contains the serious condition: we assume that it is possible to consider simultaneously evolutions of the different states of the considered ensemble, e.g., potentially, all the space of its quantum states. How it corresponds with experiments where only a singe state of the real ensemble is observable and we can trace only its evolution, not evolutions of all states? The correspondence is given by the superposition principle, which states that the operator $S$ is always linear. It means as follows. We always can repeat the evolution for the other states of our ensemble chosen from the space ${\cal H}$ and can thus check the linearity immediately. This requires the possibility to generate vectors from the space ${\cal H}$ randomly, which presumes, in particular, the generation of the same states as many times, as needed. Shroedinger equation results in that, at least in the ordinary quantum mechanics, the operator $S$ has the form $exp(iH)$ for the appropriate hermitian $H$, hence $S$ must be a unitary operator in ${\cal H}$. The requirement that all the evolutionary operators must be unitary takes place for quantum electrodynamics as well. The main algebraic object connected with the space of quantum states is thus the group of unitary operators $U(dim({\cal H})$, which dimension equals the dimension of ${\cal H}$, that is in turn the exponential of the dimensionality of the configuration space $K$. 

We can then consider the Lie algebras of such groups, defining the multiplication as the matrices commutator and the addition by the natural way, tensor products of such groups corresponding to the evolutions of independent ensembles, restriction to subspaces, etc. 

The non deterministic random operator of measurement is the random variable, which distribution depends on a given state $|\Psi\rangle$.   
If ${\cal H}=\bigoplus\limits_{k=1}^lH_k$ is the expansion of ${\cal H}$ to the orthogonal sum of subspaces then the measurement corresponding to this expansion is the random variable, each of which values is the operator of projections to subspace $Н_k$ with the probability equal the squared length of the projection of $|\Psi\rangle$ to this subspace. We will show below that by the very natural method of constructivism (grain of amplitude) the random operator of measurement can be included to the scheme joint with the time evolution operator. This method presumes that the configuration and Hilbert spaces are grained. The grain means the finiteness not only all possible positions of a system, but also all values of the amplitude. Hence, we will consider only the operators of the type \ref{evolution}. Factually, assuming the method of constructivism based on the grain of amplitude we reduce the  measurement to the expansion of the configuration space. 

\subsection{Classical algebraic systems}

We call a signature a pair: (set of symbols $\bar f$ for functions, and set for symbols $\bar p$ for predicates). Each symbol in these sets has its own quantity of places, for example, two place functional symbol, or one place predicate symbol $>0$. 

By algebraic system in this signature we mean a set $G$ and two lists $\bar F$ and $\bar P$ of functions and predicates, which correspond in their order to the symbols in the signature. The example is the group $U(dim({\cal H}))$, considered in the signature and consisting of one two place functional symbol  $\cdot$ of the group multiplication that corresponds to the sequence of the operators of evolution and reflects the real time flow. 

Classical algebraic systems based of real numbers, for example, algebras over the fields $R$ or $C$, presumes the application of the standard mathematical analysis. Factually, algebraic systems give convenient algorithms for finding solutions of quantum equations, at first Shredinger equation. Particularly, the matrices product exactly expresses the phenomenon called quantum interference. Given two matrices $U=(u_{ij})$ and $V=(v_{jk})$ we consider their product: $UV=W=(w_{ik})$. If $U$ and $V$ are the operators of sequential steps of the time evolution, then this product is the operator of the resulting evolution. What the matrices product $w_{ik}=\sum_ju_{ij}v_{jk}$ expresses ? It expresses the law of interference accordingly to which to obtain the amplitude of the passage from the state $i$ to the state $k$ we must add the deposits of all paths from $i$ to $k$ through all intermediate states $j$ where for any path we must multiply the amplitudes of its sequential parts. This law of interference lies in the basement of quantum physics. We see that the algebraic language of matrix algebra is the most precise and concise expression of this law. It substantiates the important role that algebra plays in quantum theory. 

\subsection{Constructive algebraic systems}

Resting on the constructive mathematical analysis we can make algebra constructive. As in the constructive mathematical analysis, algorithm will generate the objects of algebraic systems. It could seem that the constructive mathematical analysis is the first level of constructive algebra because we there considered the easiest algebraic system: real numbers. This is wrong impression. The constructive analysis contains the condition of passages to limit, determined by algorithms: $r\ar x,\ T(r)\ar y=f(x)$. Though it is formally reducible to constructive algebraic systems, the main in them is not the convergence but the algebraic properties. Nevertheless, if we need to make a finite algebra over real or complex field constructive, we can simply transform all the definitions inserting instead of real numbers constructive real numbers.  Since all functions used in algebraic systems have the expressions in terms of formulas (as the function of summing), all they will automatically become constructive and we remain in the frameworks of the constructive mathematical analysis. 

The single peculiarity of algebra is that it contains not only infinities resulted from real numbers, but also its "own" infinities occurring in algebraic systems. These infinities arise in the consideration of algebraic systems of infinite dimension (for example, linear spaces of groups) or in the potentially infinite complication of the internal structure of even finite dimensional algebraic systems (for example, groups with the complex system of generative equations). We have to explain how to make constructive these purely algebraic infinities. 

We call a system of generative relations for the algebraic system such a set ${\cal A}$ of equations of the form $F_1(x_1,x_2,\ldots,x_h)=F_2(y_1,y_2,\ldots,y_d)$, which contains the individual variables for elements of the system, such that any true equation between the elements of this system is logically derivable from this set. If we use the analogy with the constructive analysis, it means that we require the existence of algorithmically determined set of generative equations for any algebraic system we are going to consider. In the other words it would mean that all right equations can be enumerated by some algorithm. 

But this analogy is not really true. The point is that in the case of constructive mathematical analysis we are not given a number itself, but an algorithm, which gives us the approximations of the number. A number itself remains unknown at each finite step in course of work of the algorithm. The analogy with algebraic systems rather requires the algorithmic solvability of the false equations than of the true equations. There are no reasonable deductive rule generating all false equations, and we must state that there is no good analogy with the constructive mathematical analysis as it is. The analogy exists with the analysis restricted to the fixed grain of spatial resolution, which corresponds to the finite algebraic systems. Here the analogy is full because all equations touch only finite set of individual elements. 
  
The more refined version of algebraic constructivism follows from this case of finite algebraic systems. It results from the requirement that the set of all true values must be algorithmically resolvable. If earlier we required that this set is enumerable, than now we require the more: it must be resolvable algorithmically. The solvability is the stronger requirement than the denumerability. For example, the set of codes of Turing machines applicable (finish the work) to its own codes is enumerable (e.g., there is the evident algorithm generating it) but is not solvable due to the non-solvability of the problem of the self-applicability.

The difference between these two versions of constructivism for algebraic systems is important in case of complex algebraic objects (for example, specially defined groups). In quantum theory, which uses mainly groups of the type $U(n),SU(n)$ the corresponding Lie algebras and their matriz representations these exotic objects are rare exceptions and can appear in the specially constructed situations only. I then will not go into details of constructive algebra that became the large area. All that we need to know for the physical constructivism is that all algebraic apparatus, following to the mathematical analysis can be transferred to the constructive way. 

\section{Summary of mathematical constructivism} 

Constructive mathematics completely preserves the computational part of classical mathematics, e.g., all tricks concerning the differentiation, integration solution of differential equations, algebraic methods and the questions of stability, optimization, etc. remain valid. The difference of constructive mathematic from classical (ordinary) mathematics is in that in the constructive mathematics we require the standard condition of effectiveness to all procedures that is the more restrictive than the traditional consistency. The effectiveness does not mean the total computability. The computations with oracles determine the constructive form of the work with processes of non-computable nature. Computations with oracle directly points to the form of interactions between a human and algorithms created in the course of development of the constructivism. All this seriously distinguishes constructive mathematics from the traditional. 

Constructive mathematics excellently fits to the fulfillment of the role of formal apparatus of physics including its most fundamental part: quantum theory. Constructivism is able to establish order in the area, which now lies out of the limits of applicability of quantum theory and generates fruitless and exacting discussions between physicists who are forced to hide the defects of mathematical apparatus by the useless philosophy. We have every reason to state that constructivism represents more ripe form of mathematics in comparison with the traditional classical mathematics. Its regular application in the theoretical physics seems to me the most desirable perspective. 

\chapter{Models based on classical physics}

Classical physics forms the basis of the physics at all because it expresses the human perception and graphic representation of the world. We start from the classical physics also for the sake of preservation of the succession in science. This succession is not simply the tribute to traditions. It has the certain form towards quantum mechanics. Classical physics is the integral part of quantum mechanics, which merely cannot exist without this part. The fundamental notion of measurements of physical values, which lies in the basement of quantum theory, presumes the presence of ordinary, classical physics. Estimating the place of classical physics in the algorithmic approach, we must clearly understand that the classical character of physics does not mean the usage of classical algorithms. In particular, by means of classical algorithms we can describe both classical and quantum mechanics.

\section{Particles and elementary interaction}

The key notion in all physics is the notion of a particle, because it allows the application of mathematics to the reality. A classical particle represents an abstract object with such characteristics as the coordinates in the abstract (configuration) space and time, the mass, the charge, the spin, the dynamical parameters like the energy and the impulse, and also the possible indications to its components and how these components change in time. We see that the definition of particle turns out vague, and we can adapt to it almost any object by request. It means that the rights to determine, what a particle is, belong to a user of the model. We accept that the main attribute of a classical particle is the set of its coordinates in space and time, the law of its evolution (the speed) and the type of this particle that determines the rest attributes of it. 

The important classical concept is the representation of the identity of particles that is the agreement that all particles of the same type are identical. We devote some attention to this question in chapter 6, and now mention the following. The total number of real particles in Nature is so large that there is no possibility to address personally to each particle in the model separately. For example, if physicists divide the electromagnetic field to photons, they state that the total number of them is infinite. It appears in the usage of real numbers for physical values even if it is possible to speak about separate components of these values, as in the case of photons. 

{\bf Field representation of a system of particles is the formal encoding of it by numbers expressing the quantity of particles or the sum of some of their attributes}.

For example, instead of a system of particles $S=\{ s_1,s_2,\ldots,s_k\}$ we can consider their density in the different points given by the rule 
$$
\rho (\bar r)=\lim\limits_{dx\ar\infty}\frac{N(\bar r,dx)}{dx^3},
$$
where $N$ denotes the total number of particles in the cube with the side $dx$ and the center in the point $\bar r$. Instead of the enumeration of the set $S$ we can then use the function of the density $\rho(\bar r)$. Of course, it makes sense if only the quantity of all particles is sufficiently large. In addition, we have in any case to restrict the mathematically exact notion of the limit in this definition, because the quantity of particles is always finite. E.g., we must apply the ideology of the constructive mathematical analysis (see the section "constructive mathematical analysis"). At the same time, the application of standard analysis is restricted by the possibility to consider the approximate solution. The usage of the density, e.g., the total number of particles instead of their enumeration promises exponential profit in complexity, because instead of a number $k$ we will keep in memory something of the type $\log \ k$ or some combination of such logarithms in some degree. In any case, the benefit from the field representation will be almost exponential. 

However, this exponential advantage exists only if we fix the dimensionality of the configuration space. In the other word, for this advantage we must agree that the dynamical scenarios develop only in $R^3$! If we suppose for a moment, that the particles forming the field are able to some individual interactions, e.g., to the interaction not with all other particles but only with some selected ones, for example, are able to form pairs with one of the close neighbors, the situation changes radically. The dynamical scenario will then depend not on the density $\rho(\bar r)$, but of the more fine details of mutual positions of particles.  In the limit case, we must fix these positions with high accuracy (for example, to point to a single atom at the crystal surface) for each particle. If, for example, there are $N$ positions for each particle and the total number of particles is $m$, we have $N^m$ all possible configurations. Let us apply to this situation the same method of the passage to the field representation. Then in the best case it requires the memory of the size of the order $\log\ N^m=O(m)$ provided $N=const$ and only the total number of particles grows. Factually, we have to apss to the consideration just of the separate particles, not a field! These remarks concern the classical description of particles. In the quantum mechanical description of particles, the advantage of the passage from the field to the separate particles, will be much bigger, we consider it below. Here we only mention that with the increasing of the complexity of interaction between the particles the dimensionality of problem grows dramatically. 

The class of phenomena allowing the field representation we conditionally call the continuous systems. The condition character of this name is in that we include to this class not only media studied in hydro and aero dynamics but also solid states, electron gas, electromagnetic field and all the systems in which the individual features of the separate particles interaction are negligible. Here the numerical criterion is the possibility to obtain the right models of dynamical scenarios by the field representation of systems.\footnote{There is no any criteria depending on the total number of particles. Individual features of interactions can have radical consequences even for systems with the large number of particles that causes the limited applicability of the laws of hydro dynamical type to such systems. 
For example, the dynamical scenarios of the evolution of stars are rather the result of astronomic observations than of the direct application of field representations, and it is unknown how these scenarios depend on the individual interactions.} In this section we briefly go through systems admitting the field consideration. Its advantage is in the possibility to apply the analytical methods of computations connected with the solution of differential equations. We will not derive these equations (one can find their deduction, for example, in the book \cite{Vl}). We consider only selected questions that have attitude to the further account.

 All the physics is built on that there are only several types of particles whereas the total number of particles is huge. Coordinates of particles is just the attribute marking out a given particle among all the others. \footnote{and, mainly, among those which are in the considered system, but form the so-called reference frame.}. Correspondingly, we can divide all attributes of a particle to the absolute attributes: its type and all its derivations (mass, charge, possible values of spin, etc.), and the relative attributes: coordinates, impulse, and the concrete value of spin. It is valid also for the not point wise particles, only their configuration space has the dimensionality bigger than 3.

The division of attributes of particles to the absolute and the relative is the conditional. If necessary, we can consider some of relative attributes as merely coordinates in the thus extended configuration space. In particular, we can introduce to this space the new coordinate, called spin, which will point the direction of the spin of particle in some special spin subspace of the extended configuration space. We can analogously consider some of the absolute attributes, for example, some types of particles\footnote{Introduction of isotopic spin makes possible to treat a proton and a neutron as particles of the same type but with the different values of isotopic spin.}.

In the quantum mechanical consideration of particles, the reverse is possible: we can distinguish particles depending on their states and even temporarily include this difference in their type, for example, speaking about electrons in ground state, etc.

The total number of types will be strictly limited, for example, we can regard atomic nuclei and electrons, or molecules. Sometimes it is convenient to extend the list of these standard types adding to them some new ones, for example, we can consider as the particles quanta of the oscillatory movements of a crystal (phonons), or photons, or such exotic particles as Cooper pairs of electrons or big systems of many atoms that for some reasons behave as one particle. 

If a particle $P_0$ consists of the set of smaller particles $P_0^1,P_0^2,\ldots,P_0^k$, then the charge or mass of this particle equal the sum of charges or masses of its components. 

Interaction between particles is described by means of field. In standard mathematical formalism a field is a fundamental object and is determined by the function of the form
\begin{equation}
S\ \ar\ L,
\end{equation}
where $S$ is the configuration space of the considered system of particles, $L$ is some vector space. Here $L$ can be tensor product $L=L_1\bigotimes L_2\bigotimes\ldots\bigotimes L_h$ of different vector spaces, whose basic elements has its own sense for all $L_j$, for example, some spaces can consist of primary elements, say ordinary vectors or some real or complex functions, and they are then called contra variant, whereas the other spaces from $L_j$ can be the spaces of linear functionals over the primary spaces, and they then are called covariant. Such linear space is called a tensor space, and the field is then called a tensor field. Such type of spaces is convenient in the theory of gravity. 

For us the key point in this definition is that by a field we mean some rule, which associates to each position in the configuration space some vector from the fixed vector space. In the other words to each spatial position of a particle corresponds some object so that we can add these objects and multiply them to real numbers where the associative an distributive (of addition towards multiplication to numbers) laws are true. The notion of field is thus well fit to the consideration not only a separate trajectory of a particle in the sequential time instants $t$, but also to the consideration of the different possible trajectories of the same particle or of the different particles, and to the (simultaneous from the user's, not administrator's viewpoint) consideration of the states of the same particle in the different time instants. A field thus gives the necessary tools for the construction of quantum mechanics, when we deal with the different trajectories of the same particle, as well as for relativism, where the different time instants are considered uniformly (see the section 4.3).

\section{Differential equations}

Mathematical analysis aroused from the necessity to consider very small (in the limit infinitely small) segments of space and time and to compute big values by means of processes going in these small areas. The technique of the exact solution of differential equations is the single effective way of the construction of dynamical models without computers, and we illustrate our agreements about the models on this traditional material.

We take up the processes in the three-dimension space; hence, we will use differential equations with partial derivatives from the very outset.
Dynamical models typically are considered in terms of equations of the second order, which are reducible to the parabolic or the hyperbolic forms by the appropriate choice of variables (elliptic equations are applicable mainly for the description of stationary states). 
We limit our consideration by these types of equations, namely, we briefly discuss standard numerical methods of their solution. The main example for us will be the equation of heat transfer belonging to the parabolic type (this scheme can be apply to the equations of hyperbolic type as well).

Let we are given Couchy problem for the rectangle $x_0\leq x\leq x_m,\ t_0\leq t\leq t_n,$ of the form 
\begin{equation}
\begin{array}{ll}
\a\frac{\partial^2 u}{\partial x^2}&=\frac{\partial u}{\partial t},\\
u(x_0,t)&=u_a,\\
u(x_n,t)&=u_b,\\
u(x,0)&=u_{x,0},
\end{array}
\label{cau}
\end{equation}
We divide our rectangular area to small rectangles by the points $x_j,y_s$, forming the uniform lattice with steps $\Delta x,\ \Delta y$ on the both coordinates, and denote by $u_{j,k}=u(x_k,t_j),\ j=1,2,\ldots ,m,\ k=1,2,\ldots,n$ the values of the target function in the nodes of the lattice. By a layer we mean the set of nodes of this lattice that have the same value of the time. Ideologically the simplest method of the solution of this Cauchy problem \ref{cau} is the explicit method. It consists of that we find the values $u_{j,k}$ layer by layer, solving the evident equations of the form
\begin{equation}
\frac{u_{j,k+1}-2u_{j,k}+u_{j,k-1}}{(\Delta x)^2}=\frac{1}{\a}\left(\frac{u_{j+1,k}-u_{j,k}}{\Delta t}\right)
\label{dir}
\end{equation}
towards the unknowns $u_{j,k+1}$ sequentially for $k=1,2,\ldots,m$ for all $j=1,2,\ldots,n$. This method merely generalizes Euler method for the solution of the ordinary differential equation. It terms of a model the explicit method means that we consider only the single scenario of the evolution given by the function $u$, namely, the scenario resulted from the approximation of the function by the definition of its first derivative. The condition of stability of the explicit method is well known:
\begin{equation}
\Delta t\leq\frac{\Delta x)^2}{2\a }.
\label{equi}
\end{equation}

It follows from this condition that the building of the real picture of the evolution requires of the order of $O(\d^{-3})$ steps of the simulation of the whole process given a fixed time segment $t_m-t_0$, if $\d $ is the grain of spatial resolution. 

There is the more economical way for the solution of Cauchy problem \ref{cau}: the reverse method. Here instead of solving equations \ref{dir}, we compose the system of equations, linear relatively to $u_{j+1,k}$ starting from the equalities
\begin{equation}
\frac{u_{j+1,k+1}-2u_{j+1,k}+u_{j+1,k-1}}{(\Delta x)^2}=\frac{1}{\a}\left(\frac{u_{j+1,k}-u_{j,k}}{\Delta t}\right),
\label{dir}
\end{equation}
for $j=1,2,\ldots ,m,\ k=1,2,\ldots,n$. This system has the matrix form
\begin{equation}
\left(
\begin{array}{llllllll}
&\b &1 &0 &0 &0 &0 &0\\
&1 \b &1 &0 &0 &0 &0 \\
&0 &1 &\b &1 &0 &0 &0\\
&0 &0 &- &- &- &0 &0 \\
&0 &0 &0 &- &- &- &0\\
&0 &0 &0 &0 &1 &\b &1\\
&0 &0 &0 &0 &0 &1 &\b
\end{array}
\right)
\left(
\begin{array}{ll}
&u_{j+1,1}\\
&u_{j+1,2}\\
&u_{j+1,3}\\
&-\\
&-\\
&u_{j+1,n-1}\\
&u_{j+1,n}
\end{array}
\right)
=
\left(
\begin{array}{ll}
&\Omega u_{j,1}-u_a\\
&\Omega u_{j,2}\\
&\Omega u_{j,3}\\
&-\\
&-\\
&\Omega u_{j,n-1}\\
&\Omega u_{j,n}-u_b
\end{array}
\right)
\label{progonka}
\end{equation}
where 
$$
\b = -2 -\frac{(\Delta x)^2}{\a \Delta t},\ \Omega =\b+2.
$$

Solving the systems \ref{progonka} sequentially for $j=0,1,\ldots,n$, we obtain the values $u_{j,k}$, which dive the solution of Cauchy problem \ref{cau}. 
This method is called the sweep method. For hyperbolic equations, it looks analogously, but the difference scheme for the first derivative on the time must be replaced by the differential scheme for the second derivative. 

The sweep method has the same computational complexity as the explicit method for one step because the solution of the system \ref{progonka} requires of the order of $O(m)$ operations due to the almost diagonal form of the matrix of system. However, the sweep method is stable for such a choice of steps $\Delta x,\ \Delta t$, which guarantees the accuracy of the approximation of derivatives by the difference schemes, e.g., here for the fulfillment of the condition \ref{equi} is not necessary for the stability. It allows the choice of the step on the time relatively big, and the resulting complexity of the final state finding becomes thus radically simpler: to the linear on the time $t_n-t_0$, which permits to speak about the simulation of process in the real time mode. 

Now we look at the sweep method from the viewpoint of the simulation of the process described by Cauchy problem \ref{cau}. The solution of sequential systems of equations \ref{progonka} is the way to create the dynamical scenario for our system. It is naturally that this scenario appears sequentially accordingly with the flow of real physical time. The sweep method differs principally from the explicit method because it uses the implicit scheme for the equation \ref{cau}. It means that at each separate step we substantially use the information about the structure of the result of this step before we do it. In the other words at each time step we run a step forward and see what must result, whereas in the explicit method we merely compute the next step using only the approximation of derivative by the difference scheme. We can thus state that such looking ahead is very effective trick in the creating of the dynamical scenarios for classical systems, which have effective description in terms of systems of differential equations in partial derivatives, e.g., such systems which behavior allows the satisfactory approximation by means of classical physics. 

\section{About the scalability of classical models}

By the scalability of model, we mean the ability of this model to the expansion to the new areas of application in comparison with the initial problem for which it was created. Here the expansion does not presume the substantial change of the algorithm of the building of scenarios in the model, e.g., the scalability means the application of the former algorithm to the wide range of the different systems. By the new areas, we mean at first the spatial areas, we then speak about the spatial scalability of models. The important case of spatial scalability is the increase of decrease of the grain of spatial resolution, which is the passage to the more lengthy systems or the consideration of the system in more details. That is what we mean by the spatial scalability. In addition, we can speak about the scalability of the model on the interactions. It means that a given algorithm of simulation fits to some other interactions beyond those it was initially designed for\footnote{For example, the model of nuclear interactions of Yang Mills resulted from the adaptation of the electro dynamical formalism to the nuclear interactions.}. 

Of course, here we regard the ideal case. In the reality, the expansion of a model to the new areas or including of new interactions always requires some changes in the algorithm. It is important that this change must be local and does not lead to the revision of heuristic, e.g., the idea of the algorithm. There are no exact criteria of the locality of changes, but it does not cause any serious problem. Every transfer of computational methods from one problem to the other assumes the resolution of some obstacles arises in the adaptation of the method to the new problem. It appears, at least in the assignment the new values to the variables in formulas that is the example of local changes. We treat any change, which can be done quickly, by the local change, and this practical criterion reflects the stability of the main constructions of the model. 

It is important that the scalability is not the property of one model. It depends on the problems, which the model is applied to. It means what interactions we must account anyway, and what we can neglect. For example, in chemical reactions nobody accounts the gravitation interaction of the reagents. 

The other example comes from the celestial mechanics. The classical model of gravitation applied to these problems admit the gigh level of scalability for the extension of the considered system to $10^{20}\ cm$, because it satisfactory explains the movement of galaxies, but the processes going in the spatial scale of the order of $10^{30}\ cm$, already have not so good predictions because here the expansion of the Universe as the whole begins to play the role. On the other hand it is impossible to expand the classical gravitation theory (as the relativistic theory, which is applicable to the far cosmic flights and give the corrections to the classical model) to the scale less than $10^{-10} cm$, because here quantum processes begin to play the key role and we have no the unified field theory. 

However, it would be the error to think that classical physics permits to create satisfactory models of all phenomena concerning gravity in the spatial scale of celestial mechanics. It gives the good description of the stationary sputnik orbits and the trajectories of spaceships. But it cannot solve the question about the stability of a system of many bodies. I have already mentioned the open question about the stability of Solar system in the period of a thousand years. This question formally turns on the computational difficulties appearing in the known ways of its numerical solution. However, we have no way to determine what principally determines this stability. Hence, it would be judiciously to agree that the existing methods are not sufficient to resolve this question. 

The scalability of models represents thus the principal question because it is a matter of agreement of the languages of the different sciences on the border of their areas of applicability. Such an agreement is much more difficult than the questions inside their areas of application; the questions belonging the certain sciences typically have the regular solutions. However, more significant in all practical senses are the problems, which lie on the borders of the different sciences. We explain the significance of such boundary questions on the old example, which led to the crisis of classical electrodynamics. 

Classical electrodynamics rests on Maxwell equations for the pair $E,\ B$ of electric and magnetic fields:

\begin{equation}
\begin{array}{lll}
&rot\ B &=\frac{1}{c}(E'_t+4\pi j)\\
&rot\ E &=-\frac{1}{c}B'_t\\
&grad\ E &=4\pi\rho\\
&grad\ B &=0.
\end{array}
\label{maxwell}
\end{equation}

These equations exactly describe all electro dynamical phenomena in the macroscopic size scale that includes Coulomb, Ampere and Faraday laws.\footnote{Two lasts follow from the first two Maxwell equations, the deduction of Coulomb law given below by the book \cite{Fe}.} However, in the problem of the flight of electron in Coulomb field of an atomic nucleus these equations with the natural suppositions result in the continuous emission of energy that makes impossible the stability of atoms observed in reality. This fact establishes the limits of the scalability of classical models. We cannot expect the building of the model of chemical reactions on the basis of classical physics, because it is not able even to give the right description of the simplest chemical objects.\footnote{One could suppose that to find the electron orbit in atom we should use electrodynamics with the relativity. Nobody has passed this way to the end because it is unimaginably hard. The modern physics at all acquired the existing form only because its creators always choose the more efficient algorithms for the solution of any problem. Algorithmic criterion lied in the foundation of physics from the very beginning, though nobody mentioned it explicitly, this becomes actually only nowadays.}

\chapter{Quantum processes}

This chapter is not the introduction to quantum physics. Here the subject is only its formalism from the qubit viewpoint. I do not suppose that a reader is familiar with quantum physics, though it would be good to take a look at its basis by one of canonic books, for example, by \cite{LL}. In any case we will clarify the qubit formalism in order to give a reader the possibility to rewrite any part of quantum theory on the language of qubits. Traditional notations are commonly accepted in the physical literature, where, for example, a wave function is written as $\Psi (x)$, which leads to the collision with its value in a concrete point $x$, and therefore for the last value the notation $\int \Psi(y)\delta_x(y)\ dy$ is used, etc. These traditional notations are convenient for hand computations where the resolving of such collisions makes no serious problem for a human. Our understanding of the computer simulation requires the higher level of formalization of main notions. Moreover, the formalism must be able to deal with finite objects even if the used formulas allow the substitution of the infinite values. It especially touches quantum electrodynamics, fro which we propose here the system of formal notations of the qubit type. 

\section{Main concepts of one particle quantum mechanics}

The main axiom of quantum mechanics is that all the dynamics of any system is determined by its wave function, which is the complex function of the coordinates of all particles forming this system:
$$
\Psi (t,r_1,r_2,\ldots,r_n).
$$
Here $r_j$ are the coordinates of particle $j$ (they include not only spatial but also spin coordinates). We should treat this wave function as the vector in Hilbert space of states of the $n$ particle system. The values of this function are called amplitudes corresponding to the presence of the system in the time instant $t$ in such a state that for any $j=1,2,\ldots,n$ the $j$-th particle has the coordinates $r_j$. This treatment of a state as the vector immediately leads to the non-trivial consequence: any linear combination of states again represents some physically possible state of this system. The space of states thus possesses the linear property, which means that every equation, to which the vector $\Psi$ obeys must be linear. This principle is called the superposition principle, and it results in the existence of the special process: the amplitude interference that has no direct analogue in the classical physics (but the wave physics where the interference appears as the collective effect to which we return below). 

It is easy to demonstrate the interference of amplitudes using matrices. Let we are given a basis in Hilbert space of states of the system, and represent any vector $\Psi$ in the form of column of its coordinates in this basis. The principle of superposition then gives that the state in the next time instant $t+\delta t$ can be found applying to the state in the instant $t$ some linear operator $U$, called the operator of unitary evolution (we will see that it must be not only linear but also unitary). This fact has the following expression on the language of matrix product as follows:
\begin{equation}
\left(
\begin{array}{lllll}
&u_{1,1}&u_{1,2}&\ldots &u_{1,n}\\
&u_{2,1}&u_{2,2}&\ldots &u_{2,n}\\
&\ldots&\ldots&\ldots &\ldots\\
&u_{n,1}&u_{n,2}&\ldots &u_{n,n}
\end{array}
\right)
\left(
\begin{array}{ll}
&\psi_1(t)\\
&\psi_2(t)\\
&\ldots\\
&\psi_n(t)
\end{array}
\right)=
\left(
\begin{array}{ll}
&\psi_1(t+\delta t)\\
&\psi_2(t+\delta t)\\
&\ldots\\
&\psi_n(t+\delta t)
\end{array}
\right) .
\end{equation}
It means that any amplitude $\psi_j(t+\delta t)$ can be found by the formula 
\begin{equation}
\psi_j(t+\delta t)=\sum\limits_{i=1}^n\psi_i(t)u_{j,i}.
\label{sup}
\end{equation}
The formula \ref{sup} means that for the obtaining of the amplitude in some point in the next time instant we must sum the amplitudes in all other points in the previous time instant, preliminarily multiplying them to the corresponding amplitudes of the passage from these points to the initial point. The movement of quantum particle is then represented as the flow of some media where the amplitude in each point is the sum of deposits, which the movements from all the other points bring to this point. Here each deposit is taken with the complex weight corresponding to the corresponding passage from one point to the other. This representation of a quantum particle as a media arouses the analogy of quantum physics with hydrodynamics that we will consider further.

We now take up two sequential passages corresponding to the same evolution operator $U$: from the moment $t$ to the moment $t+2\delta t$. We then have: $\Psi(t+2\delta t) = U^2\Psi (t).$ Writing out in more details, we obtain $\psi_j(t+2\delta t)=\sum_{i,k}\psi_i(t)u_{i,k}u_{k,j}$. It means that the quantum particle can move, generally speaking along an arbitrary trajectory, not only along the line, and its amplitude in each point is the result of summing of all amplitudes along all ways leading from each point to this point. Here the deposit to the sum of each path is obtained by the multiplication of the elements of evolution matrix $U$, which correspond to the sequential parts of this path (we represent a path in the form of polyline which segments are the parts of the path). The amplitude is thus computed as the sum over all paths, and along each path as the multiplication of amplitudes of sequential parts. 

We draw attention to one small detail of this scheme which often escapes notion, but which absence makes impossible to apply this scheme practically. On the distances of the order of the spatial grain, any path must be the segment of a straight line. For the time segment of the order of the time grain, we must take into account only passages from the nearest points to the point where we find the amplitude. We will see that these suppositions are necessary in the proof of equivalence between Feynman path integrals and Shredinger equation. It means that the flow of imaginary media has some common features with the flow of real liquid. In what follows, we will take up this analogy more exactly. 

The rule about summing over all paths, formulated above, is true everywhere including quantum electrodynamics where the diagrams describe processes. It exactly corresponds to the formula of the full probability of a complex event in the probability theory with only difference that in the probability theory the values are positive and real, whereas in quantum mechanics they are complex. This analogy suggests the possible refusal from the complex numbers in the description of quantum mechanics, and we will realize this possibility below through the method of collective behavior. 

Quantum mechanics instead of dynamical values deals with the corresponding operators. The value of coordinate $x$ corresponds to the operator of multiplication to this coordinate: $x:\ f(x)\ar xf(x)$, the vector $\bar r=(x,y,z)$ corresponds to the operator $\bar r:\ f(x,y,z)\ar (xf(x,y,z),yf(x,y,z),zf(x,y,z))$, the impulse $p_x$ along the coordinate axe $x$ corresponds to the impulse operator $p_x=\frac{h}{i}\frac{\partial}{\partial x}$, the full impulse $\bar p$ corresponds to the operator $\bar p=\frac{h}{i}(\frac{\partial}{\partial x},\frac{\partial}{\partial y},\frac{\partial}{\partial z})$, the energy corresponds to the operator of energy $\frac{p^2}{2m}+V(x)$, where $V$ is the potential energy of particle. The operator of energy is called Hamiltonian. Here we assume the ordinary rules for the passage to the vector values, for example, the operator of squared module of coordinate acts as  $|r|^2:\ f(x,y,z)\ar (x^2+y^2+z^2)f(x,y,z)$, the operator of squared impulse acts as $p^2:\ f\ar -h^2\Delta f$ (e.g., as the square we treat as the scalar square), the moment of impulse $\bar r\times\bar p$ corresponds to the operator of moment, which coordinates are obtained by the rule of vector product from the coordinates of its multipliers-operators, etc.

Fourier transform from the wave function is called the impulse representation of the wave function:
\begin{equation}
\Phi(p)=
\int\limits_Re^{-\frac{ipx}{h}}\Psi(x)dx,
\end{equation}
where the reverse operator looks as
$$
\Phi(x)=
\frac{1}{2\pi h}\int\limits_Re^{\frac{ipx}{h}}\Phi(p)dp.
$$
The complete transform to the impulse representation and vice versa in the three-dimension space has the form
$$
\begin{array}{lll}
&\Phi (p)&=\int\limits_{R^3}e^{-\frac{ip\cdot R}{h}}\Psi (R)d^3R,\\
&\Psi (R)&=\frac{1}{(2\pi h)^3}\int\limits_{R^3}e^{\frac{ip\cdot R}{h}}\Phi (p)d^3p,
\end{array}
$$

 Here if the wave function depends on 3 variables we can pass to its impulse representation on each of coordinates independently of the others, for example, we can consider the function of the form $\Phi(x,p_y,z)$, or $\Phi(p_x,y,p_z)$, etc. We stress that the wave function remains unchanged in the passage to its impulse representation; it remains the same vector in Hilbert space of states. The impulse representation is merely the notation of this vector in the different basis, namely, in the basis consisting not of Dirac's delta functions, as for the coordinate representation, but of the functions of the form of plane waves $exp (ipR)$. We could take some other basis, for example, the basis corresponding to the eigenvectors of Hermitian operator of the sum $R+p$ coordinate plus impulse, and arrange the corresponding representation of the wave function if necessary. All manipulations connected with the passage to impulse representation thus follow from the simple change of basis. The usage of these passages is the integral part of the standard formalism 

The main rule of quantum mechanics is Born rule, which claims that the squared module of the wave function is the density of the spatial position of the particle:
\begin{equation}
p(x)=|\Psi(x)|^2
\end{equation}
This rule is invariant of a basis of the space of states in the sense that the density of detecting the particle with the value $p$ of impulse is the squared module of impulse representation of its wave function. 

At last, the wave function dynamics is determined by Shredinger equation that has the form
\begin{equation}
ih\frac{\partial\Psi}{\partial t}=H\Psi
\end{equation}
where $H$ is the operator fo energy of particle (or of the system of particles). In the simplest case when one particle is in the potential field, the energy operator is written above. In the more complex cases (many particles, existence of the vector potential of the electromagnetic field), we obtain the energy operator from the classical energy by the replacement of all physical values by the corresponding quantum operators. In particular, it follows from Shredinger equation that in case of stable in time potential the general solution of it looks as 
\begin{equation}
\Psi (x,t)=exp\left( -\frac{i}{h}Ht\right)\Psi (x,0)
\end{equation}
The exponential of the operator is determined as the corresponding row composed from operators. This equality remains true for a variable potential as well, only the exponential then will be the so-called chronological exponential (see \cite{BS}).

Factually, we have described all standard formalism of quantum theory. This main conceptions involve the other thesis’s, (for example concerning measurements and the possibilities of choice of a basis), which we consider in the section devoted to qubit formalism. 

\subsection{Qubit formalism}

We now take up the important moment in the passage from the wave function to a finite vector. This is the introduction of the qubit formalism of wave function. Here the qubits themselves play the decorative role but the procedure of discretization of the wave function is important. The discretization or sampling of wave functions has the deep sense because it concerns the existence of the grain in the configuration space. If we assume that the configuration space is not infinitely divisible, but there is some minimal nonzero length $d>0$, then we obtain that instead of continuous wave function we must consider its discrete approximation, which then turns not approximate, but the exact evaluation. In the other words in case of grain, just continuous representation of the wave function becomes approximation. The existence of such a discretization of space indirectly follows from the divergence of rows in quantum electrodynamics (see below), and from the properties of wave functions even of one particle when the size of grain decreases (we consider it in the section devoted to Feynman path integrals). The main aim of the discrete representation is the constructive modification of quantum theory. The necessity to deal with just approximations of the coordinates of particles instead of their real "exact" values leads us to qubit formalism.

Let we are given one particle, which coordinates takes values from some configuration space $R$ (for one-dimensional particle it is real numbers, but in our case the structure of $R$ is not important for us). We divide $R$ to the finite number of segments $D_1,D_2,\ldots,D_m$ and consider the approximation of the function $\Psi$ by the stepped function $|\Psi\rangle$, which takes on a segment $D_j$ some value $\la_j\in C$. Let $|j\rangle$ denote the characteristic function of the segment $D_j$. We then can write the formal equation
\begin{equation}
|\Psi\rangle=\sum_{j=1}^m\la_j|j\rangle
\label{expe}
\end{equation}
We consider the linear space spanned by the functions $|\j\rangle$. Introducing the dot product of functions by the standard rule $(g,f)=\int\bar f(x)g(x)dx$, we obtain that $|j\rangle$ form the orthogonal basis of this space. Normalizing them (we can do it fixing $\la_j$ and choosing the appropriate $D_j$), we make this basis orthogonal. The equation \ref{expe} will then be the decomposition of the vector $|\Psi\rangle$ on the orthonormal basis consisting of vectors $|j\rangle$. The representation of the wave function in the form \ref{expe} is correct, in contrast to the ambiguous expression $\Psi(x)$, because in the last expression the variable $x$ is free, and thus is unclear what the expression $\Psi(x)$ means: the function $\Psi$ or its value in the certain  point $x$. Such a fine point has no value in the classical mathematics, because for the analytical computations a free variable can be always tied with quantifier. In the constructive mathematics, it is important. In the building of algorithm we must clearly differ a function, which includes all its values as in \ref{expe}, from its concrete value in a fixed point.  

We must now establish the correspondence between quantum formalism and linear algebra. Let $|a\rangle$ denote a representation of vector $a$ in the Hilbert space of finite dimensionality in the form of its coordinates in the fixed basis. The action of a linear operator $A$ on this vector will be then the matrix multiplication $A|a\rangle$, where $A$ denotes the matrix of $A$ in this basis. We also agree to treat $\langle a|$ as the vector - row, obtained from $|a\rangle$ by the transposition and the complex conjugation of elements. This operation for matrices: the transposition and the complex conjugation is called simply conjugation. We also will merge any two vertical lines standing beside. The scalar product of the vectorsв $a$ and $b$ is written as $\langle a|b\rangle$. We can treat the notation $\langle a|A|b\rangle$ in two ways: either as $\langle a| (A|b\rangle )$, or as $(\langle a|A^*)|b\rangle$. By if the matrix $A$ is self-conjugated, e.g., if $A=A^*$, then the ambiguity disappears and we can use the written expression without brackets. The self-conjugated matrices are called Hermitians. The matrices of the form $\exp (iH)$, where $H$ is Hermitian are called unitary. By the reduction to the diagonal form it is easy to prove that a matrix $U$ is unitary if and only if $U^{-1}=U^*$, or, in the other words, when this matrix preserves all distances in Hilbert space. 

We can define the measurement of a state $|\Psi\rangle$ in the orthonormal basis $|\phi_1\rangle,|\phi_2\rangle,$ $\ldots,|\phi_N\rangle$ as the random variable taking any value $|\phi_j\rangle$ with the probability $|\langle\phi_j|\Psi\rangle |^2$. This is the reformulation of Born rule. The measurement thus determines the random process called the wave function collapse. In this process the passage takes place from a state $|\Psi\rangle$ to one of states $|\phi_j\rangle$, where for each $j=1,2,\ldots,N$ we now only the probability of the passage to this state and nothing more. This is the sense of Copenhagen quantum theory. It is complete in the sense that it is impossible to change it somehow, for example, to introduce some parameters besides the wave function (the so-called hidden variables, which participate in the quantum evolution). The attempts to introduce some addition constructions to the apparatus of standard quantum theory immediately lead to the refusal from its mathematical apparatus at all, and without alternative apparatus these attempts turn into nothing. 

Copenhagen quantum mechanics has the all flexibility that is needed from the complete physical theory of one or two particles. For example, any physical magnitude is associated with some Hermitian operator $A$, such that classical values of this magnitude are the eigen values of this operator. The measurement of this magnitude is the measurement of a state of this system in the basis of eigenvectors of $A$. If two operators have the same set of eigenvectors (it means that they commute) then these magnitudes are simultaneously measured. The example is the energy of particle and the projection of the operator of momentum to one of the coordinate axes (typically $z$). If the operators have no common system of eigenvectors, then the corresponding magnitudes cannot be measured simultaneously. For example, the coordinate and the impulse along the same axe cannot be measured simultaneously. Really, if we choose the segments $D_j$ as the sequential segments with the same length then within the constant coefficient the matrix of the coordinate $x$ operator has the form  
\begin{equation}
\left(
\begin{array}{lllll}
&1&0&\ldots &0\\
&0&2&\ldots &0\\
&0&0&3&\ldots\\
&\ldots&\ldots &\ldots &\ldots
\end{array}
\right)
\end{equation}
whereas the matrix of the impulse operator along the same axe has the form 
\begin{equation}
DFT\left(
\begin{array}{lllll}
&-h^2\ 1^2/2m&0&\ldots &0\\
&0&-h^2\ 2^2/2m&\ldots &0\\
&0&0&-h^2\ 3^2/2m&\ldots\\
&\ldots&\ldots &\ldots &\ldots
\end{array}
\right)
DFT^{-1}
\end{equation}
where $DFT$ denotes the discrete Fourier transform (see Appendix, $QFT$). Really, the first statement follows from the definitions straightforwardly, to prove the second statement we mention that $DFT$ transforms the differentiation to the multiplication on the imaginary unit and the argument of Fourier transform. We use this property here to make the matrix corresponding to the impulse operator diagonal. This matrix will have the diagonal form in the basis obtained by the Fourier transform from the initial basis; hence, there are no common eigenvectors of the impulse and coordinate operators. Of course, if we consider, for example, the coordinate $x$ and the impulse along the other axe, say, $p_y$, there will be the common system of eigenvectors for such two operators, which means that we can measure them simultaneously. 

The simultaneous measurement we treat here as the measurement of the both magnitudes with the same absolute accuracy. It means the possibility to know simultaneously the values of these two magnitudes exactly. If we refuse from the requirement of absolute accuracy, then, of course, it is possible to measure any two magnitudes. Here, if one of them acquires the certain value (e.g., our vector of state coincides to the eigenvector of the operator, corresponding to this magnitude) then the other magnitude, generally speaking, will have some probability distribution accordingly to Born rule. We consider for example, two dimension Hilbert space in which we measure the "coordinate" and the "impulse". I take the names of physical magnitudes in would-be sense because there are not the real coordinate and the real impulse, but their most rough approximation. We obtain this zero-step approximation if we suppose that the particle can occupy only two spatial positions, not a continual set as in the standard physical courses. Since the measurement of the "coordinate" results in the hit to one of two states $|1\rangle , |2\rangle$, and Fourier transform in the two dimensional case has Hadamard matrix
$$
\left(
\begin{array}{lll}
&1/\sqrt{2}&1/\sqrt{2}\\
&1/\sqrt{2}&-1/\sqrt{2}
\end{array}
\right)
$$
then the probability to obtain each of the possible values of the "impulse" $1$ or $2$, will be $1/2$. We mention that the so-called physical intuition could hardly give us the understanding of what "impulse of particle" means provided this particle can occupy only two spatial positions $|0\rangle$ or $|1\rangle$. Nevertheless, we can ascribe the certain understanding to the notion of impulse. We then will work with this notion further, though this understanding is not reducible to one number. We made it through the qubit formalism, e.g., by the special mathematical trick. \footnote{I suggest to a reader to consider the possibility to apply here the heuristic of standard mathematical analysis (e.g., the physical intuition), where the speed means the limit $\Delta s/\Delta t$, etc. We will do this further.} We see that the qubit formalism is more reliable than the standard formalism of wave functions. It possesses the bigger expressing abilities, because it permits to treat operators as the concrete Hermitian matrices, and thus to use algebraic methods explicitly. The qubit technique is more particular than the traditional analytic technique because it explicitly contains the grain of spatial resolution. Each defect of the formalism, which we can easily hide by the analytical technique becomes obvious when we use the qubit way. We will see this for quantum electrodynamics. Moreover, the supposition about the existence of the grain of space is much closer to the reality than the analytical axiom about the unlimited divisibility of the space. In view of it I will try to use the qubit form of notations for quantum object without the special mentioning. This situation is also instructive and typical: the shortest way to the numerical result is the simple manipulation with the mathematical object; all called the physical intuition certainly is reducible to such a manipulation. The independent existence of some physical intuition not shaped by this manipulation is the right sign of the backwardness of the used mathematics. 

We consider the capability of the qubit method on the simple example of the uncertainty relation of Bohr - Heisenberg, which has analytical and algebraical forms: 
\begin{equation}
\begin{array}{lll}
&\delta x\delta p&=2h,\\
&[x,p]&=h.
\end{array}
\end{equation}
where the commutator of two operators is $[a,b]=ab-ba$.
The analytical form we can deduce, for example, by means of path integrals (see below). The analytical uncertainty relation is derivable by the direct differentiation, using the analytical representation of the operators as $p=\frac{h}{i}grad$ and $x$ as the multiplication on $x$. We apply the qubit formalism and represent the matrix of the operator $p$ as $-ih\ F\ x\ F^{-1}$, where by $F$ we denote the discrete Fourier transform, and by $x$ and $p$, as usual, the matrices of the corresponding operators. We thus obtain the following expressions for the coordinate and the impulse: 
\begin{equation}
[x,p]=-ihxFxF^{-1}+ihFxF^{-1}.
\label{u}
\end{equation}
 The equation $F^{-1}xF=-p$ follows immediately from the definition of Fourier transform. Multiplying \ref{u} from the left to $F^{-1}$ and from the right to $F$, we obtain $F^{-1}[x,p]F=[x,p]$, and then $[x,p]=cI$, for some constant $c$, which must be proportional to $h$.  

Copenhagen quantum mechanics thus possesses the complete arsenal for the consideration of one or two particles because the case of two particles is reducible to the case of one particle, by the choice of its center of masses as the reference point of the new coordinate system. However, Copenhagen quantum mechanics is not applicable to the more complex problems. The difficulties begin already for the formally one particle problems but with the measurements, see example 2 from the section 2.4.2. Physically, the measurement means the interaction of the considered particle with its environment, which consists of the great number of the other particles, and thus has the classical properties. Factually, the problem with measurements is not one particle problem. It becomes apparent in the form of the so-called decoherence, e.g., the decay of the quantum state when the system contacts with the environmental particle, which do not contain in our ensemble. 

\section{Feynman path integral}

The remarkable principle of quantum physics is the conformity principle, accordingly to which every magnitude of classical physics corresponds some Hermitian operator acting in the space of states of the particle, and this correspondence is such that for big actions quantum dynamics completely transforms to classical dynamics. Quantum physics does not exist without classical physics in the strong sense of this phrase. There is no "pure" quantum theory, it necessarily must contain the classical physics, at least because the measurement procedure is impossible without the measuring device obeying the laws of classical physics. 

This impossibility to separate quantum theory from classical has the other sense belonging to the area of heuristic. The heuristic in the standard quantum theory (as well as in the usual physics) is the roughly formalized system of notions and agreements that determines how to apply the laws and formulas to obtain the valuable results \footnote{It is appropriate to use the analogy with the juridical system, which besides the laws contains procedures of its application. This situation does not depend on the localization. For example, it can be the sub legislative acts or precedents, or something else. It is important that this "addition" to laws is unavoidable, because without it the laws will not work.}. Classical heuristic stands behind all advantages of quantum theory including electrodynamics. This heuristic rests on the notion of point wise particle.

The simplest systems either consist of one particle, or can be reduced to one particle in sense that the approximation of their dynamics by the simple combination of one-particle systems is satisfactory. The example is the system of one particle in the potential. A real potential\footnote{We usually speak about the electromagnetic potential. Nevertheless, all formulated for this potential is true for all other potentials, including the nuclear and gravitation. For the last one the question is only how to divide the observed field to quanta.} is the sum of deposits of the quanta of this potential: photons. The representation of a particle in the potential is the approximation, which rests on that a) there is lot of photons, and b) the source remains unchanged after the emission of one photon. The second example gives us two interacting particles with the coordinates $r_1$ and $r_2$. If we ignore amplitude quanta we can introduce new coordinates $R=(r_1+r_2)/2,\ r=r_1-r_2$ and our system is reduced to the system of two independent particles, e.g., to the simple combination of one particle systems. This trick is not applicable to the case of three particles because the case of three particles is a kind of model system for the checking of the hypothesis in quantum informatics. 

In this paragraph, we consider systems reducible to one particle. For these systems in quantum case the classical heuristic is valid which permits to reformulate all results of quantum physics on the language close to classical (further we call this language quasi classical). It permits to use the phrases of the type "trajectory", "movement of particle along the trajectory", "deposits from different trajectories", etc., despite of that there is no any trajectories or deposits in the formalism of quantum theory. This powerful heuristic is the source of success of quantum theory up to nowadays. In the planned constructive reformulation of quantum theory we have to use this heuristic because we have no other way. The first step in this direction belongs to Feynman who formulated quantum mechanics on the language of path integrals. 

The idea of this language is simple. We consider the movement of quantum particle from the point 1 to the point 2. It can fulfill this passage by the different ways, e.g., along the different paths. Let we be given an algorithm, which permits for a given path $\g$ to determine the number $A(\g )$, and for the different such numbers $A_1,A_2,\ldots, A_h$ to determine their result $K(A_1,A_2,\ldots,A_h)$, which is the number. Then the more the module $|K|$ is, the more probable will be the passage to the point 2 of the particle, which initially was in the point 1. This idea does not touch numbers, but only estimations. Nevertheless, it points to some algorithmic scheme permitting to give the rough answer to the question: where will be the particle provided it is initially in the point 1 ? This scheme implies the principal conclusion that we must consider a quantum particle not as one point travelling along a path, but as a set of points, each of which equally represents the initial point an moves along its own trajectory. The equality of these points makes possible to name them the samples of the initial particle. 

We could liken a quantum particle to a whole galaxy, which stars are its samples. The carrying if this analogy requires some efforts in th edescription of interactions. For the method of collective behavior this analogy will become evident from the chapter 5. 

This situation influence to the mixed consideration of the many body systems by the Born Oppenghaimer method. This method considers the heavy particles as atomic nuclei as classical objects whereas electrons as quantum objects. 

In view of the above mentioned I will call thus ensemble of samples of the particles the swarm to stress its unusual character, when each its sample represents the whole swarm. 
It brings the question: what forces the samples to keep together, or what would happen if some sample flies too far from the main part of the swarm ? 
The answer to this question requires the recognition of samples as the non-erasable, and the explicit mechanism of the return to the swarm the too far samples, that we take up in the chapter 5.

It would be logical to consider each sample as the non-erasable, e.g., to ascribe to each of samples its own history. We will return to this further. Now we agree to treat the samples as the auxiliary tool for the description of the wave function $\Psi$, and to redefine them in the short time frame $\delta t$, on basis of the wave function $\Psi(t)$. We call such a swarm a wave swarm as a sign of its sample have only the history restricted by the duration of the time $\delta t$. Our scheme then looks as the iteration of three main steps:  
\begin{itemize}
\item computation of the wave function on basis of the states of all samples of the wave swarm, 
\item new definition of samples of the wave swarm,
\item free flight of samples and the change of their amplitudes.
\end{itemize}

What exactly is the probability of the passage $1\ar 2$ ? This question concerns the interaction of the samples with each other. We know the answer: the probability equals the squared module of he value $K$, and further, we establish why it must be so. We are not restricted with the necessity to consider the history of each sample separately and to treat them as non-erasable, we agree that each sample carries some special complex number associated with it and called its amplitude. To specify the details of this scheme we must define what means the expressions $K(A_1,A_2,\ldots,A_h)$ and $A(\g )$. The first expression we define merely as the sum
$$
K(A_1,A_2,\ldots,A_h)=\sum\limits_{j=1}^hA_j
$$
The second we define as 
$$
A(\g )=\frac{1}{A}\exp (\frac{i}{h}S[\g ]) 
$$
where $A$ is some constant, $S[\g ]$ is the action along the path $\g$, defined as  
$$
S[\g ]= \int\limits_{t_1}^{t_2}L(x_t,x,t)dt
$$
where the Lagranjian of the system is $L=E_{kin}-E_{pot}$ the difference between the kinetic and potential energies.

The function $K$ is thus complex valued and we can express it by the following formula:

\begin{equation}
K(2,1) = \frac{1}{A}\int_{T(2,1)} \exp (\frac{i}{h}S[\g ]) D\g
\label{K}
\end{equation}
where the integration on $D\g$ means the summing on the set of all trajectories $T(2,1)$, leading from the point 1 to the point 2. We can then express the wave function in the moment $t_2$ through the wave function in the moment $t_1$ by the formula 
\begin{equation}
\Psi(x_2,t_2)=\int_{x_1}K(x_2,t_2;x_1,t_1)\Psi(x_1,t_1)dx_1
\label{ps}
\end{equation}
for any $t_1, t_2, x_2$. Here the point 1 has the coordinates $x_1,t_1$, the point 2 has the coordinates $x_2,t_2$. The formula \ref{ps} expresses the fact that the wave function is the magnitude which square is the probability to find the particle in this point, and this value equals $K$ provided the particle initially was in the point 1 (in this case we simply obtain on the right the action of delta-function to the wave function that immediately gives the desired). The formula \ref{K} formalizes the computation of the value $K$, given above.

Resuming all the preceding we can define all three points of the evolutionary step of the wave swarm. The computation of the wave function in the given point is the summing of the amplitudes of all its samples occurred in some cube with the centers in this point. The redefinition of the swarm means that we split the value of the wave function in this cube to many equal parts, e.g., represent it as $n\a$, where the natural $n$ is large, and create $n$ new samples, ascribing to each of them the random speed from the uniform distribution in some big cube. At last, the free flight of samples goes by Galilean law, when the amplitude of each sample is multiplied to $e^{-\frac{i\Delta S}{h}}$, where $\Delta S$ is the action of this sample along the straight path provided the time $\delta t$ of its flight is fixed. To simplify the computations we assume that $\a$ is the same for all points, and $n=n(x)$ is proportional to $|\Psi (x)|$ in this point $x$. 

We make certain in that the formalism of path integrals is equivalent to the standard quantum mechanics. For this, following \cite{FH}, we find the value of wave function defined accordingly to \ref{ps}, in the next time instant. We need to represent somehow the kernel $K$. We suppose that for the small values of period $\e$ exists only one path of the form $1\ar 2$, such that the integration on \ref{K} is reducible to one summand. This trick needs the substantiations from the viewpoint of standard mathematics, whereas it is legal in the constructive mathematics, because the entire path become of straight lines provided we limit the grain of spatial resolution by some lower bound, and we obtain what we wanted. Analogously it is possible to regard that the action is limited by the multiplication of $\e$ to Lagrangian taken in some intermediate point. We then have:
\begin{equation}
\Psi(x,t+\e )=\frac{1}{A}\int\limits_R\exp(\frac{i\e}{h}L(\frac{x-y}{\e},\frac{x+y}{2}))\Psi(y,t)dy
\end{equation}
Substituting here $L=mx_t^2/2-V(x,t)$ and $y=x+\delta$, we obtain 
\begin{equation}
\Psi(x,t+\e )=\int\limits_R\frac{1}{A}\exp(\frac{im\delta^2}{2h\e})\exp(-\frac{i\e }{h}V(x+\delta /2,t))\Psi(x+\delta ,t)d\delta
\end{equation}
We see that the maximal deposit comes from $\delta$ of the order $\sqrt{\e h/m}$. E.g., if we decompose $\Psi$ on the degrees of $\e$, preserving the summands of the first order only, we must keep the summand of the second order on $\delta$. With this accuracy we find
\begin{equation}
\Psi(x,t)+\e\frac{\partial\Psi}{\partial t}=\int\limits_R\frac{1}{A}e^{im\delta^2/2h\e}[1-\frac{i\e}{h}V(x,t)][\Psi(x,t)+\delta\frac{\partial\Psi}{\partial x}+\frac{\delta^2}{2}\frac{\partial^2\Psi}{\partial x^2}]d\delta ,
\end{equation}
then in our approximation in view of the known equality 
\begin{equation}
\int\limits_R e^{im\delta^2/2h\e}d\delta=\left(\frac{2\pi ih\e}{m}\right)^{1/2}
\end{equation}
we have 
\begin{equation}
A=\left(\frac{2\pi ih\e}{m}\right)^{1/2}.
\end{equation}

Now applying the known integral
\begin{equation}
\int\limits_R\frac{1}{A} e^{im\delta^2/2h\e}\delta^2=\frac{ih\e}{m}
\end{equation}
we get 
\begin{equation}
\Psi+\e\frac{\partial\Psi}{\partial t}=\psi-\frac{i\e}{h}V\Psi -\frac{h\e}{2im}\frac{\partial^2 \Psi}{\partial x^2}
\end{equation}
which immediately gives Shedinger equation. 

The path integral method is thus the version of standard quantum mechanics. Its practical application presupposes masterly computations with integrals and algebra, but it also gives some new view to the standard quantum theory comparatively to Shedinger equation. For example, we can compute the kernel $K$ for a particle in the different potential fields (see \cite{FH}), for a free particle it will be 
\begin{equation}
K(x,t,0,0)=\left(\frac{2\pi iht}{m}\right)^{-1/2}e^{imx^2/2ht}
\label{ker_free}
\end{equation}
that gives us the character of its samples movement in the point with coordinate $x$: they fly with the speed $v=x/t$, that exactly corresponds to the classical picture of the dispersion of samples in the "explosion" of a particles which was initially in the reference point of the coordinate system. 

With the help of path integral method we can obtain the uncertainty relation "coordinate-impulse", if we analyze the passage of a particle through the narrow slit of the width $2b$. Some bulky computations of the kernel of the passed particle (see \cite{FH}, pages. 63-64) show that after the passage through the slit the support of wave function obtains the widening $ht/mb$, which means the addition of some uncertainty in the impulse equal to $h/b$, from which we obtain the uncertainty relation of Bohr-Heisenberg:
\begin{equation}
\delta p\delta x=2h.
\end{equation}

In the formalism of path integrals the notion of sample plays auxiliary role because a sample of real quantum particle preserves its history during time segment $\delta t$ only. The wave function plays the main role here which factually determines how many and what samples must be created in each small cube of space. Nevertheless just free movement of the samples is the step of evolution, which creates the unitary dynamics. The secondary character of the notion of sample in the wave swarm is evident from that the density of this swarm is proportional to the module of wave function, not to squared module. The density of samples in the wave swarm is not the probability density to find the particle in this point. Just this results in the brief history of each sample that means the absence of the dynamical characteristic of samples. The mass of sample (in this approach it is the mass of the whole particle) occurs in the evolution of its amplitude only through the action. In spite of the ephemerality of samples of the wave swarm, its application gives much. In particular, we can formulate in terms of path integrals the criterion for the determining which type of mechanics is applicable to a given particle: quantum or classical. 

 Really, we consider the passage from the point 1 to the point 2 along two paths:
 $\g_{cl}$ the trajectory , representing the solution of the equation of classical mechanics for this particle, and $\g_{noncl}$ some other trajectory. Without loss in generality we can assume that samples preserves their history during the travelling from the point 1 to the point 2 , which means that these two points are sufficiently close. We compare two deposits to the wave function: 
\begin{itemize}
\item  the deposit from amplitudes of the samples moving along the paths close to  $\g_{cl}$, and
\item the deposit from amplitudes of samples moving along the paths close to $\g_{noncl}$. 
\end{itemize}
We denote them by $K_{cl}$ and $K_{noncl}$ correspondingly. Let $S_{cl}$ be the action of the particle along the classical trajectory. We suppose that 
\begin{equation}
|S_{cl}|\gg h.
\label{crit}
\end{equation} 
We can assume that the change of the action in the order of magnitude equals the action itself if the trajectory has the sufficiently general form. Since the classical trajectory $\g_{cl}$ is the solution of the equation $\frac{\delta S[\g ]}{\delta \g}=0$ (least action principle), then the change of the phase on all trajectories close to $\g_{cl}$ will be small. At the same time, the change of the phase on the trajectories close to $\g_{noncl}$ will be large in view of the inequality \ref{crit}, since for that trajectories the equation $\frac{\delta S[\g ]}{\delta \g}=0$ is not true. It means that the deposit $K_{cl}$ on absolute value will be much more than the deposit $K_{noncl}$, because in the first deposit the summands have almost the same phase, whereas in the second deposit the summand have the different phases. We obtain that all amplitude is concentrated along the classical path of the particle, e.g., it behaves as classical particle. 

If now \ref{crit} is not fulfilled the situation changes, and the deposit $K_{noncl}$ can compete with $K_{cl}$, that is the real particle will not yet move along the classical path only, and will thus show the quantum properties (for example, will interfere after the passage through the slits etc.) We thus can treat the inequality \ref{crit} the criterion of the applicability of classical mechanics. We note that the smallness of the action can be reached by the small mass of the particle, small speed or minor period. Even the massive particles, moving slowly in the short time segments show quantum properties. In the practical description the time segment $\delta t$ is chosen with the aim to obtain the substantial picture of the evolution of the system (not of the dynamic of the separate particle). This is why electrons typically have not the shape of point wise particle, but has the shape of a cloud (wave function) whereas the nuclei should be treated as point wise, e.g., classically. This description is called the model of Born Openghaimer. It is convenient for the atomic physics where the subject is the electron states, are molecular dynamics, investigating the stretching and rotation of molecules. This representation is not appropriate for the description of chemical reactions, where the quantum character of the nuclear movement is important. In general, in course of the same reaction typically arises the necessity of the classical as well as quantum consideration of the reagents, and the model of Born- Oppenghaimer is assumed mainly due to its simplicity, based on the large in 1000 times) difference in the masses of proton and electron. 

Path integrals represent the important method showing the necessity of the heuristic  of the swarm of samples, ard we will develop this heuristic further. 

\section{Formalism of the many body quantum theory}

Ultimately, the aim of quantum theory is to explain the dynamics of complex systems, which have, in general case, macroscopic size. The way to it lies through the theory of atoms and molecules, because we think that their behavior has the simpler explanation. To tell the truth, we factually have no theory of many body systems, which completely accounts quantum effects. We speak about the entangled states of many particles that certainly detected in the numerous experiments, for photons and for massive particles, like complex molecules. To create such theory is the aim of constructive physics. 

Quantum formalism for many bodies rests of the simple algebraic reasoning. If the description of one particle requires the wave function of one argument: coordinates and time, then the description of the system of $n$ particles requires the wave function of $n$ arguments: coordinates of all these particles and the (common) time. We write it as 
\begin{equation}
\Psi (x_1,x_2,\ldots,x_n,t).
\label{wave}
\end{equation}

 The commonness of the time follows from the non relativistic approach when there is the absolute time, which makes possible to consider the evolution of system substituting to the wave function the sequential values of the time: $t=t_0,t_1,t_2,\ldots$. 

Here we assume that the coordinate $x_j$ is the concrete point in the configuration space (for example, for a particle in three dimension space without spin $x_j\in R^3$), occupied by the particle $j$. We thus mean by this notation of the wave function the possibility to distinguish particles, for example on their spatial positions. 

Algebraic representation \ref{wave} of the system of $n$ particles is compact and beautiful in its appearance. With the help of it we can prove the different useful properties of operators and the following properties of solutions of the equations which this function satisfies to.\footnote{For example, if the function \ref{wave} possesses for $t=0$ the property of the anti symmetry or the symmetry on the coordinates of particles and is the solution of Shedinger equation, then in the other time instants $t>0$ it will satisfy these conditions as well.}.

However, the algebraic representation \ref{wave} harbours the serious defect. This defect has the different manifestations, for example, it does not allow the computation of such functions already for small number of particles, for example 10, let alone for the more complex systems. This complexity barrier is principal and we will return to it. Moreover, this defect makes impossible to apply in quantum physics the natural heuristic of point wise particles we discussed in the previous section. At the same time, without this heuristic we cannot penetrate into the many body quantum world. 

The heuristic corresponding to the formula \ref{wave} can consists in that the different samples of swarms corresponding to particles $1,2,\ldots,n$ join in the cortege and the wave function just determines the distribution of such corteges. The destination of heuristic is in to find the idea of the building of effective simulating algorithms. In our case, its purpose is to reach the radical economy of the computational resources needed for the work with objects of the type \ref{wave}. It means that we have to motivate why one or the other sample is included to this cortege. This heuristic is irreducible to the one particle case we already studied. This heuristic I will name the many body heuristic. It must help us to clarify the events going in the complex systems, including the fundamental phenomena that are still hidden from us. We yet have no such heuristic, but we possess some its fragments, which we take up in the next chapter. The standard quantum theory proposed in the many body case to rely on the standard formalism of the object of the type \ref{wave}, which we now study. Here we must only keep in mind that in view of the absence of many bodies heuristic in standard quantum theory at all, its formalism turns much more unsteady support than in the one particle case, which we studied earlier. We should not think then that the quantum theory for many bodies gives such reliable answers to questions, as it was in one particle case. 

Quantum theory satisfactory explains the behavior of one particle or a few (2, at most 3) particles provided all external conditions are stable and simple. These methods of a few particles were elaborated in the first half of 20-th century, and from that time all quantum theory investigates many body systems only by application of one-particle methods. The separate particles in the gigantic ensembles (for example, a gas, or a crystal lattice) were treated as the isolated particles in the common potential created by all the other particles. This is, however, not quantum, but the semi classical way. The typical example of this approach is the statistical quantum theory, which uses the notion of quantum states of separate atoms for the computation of the thermodynamic characteristic of a gas. The other example is quantum theory of solid states. Quantum theory of liquids exists as well but it contains more problems that are open. Yet less elaborated is quantum theory of phase transitions or chemical reactions. By the same manner quantum theory works on the molecular simulation. For example, in the investigations of the polypeptide conformations it serves for the obtaining of the constants of bounds between the separate amino acid units, whereas further only classical methods work. At last , for the complex systems as bimolecules or for the dynamical problems as chemical reactions quantum theory is not applicable but the computation of stationary states and the trivial statements of the type "it works in dynamical processes, but we still do not know how, because it rests again electrodynamics, which immediately creates the huge computational difficulties."  

Hilbert formalism of quantum theory for many bodies is factually the program for the studying of many body problems. We are going to discuss this program in more details than it is done in the standard courses on quantum mechanics. In particular, we will discuss the problem of a quantum computer, which is the core of this program. We also discuss the interpretation of experimental results on this area, the fundamental phenomenon of decoherence and its model in the standard formalism. In the conclusion, we resume the practical recipes that this way gives for the elaboration of the theory of many body systems. 

\section{Unitary dynamics and measurements}

The most intriguing feature of quantum theory is the dual description of the dynamics of all objects that is divided to two types: unitary dynamics and measurements. This feature is so fundamental that apparently must remain for any new approach to quantum theory and even for any other theory, which could replace it in future. In the full measure it concerns algorithmic approach where the model consists of two parts: user and administrative. 

We consider separately these two types of dynamics.  

{\bf Unitary dynamics}. As we know, all possible states $|\Psi\rangle$ of quantum system form some Hilbert space of states ${\cal H}$, so that the evolution of each state when the time flows from the instant $t_0$ to the instant $t_1$ is determined by the unitary operator 
\begin{equation}
U_{t_0,t_1}:\ {\cal H}\ar{\cal H}
\end{equation}
We can still specify this formula by the representation of the time evolution operator as $U=exp(-\frac{i}{h}H(t_1-t_0))$, where $H$ is called Hamiltonian of the system and it equals teh operator of its full energy (which depends on this system and the external potentials) that is reformulation of that the evolution of state $|\Psi (t)\rangle$ is given by Shredinger equation. The important is that if we take instead of $|\Psi\rangle$ any other state $|\Psi_1\rangle$ in the space ${\cal H}$ and put it in the same conditions as $|\Psi\rangle$, we would then obtain the result of application of the same operator  $U_{t_0,t_1}$. 

How can we verify this statement? There is the single way for this. We have to take all possible states $|\Psi\rangle$ as the initial conditions and put them to the action of the same potentials to see the results of their evolutions. After this we have to compare somehow these results, e.g., the states of the form $U_{t_0,t_1}|\Psi\rangle$, to make the conclusion about the validity of this axiom.  

But it requires to compare the different quantum states. It turns that it is impossible to do by the unitary evolutions only. It requires the completely different type of the evolutions, called measurements. The definition of unitarity itself means nothing up to the moment when we apply the measurement to the system. This principal fact witnesses that {\bf there is no pure quantum mechanics, which does not rest on the classical mechanics describing the process of measurement.} 
\nn

{\bf Measurements}. We fix some basis $\{ |\psi_1\rangle,\ldots,|\psi_n\rangle$ in the space ${\cal H}$, and assume that it is orthonormal. Then the measurement in this basis is the random variable taking values $|\psi_j\rangle$ with probabilities $|\langle\psi_j|\Psi\rangle |^2$. This axiom of quantum mechanics is known as Born rule of computation of quantum probabilities. There is no other way to extract the information about quantum state but to fulfill the measurement. The single choice here is the choice of the basis $\bar |\psi_j\rangle$; it depends on the experimental device on which we fulfill the measurement. Typically, there are two main choices of the basis: coordinate basis (the measurement of the coordinates) and the impulse basis (the measurement of the impulse). It is impossible to measure these two values simultaneously because there is no the common basis of the eigenvectors for these two operators. 

The measurement factually means the physical contact of the considered system with the classical object consisting of many particles. Since the unitary character of dynamics disappears in this process only probabilities remain after the measurement. We stress that there is no way to extract any information from quantum system behind the measurement. Therefore, classical physics is necessary to use quantum theory. There is no way to avoid this fundamental limitation. 

These two postulates of quantum theory imply the serious corollary. The statement about the unitarity of quantum evolution touches not a separate evolution but the big number of experiments of the same type. Only after the statistical processing of its results the properties like linearity, conservation of the norms, etc. become explicit.  Here even the processing of one separate experiment (one evolution) requires the attraction of many body systems, for which we cannot apply the laws of quantum mechanics, but will have to use the classical mechanics to establish what is the result of the experiment:   
$|\psi_{j_1}\rangle$ or $|\psi_{j_2}\rangle$. If we for some reason cannot ensure the identity of conditions for the different experiments (including the coincident tuning of measuring devices and the possibility to use the macroscopic devices at all) the statement about the quantum character of the evolution loses any sense. 

Quantum theory thus from the very beginning requires the existence of many body systems and the possibility to involve them in experiment directly. It is required the presence of macroscopic device with many particles simultaneously for the measurement of a state and many particles for the choice them as the measured quantum system which state $|\Psi\rangle$ we investigate. How much time will the measurement take? If we suppose that the particles of the same type are identical, we can fulfill simultaneously many experiments (this is the case in the statistical quantum theory). Alternatively, we can try to check the hypothesis about the identity of elementary particles of the same type. For this, we need to fulfill the row of sequential experiments on the same particle, then the analogous row with the other particle, etc. The modern devices, like scanning tunneling microscope permit to address immediately to the individual atoms, hence for the fulfillment of such experiment does not meet the serious difficulties but the time. 

\section{Many world interpretation of quantum theory}

Hilbert formalism is the basis of the standard mathematical apparatus for quantum theory. The first corollary from this formalism is the principle of superposition, e.g., the possibility to fulfill the linear operations on the state vector. The second important corollary is the method of the consideration of many body ensembles, which is called the many world picture of quantum theory (term of Everett). The many world picture of quantum theory 
\footnote{It is often called many worlds interpretation, but this is not exact term. Many world picture is the integral part of Hilbert formalism and thus cannot depend on any interpretation.} consists in the following. The entire Universe has some wave function  
\begin{equation}
|\Psi_{global}\rangle .
\end{equation}
This function depends on the coordinates of all particles in the Universe and exactly determines its state. In what follows in the study of the formalism of quantum electrodynamics (QED) we will understand that $|\Psi_{global}\rangle$ must depend also on the times, which generally speaking must be different for all the most of particles, at least slightly. Nevertheless, if some particles interact with each other their time must be the same. We will show it further, and now assume without the proof. We then can write for the wave function of the Universe the representation of the form 
\begin{equation}
|\Psi_{global}\rangle=\sum\limits_- \la_-|\bar R_-,t_-\rangle
\label{uni}
\end{equation}

where the index $_{}$ denotes the so-called "world", e.g., the maximal set of particles which either interacted some time in the past, or will interact in future. The index of the world $-$ take the values from some set ${\cal T}$. The different worlds have the different values of the time 
 $t_-$, and the worlds thus do not intersect. The notation of the wave function in the form \ref{uni} thus has the decorative character. In the standard formalism based on classical mathematics the worlds cannot transform to each other and even to interact somehow with each other just because they have the different time $t_-$, and, correspondingly, the different basic states. The history of each world in the result of the grouping in \ref{uni} the summands with the same value of the time $t_-$. We now ascribe to the parameter $t_-$ the physical sense of the time. Then the structure of the set ${\cal T}$ becomes important. If it is the linearly ordered set, we obtain the deterministic picture when in each time instant there is the wave function of the Universe and the next its state is predefined completely. This picture prevents us from any possibility to influence to the events and thus it lies besides any reasonable human activity. We thus must accept that ${\cal T}$ is the set with the partial order.. The most natural partial order is the tree like structure. Just this is typically accepted in the standard quantum theory. Then in some instant, the "fork" is possible in the evolution of the Universe, so that one of the words evolves along one way, the other along the other way. Since the unitary evolution is deterministic on the level of wave functions, we conclude that the fork concerns only the different results of the same measurement of the same state. If the universe has many "observers" then each of them of course, cannot detect was its own observation (the initiation of the measurement) the cause of the fork or it is the result of activity of some other "observer". Moreover, he cannot even detect the presence of the other "observers" up to the moment when they exchange the information directly. It means that we can determine that the fork has happened with the high probability gathering the statistics of many states and discovering that the dynamics is not unitary. In the reality the forks 	permanently happen and compose the phenomenon called decoherence. 

The measurement results in the transformation of the state vector of Universe to the new state vectors, to each with the corresponding probability. These new states will be the different worlds. Since the measurement of the system presupposes the contact of the system with some bigger system, called the measuring device, then in the standard quantum mechanics we conclude that it is necessary to introduce something more than the Universe. For example, it can be the some non-material entity as the Consciousness, or something like this. If we treat this entity, the material as well we would have to extend the Universe including this entity to it, and all repeats again.  E.g., to avoid the loop we must accept that the observer is not material; this means that it is out of the scientific knowledge and we will never extend the scientific knowledge to the whole world. One could object that the Universe is infinite and we always can imagine that there is some part of it lying beyond our perceptions, etc. This assumption does not lead to any contradictions but contains the reach possibilities for the suspicious manipulations.\footnote{Like the paradox of a self-murderer, who always meets the misfire, because he lives only in that world where it happened, in the other worlds he is simply absent, etc.}

The single reasonable outlet from this difficulty is to agree about the following: 
\begin{itemize}
\item decoherence is the property of the system itself, not of its contact with the environment (this contact only creates the conditions of it)
\item the evolution of state vector in each world must have the description in the exact terms without the attraction of non material entities,
\item the different worlds exist in the administrative part of the model only, and serve exclusively for the debugging of the main scenario which we observe in the real world.
\end{itemize}

The random factors we observe result from th ework of administrative part of the model; their source is inaccessible for a user because of the construction of the model. The randomness’s must ultimately factorize so that the result is independent of them. Accepting these thesis’s we can go further, on the way to physical constructivism. 

\section{Quantum computer}

In this paragraph, we describe the fundamental device which existence directly follows from the many body Hilbert formalism in the framework of classical mathematics. This device, makes possible to solve the row of computational problems much faster than any classical computer, provided this consideration adequately describes the real physics. This device is called a quantum computer (QC). The limited quantum computer (on several qubits) is already built on the different technologies\footnote{Nuclear-magnetic resonance, Josephson junctions, ion traps.}. The main question concerning QC is to what extend in memory is it scalable. 

{\bf The question about the physics of quantum computer unavoidably leads us to the necessity to investigate complex quantum phenomena, and thus to the constructivism in quantum theory.}

In this section, we show the brief description of the general scheme of quantum computer and quantum computations. In view of the importance of the theory of quantum computer I give the more extended description of its formal side in the Appendix. One can read about the technological aspects of QC in \cite{VK}. 

\subsection{Idea of quantum computer}

At first we give the general idea of a quantum computer. We imagine that we need to find some unknown integer $|l_t\rangle$. We consider Hilbert space of states of quantum system with the basis $|0\rangle,|1\rangle,\ldots,|N-1\rangle,$, so that $l_t$ contains among the numbers $0,1,\ldots,N-1$. We prepare the quantum state of the form 
\begin{equation}
|\tilde 0\rangle=\frac{1}{\sqrt{N}}\sum\limits_{j=0}^{N-1}|j\rangle ,
\end{equation}
in which all amplitudes are equal. The probability to find the unknown number by measuring $|\tilde 0\rangle$, equals $1/N$ that is equivalent to the bruit force. 

\begin{figure}
\centering
\caption{Quantum computation}
\vspace{250mm}
\makebox[180mm][l]{\includegraphics{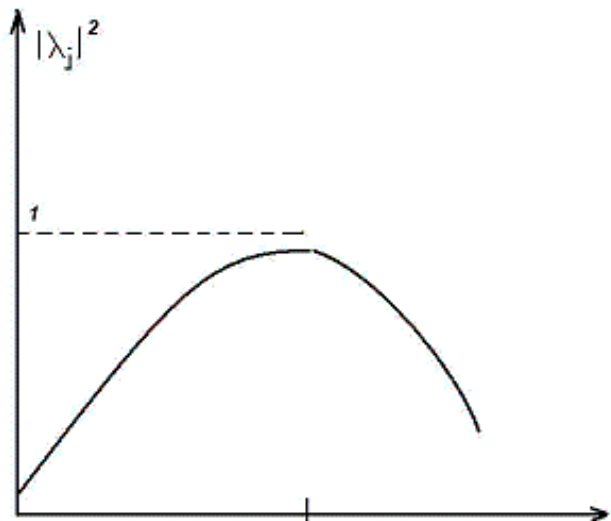}}%
\end{figure}

We suppose that by using the information available about the unknown number $l_t$, we can build some unitary evolution $U_t$ of the quantum state of our system $|\Psi (t)\rangle$, where $|\Psi(0)\rangle =|\tilde 0\rangle$ such that $|\Psi(t)\rangle = U_t|\Psi(0)\rangle$. For example,  it may be the evolution induced by a constant Hamiltonian $H=const$, and then $U_t=exp(-iHt/h)$, or the Hamiltonian depends on the time and then we must treat the exponential as the chronological exponential. The choice of Hamiltonian is the question of the control on the quantum computer that the classical system fulfills. This controlling classical system can have the form of laser impulses, or the voltage on the electrodes. The evolution of quantum state then gives us in each time instant the state 
$$
|\Psi(t)\rangle=\sum\limits_{j=0}^{N-1} \la_j(t)|j\rangle ,
$$ 
where complex numbers $\la_j(t)$ depend on the time $t$. If we measure tis state, it gives the target number $l_t$ with the probability $|\la_{l_t}|^2$ accordingly to Born rule. 

Let an evolution of the system be chosen so that the function $|\la_{l_t}|^2$ takes the value close to 1, in the time $T_{qua}$ (see the picture). Let $T_{class}$ be the time of search of the unknown number $l_t$ on a classical computer. If $T_{qua}<T_{class}$, we say that the quantum speed up of this classical search algorithm takes place. Really, if we create a device realizing this evolution and fulfill the measurement of the state of the quantum system in the instant $T_{qua}$, w eobtain the desired result $l_t$ with the high probability in the time $T_{qua}$, e.g., faster than if we search $l_t$ on the classical computer. 

The method of determining the evolution $U_t$ is called the control on the quantum computer. The classical device fulfills the control, this device acts independently of the quantum state. This evolution we call the quantum computation. At the end of quantum computation, we measure the final quantum state that results in the answer with some probability, which must ne close to 1. It follows from the definition that any classical algorithm we can treat as quantum. E.g., the computable on classical computers functions will be also computable on quantum computers. The reverse is true as well: a function computable on a quantum computer will be computable on some classical computer. Really, we can, in principle, simulate any quantum computation by the appropriate classical computer, given a law $U_t$, because this law means the possibility to write the unitary matrix corresponding to this operator, and thus to restore all states of the form $|\Psi(t)\rangle$.\footnote{This simulation is hypothetical because the required classical memory grows as exponential. We here reason from the viewpoint of classical mathematics where one does not pay attention to the complexity.} The stock of computable functions is thus the same for quantum and classical computers. The advantage of quantum computers is that it makes possible to compute some important function substantially faster than classical, provided we accept quantum theory in the area of many body systems that we do in this chapter. 

For the same problem of finding some set of numbers $l_t$, there are, generally speaking, many quantum algorithms as well as many classical. We say that the problem admits the quantum speed up, if there exists the quantum algorithm, which solves this task faster than any classical. 

It turns that the total number of problems admitting quantum speedup is small (see \cite{Oz2}, \cite{Oz3}). Nevertheless, there are important problems among them: direct search and the problem of factoring integers. For the description of them we need the more detailed model of quantum computer.

\subsection{Abstract model of quantum computer}

We now describe the abstract model of quantum computer, which already can serve for the estimation of its complexity, e.g., the time $T_{qua}$. 

The model consists of two parts: classical and quantum. The classical part consists of registers in which the codes of elementary unitary operators from some list $U_1,U_2,\ldots$ stand. This list consists of simple 1-2 or 3 qubits unitary operators and the pointers, e.g., the arrows that point to which qubit we must apply this operator. Besides this, the classical part contains two special registers: the register of the end of computation and the register of the query to an oracle.

\begin{figure}
\centering
\caption{Abstract model of quantum computer}
\vspace{250mm}
\makebox[180mm][l]{\includegraphics{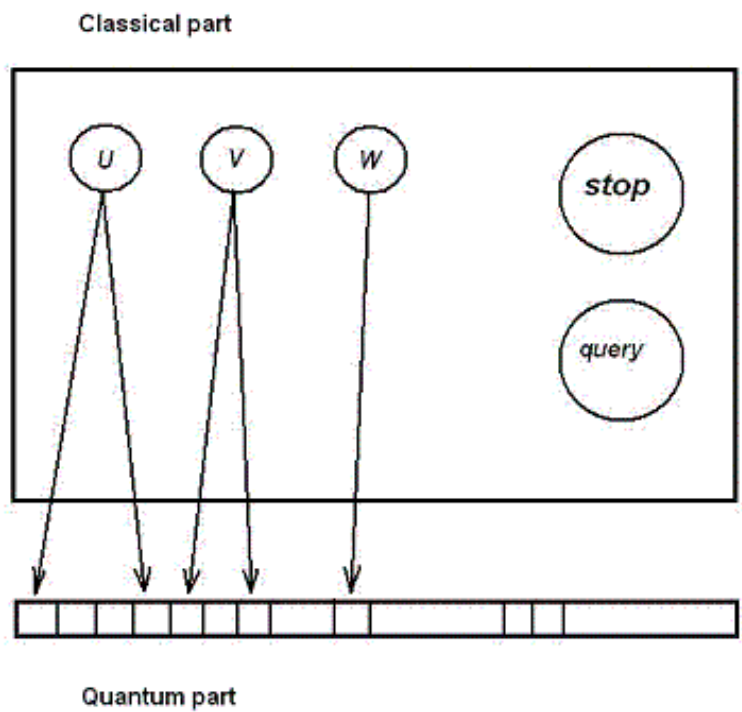}}%
\end{figure}

Quantum part of computer is the tape in which cells stand qubits. Potentially the tape is not bounded in the sense that we always can add new qubits initialized by the state If the tape contains $n$ qubits, its quantum states fill $2^n$ dimensional Hilbert space of states. The general form of state fo quantum part is:
\begin{equation}
|\Psi\rangle = \sum\limits_{j=0}^{N-1}\la_j|j\rangle
\label{qubi}
\end{equation}
where $N=2^n$, and the coefficients $\la_j$ are the complex numbers called the amplitudes corresponding to the states $|j\rangle$. We thus can treat the evolution of the state of quantum part as going in the finite dimensional Hilbert space which dimension $N$ is inaccessible for us despite of the total number of qubits $n$ is an accessible value. We see that \ref{qubi} coincides to what we call a qubit representation of wave function. Quantum computer thus expresses the standard form of many particles Hilbert formalism. 

Quantum algorithm is the classical algorithm determining the evolution of the state of the classical part of computer. A computation on quantum computer is a sequence of unitary operators determined by the classical part and applied to the state of quantum part. On each time step $j$ on the state $|\Psi_j\rangle$ we fulfill the transformation of the form
\begin{equation}
U\bigotimes V\bigotimes W\bigotimes\ldots
\end{equation}
where the elementary operators $U,V,W\ldots$ which codes stand in the registers of classical part are fulfilled on that qubits to which the pointers outgoing from the register point. 

The purely classical law of the evolution of classical part thus induces the unitary evolution of its quantum part of the form: 
$$
|\Psi(t)\rangle=\sum\limits_{j=0}^{N-1}\la_j(t)|j\rangle
$$
e.g., it is reducible to the change in time of the basic quantum states amplitudes. 

We gave the definition of a computation without oracle, or an absolute quantum computation. Analogously to the classical case we can introduce the notion of quantum computation with oracle. Let a unitary operator $U:\ H_1\ar H_2$, be given where $H_1$ - $m$ and $k$ are the qubit Hilbert space of states. We We set up on the tape the special place: the set of qubits (register) of $m$ qubits, and the special register in the classical part named query register. We agree that if the query register contains $0$, the computation goes in the usual order. If the query register contains $1$, we instead of usual unitary operator induced by the classical part, apply the query to the oracle, which transform the basic state $|j\rangle$ to $I\bigotimes U$, where $U$ is applied to the separated $m$ qubit register, and the identical transformation to all the rest. This definition is the straighworward extension of the notion of computation with oracle to the quantum case. 

We concretize the notion of quantum oracle to the case of the usual function of the form 
$$
f:\ \{ 0,1\}^m\ar\{ 0,1\}^k
$$
We separate on quantum tape two registers of $m$ and $k$ registers, and call the first query register, the second - the register of the reply. Let $a$ and $b$ be corteges of ones and zeroes containing in these registers. We introduce the unitary operator by its definition on the basic states as:
\begin{equation}
Qu_f|a,b\rangle \ar |a,b\bigoplus f(a)\rangle
\label{oracle}
\end{equation}
where $\bigoplus$ denotes the bitwise addition modulo 2. This operator is merely the permutation of the basic vectors, hence it linearly continues to the unitary operator on the whole space of quantum states. It is involutive, e.g., $Qu_f^2=I$. We describe in the Appendix how by this trick to build the fast quantum search algorithm.

\section{Role of entanglement}

As we know, an entangled state of a system consisting of two quantum subsystems $S_1$ and $S_2$, is a state which has not the form $|\Psi_{S_1}\rangle\bigotimes |\Psi_{S_2}\rangle$ for any choice of $\Psi_{S_1},\ \Psi_{S_2}$. It corresponds in the analytical notations to the impossibility to represent the wave function of the whole system as the product of wave function of its parts $S_1$ and $S_2$. We can define the measure of entanglement for entangled states by the different ways. For example, using the relative density matrix $\rho_{S_1}=tr_{S_2} \Psi$. That is we define the entropy of entanglement as $H=tr(\rho_{S_1}\ln\ \rho_{S_1})$, which equals to the analogous expression taken for $S_2$. This measure of entanglement often works in the theory of quantum information. 

It straightforwardly follows from the definition that the evolution of quantum system which has the for of the evolution of non entangled states (e.g., tensor product of one particle states) can be simulated on a classical computer in real time mode, that is with the complexity proportional to the real time. It implies that there is no fast quantum algorithm on non-entangled states. In the other words without the entanglement, a quantum computer is reducible to the classical part.

Nevertheless, it would be wrong to think that it is possible to identify the set of all entangled states with the states with the maximal entropy of entanglement $H$. The possibility to generate many particles entangled states even with the maximal degree of entanglement does not mean that we create a quantum computer. For example, we consider two classes of entangled states that were really detected in experiments on ion traps of Paul. There are $GHZ$ and $W$ states of the form

\begin{equation}
\begin{array}{ll}
&GHZ:\ \ \la_1|11\ldots 1\rangle +\la_2|22\ldots 2\rangle+\ldots +\la_k|kk\ldots k\rangle,\\
&W:\ \ \la_1|100\ldots 0\rangle+\la_2|010\ldots 0\rangle+\ldots+\la_k|00\ldots 1\rangle .
\end{array}
\label{GHZ_W}
\end{equation}

We see that to keep such states in the memory of computer it is sufficient to reserve of the order of $n$ cells where $n$ is the number of qubits. It means that any computation with such states can beb repeated on classical computers, and with these states it is thus impossible to realize fast quantum algorithms. (see \cite{Ak2}). 

Factually, the states of the type $GHZ$ and $W$ are reducible to one particle states. We consider at first the state $GHZ$. Its analog for $n=2$ is the so-called Shmidt state of two particles, that is the state 
\begin{equation}
\sum\limits_j \la_j|\psi^1_j\rangle\bigotimes |\psi^2_j\rangle
\label{shm}
\end{equation}
where $\{ |\psi^1_j\}$ and $\{ |\psi^2_j\}$ are orthonormal basis’s in the spaces of states of the first and the second particles correspondingly. The state \ref{shm} is determined by $N$ numbers, where $N$ is the dimensionality of Hilbert spaces of one particle. To hold such bipartite state the same volume of memory is necessary as for one particle state. 

The state $GHZ$ is the form of Schmidt state for several particles. This state means the following. We really have only one particle, but it consists of a several parts, which behave as the unit. We consider $GHZ$ state for $k=2$: $|\Psi\rangle=|00\rangle+|11\rangle$. We treat $0$ and $1$ as the positions of one particle in two different fixed points. The measurement in the standard basis then results in the presence of the both particles in the same point. If we want to measure the impulse of particles we have to apply to the corresponding qubits quantum Fourier transforms, which in case of one qubit is merely Hadamard operator. Applying this operator to the both qubits, we pass to the basis in which the values of qubits will be he impulses of the corresponding particles. The straightforward calculation shows that the result will be the same: $|00\rangle+|11\rangle$. It means that the impulses of two particles will be the same if we measure them. It explains the fact that this state is factually the state of one particle. To detect such state it would suffice to find the interference properties of the object consisting of several parts, for example, the interference of a molecule of hydrogen on two slits. The presence of interference picture means the existence of the entanglement of the type $GHZ$. This type of entanglement thus is widely spread. The more interesting is to detect this type of entanglement on the large distances. For ions in Paul traps, it is about a few millimeters, for photons up to several tens of kilometers. 

Independently of the number of points in the configuration space (at least 2) the following statements take place:

1) Any quantum state of two particles can be represented as Schmidt decomposition.

2) There exist the states of three particles which cannot be represented in the form of Schmidt decomposition.

The easiest example for 3 qubits is the state $|100\rangle+|010\rangle+|001\rangle$, that is the example of state of the type $W$.

We now turn to the general form of $W$ state. This state is irreducible to $GHZ$ states by any local operations (e.g., unitary operators on the separate particles and local entangling and measurements, the so-called LOCC operations). Nevertheless, we can treat it as the one particle state as well. Let us consider each qubit as the element of configuration space. We remember that in the qubit representation of the wave function we agreed to encode by qubits the points of configuration space in the sense that each next qubit is the specification twice the spatial position of the particle. Temporarily we accept the other agreement: each qubit we identify with the point of configuration space itself, taken with the highest accuracy. This new agreement is completely different. Than say, the basis state $|100\rangle$ denotes the presence of the particle in the point $1$ (whereas the points 2 and 3 are free), the state 
 $|010\rangle$ means the presence of the particle in the point 2 (whereas the points 1 and 3 are free), etc. The state $GHZ$ of the form \ref{GHZ_W} corresponds then the wave function of one particle of the form $\sum\limits_{j=1}^k\la_j|j\rangle$.  

\section{Formalism of quantum electrodynamics in qubit form}

We are now ready to describe the formalism of quantum electrodynamics. The reason why we delayed this description is technical. We are going to look at the many particle quantum electrodynamics, and for this the standard form of representation of the wave functions as $\Psi (r_1,r_2,\ldots )$ is not appropriate. This is because in QED it is convenient to represent the main operator: emission and absorption of a photon by a charged particle in the impulse-energetic basis of theh space of states, whereas for us the important will be the for of this operator in the spatial-time basis, and the passage between basis’s is convenient to represent in the qubit form. Moreover, we interest in the arbitrary vectors of state of the many particle system that may be entangled, and it is thus convenient to investigate them in the qubit form.  At last, the representation of the main operator of QED requires the notion of ancillary qubits (ancilla) which came from quantum computing. Since, the qubit form for QED is the best. Its advantage is also in that the defect of standard mathematical apparatus as applied in QED, becomes evident. 

QED represents the result of application of quantum thoery to the electrodynamics, e.g., to the interaction between the charged particles and the field described by Maxwell equations. The natural way for this at the first glance would be to represent the basic states of $n$ charged particles and $k$ photons as  $|x_1,x_2,\ldots,x_n,f_1,f_2,\ldots,f_k\rangle$, where $x_j$ and $f_i$ are the coordinates of particle $j$ and photon $i$ correspondingly. However, this way is difficult in QED due to the peculiarities of photons. These particles move only with the speed of light and have zero mass. Therefore, we cannot associate any reference frame with a photon, because the reference frame requires the presence of the nonzero mass in the reference point. Moreover, the photons show only in their interaction with the charged particles and do not interact with each other (but the negligible higher degrees - through electro-positron pairs). Generally speaking, we cannot write any analog of Shredinger equation with some potential for them. Photons themselves create the potential. At last, photons appear and disappear only in the interaction between charged particle and field. It can happen also in vacuum where a photon can create elector-positron pair. Hence, in QED the full formalism of Hilbert spaces is not completely applicable and we cannot speak about the completeness of QED as for usual quantum mechanics. 

QED consists of the description of fundamental processes of interaction between one or two charged particles with photons and the computation of amplitudes of these reactions of scattering, and the rules permitting to derive classical Maxwell equations from these amplitudes. 

Nevertheless, in this section we try to describe the formalism of qubit type for QED in the full form. It has three principal differences from the considered case of usual quantum mechanics. At first, the space of states for QED explicitly contains the time. At second, this time will be in general case the different for the different particles, and at third, the number of particles in the system will not be stable. 

Let we be given a set ${\cal X}$ of spatial states $x$ and a set ${\cal T}$ of the time instants $t$ for one point wise particle. These values belong to the user part of the model, e.g., they will describe the film which action will develop on a given set of points in the space for this user and which cadrs are enumerated by a given time instants. 

We consider Hilbert space of states for one poin wise particle which is spanned by the basic states of the form $|x,t\rangle$, where $x\in {\cal X},\ t\in {\cal T}$.
The general form of this state is:
\begin{equation}
|\Psi\rangle = \sum\limits_{x\in {\cal X},\ t\in {\cal T}}\la_{x,t}|x,t\rangle
\label{seq_scattering}
\end{equation}
The state of a system thus depends on the user time: the time occurs in the basis of space of states. Nevertheless, the state depends on the administrative time $\tau$, e.g., we can treat the sequence of states of the form
\begin{equation}
\Psi(\tau_1)\rangle, \Psi(\tau_2)\rangle,\ldots
\end{equation}
where each passage $\Psi(\tau_j)\rangle\ar \Psi(\tau_{j+1})\rangle$ we call the reaction of scattering. A user investigating the process of QED will observe the film devoted to only one reaction inside of which its user's time will be $t$, which really varies not in the infinite limits as in the analytic formalism, but in the finite limits though as large as it is needed to a user. At the same time, the other user, who investigate, say, chemistry and who needs QED only for the substantiation of the going association or dissociation of molecules will see the different film, which has the form 
\ref{seq_scattering}, and for it the user's time will be $\tau$. These scales differ in the orders of units.  

We then consider the system of several particles, enumerate them by the natural numbers: $ 1,2,\ldots, n$, and agree that the attributes of a particle number $j$ have the upper index $j$. For the simplicity of notations we will also write $\sum\la_{-}f_{a}$ instead of $\sum\limits_{a}\la_{a}f_{a}$ for any ensemble of objects $a$.

The space of quantum states of $n$ particle system consists of the formal linear combinations with complex coefficients of the form
\begin{equation}
|\Psi\rangle = \sum\la_{-}|x_1,t_1,\bar\e_1,x_2,t_2,\bar\e_2,\ldots,x_n,t_n,\bar\e_n\rangle
\label{QED_state}
\end{equation}

We note the difference of QED from the usual quantum mechanics, shown in that in QED states there are the summands of the form $|x_1,t_1, x_2,t_2\rangle$ e.g., we can speak about the simultaneous treatment of particles in the different time instants. 

Factually, a state \ref{QED_state} we can consider as a scenario because there contains the information about the states of the system at hand in all possible time instants. 

We remember that in QED $p=(\bar p,E)$ is the generalized impulse of charged particle, the photon amplitude is 4-vector $\e_\mu$, such that $A_\mu=\e_\mu\ exp(-iKx)$, where $K$ is the 4-vector of the impulse of photon of the form $(\bar p,\w)$. Dot product of 4-vectors we define as 
$$
a_\mu b_\mu=a_4b_4-a_1b_1-a_2b_2-a_3b_3,
$$
e.g., there is silent summing on the repeated indices. We can assume that (see \cite{FH}) for the detected (free) photon $\e_4=0$. 

Theh main operation determining the evolution of state \ref{QED_state} is the operator of elementary scattering, which describes two processes: the emission of a photon by a charges particle and the absorption of a photon by it. The following conditions determine these processes in QED:

1) they go only whith the conservation fo energy and impulse: $\bar E_1$ and $\bar p_1$ of the system, and 

2) the amplitude of the first is proportional to $\la_{\bar E_1,\bar E_2, \bar p_1,\bar p_2}=i\ e\ (p_1+p+2,\epsilon)$, where $\epsilon$ is the vector of polarization of the photon, $p_1,\ p_2$ are impulses of the charged particles before and after the reaction, and the amplitude of the reverse process is complex conjugate to the amplitude of the direct process. 

Here we accept that each emission or absorption of photons conserves the energy and impulse, so that if
$\bar E_1,\bar p_1$ are the energy and impulse of particles before the reactions, and $\bar E_2,\bar p_2$ ,$\bar E,\bar q$ are the energy and impulse of the particles and photons after the reaction , then $\bar E_1=\bar E+\bar E_2,\ \bar p_1=\bar q+\bar p_2$.

To transform these conditions on the language of qubit formalism we must introduce one agreement about the notation of non-existing photons: they must have the zero energy and impulse, and the random polarization. We thus introduce to the basic elements of our space all photons: really existing and still not emitted in course of reaction, and put to zeroes the impulses and energies of the last. The reaction of scattering has the form of unitary operator, which matrix we denote by $S$ accordingly to the tradition od QED. 
Since the scattering is typically defined in terms of energies and impulses we pass to this basis and find the matrix $S$ in it. The passage from $(\bar x,\bar t)$ representation of matrix to $(\bar p,\bar E)$representation is given by the direct Fourier transform along each coordinate $\bar x$ and by the reverse transform along each time $\bar t$. 

The operator $S$, as any unitary operator we can represented in the form $S=exp(iH)$ for some Hermitian operator $H$. We accept that the scattering goes on the short frame of the physical time $t$: $[t_0,t_0+\Delta t]$, so that all times $t_j$ lye on this frame. It means that we can use the approximate equality 
$$
S=exp(iH)\approx 1+iH. 
$$
We describe the matrix $H$. In $(\bar p,\bar E)$ representation the matrix $H$ has the following form. In the column corresponding to the state with charged particles with impulses $\bar p_1$ and energies $\bar E_1$, in the place corresponding to the state of charged particles where they have impulses $\bar p_2$ and energies $\bar E_2$, and the corresponding photons have impulses $\bar q$ and energies $\bar E$, provided $\bar p_1 = \bar p_2+\bar q$, $\bar E_1=\bar E_2+\bar E$, stands the number $\la_{\bar E_1,\bar E_2, \bar p_1,\bar p_2}$, the elements of $H$, standing on places symmetric to the described relative to the main diagonal equals complex conjugations of the above described elements, on the rest places stand zeroes.

We consider the result of application of $H$ to the state of one charged particle. Let its initial state have the form
\begin{equation}
|\Psi\rangle=\sum\limits_{-}\la_{\_}|x_1,t_1\rangle
\end{equation}
We pass to the impulse-energy basis and add the ancilla initialized by zeroes $|\bar 0\rangle$: 

\begin{equation}
|\Psi\rangle\ar\sum\limits_{-,=}\la_{\_}e^{i(\bar x_1\bar{\bar {p_1}}-\bar t_1\bar{\bar {E_1}}}|\bar{\bar {p_1}},\bar{\bar {E_1}}\rangle |\bar 0\rangle
\end{equation}

This ancilla corresponds to the non-appeared photon associated with this charged particle; it will emit this photon at the next step.
Now we fulfill the operator of emission of the photon. It results in the state:
\begin{equation}
\sum\limits_{-,=,\tilde\ ,\\
\tilde p_1+\tilde p_2=\bar{\bar{p_1}},\\
\tilde E_1+\tilde E_2=\bar{\bar{E_1}}
}
\la_{-}
\mu_{\ _{\tilde\ }}
e^{i(\bar x_1\bar{\bar {p_1}}-\bar t_1\bar{\bar {E_1}})}
|\tilde p_1,\tilde E_1,\tilde p_2,\tilde E_2\rangle
\end{equation}
and now we have to return again to the spatial-time basis:
\begin{equation}
\sum\limits_{-,=,\tilde\ ,\\
\tilde p_1+\tilde p_2=\bar{\bar{p_1}},\\
\tilde E_1+\tilde E_2=\bar{\bar{E_1}},\\
X_1,X_2,T_1,T_2
}
\mu_{\ _{\tilde\ }}
e^{i(\bar x_1\bar{\bar {p_1}}-\bar t_1\bar{\bar {E_1}})}
e^{i(X_1\tilde p_1+X_2 \tilde p_2-T_1\tilde E_1-T_2\tilde E_2)}
|X_1,X_2,T_1,T_2\rangle
\end{equation}
We substitute here $\tilde p_2=\bar{\bar {p_1}}-\tilde p_1,\ \tilde E_2=\bar{\bar {E_1}}-\tilde E_1$, and then find the coefficient at the fixed basic state $|X_1,X_2,T_1,T_2\rangle$. It equals 
\begin{equation}
\sum\limits_{-,=,\tilde\ }\la_{-}
\mu_{\ _{\tilde\ }}
e^{i(\bar{\bar {p_1}}(\bar x_1-X_2))+i(\tilde p_1(X_2-X_1))+i(\bar{\bar{E_1}}(\bar t_1-T_2)+\tilde E_1(T_2-T_1))}.
\end{equation}

It means that the coefficients at the basic states for which $\bar x_1$ differs substantially from $X_1$, or $X_1$ from $X_2$, or $\bar t_1$ from $T_2$, or $T_2$ from $T_1$, are close to zero. We thus conclude: in the emission of photon we obtain the state in which for each summand the times of photon and charge particles are the same. 

We consider the basic state in the coordinate-time basis $|x,t\rangle$, add to it the ancilla initialized by zeroes and apply to this state our operator $H$, and then return back to coordinate-time basis:
\begin{equation}
\begin{array}{ll}
&|x,t\rangle=\sum\limits_{p,E}e^{ipx-iEt}|p,E\rangle\ar\sum\limits_{p,E,p_1+q=p,E_1+E_q=E}e^{ipx-iEt}(p_1+p)\cdot\e |p_1,E_1,q,E_q,\e\rangle\ar \\
&\sum\limits_{p,E,p_1+q=p,E_1+E_q=E,x_1,x_q,t_1,t_q}e^{ipx-iEt}(p_1+p)\cdot\e e^{-ip_1x_1-iqx_q+iE_1t_1+iE_qt_q}|x,x_q,t_1,t_q,\e\rangle=\\
&\sum\limits_{p,E,p_1+q=p,E_1+E_q=E,x,\Delta x_1,t_1,\Delta t}e^{ipx-iEt}(p_1+p)\cdot\\
&\cdot\e e^{-ip_1x_1-iq(x_1+\Delta x)+iE_1t_1+iE_q(t_1+\Delta t)}
|x_1,x_1+\Delta x,t_1,t_1+\Delta t,\e\rangle =\\
&=\sum\limits_{p,E,p_1+q=p,E_1+E_q=E,x,\Delta x_1,t_1,\Delta t}e^{ip(x-x_1)-iq\Delta x}(p_1+p)\cdot\\
&\cdot\e e^{iE(t_1-t)+iE_q\Delta t}|x_1,x_1+\Delta x, t_1,
t_1+\Delta t,\e\rangle
\end{array}
\end{equation}
We now straightforwardly convince that the coefficients at the states for which one of the inequalities $x\neq x_1$, $t\neq t_1$, 
$\Delta t\neq 0$, $\Delta x\neq 0$, is true are close to zero for each vector of polarization $\e\in R^4$. It follows from that such coefficients equal to linear combinations of Dirac delta-functions $\delta_{x-x_1},\ \delta_{t-t_1}$, $\delta_{\Delta t}$, $\delta_{\Delta x}$ and their derivatives. The closeness to zero instead of the exact equality appears from the discrete character of states that is the result of its qubit representation. 

We can conclude the following: the fundamental interactions of QED bear the local character in the sense that they happen in one point of space-time, in which the charged particle stands. This conclusion is not trivial because in the standard formalism of wave functions there are unitary operators for which the final sum contains the summands with $x\neq x_1$ or $t\neq t_1$, or one of $\Delta t$, $\Delta x$ differs from $0$ and the interaction then would not be the local. The locality of interaction means that it results in the state where the position of the initial charged particle does not change in comparison with the initial: $x=x_1,\ t=t_1$ and the photon "occupies" the same point. If the equalities are exact we would obtain that the operator $H$ is the identity. Of course, here we could postulate that one must consider the fundamental interaction of QED in the impulse-energy basis only, and we then could write the law of the building of matrix $H$ and to work with it operating with this law instead of direct manipulation with the matrix itself. This agreement is acceptable in the constructive physics, but for one particle only which interacts with the filed it creates. This case typically considered in QED (see \cite{Fe}). We here can avoid the divergence of rows, introducing the renormalization of charges and masses depending on the environment, and then compute the cases of 1-2 particles to the end, which brings, for example, the value of magnetic momentum of an electron or the correction to the energy of the ground states in atoms, resulted from the finite speed of vector photons. 

However, the situation differs radically if we intend to develop QED in the many particle area. Then the necessity arises to work with states in which every particle has its own time: $|x_1,x_2,\ldots; t_1,t_2,\ldots\rangle$. This is still the surmountable obstacle. The real problem arises from that the notation of the operator $H$ in КЭД turns depending on the chosen grain of space-time. We saw that the representation of the operator $H$ in qubit form not only determines its accuracy as it takes place for Fourier transform, but factually determines its action for the chosen space-time basis in the configuration space. It makes the question about the choice of basis principal for computations, that closes the door to use the qubit representation for QED, and, consequently, to use computers for the computations of quantum electro dynamical systems\footnote{It makes sense to apply a computer only if it can fulfill the whole volume of the manual computing. If the choice of basis begins to depend on a user in each time instant the computerization becomes unreal.}. 
The hypothetical scalable quantum computer does not save the situation because every its application to the simulation of quantum physics begins with the choice of discretization, e.g., requires the qubit representation of the wave function.

\subsection{Review of standard formalism for QED}
 
Quantum electrodynamics is the core of the whole quantum theory. It makes possible to reach numerical results that is the success criterion of any physical theory. However, we treat not the physical theory as the whole, but its mathematical formalism in respect to possibilities it gives. Namely, we are interested in the application of this formalism to the analysis of the dynamics of complex many particle systems.  Investigation of such systems always requires the application of computers for which we must represent QED in the qubit form. We stress that it does not depend on the degree of scalability of quantum computer: in any case it is necessary to apply the qubit representation. 

At the same time the formalism of wave functions in QED as we saw does not admit the complete qubit realization, because we can treat the main operator of emission - absorption of a photon by a charged particle independently of the step of discretization in the impulse-energy representation only. If we pass to the space-time representation, we obtain the dependence of this operator on the grain of space-time resolution. This is the main problem of the formalism of QED. 

The next peculiarity of the QED formalism is that we cannot find the amplitude of processes concerning more than one photon , for example, as shown at the picture, by the standard mathematical analysis, because it gives the divergent row. In means that the value of this amplitude (it is one of the elements of the matrix of the main process) depends on the grain of spatial resolution. We note that the necessity to compute to the number leads to the application of the special tricks of QED like the special methods of diagram summing or the change of a charge and mass of the particle depending on its environment called renormalization. We cannot consider these tricks as the part of QED formalism we are speaking about. In the formalism, these tricks have no serious status. However, just the method of renormalization and principles of summing of Feynman's diagrams contain the main value of QED, because these tricks only make possible to obtain the numerical values of physical magnitudes with the surprising accuracy!\footnote{The computation of the magnetic moment of electron serves as the good example.} Al the rest is the routine computation where we can apply the standard computer programs. 

This is the evidence of the serious defect of the mathematical formalism of QED, and of the whole quantum theory as well. The lack of integral formal apparatus for the most efficient of physical theories\footnote{QEDis the standard for the creation of the theory of nuclear forces and probably for the quantum gravity theory.} is the sure sign of the drawbacks in our knowledge about the microcosm. If we can somehow cope with this for the simple problems (for example, we can estimate Lamb shift of the atomic levels), then for the more complex systems, like chemical reactions these gaps turn fatal.

We thus can conclude: 

{\bf The problem of scalability for QED lies in the area of its mathematical apparatus, and its solution is only possible on the path of the modification of this apparatus.}

Further we see that the constructive mathematics gives us just the required mathematical apparatus. 
 
\section{Simulation of quantum systems}

We now turn to the problem of the simulation of quantum systems. Just this problem was the main motive for R. Feynman when he proposed the idea of a quantum computer.\footnote{Earlier we saw the universality of the states of QC in the many particle area.} We can describe a state of many particles system by the set of numbers containing the values of physical magnitudes as mass, coordinates, speed, time etc. These numbers (in contrast to amplitudes) are real. Moreover, if we limit the area and the resolution of measuring device we can assume that all of them have the form $\frac{l}{2^n}$, where $n$ is the number of the reasonable value. We can then encode a basic state of the system at hand as a basic vector of quantum memory of $n$ qubits. Correspondingly, a linear combination of basic states of the system we associate with the state of quantum computer with the same amplitudes. The qubits for our simulating quantum computer have the virtual character, e.g., we cannot ascribe to them any natural physical sense. However, in our simulating quantum computer these qubits are real. This approach to the description of physical systems we call "qubit" approach. We will see that this approach to the description of physics is more effective than the traditional "bit" approach, used for the numerical simulation of many particle processes of classical computers. For this, we try to solve on a quantum computer Shedinger equation. Again, here the key role belongs to quantum Fourier transform, but we will use the other its property, than earlier. This property is that Fourier transform change the operation of differentiation into the operation of multiplication to independent variable with imaginary unit. Hence, if we apply it to a wave function, the operator of double differentiation turns for the Fourier image of wave function into the multiplication to the new variable squared with some coefficient. This new variable is the impulse. This idea is well known to everybody who often solves Shredinger equation at hand, works well for a quantum computer. We must make sure only that quantum Fourier transform possesses the analogous property concerns to the differentiation (for our quantum computer the role of differentiation belongs to the finite difference). It follows from that QFT is the approximation of the continuous Fourier transform in the passage to the qubit representation of a wave function. In the Appendix is written how to realize QFT on a quantum computer.

Our aim is to obtain the state of our computer corresponding to the state of the system at hand in some time instant $t$. We will approximate by the working operators the action of the evolutionary operator $e^{-iHt/h}$ on the wave function $\psi_0$, where $H=H_p +H_q$, $H_p =\frac{p^2}{2m}$, $H_q =V(q)$, $p=\frac{h}{i}\frac{\partial}{\partial q}$ and the potential $V(q)$ is the real function. For the simplicity we take the time $t$ equal unit. It is easy to realize on the quantum computer the action of $H_q$. Since the matrix of this operator (and then $e^{iH_q}$) is the diagonal, we need only to change the phases depending on the form of basic states; we can do it as for the inversion of states of zeroes in Grover algorithm (see Appendix). This simple way is impossible for the second summand in the Hamiltonian. The difficulty is that the operator $H_p$ is not diagonal in the chosen "coordinate" basis. We already know what to do: we must pass to the "impulse" basis, by the application of Fourier transform, which we did earlier. For this we choose the small time interval $\D t$, and represent approximately our evolutionary operator by Trotter formula:
\begin{equation}
e^{-iH}\approx (e^{-iH_q \D t}\ e^{-iH_p \D t})^{1/\D t}.
\label{Ev}
\end{equation}
It is easy to get this formula by the expansion of exponential to the row. We choose the coordinate basis where $H_q$ has the diagonal form. We apply quantum Fourier transform: $\QFT :\ f\ar\int_{-\infty}^{+\infty}e^{-ipq}f(q)\ dq$ and its property to turn the differentiation $\partial /\partial\ q$ into the multiplication to $ip$, we represent the action of impulse part of the operator as $e^{-iH_p}=\FT^{-1}\ e^{-ip^2 \D t/2m}\ \FT$, where the middle operator has the diagonal form. Now the sequential applications of QFT and phase shift to $-p^2 /2m$ in the sequence (\ref{Ev}) give the required approximation. 

Fourier transform from this method is applied towards each coordinate independently, and in case of several particles on each coordinate of each particle separately. The realization of the unitary operator $e^{-iH_q}$ on a quantum computer represents some separate task depending on the form of the potential energy. If the potential energy equals $q$, we can realize this diagonal operator by the sequential turns of the phase of the form $|0\rangle\ar |0\rangle$, $|1\rangle\ar e^{i\phi}|1\rangle$, depending on the place of the qubit in the register containing the value of coordinate. In case of the arbitrary potential energy, we suppose that it is possible to decompose it to the Tailor row with the coefficients, which are defined by some fast algorithm. For example, if this potential is the sum of Coulomb potentials of $n$ different particles, this algorithm will have the linear complexity depending on $n$ provided the coordinates of particles are given by some fixed algorithm (which we can treat as the oracle). The operator $e^{-ihH_q}$ we can then realize in the form of quantum gate array of the linear size of $n$. 

The complexity of this method depending of the time $t$ of real physical system is $O(t^2)$. It follows immediately from that the accuracy of Trotter formula has the second order, since it results from the Tailor decomposition of exponential up to the first summand. We can decrease the accuracy to the value $O(t^{1+\e})$ for any $\e >0$, if instead of Trotter formula we use Tailor decomposition of the higher degrees (see \cite{Oz4}). 

We also note that analogously it is possible to simulate the unitary quantum dynamics of the moving charged particles in non relativistic approximation taking into account the electromagnetic field with the vector potential $A$. For this we have to replace the impulse operator $p$ of any particle by $p-\frac{e}{c}A$, where $e$ is its charge, $c$ is the speed of light. Tracing the reasoning from above, we see that it does not touch the final result. We not that this is not the consideration of quantum dynamics, but its non relativistic approximation only, for which we can account the field effects introducing this correction to the Hamiltonian. E.g., here we treat the field as classical that accordingly to the definition means that we can include it to the Hamiltonian as the potential energy, or as the addition to the impulse. In the general case, the field in many particle quantum systems is quantum as well, e.g., it breaks up into photons and it is possible to speak about the history of each photon. The substantiation of the possibility to consider the field of many particle system as classical is the advantage of many body Hilbert formalism;; we now show it, repeating the reasoning from the book \cite{Fe}. 

Let one electron emit a photon and the other electron absorb this photon (instead of electrons one can consider atoms, or atomic ions). The electron, emitting this photon meets the return (for example, it can change its spin). It means that the system creating the field (the first electron) changes and the emission of the next photon will be emitted in the different conditions. It means that this field is quantum, because the photons are distinguishable, because the corresponding states of electron will be different for them. We then must sum the probabilities of the emission of the first photon, the second photon, etc., for the probability of the emission of many photons. The resulting probability will be $n|\a |^2$ where $\a$ is amplitude of the emission of one photon (let for simplicity it is the same for each case), $n$ the total number of photons. We then cannot speak about the vector potential $A$ of the field. The field here is the set of the separate photons. We can differ them, measuring the state of the electrons, which emit them. 

Let us now consider the different situation: we have not a single atom or electron, but a whole peace of material. The return from the photon emission will be then distributed on the whole peace of many atoms, and we cannot detect which of them really emitted this photon! In this case we have to sum not the probabilities, but the amplitudes, because the peace behaves as a whole (see below, in the chapter 5). The phases of summed amplitudes will be close because we cannot detect the atom emitted the photon (we assume that the photon has some fixed impulse and polarization). This summing of amplitudes gives $n\a$, and the probability $n^2|\a |^2$ will be $n$ times more than the probability of the photon emission in the case of distinguishable emissions. Here we cannot distinguish the photons, and it makes possible to speak about the vector potential, which results from the summing of the wave functions of the different photons. This is just the summing of amplitudes, not probabilities, as in the first case. The field with the potential we call the classical field. We see that a field is classical if the big ensemble of particles situated in entangled state generates it. Only the set of entangled particles can create the classical field. Just this entanglement of the particles inside the source of the field causes the distribution of the return from the emitted photon among the atoms of the source. It makes possible to factorize the state of source and the photon state, where the common wave function of the system "source + photons" has the form $|\Psi_{source}\rangle\bigotimes (|\psi^1_{photon}\rangle+|\psi^2_{photon}\rangle+\ldots +|\psi^n_{photon}\rangle)$. In this case we can speak about the vector potential of the field, which equals $A=|\psi^1_{photon}\rangle+|\psi^2_{photon}\rangle+\ldots +|\psi^n_{photon}\rangle$. 

Given a separate charged particle, say, electron, we cannot simply speak about the classical potential it creates. For a separate electron we should use quantum electrodynamics, that is to consider the emission of separate photons, entangled states of these photons and this electron (or atom), and to obtain the field by this consideration only; we take up this in the chapter 5. For example, for the electron in the quantum dot this electron will be in entangled state with the particles forming this dot, that results in the redistribution of the return from the emitted photons, which ensures the classical character of the field created by the electron. 

The consideration of quantum computations on the separate particles (for example, on the separate electrons) by the standard quantum physics requires the care about the classical character of the field generated by them. Otherwise, it is impossible to count that there is a potential. In the practical sense in the simulating of quantum systems, we must regard as quantum all the particles contained in the model. The extension of the system be one by one. If we introduce the bound conditions as the external potential, we must guarantee that it is created by the great number of particles so that the condition of classical character is satisfied. 

We compare the quantum method of solution of Shredinger equation with the traditional method of finite differences. The method we have described above gives the exponential win in the memory, because for the differences method we must keep in memory all the wave function whereas the quantum memory permits to manage with only logarithmic number of qubits. The win in the time will be about this order. The more accurate analysis (see \cite{Oz4}) shows that the evolution on the time segment $t$ can be simulated quantumly in the time of the order $t^2$. It is important that with the growth of the number of particles in the simulated system ($q$ can denote the set of many coordinates) provided the accuracy is fixed, the number of the required qubits grows linearly. The effective simulation of many body systems on a quantum computer is thus possible. 

It is interesting that in some cases we can even to predict the state of the simulated system beyond the simulation of its evolution. In the other words, to simulate in advance, when to obtain the state corresponding to the state of real system in time instant $t$ we need the less time $t'<t$. This possibility appears in case when the Hamiltonian of the real system has the sparse spectrum, that is the eigen frequencies group near few typical frequencies. The corresponding method is described in the work \cite{Oz5}.

\section{Ensembles of identical fermions and bosons}

The single case when we can forsee some certain for complex systems by means of the wave function is the ensemble of identical particles: bosons or fermions. At first we consider the case of two identical particles. Their common state is given by the wave function of two variables: $r_1$ and $r_2$, which are the coordinates of the first and the second particle correspondingly. Since the particles are identical, we cannot distinguish them. The wave function must show the symmetry properties in the replacements $r_1$ to $r_2$ and vice versa. Because Pauli principle is valid for fermions, it is accepted, that its wave function is anti symmetric:
$$
\Psi_{fer} (r_1,r_2)=-\Psi_{fer} (r_2,r_1),
$$
for bosons, where there is no Paili principle, it must be symmetric:
$$
\Psi_{bos} (r_1,r_2)=-\Psi_{bos} (r_2,r_1).
$$
We suppose that each of separate particles can be in one of two states: $\psi_1(r)$ or $\psi_2(r)$, which are mutually orthogonal. What symmetric and anti symmetric combinations can we compose from these two functions? There are two such combinations. At first we can build he determinant and permanent of matrix on these functions and coordinates of the both particles:
$$
\begin{array}{ll}
\Psi_{fer} &=\psi_1(r_1)\psi_2(r_2)-\psi_1(r_2)\psi_2(r_1),\\
\Psi_{bos} &=\psi_1(r_1)\psi_2(r_2)+\psi_1(r_2)\psi_2(r_1).
\end{array}
$$
At second, we can consider the combinations:
$$
\begin{array}{ll}
\Phi_{fer} &=\psi_1(r_1)\psi_2(r_1)-\psi_1(r_2)\psi_2(r_2),\\
\Phi_{bos} &=\psi_1(r_1)\psi_2(r_1)+\psi_1(r_2)\psi_2(r_2).
\end{array}
$$
The first way is better because we obtain the formulas scalable to many particles. It is simply to see through the qubit representation.

We represent the one-particle states $\psi_1$ and $\psi_2$ in the qubit form as $\psi_1=\sum\limits_j \la^1_j |j\rangle ,$ $\psi_2=\sum\limits_j \la^2_j |j\rangle $. The determinant representation of two-particle state has the simple form $|\Psi\rangle =|\psi_1\rangle\bigotimes |\psi_2\rangle -|\psi_2\rangle\bigotimes |\psi_1\rangle$. With this definition of two particles wave function we have $|\Psi\rangle =\sum\limits_{j_1,j_2}(\la^1_{j_1}\la^2_{j_2} -\la^2_{j_1}\la^1_{j_2})|j^1_{j_1}\rangle\bigotimes | j^2_{j_2}\rangle$. It exactly corresponds to the relation between functional and qubit forms of quantum state notation, which we defined earlier. We can generalize this consideration to the case of $n$ identical fermions, when we have the determinant of matrix. The case of bosons is analogous; it differs only in that the permanents stand in place of determinants. 

For the second form of two particles wave function, there is no such correspondence. 

Of course, not all anti symmetric function we can represent in the form of determinant. However, we always can represent it as the row composed from such determinants. Determinants of the considered form we can treat as the basis of some Hilbert space, which we call Fock space of occupation numbers. In Appendix, we consider it in more details. 

One could treat the anti symmetry property as the real entanglement of many particles. Formally, it is right, since $\Psi$ is the entangled quantum state in Hilbert space. However, this entanglement bears the special character, and we illustrate it by the simple reasoning. Let us suppose that the basic states $|j\rangle$ are step like functions with the supports concentrated in the segments $[ x_j,x_{j+1}]$ of the configuration space. We consider two states $|\psi_1\rangle\bigotimes |\psi_2\rangle ,\ |\psi_2\rangle\bigotimes |\psi_1\rangle$, which form the determinant. We can slightly change the values of complex functions $\psi_1(r_1),\ \psi_2(r_2)$ so that their supports become non overlapping sets. The determinant representation  $\Psi$ is then reduced to the single summand, and represent the non entangled state. The entanglement resulted from the anti symmetry of the wave function is then degenerated. We can replace such entanglement by the requirement that the supports of wave function are none overlapping. This requirement expresses Pauli principle.

\section{Spin and spatial coordinates}

The main peculiarity of Hilbert formalism in the area of many particles consists in the exponential growth of the space of states dimensionality. It is the cause of the quantum computer project. On the other hand, this peculiarity makes Hilbert formalism inconvenient for systems with many particles. It reveals in the absence of the good model of decoherence, e.g. of the quantum state decay. However, there are the other difficulties of Hilbert formalism, which reveal in many body systems. There touch the passage from the ordinary representation of wave functions to the qubit representation, and they reveal in the choice of the space of states as it is. We have already met them in the consideration of entanglement in the space of occupation numbers. Here we consider the other example of the same type.

This example concerns the division of the configuration space to spin and spatial variables. Corresponding to the traditional formalism, the configuration space for one particle is divided to two non-overlapping parts: coordinates and spin. It means that the state of ensemble of particles with spins has the form 
\begin{equation}
|\Psi\rangle =\sum\limits_{j_c,j_s}\la_{j_c,j_s}|j_c\rangle\bigotimes |j_s\rangle
\end{equation}
where $|j_c\rangle$ and $|j_s\rangle$ are coordinate and spin basic vectors. Let in the initial instant spins and coordinates be non-entangled: 
$|\Psi\rangle_{ini}^{nonent}=|\Psi\rangle_{ini}^c\bigotimes |\Psi\rangle_{ini}^s$, where  $|\Psi\rangle_{ini}^c,\ \  |\Psi\rangle_{ini}^s$ - are the coordinate and spin wave functions. Then the application of the operator acting to coordinates of particles only leaves spin state unchanged. 

Now we consider the following practical situation. Let we be given a magnetic peace of metal, which we keep in the stable position relatively to some inertial reference frame. We begin to turn it so that it occupies the other position. The direction of magnetic field induced by this peace will change correspondingly to its new position. We try to analyze this simplest experiment from the viewpoint of Hilbert formalism. We suppose that this formalism is completely scalable, that is it is applicable to the random numbers of particles so that it conserves all classical limits. We will account only coordinates of all nuclei and electrons, which form this peace, and electron spins (without nuclear spins). The state of spin wave function of all electrons determines the direction of classical magnetic field. In the magnetic material in the rough approximation, we can treat spin and coordinate parts of the wave function non-entangled. It means that we have the quantum state of the form $|\Psi\rangle_{ini}^{nonent}$ in which spin wave function  $ |\Psi\rangle_{ini}^s$ determines the direction of classical magnetic field induced by the peace.  

It is naturally to accept, that the mechanical turn of the peace touches the spatial coordinates of all particles only, hence, the spin remains unchanged. If quantum standard formalism is completely scalable, the turn must preserve the magnetic field induced by the peace that contradicts to the experiment. We thus conclude that the standard formalism cannot be scalable. If we consider the single electron, its spin will be unchanged when it moves inside of the atom. For the single electron the division of spatial and spin coordinate takes place. Nevertheless, for the large ensembles it is wrong. The standard formalism is not applicable to the big peaces of material.\footnote{The situation is standard. One could accept that quantum physics is not applicable here. However, this answer cannot satisfy us because we take up not the standard quantum theory, but its extension to the complex systems. Our tool is the computer. The program must correctly process all exceptions. Where the traditional physicist refers to the tradition and intuition, we must only apply the algorithm. Hence, we must explicitly determine how to act in all cases, not only for the narrow range of traditional problems.} Further, we discuss this example from the positions of constructivism.

\section{Problem of decoherence or why to reconsider the basement of quantum theory}

We return to our main subject about the adequateness of the mathematical apparatus of quantum theory. From the previous paragraph, we see that there is the principal difference between two types of the evolution of quantum systems: unitary evolutions and measurements. The procedure of the measurement has no description in terms of standard quantum formalism, but Born postulate containing only the rule of computation of the quantum probability. 

We thus can look at the measurement in two ways: we can treat it as the kind of friction, which acts on the considered system from the environment, or otherwise, count that the ability to fulfill measurements is the internal property of quantum system itself. In the first case we can in principle remove measurements at all, or fulfill them at our own discretion, because the friction is removable, for example, by the creation of the deep vacuum. In the second case we should refuse from the standard formalism of quantum theory as from the non appropriate, because it does not reflect some fundamental property of quantum systems. 

Experiments carried on quantum systems aimed to the investigation of quantum processors, witness that decoherence could hardly be treated as the sort of friction. The entangled states detected in the experiments on ions in Paul traps serve the evidences of it. These experiments show the decay of quantum states despite of the appearance of states of the forms W and GHZ. There are just the states with the good description in the collective behavior method (see chapter 5). 

The difficulties with decoherence lead us to the problem of the applicability of quantum theory to many particle systems, because decoherence turns to be the fundamental property of such systems. There is no criterion in quantum theory for the distinguishing between the external potential (in Hamiltonian) and the so-called harmful influence of environment to the system inducing decoherence. 

The area of applications of quantum theory is thus strictly limited by the problems reducible to one particle. The example is quantum chemistry with the problems of finding the stationary states of electrons in the field of many Coulomb centers (see \cite{Novo}). The problems, where the quantum behavior of many particles is important, as for example, chemical reactions, lie besides the frameworks of applicability of the standard quantum theory.\footnote{The attempts to create semi empirical methods of Landau-Ziner type do not change this thesis.} We must stress that this limitation is principal and cannot be changed by the addition of some new parts to quantum theory. The obstacle for the expansion of this theory to complex systems is not its imaginary incompleteness, it is complete. The main and the single obstacle is its mathematical apparatus. The application of standard analysis is constructive in its computational part only, e.g., when we use integration and differentiation, and solve differential equations. As for the ideology of the actual infinity contained in standard analysis, in the form of the unlimited extrapolation of formulas is not adequate to the real nature. This is the irremovable defect of standard quantum mechanics. 

Decoherence is not a friction. It is the evidence of the inaccuracy in the description of real systems following from the standard formalism. The so-called contra intuitive character of quantum mechanics follows from this drawback. It is the rare success that this rough apparatus gives the good description of simple systems, like the electron states in hydrogen atom. Undoubted success of quantum theory in such cases results from the power of analytical technique, and it is not the merit of the ideology of standard analysis with its actual infinities. When we pass to slightly more complex systems the visible simplicity of quantum theory turns to the insurmountable obstacle in any computations at all. The states of hydrogen molecule with the complete account of quantum character of the movement of nucleus and electrons, e.g., the consideration of the states of this object in terms of compelte wave function $|\Psi (r_1,r_2)\rangle$, where $r_1$ and $r_2$ are the coordinates of protons and electrons have no good exact description as the states of hydrogen atom. Even the attempts to describe the states of real hydrogen atom with the possibility of quantum smearing of its complete wave function and the computational difficulties arising if one tries to reach numerical results, reduce to zero the beauty and power of quantum theory. 

We consider the following model problem in which the numerical answer cannot be obtained even approximately. It is the problem of dissociation of molecular hydrogen ion. In consists of two positive charged protons and one electron in the Coulomb field of two centers. If $r$ is the distance netween them the main state depends on $r$. The potential energy of the system of two protons provided we treat the electron as the bounding force, has the form shown at the picture. It means that the ground state of electron is bounding for all distance between the protons. Now we assume that we give the initial speed to the protons sufficient to overcome the potential barrier. The problem is to determine in which instant the electron transforms from the ground state in the field of two protons to some stationary state near some of them. Of course, there are the different channels: electron can pass to the state of continuous spectrum; protons are quantum as well, etc. We can concretize this model problem, assuming that, for example, the electron passes to the ground state near one of the protons. 

\begin{figure}
\centering
\caption{Potential energy of two protons in $H_2^+$.}
\vspace{100mm}
\makebox[180mm][l]{\includegraphics{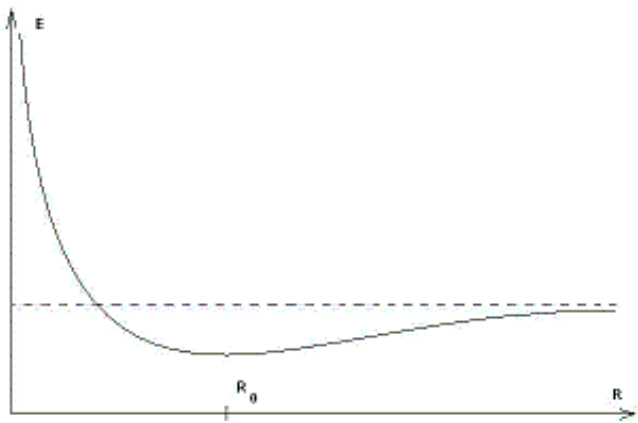}}%
\end{figure}

\begin{figure}
\centering
\caption{Dissociation of the molecular hydroden ion}
\vspace{100mm}
\makebox[180mm][l]{\includegraphics{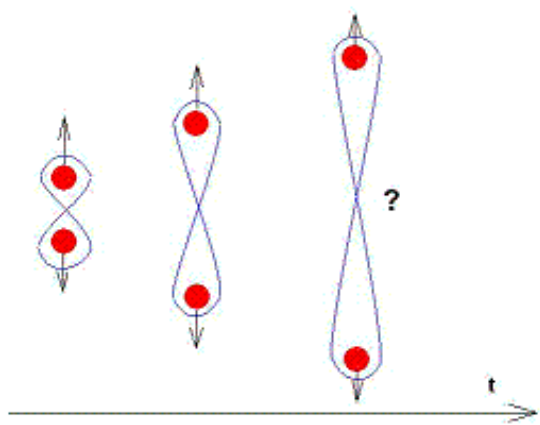}}%
\end{figure}

This is the problem about the decay of the ground state in two potential holes, and it in principal cannot be solved by means of quantum theory. The cause is not the requirement to use electrodynamics (in QED this difficulty would grow many times). The cause is that here we deal with the collapse of wave function, e.g., with the procedure of measurement. This procedure lies besides the frameworks of quantum mechanics, and we can do nothing with it. 

The basement of Copenhagen quantum theory is the wave function of many particles $\Psi(r_1,r_2,\ldots,r_n)$. This object from the formal view point is very hard. It directly presumes the usage of Riemann scheme of integration of Shredinger equation, that is the processing of all the configuration space for $n$ particles, that is inacceptable for the complex systems. Only Born rule for quantum probabilities represents the connection of wave function with the reality, and this rule cannot be checked immediately in the many particle case, because the amplitude turns to be smeared among incredible quantity of states. We meet the necessity to concentrate amplitude on the so-called substantial states and ignore all the rest, just by this reduction all results about many particles are obtained in quantum mechanics (for example, Feynman's reasoning about the classical character of the field created by the peace of material). In the general case the concentration of amplitude is given by quantum computer, which in turns becomes a hostage of the fundamental property of decoherence. The wave function of Copenhagen quantum theory turns to the object which verification is doubtful. We have no a reliable method of the checking of the existence of such wave function for many particles. It is impossible to build the quantum physics of many particles on such a shady object. We must get rid of it.

If all is limited by the difficulties in the obtaining of exact estimations, it would not be so critical. However, the quantum formalism is such that the main scenarios determining the behavior of complex systems, like chemical reactions, completely depend just on the accuracy of these estimations. We thus deadlock: for the description of complex systems, we need a scalable quantum computer, whereas its building in turn, meets decoherence as the main obstacle.

This is one more argument for the reconsideration of quantum theory on the basement of constructivism. The essence of it is that instead of the standard mathematical analysis we must apply the constructive mathematical analysis. Just constructive functions represent the adequate description of physical processes in the many body systems, because they require the explicit determining of the technique of the concrete computations that is ignored in the standard quantum mechanics. 

This rebuilding of quantum theory represents the hard but necessary aim. Its significance follows from the impossibility to ensure otherwise the exact analysis of many particle quantum scenarios with the separation of the role of individual particles. The further chapters show how this rebuilding can look.

\newpage
\section{Resume of standard quantum mechanics}
\ \ \ \ \ \ \ \ {\it "I bet that the superposition principle will stay till the end of time."}

\ \ \ \ \ \ \ \ \ \ \ \ \ \ \ \ \ \ \ \ \ \ \ \ \ \ \ \ \ \ \ \ \ \ \ \ \ \ \ \ \ \ \ \ \ \ \ \ \ \ \ \ \ \ \ \ \ \ \ \ \ \ \ \ \ \ \ \ \ \ \ \ \ \ \ \ \ \ \  \ \ \ \ {\it R. Feynman}
\nnn

What must remain in quantum theory after the unavoidable change of its mathematical apparatus? 

The short answer is that all its results checked in experiments must remain. It concerns one particle problems which solution rests on the superposition principle, e.g., on the interference of amplitudes. It does not mean that it is obligatory to use for the computations just the technique of the solution of differential equations and summing of diagrams from the computational arsenal of quantum theory. Conversely, where this technique requires the inacceptable computational resources we must change it by the more economy methods. For example, it may be the method of collective behavior, which is able to change Shredinger equation, and the genetic method of the building of quantum scenarios for the scattering problems. 

In any case, the application of new computational tricks must give the same asymptotic results than the standard methods where the last are applicable and give the concrete results. For example, the full correspondence must be for the stationary states of several electrons with spins in the filed of many Coulomb centers. The converse example, where the standard methods do not give the results is a quantum computing. Constructive physics must not simulate the work of an ideal scalable quantum computer without decoherence, because the standard quantum mechanics cannot do it as well. The project of a scalable quantum computer is the project of just standard quantum physics, and its realizability is not yet confirmed by experiments. Therefore, the constructive physics will not give us the solution of factoring problem by Shor algorithm. However, the constructive physics must answer what will happen if we try to realize practically quantum computer, say, on ion traps. The realization we will explicitly see how namely the influence of decoherence destroys quantum computation, because decoherence must be the inbuilt property of the simulating algorithms used in constructive quantum theory. 

At last, the many particle Hilbert formalism must certainly survive. It will only acquire the special form determined by the method of collective behavior (see the next chapter). The point is that Hilbert formalism is the completely constructive methodology for the consideration of many particle systems, which applications do not depend on the sort of analysis we use: traditional or constructive. This methodology is inseparably linked with the computational methods like the differentiation and integration, about which we spoke above and which simply will not work without Hilbert formalism.

Hence, the main intrigue of quantum theory completely remains, that is the many world interpretation and the possibility of the mutual influence of these "worlds" to each other. Still this intrigue is resolved partially, by the described below genetic method of the building of dynamical scenarios, and partially, by the presence of a user in our models, who has the rights of expert. This solution in contrast to the other approaches is completely constructive. However, it does not contradict to the influence to the experiment of the volutional impulse of experimentator; only such impulses must be always treated through the inclusion in the system at hand the particles connected with the experimentator. The potential realizability of this inclusion results from the fundamental principle of constructive physics requiring no more than the linear growth of the computational resources for the simulation. 

The preserving of Hilbert formalism does not mean the obligatory using in the simulation of the wave function for $n$ particles of the form
$$
|\Psi(r_1,r_2,\ldots,r_n)\rangle .
$$
The usage of such function, as we saw, leads to the unacceptable growth of computational resources during the simulation. Therefore, the more practical methods must replace the usage of such objects, for example, the method of collective behavior, presuming the grouping of particles to corteges and the application of Lebesgue scheme of integration instead of Riemann. 

Resuming the standard quantum theory, we declare that it must survive as the whole, but the part of it, inseparable from the abstraction of potential realizability, which thus cannot be constructive. I know only one such construction in quantum theory: a scalable quantum computer. This project cannot transform into constructive form and thus we must sacrifice it to the planned reformation of the theory. I stress, that the matter concerns just an unlimitly scalable quantum computer. The limited versions of QC already work in reality, and we hope to obtain the advantage from the usage of these processors in the simulation of physics (see the works \cite{Va},\cite{Va2}, \cite{VK}). Moreover, the experiments on quantum computing represent the necessary foundation of the renewed quantum theory, because they are able to clarify the principle question about the computational resources required for the simulation of complex systems of many particles. 
 
\chapter{Algorithmic modification of quantum theory}

We saw that the constructive mathematics appeals to experiments for the checking of its axioms in more degree than the classical mathematics. This is why it is more dependent on physical experiments. It shows, for example, as the property of pluralism, which is the feature just of constructive mathematics. Logic evaluations in constructivism are not so deterministic as in classical mathematics, we made certain of it by the example of Kripke models. The classical mathematics contains the absolute truth: given an axiomatic theory and a concrete model (an algebraic system) for any formula, it will be either true, or false. It is not the case in constructivism. Though constructive system of logical deduction is of the same reliability as classical, it contains more restrictions than classical deduction. The restrictions of constructivism follow from its procedure character, and the assumption of these restrictions compels us refuse from the classical understanding of the notion of "absolute truth", which role is expressed by Goedel theorem of completeness for the predicate calculus. The attempt to build the analogue of Goedel theorem for constructive logic leads to Kripke model, e.g., to that pluralism of true values, which is the gender sign of the constructivism.

As applied to physics it means the serious conclusion. Constructive mathematics merely cannot exist without its physical part. The attraction of physical experiments in constructivism is mandatory for the preserving of its integrity. If the classical mathematics can exist in the so-called pure form e.g., in the form free of any (including physical) applications, this is impossible for the constructive mathematics. In is created essentially for one aim: to serve as the formal apparatus for physics. This is why the algorithmic version of physical theories, and first, the most advanced from them: quantum theory is the integral part of constructivism. 

The constructivism in quantum theory requires the certain heuristic for the building of simulating algorithms. In this chapter, we investigate such heuristics, based on the notion of point wise particle: the heuristic of the collective behavior. This approach, strictly speaking, is not the single possible for the physical constructivism.\footnote{For example, one could try to build the simulating programs starting from the notion of wave functions for one or two particles.} The advantage of heuristic of point wise particles is that it is vivid and thus is widely used in the other branches of natural knowledge. In the physics of microcosm, the real particles behave not as classical point wise particles and due to this the wrong opinion that the heuristic of particles must be replaced by the heuristic of wave functions and operators became widely spread. \footnote{Instead of the senseless polemics with the supporters of this opinion I can propose to them to describe the process of DNA replication by the language of wave functions.} 
In reality, if we only want to go forward we need the quantum heuristic on the language of particles, but not of wave functions, because wave functions are not applicable for systems with many real particles.

The keynote of the heuristic of particles is the representation of a real particle as the ensemble of point wise imaginary particles, called its samples. This is the idea of the method of collective behavior. We show how to reproduce the conclusions of standard quantum formalism via this method. I note that even here our approach is not the single possible. For example, if we preserve the complex numbers in the basement of formalism, we could make our work as so easier using Feynman path integrals. However, this way is already traced below. It exactly corresponds to the standard quantum formalism. Its drawback is the absence of the individual histories of the samples that sharply weaken the heuristic: it becomes inapplicable to many particle systems. The presence of individual histories of all samples is the element of heuristic important for its scalability.\footnote{It is also concerns the possibility of the genetic memory of particles, admissible in constructivism, that is discussed below.}. The main conclusion is: the constructivism in quantum theory can be fruitful only if it is sufficiently radical.

Practical question in the physical constructivism is the algorithmization of quantum mechanics. We assume without discussions that the right description of the elementary reactions with $n$ particles is sufficient for the building of actual models of the complex processes, including the simple forms of the life, as viruses and bacteria. Namely, the routine generalization of the scattering processes with many particles gives us the picture of behavior of very complex objects. 

We saw in the previous chapter that the effective classical algorithms could not simulate Hilbert formalism of quantum theory. Nevertheless, the impressive successes in the application of computers in quantum physics prove the perspectiveness of the simulation of real processes. This brings the new viewpoint to decoherence - it corrupts the unitary evolution, but at the same time makes the states simpler, and thus facilitates the computer simulation. 

Decoherence, as we saw above is not a part of the "bible" of standard quantum theory and physicists regard it as the kind of friction, or outside influence corrupting the ideal picture. Our treatment of decoherence is completely different. The divergence of quantum evolutions from the unitarity, from the algorithmic viewpoint, results from the lack of classical memory for the placing of complex quantum states. It can look unusually, but one should understand that we give to algorithms the status of the element of new mathematical apparatus, whereas the description of any physical phenomena given by the apparatus is exhaustive and does not anticipate the search of any more fundamental "mechanisms".

The principle of correspondence formulated by Bohr for the relation between quantum and classical mechanics has the certain sense for the relation between standard quantum theory and constructive quantum theory. This principle of correspondence, reformulated for our case acquires the following form. 

{\bf In the algorithmic simulation the resources of computer must be distributed for the achievement of the best approximation to the standard formalism of Hilbert spaces.}

It means that in case of potentially unlimited computational resources the algorithmic description turns to the standard formalism of Hilbert spaces. 

This principle of correspondence for constructivism has one feature: it is unclear, which approximation to standard formalism is the best in case of complex systems? (For one particle, we have the possibility to compare the found numerical values with experimental data.) The solution of the question about the quality of approximation is the prerogative of a user; he solves it in form of an expert estimation. We will see that in the wide class of problems is reducible to the choice between two answers: "yes" and "no", e.g., can be exactly formalized. It will be the practical advantage of constructivism over the standard approach where decoherence has no first principle model, because the status of user is there undefined. 

\section{About physical sense of algebraic operations}

Conctructivization of quantum theory presumes the passage from the mathematical constructivism to the real work on the transformation of quantum theory. We sequentially approach to this aim but have to devote still some attention to the general questions without which the effective practical work is senseless. We have already seen the fallaciousness of the standard formalism, based on analysis, and we have the alternative: constructive analysis based on algorithms and approximations. However, this is too less. We need the heuristic of the building of algorithms without which the definition of Turing machines and general theorems are the useless toys. We still have no such heuristic but we can build it accounting the existing heuristic of standard algebra. We have to analyze the sense of algebraic operations more carefully in comparison to their meanings accepted in standard physics. The heuristic outlook that we are doing will help us to understand the method of collective behavior. The reader must be ready to meet the divergence between the algebraic technique and our problems, and our aim is not to describe this technique, but to change it towards our subjects. 

What is the nature of a number? This is a quantity of objects belonging to the same type, and containing in some set $M$. We call these objects samples, and treat the set $M$ as a sack. Imagine a human who does not know arithmetic, and who needs to operate with quantities. If he join two sets it results in the sum of numbers of samples. There are two ways of summing quantities. We can move them from one sack to the other sack one by one. Alternatively, we can pour the samples from one sack to the other simultaneously. Now it does not play a role. However, the subtraction meets the certain difficulty: we can fail the samples. Here we can make this: get some storage of such samples and add them into the sack in which there is the shortage of samples, simultaneously memorizing where and how many samples we added. 

It bring the serious problem about how we will record the quantities. Without the subtraction, all was simple: we stores samples in sacks and took them out when needed. Now we must encode the quantities. The simplest way here is: put lines in the raw, $|$ means one sample, $||$ means two samples, $|||$ three samples, etc. If we need to encode twelve samples, we write the figure $||||||||||||$. Here one could guess that this is too hard way. We can act easier: agree to represent some quantity of samples, say ten, as one sample of new sort, and represent this number as ten plus two. Then the notation wil be shorter: $I ||$, where $I$ denotes this new sample. Continuing improvements in this style we quickly reach the arithmetic notations of quantities, e.g., numbers. It is easy to understand that the gain from such modernization will be exponential in memory as well as in time. Now to sum we do not need to search through all the samples in the sacks. To apply the rule for addition of numbers would suffice. 

How could we represent the operation of multiplication? We can define it as in is done in algorithm theory, e.g., by the primitive recursion. At first, we explain it on the operation of addition. The basis is $a+0=a$. The step of recursion is simple: $a+(x+1)=(a+x)+1$. The elementary operation of recursion is the addition of unit. Because of $a+x$ at this step is defined we can define $a+(x+1)$ by the given formula. 

Analogously, having the addition we can define the multiplication. Here the recursion looks as follows. {\bf Basis}. $a0=0$. {\bf Step}. $a(x+1)=ax+a$. As we see, all is easy! If we remember the method of multiplication of numbers, it requires of the order $O(n^2)$ operations on Turing machines, where $n$ is the length of the notation of numbers. The quantity $N$ of the samples is connected with the length of arithmetic notation as $N=10^n$, hence we conclude that the operations on quantities in the arithmetic notation gives exponential gain in comparison with the operations with sucks. 

However, there is no free of charge. For this quickness, we have to pay. Here we pay by the individuality of samples. If there are the equivalent coins in the sacks, this method fits. If there are potatoes, which slightly differs one from another, the arithmetic method is also not bed, because we can treat them as approximately equal. The difference between samples becomes principal if there are living things, each of which possesses its own individuality. It determines the border beyond which algebraic tricks cannot be efficient. Nevertheless, the fundamentality of physical laws means that they are equally concern the living world. This is especially actual for us because our aim is the simulation of complex systems. Therefore we owe to trace how algebraic operations looks as applied to such delicate objects which must preserve their individuality. Of course, it in turn requires some pay. 

We have already seen that the summing of samples in sacks does not cause any problem. What we have to do with the multiplication? We can act accordingly to the recursive definition. It means that we take in the order each sample from the first sack, and fulfill the following procedure: in the order take for it the new sample from the storage, and after all sum all quantities. We then preserve the individual memory about each of samples and can transmit it at each step from the initial samples to the new. However, it is not easy to understand that the building of Egyptian pyramids would be more productive activity than this way. Hence, the recursive methods, which serve in standard mathematics as the definitions of algebraic operations, are good for these aims, but cannot work in practice. 

We have reached the moment when we have to sacrifice something. If earlier we had the full correspondence between quantities and numbers, now we must violate the strictness of this correspondence. Further, this will have the other manifestations, but we fix it just for the multiplication as the main algebraic operation, because the simulation of physics is not possible without it. 

How do we represent the multiplication? We agree, that the quantity resulted from the multiplication of two real quantities is not as real as they are, but represents the indication to some virtual process. Let we have to multiply the quantities of samples in two sets $A$ and $B$. We represent the step of this virtual process as the forming of the pairs of the form $a,b$, where $a\in A$, $b\in B$. There are exactly $|A|\ |B|$ ways of such forming. We will not enumerate them all, but content ourselves with only one representative from this set. Still leave apart the way to choose this representative; it is important that it is only one, because it gives us the huge economy comparatively to the method of Egyptian pyramids. We can fix only one pair $a,b$. But without big loss in memory we also can fix some set of pairs $a,b$ such that each $a$ and each $b$ occurs only in the single pair. We thus fulfill the multiplication of quantities. 

Let us go further and look how we can introduce the notion of complex number. Here we have to pay yet more than for multiplication because the operations on complex numbers are not so easy than on real numbers. The simplest way here is the attraction of physical notions. The phase of complex number is connected with the impulse of quantum one dimension particle by the simple rule: the wave function for impulse $p$ has the form $exp\ (ipx/h)$. We thus can write that the phase of wave function $\phi$ is connected with impulse by the equality
$$
\frac{d\phi}{dx}=\frac{p}{h}.
$$
The physical sense of the phase is thus that the speed of its change along some direction is proportional to the projection of impulse to this direction. Only relative phase has thus the sense, and the phase changes with the speed with which the particle moves along this direction. We thus need to determine somehow the spatial positions of the samples. It can be done within the common shift, e.g., only the relative spatial positions can be established. Keeping in mind that the speeds of samples are proportional to the change of the phases we can do the following trick. We change the sign of a phase to the opposite that gives the change of the speed to the opposite, and then consider the collision of samples in the imaginary space. If we agree that this procedure makes possible to find the positions of samples, then their density we must define as $\bar f\ f$, where $f$ is the complex number corresponding to one sample, the line denotes complex conjugation. Really, if we associate with each sample $j$ some complex number $f=p\ e^{i\phi_j}$ with the same module $p$ and individual phase $\phi_j$, then the change of speed can be caused by the complex conjugation. 
The collisions of samples can be thus caused by the change of the speed of one of them to the opposite and by the coupling of these samples. This procedure can give us the rough way to determine the spatial positions. Here we must compute the density of samples not as their quantity, but by peculiar procedure expressed by the production of complex numbers $\bar f\ f$.

There is the different way to define the speeds of samples. We can regard the point of the imaginary space in which a sample will be in the fixed time frame, and suppose that there is already some other sample there. The determining of speed means the fixation of some pairs of samples $a_1,\ a_2$.

It brings the conclusions. The multiplication of quantities arises in connection with the presence of speeds of samples. The multiplication is represented as the forming of pairs of samples. The multiplication determines the relative spatial positions of samples. These conclusions are still uncertain, but we will specify them further. Nevertheless, just these conclusions lie in the basement of collective behavior heuristic. 

\section{Amplitude quanta and Born rule}

Many equations of mathematical physics describing the dynamics of classical systems (heat transfer, oscillations, diffusion) arise in the limit process in the problems about the dynamics of huge number of small bodies, or quanta of matter. Correspondingly, the area of applications of such equations is limited by the size of these small bodies. These equations follow from the more fundamental laws, or mechanisms of interaction between the small bodies (for example, for the equation of oscillation it is Guk law). In QED this mechanism is determined by Feynman diagrams for fundamental processes (see \cite{Fe}). Diagrams of fundamental processes help to make QED constructive, but do not give the good starting point for building algorithms, because they admit the operations with infinitesimals. Algorithmic approach means the complete transfer to the operations with finite quantities\footnote{Illusion of some that if would restrict the capability of the theory is wide spread and has the same origin as the "algorithm phobia". Nobody can operate with the infinite values, as well as with the non-computable procedures. The question is only in the inaccessibility of some elements of such procedures (as the administrative parts of the quantum model). As for the internal beauty - in the world of algorithms it is not less than in the part of this world represented by formulas. There is one principal advantage of algorithms: the possibility of visualization. Mathematics, which in principle does not allow visualization, can rest on the deductive method only, but it is not reliable basement because of Gedel theorem of incompleteness. The reckless using of such mathematics is similar to the walk on thin ice.}. The method of collective behavior gives the concrete way to the algorithmic form of theory. We now take up on eside of this method: amplitude quanta. 

We now derive Born rule from the concept of amplitude quanta (see \cite{Oz1})\footnote{The similar reasoning are in the work \cite{Zu}, but we will obtain Born rule directly from the algorithmic concept without additional suppositions.}.

The consideration of quantum evolution from the viewpoint of Hilbert formalism for many particles gives the states of the form

\begin{equation}
|\Psi\rangle=\sum\limits_j\la_j|e_j\rangle,
\end{equation}

where the summing goes on the infinity set of basic states of the system $|e_j\rangle$. Algorithmic approach reduces this row to the finite sum, which arises if we omit all summands with coefficients $\la_j$, which modules are less than some fixed border $\e>0$. Such a sum contains no more than $1/\e^2$ summands. Let $N$ be the number of basic states for one particles. We then can treat that $\e=\frac{1}{\sqrt{N}}$. The state will thus have the form

\begin{equation}
|\Psi\rangle=\sum\limits_{j=1}^N\la_j|e_j\rangle,
\label{limstate}
\end{equation}
where some of summands can be zeroes. 

We call the (algorithmic) reduction the elimination of all summands which amplitudes modulo are less than $\e$, and the following renormalization of the quantum state. This constant we call the amplitude quantum. Without special mentioning we will fulfill the reduction of any quantum state in our model. The states obtained by the reduction are called admissible states. 

We now demonstrate how the reduction procedure consisting of the elimination of small amplitudes leads to Born rule for the quantum probability. Our aim is to transform the coomputation of probability to obtain the certain basic state $A$ in the measurement of initial quantum state $\Psi$ to the application of classical rule of probability theory 
$$
p(A)=\frac{N_{suc}}{N_{tot}}
$$
where $N_{suc}$ is the number of successful elementary events (e.g., events, for which the target event $A$ is realizable) to the number of all elementary events $N_{tot}$. This rule is called the classical urn scheme. We have to define the set of elementary events and establish the correspondence between elementary events and basic states of the measured system. We call elementary events those states of the extended system (the measured system and the measuring device) which module of amplitudes in the current quantum state equal amplitude quantum $\e$. The set of elementary events will thus depend on the quantum state of extended system.

We denote by $|\Psi_j\rangle$ basic states of the measured system, by $|\Phi_j\rangle$ basic states of the measuring device (which can be the user's eye), and obtain after the contact of these two objects the state of the form 

\begin{equation}
\sum\limits_j\la_j|\Psi_j\rangle\bigotimes |\Phi_j\rangle
\label{meas}
\end{equation}

Due to the big mass of the measuring device in comparison with the measured object, the attempt to describe its quantum state requires the division of all states from (\ref{meas}) to the sum of $l_j$ basic states (we need to account, for example, the states of all nuclei, electrons, atoms and molecules in the measuring device). It means that even in the very contact we had the state $|\Phi_j\rangle$, the evolution quickly gives instead of it the state $|\Phi'_j\rangle =\sum\limits_{k=1}^{l_j}\mu_{j,k}|\phi_{j,k}\rangle$, where the values $l_j$ will rapidly grow up to the instant when modules of amplitudes reach the value of amplitude quantum $\e$, after that they will be nulled. Therefore, all modules of amplitudes $\mu_{j,k}$ are approximately theh same. Substituting the espression for $|\Phi'_j\rangle$ instead of $|\Phi_j\rangle$ to the expression (\ref{meas}), we find that the amplitudes of states $\phi_{j,k}$ equal $\frac{\la_j}{\sqrt{l_j}}$ due to the unitarity of quantum evolution. 

We owe to fulfill the reduction, that is to omit the states $\phi_{j,k}$ which amplitudes are small. Since the time frame when the considered division of state to the huge number of summands goes is very small, in the computations it factually means that we split every summand in (\ref{meas}) into $l_j$ new summands so that all the arisen amplitudes are close to the amplitude quantum and approximately equal. This makes all the states equitable on the eve of reduction, which makes possible to use classical urn scheme we have defined above. Here the quantity $l_j$ of summands with the first multiplier $|\Psi_j\rangle$, which is the total number of favorable events will be proportional to $|\la_j|^2$, and if in the reduction only one state survives we obtain exactly Born rule for quantum probability.

The probability space of elementary events is thus determined by the wave function $|\Psi\rangle$. We consider factually the conditioned probabilities to obtain in the measurement one or the other result provided the system is in state $|\Psi\rangle$. 

This deduction of Born rule of the computation of quantum probabilities rests on our definition of algorithmic reduction of quantum state as the elimination of small amplitudes. Such reduction must be fulfilled at each step of the simulation of unitary evolution merely because without this reduction the simulation becomes impossible. This form of representation of quantum dynamics differs from unitary evolution only quantitatively: the measurement happens in the moment when the considered system comes into contact with the massive object which we can call the environment, and this contact leads to the splitting of summands in (\ref{meas}) into many new summands. Besides this natural supposition, we use the normalizing of wave function which conservation follows from Shedinger equation. Our explanation of Born rule thus uses nothing outside the frameworks of standard quantum mechanics but the reduction of state vector that is the elimination of too small amplitudes. We note that just this procedure of reduction turns the set of Feynman paths to the classical trajectory for a massive body. (see \cite{FH}). 
We treat decoherence as the forming of entangled states of the form (\ref{meas}) with the environment, e.g., do not make any difference between it and the measurement of the considered system. Born rule and the irreversible corruption of quantum state in decoherence are thus the corollaries of the grain amplitude. 

At last, we note that in the deduction of Born rule we used only the entanglement of the states of the form (\ref{meas}), which belongs to Schmidt type, e.g., is the natural generalization of the entanglement of EPR type.  

The fact that the main axion connecting quantum mechanics with classical - Born rule has the simple treatment in terms of amplitude quanta, is not casual. Below we show the concretization of the amplitude quanta method: the method of collective behavior, which makes possible to build the dynamical picture of quantum evolution practically. 

\section{Absolute model of decoherence}

Algorithmic approach proposes the following model of decoherence:

{\bf Decoherence is the divergence of the real evolution of quantum $n$ particle system from unitary solution of Shredinger equation resulted from the lack of classical memory for the storage of the states of this solution in Hilbert space {\cal H} for $n$ particles.}

Decoherence of quantum states is thus not a kind of friction as in standard quantum theory. Decoherence is the property peculiar to quantum dynamics from the very beginning. It is the measure of divergence of this dynamics from the "ideal" unitary evolution, which is expressed by algorithmic formalism itself: the requirement of the existence of the classical effective algorithm determining the real evolution of quantum system.

This algorithmic model of decoherence is too general to be applied to a concrete situation. We will specify it taking as the basement the constructive mathematical analysis. In it every function $f(x)$ is a constructive function of constructive real variable. It means that speaking about a function we keep in mind that for each approximation of the value of its argument, knowing only this approximation we can effectively build the approximation of the value of function on this argument. Given the more exact approximation of argument, we must give the better approximation of the function value, etc. At each step we know only the current approximation of $x$, and nothing more. If we break this process of the permanent specifications on some step, we should agree that the value of function computed to this instant is exact. It makes us to think that all procedures of building of the functions in reality must stop, but we do not know on which step. The notion of constructive function reflects our measure of ignorance of the real mechanisms of this break and the exact moment when it happens. Though, the simplest supposition is that the considered values $x$ and $f(x)$ in reality are grained, e.g., they take values of the form $\e, 2\e, 3\e,\ldots,$ for very small quanta $\e$ (separate quanta for the different magnitudes).

In the preceding section we saw that the simple assumption of the grain of amplitude gives us Born rule for quantum probability. This observation brings us the following more detailed model of decoherence:

{\bf Decoherence comes out from the existence of the minimal nonzero module of amplitude value $\varepsilon$ for any quantum state.}

This model of decoherence wee call the absolute. The sense of this name is the following. The physical dimension of an amplitude is $1/\sqrt{\bar r}$, where $\bar r$ is the dimensionality of the considered configuration space, because the rule of normalization $\int |\Psi |^2\ dr=1$ must give the non-dimensional value of probability.  However, it makes sense to speak about the dimension only if we can infinitely expand or divide the corresponding magnitude. E.g., this magnitude must be potentially unlimited, at least unlimited in the considered area. For example, the physical dimension of a length makes sense if we consider the movement of a particle along classical path and can treat that any segment in space is potentially divisible to many parts. Just this potential plenum allows the usage of the different unit systems, which gives the sense to the physical dimension. Of course, the length in space is not unlimitly divisible, but the elementary length is so small that for one quantum particle or a few classical particles, the presence of this elementary length has no effect to the model and we can think that the length is divisible unlimitly.

Is we consider one quantum particle we can trast the amplitude as continuous, and correspondingly it has the dimension. When the number of particles grows the situation changes radically. Since the quntity of all possible states groqs as the exponential of the tumber of particles we will guickly reach the situation where to one basic state falls the amplitude less than $\varepsilon$. In this case we have to choosde, which basic state really occurs in th esuperposition and ignore all the rest. Further we define this procedure of the quantum state selection. When the amplitudes aproach to the crytical value $\varepsilon$, and besides it is only zero, we must agree that the ascribing the physical dimension to the amplitude looses the sense because now we must measure the amplitude in the quantity of quanta of the elementary value $\varepsilon$, that is in non dimension values. 

We note that in this situation the representation of wave function in the different basises, for example, in coordinate or impulse bases, looses sense as well. The transitions between these basises given by Fourier transform gives the right answer for the continuous spectrum of values of the amplitude. Therefore, in the area where amplitudes becomes comparable to $\varepsilon$, we loose the possibility to use the representation of standard quantum theory! This is the consequence of thee absolute model of decoherence.

The absolute model of decoherence will be natural if we use constructive mathematical analysis instead of standard. Indeed, th econstructive function $f(x)$ of the constructive real variable $x$ requires to point the approximation of the function given an approximation of its argument. We have to be ready that the value of argument $x$ will be pointed very exact, and then our approximation of the function $f(x)$ must be exact as well! It means that for the constructive analytical technique any value of $\varepsilon$ is acceptable. As we reach this accuracy for $x$, the simulating algorithm gives us the exact value of $f(x)$. In the following section we show the clear interpretation of it for the uncertainty relations of Bohr-Geisenberg. As we reach the absolute accuracy in the determining of the spatial position we obtain the maximal uncertainty in the impulse: it can take values in the diapason from zero to the maximal possible value which always exists in constructivism.

It brings the question about the real value of $\varepsilon$ in theh nature. Our algorithms give us no information about this value. The estimation of it is the oprerogative of a user. 

Absolute model is more concrete in comparizon with thee general algorithmic, following from our approach as it is, and it pertims to count something. But it compel us to use the standard algorithms based on matriz algebra. We would never use these algorithms for the simulation of cllassical dynamics. Since we agree to treat decoherence th einbuilt property of our formalism, our algorithms must be such effective on classical systems as the methods of classical modeling, because the classical trajectories of objects represent the result of the complete decoherence of quantum states arising at the small time frames(see \cite{FH}). Hence, we must build algorithms not basing exclusively to the matrix algebra, but using the more classical representations. This slighly vague reasonings will become more clear in the next section. 

\section{Method of collective behavior}

Standard method of quantum theory consists in the representation of the state of $n$ particles in the form of wave function $|\Psi (\bar r_1, \bar r_1,\ldots,\bar r_n)\rangle$ unavoidably leads to the exponential expenses of computational resources. Hence, for the constructivization of quantum theory of many particles we need the different approach to the description of many body states. The method we show now is based on the collective behavior. A real quantum particle is represented as the ensemble of classical particles, which are its samples. At first, we do some introductory remarks about the place of this method in the physical constructivism.

\subsection{What is constructivism in practice}

We underline that the inevitability of constructivization of quantum theory arises from its application to complex systems. Factually, we speak about the new area: quantum theory for many particles, which still does not exist. Its creation is possible exclusively on the ways of the modification of quantum theory formal apparatus, namely, by the constructive quantum mechanics. 

The way to physical constructivism practically lies through the modification of its methods by means of algorithms. Despite of unconditional correctness of the principle of correspondence, which is the conservation in constructivism all advantages of quantum theory, its constructive form differs from the standard. Its difference lies in the kit of mathematical instruments. The place traditionally occupied by the formulas will pass to algorithms. A reader should think about this thesis, because it differs strikingly from the common understanding of mathematics in physics. Moreover, the wrong treatment of this thesis can bring us back to the old discussions about the hidden parameters, the nature of photons and the other things of this type. The answer we gave to these questions in the previous chapters is factually traditional: it means the using the formal apparatus of algorithm theory. Just this approach results in the division of the model to the user and administrative parts, and the identification of a user as an oracle in this model. 

What practically means the constructive modification of mathematical apparatus? There is no answer to this question in the earlier works devoted to the logic of quantum theory; it requires all arsenal of constructivism and, first of all its practical part, that is the building of computer programs. Today the computer simulation of physics looks as follows. Given formulas, for example, differential equations, the task is to build its solution. In the other words, we have to find the way to approximate it, then to write the computer program and launch it on a computer. We have already seen that one formula cannot describe all quantum mechanics. Therefore, one must use the different formulas, for example, one for classical movement, the other for quantum, or one for one particle, he others for many particles, etc. This is the simulation in its typical sense. Of course, in course of computations it turns out that there exists the bound of accuracy determined by the grain $\e$. However, this causes no consequences but the interference in computations. By this scheme is good to specify the solutions of already solved problems. On the other hand, there is no the slightest hope to create quantum theory of many particles, in our meaning of the theory, by these methods. Standard scheme compels us to use quantum computer for the simulation of dynamics of more than 2 particles, without any guarantee and even no recipe of its practical application. In the other words the described standard scheme of the computer simulation of quantum physics of many particles is blind.\footnote{It does not mean the impossibility to obtain new results on standard way, for example, to find molecular spectra. Merely the standard scheme does not fit to the simulation of complex problems and we cannot thus expect to find or to predict new effects in the complex systems by this scheme. At the same time, we must not treat the standard methods useless in constructivism. Only their role must be secondary, they can serve for the debugging of programs, and no more.}.

What constructivism proposes to us? It proposes the scheme of sequential approximations which accuracy is not known in advance! It means that the dynamics of many particles must be described completely with the given grain of spatial resolution, as if this grain represents the finals step of all simulation. Our model must be plausible at each step and must not refer to the process as the whole, because we have no actual infinities. To pass from one value of grain to the other we need the heuristic, e.g., the ideology of algorithm. The method of collective behavior rests on the classical heuristic, e.g., on the notion of particle. The passage to the lesser grain of spatial resolution then means the addition of new (imaginary) particle, as the additional sample of the real particle. The real particle will be then the ensemble (swarm) of its imaginary samples. This is the heuristic of collective behavior. 

Swarm heuristic is similar to the ideology of Feynman path integrals. However, it lies farer from the wave function. Path integrals represent the tool for computation of the wave function $\Psi$, and they presume the limit process to the "exact" wave function, typical to the classical mathematics. Here classical mechanics appears in the limit $h\ar\infty$. Swarm heuristic does not rest on the limit process. We use it on purpose to show ins correspondence with Shredinger equation only, provided the computational resources are unlimited. This will be the substantiation of correspondence principle, which must be stable in the passage to constructivism. In its technique swarm approach needs in no limit process. In particular, classical mechanics contains in it, when all the swarm consists of only single particle this is the classical mechanics. It imposes to us the duty to explain photon emission on the level of small swarms, e.g., to have the concrete mechanism of quantum electrodynamics in terms of separate samples of the real charged particle and its interactions with the other samples of the same real particle. 

This statement of the problem would be unjustified in classical mathematics because it directly contradicts to the ideology of algebraic formulas at all. The usage of formulas presumes that the summing of deposits from virtual processes has been already fulfilled, as for path integrals. Whereas the constructivism requires the explicit consideration of these processes, in the form of individual histories of the samples of real particle.  It does not permit to apply formulas easily (we will make certain of it on the example of Shredinger equation fro one particle), for this at each step of simulation we need to build the new swarm. However, it makes possible to build the simulation algorithms without support from formulas, e.g., without the abstraction of the potential realizability. The cost we pay corresponds to what we get back: the possibility to consider complex systems. 

Going slightly deeper to the general questions we could say that the different role of formulas and algorithms is connected with the different human perception of these types of mathematical apparatus. If formulas represent relatively narrow stripe of the perception and the processing of information connected with the left cerebral hemisphere, then algorithms affect a person of user deeper because they used also his creative thinking. It is not accepted to speak about such things in mathematical literature; it belongs rather to biology than to physics or mathematics. Constructivism gives us the possibility to discuss these things because here a human as user of the model has the concrete place in the formalism: this is the place of an oracle in the computations. In what follows, we will illustrate the difference between the constructive and standard approaches fro the examples we consider. 

What means the passage from the spatial representation of the wave function to its impulse representation from the swarm viewpoint? A swarm consists of separate particles, each of which possesses the coordinates and the impulse. At the same time the wave function $\Psi (x)$ determines the distribution of the real particle only on the coordinates: the density of this distribution is $|\Psi (x)|^2$. The information about the speed of particle is contained not in the module of the wave function but in its phase $\phi (x)$, where $\Psi = |\Psi |e^{i\phi}$. To see the distribution of real particle on impulses we need to pass to the impelse basis in the Hilbert space of states. Analytically it means the application of Fourier transform:
\begin{equation}
|\tilde\Psi (p)=\int\limits_R\Psi (x)e^{-ixp/h}dx.
\end{equation}
Fourier transform says that we decompose our state vector $\Psi$ on basic plane waves $e^{ipx}$. A plane wave from the viewpoint of the swarm is the subset of samples with the uniform density which move with the same impulse $p$.
The separation of such a subset in the swarm is problematic because we have only finite set of samples and all values of the impulse must be factually represented. Given a value of impulse $p$ we separate the subset of samples with this impulse $p$ that gives the relatively sparse ensemble. If we want to make this ensemble representative, we must make the total number of samples huge. However, in this case the swarm representation looses its sense at all, because the representation of the squared module of wave function as particles really stored in the computer memory leads to the exponential expense of it: this is just the gain from the arithmetic notation of a number! E.g., if we want to compute the impulse representation of one particle by means of swarm within the accuracy of analytical method we meet the certain failure!

For what do we need the method of collective behavior in this case? We need it only for the simulation of processes with many particles. High expenses and bad work of it for one particle is the necessary cost for the possibility to simulate many particle problems, which is the single aim of it. It would be naive to think that we can master the theory of the world of many particles by means, which are good for one or two particles. Similarly, we cannot expect that one can do it without giving up some conveniences from the already reached by us "comfort level" in mathematical formalism. 
Of course, it concerns to qubit formalism of quantum informatics, which represents the discrete version of Hilbert formalism for many bodies.
 {\bf Constructivism is not reducible to the discrete (or qubit) version of standard Hilbert formalism!} We saw (in Appendix it is illustrated completely) that this formalism works well in the non-relativistic case. In the relativistic case, we immediately meet the defect of this formalism when trying to pass from the impulse representation of wave function to the coordinate in the notation of fundamental process of quantum electrodynamics. It is demonstrative that this difficulty immediately arises in the consideration of three particles instead of two (emission of two photons by an electron). In any case, as we add the new particles to the system, even if a classical supercomputer can cope with the computational difficulties, the advantages of standard formalism disappear as a mirage in a desert. 

I think that this is not the simple fortuity, and the new serious physics hides behind this situation. We go to it starting from the constructivism, which is the algorithmic character of all procedures but the expert estimation of a user. This role of algorithms brings the new strict requirements to the heuristic, as the general scheme of this algorithm. Here we owe to go to complete breaking off with the standard approach because the most important for us will be then the severe economy of the computational resources instead on the concise language of formulas as in standard quantum mechanics. Swarm approach just gives us such a core - heuristic, which we will rest on in the path to the model of complex systems.

\subsection{Statement of problem}

Computational difficulties arise in standard theory even for the case of one particle for the matrix method. Let its configuration space be divided to $N$ elements. It means that the space of quantum states has the dimensionality $N$. As we apply the matrix algebra for the computations in any form we compel a computer to process all the trajectories of the system passing through all $N$ basic states. In the simplest case, it is expressed in the computation of the product of the unitary matrix of the time evolution. The mean value of the module of matrix element of a unitary matrix is $1/\sqrt{N}$. If all the interference resulted in the evolution of the system is constructive we would obtain the matrix all elements of which are approximately equal 
$N\ 1/(\sqrt{N}\sqrt{N})=1$, whereas they are of the order $1/\sqrt{N}$. It means that the bulk of interference is destructive and the main part of the computational resources goes to the determining that there is no particle in the considered point at all. The computational methods of quantum mechanics based on constructive algebra presume the big and non-efficient expenses of computational resources from the very beginning. These methods factually realize Riemann scheme of integration for the Shredinger equation based on the division of configuration space to finite elements. 

Collective behavior represents the alternative approach corresponding to Lebegue scheme of integration. We start with the samples of the real particle. The total number of them must be large, and their dynamics must well approximate quantum dynamics. We want to avoid the big and non-efficient expenses of computational resources typical for matrix algebra. Developing this technique for one particle, we can wait that such a method is applicable for many particles as well. 

The representation of quantum particle through its samples we call the method of collective behavior. It gives the severe economy of computational resources that is the basic requirement of algorithmic approach. This requirement is not purely esthetical. It makes possible to build models in which decoherence is the inbuilt property, but not an axiom, as in Copenhagen quantum mechanics. We have shown that this requirement gives us the classical urn scheme fro the results of quantum states measurements. 

At first, we give the interpretation of quantum dynamics of one particle in terms of collective behavior, and then pass to the case of several particles. The cost we will have to pay for the economy of computational resources is the necessity to build algorithms through the mechanism of interaction between the samples, instead of the differential equations. These equations for the method of collective behavior do not exist. Nevertheless, the situation has the positive sides besides the economical computations. The model of dynamics turns closer to classical than in the standard way, which allows its visualization.

Our methods are the direct generalization of the diffusion Monte Carlo method (DMC) to the general case of Shredinger equation. The fact that DMC gives the most exact value of ground state energy in comparison with the other methods inspires us with optimism towards the practical using of the collective behavior in the more complex cases. 

The method of collective behavior for one particle does not possess the accuracy peculiar to the analytic solution of Shredinger equation. Its advantage is the other. It economizes computational resources, and makes possible to build the scalable models of many particle systems. In the evolution of such complex systems, the deposit of each particle is individual. The evolution of a separate particle in the complex ensemble cannot be repeated. Even in the attempt to reproduce the evolution of the whole system, there will be the little difference in the positions of separate particles that brings to individuality the huge role. It means that the evolution of a separate particle in the complex ensemble is typically unique. At the same time, the wave function results from the huge number of repetitions of the same experiment with the identical initial conditions. This is why the usage of the wave functions of separate particles in the complex ensemble is justified in that case only if we especially roughen the behavior of such ensemble by averaging it over the different degrees of freedom of the separate particles. This is the way of statistical physics. At the same time, this way is inappropriate for substantially complex systems like living things. 

It is impossible to investigate a living organism by the methods of statistic physics because the averaging ignores specific physical mechanisms of the fundamental nature, which determine what we mean by a life. It is important that these mechanisms have a fundamental character, namely they are connected with the entanglement of quantum states. Entanglement is irreducible to the interaction, say, to photon exchange. Entanglement is the fundamental phenomenon, which reveals in the specific bounds between particles, and just these bounds permits us to build scalable models of constructive physics. We need some special computational resource for the reflection of the entanglement dynamics, and we can take this resource only saving on the description of separate particles. Collective behavior method just gives us this redistribution of resources.

Why in the simulation of complex systems we do not need to pursue the accuracy of the wave functions for separate particles? Because the wave function is the result of statistical processing of the massive of uniform experiments which cannot exist in the evolutions of complex systems. If we speak about the coordinate or impulse representation of the wave function, about the matrix algebra etc., we mean only the properties of gigantic virtual statistic ensembles consisting of one particle taken sequentially in the instants of time: $t_0,t_0+\Delta t, t_0+2\Delta t,\ldots$ in the same initial state. Moreover, all real experiments contain also the averaging over the different real particles so that if we, for example, measure a spatial position of electron, this measurement gives it a huge speed, which is the more, the more accurate we measure its position. It means that for the next measurement we will take another electron etc. We will return to this question further. Even if the electron is the single, its repeated measurements have only remote relation to the behavior of a many electron system. We can apply one-electron results only if entanglement plays the negligible role, like in ideal gas. If entanglement plays the role, we must provide the computer resource for the reflection of entanglement. If we use the time and memory to the exact reconstruction of the one-particle wave functions, we will not be able to do anything with the complex systems!  All the more this becomes impossible if we begin to reconstruct the wave functions of the several particles. In this case our computer will be occupied by the processing of extremely large mass of senseless combinations of states of several particles, which we can compare with the attempt to reproduce the literature poem by the team of monkeys chaotically biting on keyboards. 

One could attempt to defense wave functions of separate particles, intending to build not entangled states or the states close to not entangled (for example, Schmidt states for pairs of particles). It is not a stupid way, but for 3-5 particles only. If we ignore the states of 3 particles irreducible to Schmidt states (such states exists yet for 3 qubits) we can obtain some agreement with experiments, for example for the potential of chemical bounds in the stationary states (such computations are done in chemistry for several tens of particles). However, even for 3-5 particles this way will not give us the right picture of the dynamics, for example, for the association of molecules. All the more this patchwork way is inappropriate for the simulation of complex systems for which the behavior is determined rather by entangled states than by the accuracy of the computation of potentials. 

We need the integral heuristic to make our model scalable. The description of classical and quantum evolutions and measurements must be uniform. A wave function successfully manages with the statistics but it is one particle statistics and it cannot be generalized to complex systems. This is why we refuse from the traditional description of quantum states by wave functions and replace it by the swarm representation of elementary particles. We will see that this representation is on friendly terms with the standard approach in case of unlimited computational resources. As fro the complex systems where the standard approach is not applicable at all the method of collective behavior gives us that Ariadna thread, which must lead us to the target.

\subsection{Diffusion Monte Carlo method}

For the search of the main state (eigen state with the minimal energy) the easiest way is the probabilistic Monte Carlo method, which we call the method of the stationary diffusion. In this method, the swarm of its samples represents the real quantum particle, where the density of swarm equals to the module of the wave function of the main state: $\rho =|\Psi_0|$. It shows that this method is narrow, and we cannot apply it for the simulation of dynamics. The dynamics requires the description of excited states, not only the main state. However, Monte Carlo method is simple in the usage and it represents the good starting point for the building of the real dynamical model.

We consider the following process of the evolution of this swarm in the time. Let for each sample with some small probability $p$ we shift it to $\Delta x$ along one of the coordinate axes to the positive or negative direction with the equal probabilities $p$. With the probability $1-6p$ each sample stands at its old place. Let we be given a scalar function $V$ on the configuration space. We introduce the following rule of creation and annihilation of samples. Let for each sample with the probability proportional to $V$, the following process takes place. If $V<0$, this sample generates the new sample located in the same point, if $V>0$, this sample eliminates some other sample located at the distance less than $\Delta x$ from it. We call this process the static diffusion. There are no speeds of samples in this process, but only the coordinates.  

We make in the Shredinger equation 
$$
ih\frac{\partial\Psi}{\partial t}=-\frac{h^2}{2m}\Delta\Psi +V\Psi
$$
the formal replacement $t=-i\tau$. The equation thus acquires the form of diffusion equation:
$$
h\frac{\partial\Psi}{\partial t}=\frac{h^2}{2m}\Delta\Psi -V\Psi
$$
which just describe the static diffusion process. If we order the energy levels by the increasing of their energies $E_n$, the evolution of the state vector expanded on the eigenvectors $\phi_n$ will look as:
$$
\Psi=\sum\limits_n \la_n e^{-\frac{i}{h}E_nt}\phi_n .
$$
The expansion of the solution of the diffusion equation has the form
$$
\Psi(t)=\sum\limits_n \la_n e^{-\frac{1}{h}E_nt}\phi_n .
$$
We see that for $t\ar\infty$ the diffusion process converges to the main state, because the deposit of the rest states will be suppressed by the rapidly decreasing exponentials. It is known that the main state contains no differences in the phase, e.g. it coincides with its module. To find the main state we thus have to take the arbitrary initial distribution of the density and to launch the static diffusion process. It stabilizes on the distribution proportional to the main state. Of course, we must take care of the total number of the samples, which must not go out of some limits. We can guarantee it by the addition or elimination of samples uniformly accordingly to the existing density. 

The DMC method can be straightforwardly generalized to the case of $n$ real quantum particles. Here the sample will be the cortege of $n$ samples of each of real particles. The configuration space will be $R^{3n}$. 

We underline once again, that the described static diffusion method is aimed for the search of the main state exclusively, that is for the static simulation. To investigate the dynamic picture we need the dynamical diffusion method, which we consider further.

\subsection{Known ways of approximation of Shredinger equation by quasi-classical ensembles}

Attempts to reduce Shredinger equation to quasi-classical ensembles have some history. Our method proposed here, differs from the previous attempts in that we want to obtain not differential equations, but the mechanism of interaction between samples, which leads to quantum dynamics by the most economical way from the view point of the simulation on classical computers. We show two tricks similar to our approach: Bom approach and diffusion Monte Carlo method. 

We begin with Bom method, which uses the notion of pseudo potential. 
We identify the module squared of the wave function with the density of the imaginary particles, and its phase $\phi$ with the classical action. We then have: $\Psi = \rho^{1/2}e^{i\phi /h}$, and $1/m\ grad\ \phi (\bar r)$ can be regarded as the density of the flow of imaginary particles. Shredinger equation then becomes equivalent the system of equations:\footnote{These equations were derived by Madelung. Bom laid them in the basis of his interpretation of quantum theory. This method together with Feynman path integrals can be treated as the prototype of our method of collective behavior.}
$$
\begin{array}{lll}
&\frac{\partial\rho}{\partial t}+div\ (\rho/m\ grad\ \phi )&=0,\\
\frac{\partial\phi}{\partial t}+&\frac{1}{2m}(grad\ \phi )^2+V+V_1&=0,
\end{array}
$$
where the quantum pseudo potential $V_1 = \frac{h^2}{m}(\Delta\rho/\rho+(grad\ \rho )^2/\rho^2 )$ depends on the density of particles which has the singularity in the zero point. These equations coincide with the equations of the flow of classical particles provided $V_1$ has the physical sense of some pseudo potential. Bom's approach contains one serious drawback. The mechanism of pair wise interaction between imaginary particles needed for the creation of pseudo potential remains unclear. Moreover, the singularity in the zero point means that there is some force with the unpredictable direction which acts on particles in the zones with small densities. Without the mechanism of such a force, application of these equations cannot give more than Shredinger equation itself.\footnote{D.I.Blohintsev in \cite{Bl} marked the lack of simple interpretation of linear property of the solutions of Shredinger equation as the main drawback of Bom's approach. I think, this is not a drawback. The contradiction between the stochastic nature of a wave function and the linear property of Shredinger equation remains insuperable in any ensemble interpretation of the wave function. However, it is not needed for the creation of the economic simulation. All methods of linear algebra suffers from the inbuilt inefficiency of the expenditure of the computational resources; in the problems of numerical simulation of quantum dynamics, especially for many particles, in any case we need to choose between some type of ensemble method or quantum computers. This is why the sacrifice of the mathematical beauty here would be well taken.}
The deep cause of this difficulty is that the real wave function has the statistical nature. Its experimental definition always requires the big number of repeated trials, because the linearity reveals only for the result of the limit process on the infinite time. Therefore, the refusal from the simple description of linearity causes no defect to the simulation. For the effective simulation, we need to refuse not from the linearity only. We need the simple mechanism of interaction of samples in the dynamical diffusion swarm. It turns that the obtaining of such a mechanism requires the refusal from the relatively concise system of differential equations, like the represented above. We will be able to write the system of equations for a fixed spatial resolution only. This is the cost for efficient algorithms.

Now we glance at the diffusion Monte Carlo method. DMC deals with the ensemble of point wise samples without speeds. The law of movement of such samples is simple. Each sample with some probability $p$ shifts to $dx$ to each of 6 possible directions along the coordinate axes (with the same probability) or stays in the current point with the probability $1-6p$. The potential energy with the opposite sign plays the role of intensity of the creation (annihilation) of the particles in this point. Here the creation means the emission of new sample by one of already existing, as well as the annihilation means the absorption of some (close) sample by one of already existing. The presence of the simple mechanism of movements is the strong side of Monte Carlo approach. However, samples in DMC ensembles have no history. Only the density function $\rho (x)$ characterizes this mechanism.
Samples from DMC ensemble have no speeds and these ensembles can serve to the computing of the shapes and energies of stationary states, for which they give the most exact results among all computational methods. Namely, the evolution of DMC ensemble stabilizes on the density $\rho(x)$ equal module of the wave function of the main state (the eigen state with the minimal energy) of the considered system. If we acquire speeds to samples we obtain that they vary chaotically as in Brownian motion. Here the average speed of samples in each small cube will be about zero. 

We mention that DMC is closer than the other approaches to our method. It gives the most effective way of finding the main state. Nevertheless, it is impossible to simulate the dynamics with DMC, because one scalar function $\rho (x)$ is not sufficient to it. The indirect evidence of the impossibility of such generalization of DMC is that we must compare the density of samples not with the module of wave function but with the square of it. It means that in the framework of DMC we cannot reach the uniform description of classical and quantum dynamics. There is one more serious drawback of DMC. The creation and annihilation of samples makes impossible to trace their individual history. This drawback does not directly influence to the complexity of computation of the main state, but it is principal fro the description of the dynamics, because the dynamics is just the resulting description of the history of one particle. 

At last, recall what we obtain starting from some classical Hamiltonian of interaction between particles $H(r,p)$. The density of samples $\rho$ for the ensemble with classical Hamiltonian of interaction will depend on the coordinate $r$, as well as on the impulse $p$ and obeys Liouville equation
$$
\frac{d\rho}{dt}=-\{ \rho, H \} .
$$
The behavior determined by this equation cannot simulate quantum evolution with the admissible accuracy since it does not give some principal quantum effects, as for example, Rabi oscillations or quantum spectra. It means that for the simulation of quantum dynamics we need to admit some elements in the behavior of samples, which do not follow from classical physics. 

What is the physical nature of this behavior? We do not know the answer. In the work \cite{Kh} it is proposed to treat such a behavior as the result of action of the so-called pre-quantum fields. For the dynamical diffusion swarm we can assume that its samples exchange with the virtual photons that results in the impulse exchange in the definition of dynamical diffusion swarm (see below). It does not clarify the situation and we leave the question about the nature of this exchange without the answer. It does not prevent us for building algorithms, and we can return to it when necessary.

One more far analogue is the model of Calder Leggett for the partial decoherence of a quantum particle, in which this particle interacts with the bath consisting of harmonic oscillators. This interaction gives to a particle the random speed. This model starts from the standard quantum formalism whereas the dynamical diffusion swarm is designed to serve instead of this formalism.

\subsection{Dynamical diffusion swarm}

 We describe the main instrument of quantum simulation: the dynamical diffusion swarm. This object generalize two known objects: the ensemble of classical particles from DMC method and the ensemble of particles with the classical Hamiltonian $H(r,p)$.

   Why the dynamical diffusion swarm is better then the explicit solution of Shredinger equation? In the solution of Shredinger equation, we use Riemann scheme of integration. It requires the computation of the wave function in the whole configuration space, independently of the degree of constructiveness of the interference. On the main area of space where the interference is destructive and the wave function factually equals zero we are compelled to expense the computational resource only to check it. In contrast to it, the dynamical diffusion swarm realizes the more general, Lebesgue scheme of integration. In this scheme, the diffusing samples will concentrate in the areas of constructive interference as the result of diffusion dynamics and we avoid the huge non effective expenditures of the computational resources. This is the fundamental advantage of the diffusion dynamics. As we see in further, the price for this advantage is the non-uniform diffusion intensiveness on the element of the length. This intensiveness will depend on the chosen grain of spatial resolution $\d x$, in contrast with the ordinary diffusion where there is no such dependence. 

  We proceed with the definitions. We call a swarm a finite set $S$ consisting of $n$ classical point wise particles of the same type, each from which $s\in S$ possesses its own coordinates and impulse $x(s),\ p(s)\in R^3$. In the method of classical behavior one quantum particle of the mass $M$ and charge $Q$ is represented as the swarm $S$, each sample of which $s\in S$ has the mass $m=M/n$ and the charge $q=Q/n$. The elements of this swarm are called the samples of this quantum particle. We will suppose that the number of samples $n$ in the swarm is so large that it can serve as the approximation of the continuous media, e.g., for our aims we need to pass to more and more fine division of the configuration space to cubes so that at each step every cube contains some samples. We will see that the dispersion of speeds will grow with the decreasing of the grain of resolution. It means that we have the separate swarm for each value of this grain.

 The methods of definition of particles depend on the concrete problem because particles are not necessary elementary in the exact sense. The definition of what object we treat as a particle presumes the determining the typical length $\Delta X$ and the time $\Delta T$, so that the size of particles is much less than $\Delta X$, and we can treat them as point wise, and the time interval $\Delta T$ is not less than the time of processes interesting for us.   We assume that the typical average speeds of considered shifts are much less than some limit speed of the movement of material bodies $c$. For example, we can treat an atom as a point wise particle in processes with $\Delta X > 10^{-8}m$ and $\Delta T>10^{-10}s$. If we decrease the value of typical lengths and times, then to obtain the right (corresponding to experiments) picture of the dynamics we must consider the different set of "elementary" particles, for example, we must consider the nucleus and electrons inside the atom separately. If we fix $\Delta X$ and $\Delta T$, we have to determin lesser segments $\d x$, $\d t$, which will represent elementary steps of the future video film, though they must be much bigger than the typical lengths and times $\tilde\Delta X,\ \tilde\Delta T$ for more fundamental processes than the considered (the gap between such values may be of the order $10^{-20}$, that always allows such a distinction). In addition, in one process with the fixed energy the length and the time frames depend on the mass of particles. The difference in mass makes possible to consider for the bulk of QED processes only (identical) electrons, because the typical lengths of the spatial shift of nuclei will be $1800$ and more times lesser. We thus can assume that the chosen values $\Delta X$, $\Delta T$ are the size of imaginary screen and the length of the film, and $\d x$, $\d t$ are the screen resolution and the time of one card in the video film. We choose $\d x$ and $\d t$ as much as possible so our film will be informative. 

After this choice, we can make a conclusion, which particles we must treat as quantum, and which as classical. For this we must compare its actions $a=M(\Delta X)^2/\Delta T$ with Planck constant $h$. If $a<h$ we treat the particle as the quantum, in the opposite case as the classical. Further, we will see that in the method of collective behavior the passage from one type of consideration to the other means merely change of the size of swarm, and does not require the introduction of the different type of dynamics. We have already convinced that in course of the film preparation we always can decrease yet more the values $\d x$ and $\d t$ in order to make a picture better, for example, by the splitting of these intervals to the lesser parts. We treat that all the space $R^3$ is divided to the equal cubes with the side $\d x$, and the time is divided to equal time frames of the duration $\d t$. 

We introduce some value of speed $c$, which we treat as the limit speed of the movement of samples. We choose intervals of the divisions such that $\d x \gg c\d t$. It guarantees that at each step of the evolution the values of magnitudes resulted from the averaging on cubes with the side $\d x$ will change very small that is needed for the obtaining of the asymptotic approximation.

The density of swarm in the point $x$ is determined by the expression
\begin{equation}
\rho (r, t) = \frac{N(r,t)}{(\d x)^3},
\label{density}
\end{equation}
where $N(r,t)$ denotes the number of samples contained in the instant $t$ in the same cube as the point $r$. To compare this with the solution of Shredinger equation we must limit in this definition $\d x \ar 0$, which means that we consider not one swarm, but the sequence of swarms with the densities $\rho_n$ for increasing $n$. We will not do that to avoid the complex notations; instead we assume that whenever required we can divide our division to the smaller cubes so that $\d x$ decreases in the admissible limits. We write $\rho(x)=|\Psi(x)|^2$, which means that 
\begin{equation}
\rho_n(x)\ar |\Psi(x)|^2 (n\ar \infty ),
\label{asympt}
\end{equation}
where the convergence is uniform without the special mentioning. Such a sequence of swarms realizing the approximation to the wave function density (for Shredinger equation) we call the admissible approximation to quantum (unitary) evolution. 
    
    Our main aim is to define the behavior of samples in the swarm, which gives the admissible approximation to quantum evolution. 

The main requirements for the simulation of quantum dynamics via collective behavior are the following. 
\begin{itemize}
\item Quantum dynamics simulation uses the dynamics of the swarm of samples so that in each time instant $t$ the quantum probability equals the density of swarm 
\begin{equation}
|\Psi(x,t)|^2=\rho(x,t)
\label{swarm}
\end{equation}
 in each spatial point $x$ within the error of the order of $\d x$. 
\item Each sample of the swarm has its own history, e.g., it preserves its individual number in course of the whole process of the simulation in all transformations. The types of samples exactly correspond to the types of real particles.  
\item The behavior of any sample is completely determined by its own state, the states of all samples in its vicinity and some source of random numbers. 
\end{itemize}

The swarm satisfying these conditions we call the quantum swarm of one quantum particle. 

We determine the behavior of samples such that these conditions are satisfied. For this it would suffice to show that the appropriate local shifts of samples give us the needed approximation of the solution $\Psi(x,t)$ of Shredinger equation, which satisfies (\ref{swarm}). Such shifts will be a priori not natural in the dynamical sense, but we then show that the natural character of these shifts follows from the special diffusion mechanism of the movement of samples. 

 The second rule means that we refuse from the usage of complex numbers, and simultaneously want to make the model convenient for the inclusion of QED. The last requirement means the locality of all interactions. We agree that the behavior of sample is the rule determining its state (impulse, moment of impulse, and type) and the spatial position. In view of the premises, the behavior cannot be determined by classical physics. 

We define the quasi-classical behavior of samples called the dynamical diffusion mechanism. Then we show that the swarm with this mechanism satisfies our conditions.

We agree that each sample in each time instant can either stand at place or move with the speed $c$ along one of the coordinate axes $OX,OY,OZ$. 

We call the reaction of exchange the sequence of the following actions:
the choice of a pair of samples $\a, \b$, which lie at the distance no more than $\Delta x$, which speeds are the opposite: $v(\a )=-v(\b )$ and either the simultaneous change of their speeds to zero (if they are nonzero), or ascribing to them mutually opposite speeds, which modules equal $c$, and directed along one of the coordinate axes (randomly chosen).

The exchange of speeds changes neither the summary impulse of the swarm, nor the summary momentum of impulse provided $\Delta x$ is small. 
We denote by $N(r)$ and $N_s(r)$ the set of all samples in the cube containing the point $r$ and the set of samples from this cube with nonzero speeds, by  $N^+_x(r), N^+_y(r), N^+_z(r)$ we denote the sets of samples from the cube with $r$, which move along the corresponding axes in the positive direction, and by the analogous symbols with the sign $-$ the samples moving in the negative direction. By $|g|$ we denote the number of elements in the set $g$. We also agree to denote the number of elements in a set by the same sign but with the replacement $N$ by $n$. We call $r$ stationary each subset $S\subseteq n(r)$, consisting of elements with nonzero speeds for which $\sum_{\a\in S}v(\a )=0$ and $S$ is the maximal on the inclusion with these properties. The number $|S|$ of elements in $r$ stationary set (which does not depend on its choice) we denote by $s(r)$. Let $d>0$ be a chosen constant such that the diffusion coefficient is proportional to $d$, $V(r)$ is a scalar field proportional to the external potential energy with the multiplicative constant $grad\ V(r)=(V_x(r), V_y(r), V_z(r))$.  

We also agree to consider non-relativistic swarms only, e.g., such swarms for which $n_s(r)/n(r)$ is close to $1$ for all $r$. It means that the main part of samples in each cube has the zero speed. This requirement is incompatible with the point wise approximation \ref{asympt} by swarms of the exact wave functions for external potentials of Coulomb type $1/r$ because the average speed of samples near zero point goes to infinity. To obtain the asymptotic convergence \ref{asympt} we would have to assume that the speed $c$ can be chosen as large as needed for the regular swarm number $n$. In the reality $c$ cannot exceed the speed of light that establishes the natural limit of the accuracy of the swarm approximation of the solutions of Shredinger equation.

We call the dynamical diffusion mechanism of evolution the sequence of the following actions on the swarm:

\begin{itemize}
\item 1) The sequence of random exchanges with the uniform distribution leading to the distribution of speeds with the property $s(r)/n_s(r)=d$ for each point $r$. If $n(r)$ is small, this equation must be fulfilled with the highest possible accuracy (see the agreement about the accuracy from above). The highest accuracy we reach if we fulfill exchanges randomly. 
\item 2) The ascribing of speeds to some samples from $N_s(r)$, chosen randomly from the uniform distribution such that the signes of new acquired speeds along each axes are the same and if $v_u(r)$ is the summary speed ascribed to samples from $r$- cube along the axes $u$, $u=x,y,z$, then for all such $u$ the equation $v_u(r)m=-V_u(r)\Delta t$ is fulfilled with the highest possible accuracy. The highest accuracy we reach if we ascribe the speeds randomly with the probability density proportional to $V_u(r)\Delta t$ with the account of the sign. 
\item 3) The change of coordinate $r(\a )$ of each sample corresponding to the law of uniform movement:
$r_{new}(\a )=r(\a )+v(\a )\Delta t$.
\item 4) The computation of $V(r)$ accordingly to the new positions of samples.
\end{itemize}

We do not concretize the method of computing the potential energy. It can be done by the formula for Coulomb law or by the diffusion mechanism proposed in  \cite{Oz1}.

The swarm with the dynamical diffusion mechanism we call the diffusion dynamic swarm (DDS). We note that such a swarm cannot be represented as the ensemble of point wise particle with the classical interaction. 
The point 1) says about twp things:
\begin{itemize}
\item there is the random force of attraction or repulsion between samples, which does not change the summary impulse of the swarm (ср. \cite{FH}), and
\item the averaging of speeds of samples takes place within the accuracy determined by $d$ (the smaller is $d$ the more strict averaging).
\end{itemize}

We impose the following important requirement: 
{\bf Exchange of impulses between samples must be much more intensive than the exchange of samples themselves between the neighbor spatial areas.} It means the following. The change of impulse in the small cube goes mainly by the impulse exchange, not by the penetration of samples through its borders. It is important for our aim. It is expressed in that the intensiveness of impulse exchange (diffusion intensity) will grow rapidly with the decreasing of the grain of spatial resolution $\d x$: as $(\d x)^{-3}$. 

For each moment of time $t$ if $\Delta x$ is small enough, the density of swarm $\rho (r,t)$ for any point $r$ does not depend on the orientation of the coordinate axes.
Really, let $\d_1$ be such that $c\d t\ll\d_1 x \ll \d x$, $v(r)$ denote the average speed of samples in the point $r$, found by the averaging on samples with coordinates $r_1:\ \| r-r_1\| < \d_1 x$. The total number of samples passed in the unit of time from the vicinity of the point $r_1$ to the vicinity of the closed point $r_2$ will be then proportional to the scalar product $v(r_1) (r_2-r_1)/\| r_2-r_1\|^2$, which does not depend on this orientation.

We now compare the dynamical diffusion swarm we defined with the ensemble used in the diffusion Monte Carlo method (we call this ensemble the stationary diffusion swarm). A state of the dynamical diffusion swarm is determined by the fixation of coordinates and speeds of all its samples, whereas its density determines a state of the stationary swarm only, because the average speed of samples is zero. Indeed, in the step of evolution of the stationary swarm for each sample we choose either its shift to one cell with the equal probability, or it remains on its place. This choice does not depend on the initial speed of this sample, e.g., impulse is not conserved and the summary impulse in each cube turns to be zero. This model corresponds to the dispersion of the molecules of some substance in the media of the other substance, for example, a color in water. If we pay attention to the molecules of a color, its impulses change: it passes to molecules of water. Stationary model of diffusion we can thus call the model with the friction. 

In contrast with the stationary swarm, the dynamical diffusion swarm evolves with the conservation of the impulse in each interaction of pairs of samples in the exchange reaction. Not only has its density thus determined a state of the dynamical swarm, but also by the summary impulse in each spatial cube. E.g., a state of the dynamical swarm is given by the pair of functions $(\rho (\bar r, t),\ \bar p (\bar r, t))$.

However, the main advantage of DDS in comparison with the known ensemble methods is that each sample has its own history. It justifies the name of samples as the representatives of the initial real particle. DDS makes possible of the uniform consideration of a particle: from classical as well as from quantum viewpoint. To pass from the classical description to quantum and vice versa we need only to replicate (or, correspondingly, to join) samples of the swarm. Here in the swarm corresponding to the classical particle will be the single sample: this particle itself. We can also regulate the allocation of the computer memory in the simulation taking into account that the accuracy of the quantum description will grow with the increasing of the number of samples in the swarm. 

\subsection{Differential equations for the dynamical diffusion swarm}

To prove the appropriateness of the diffusion dynamics for the approximation of Shredinger equation we must pass to the differential equations for the real functions. The difficulty is that these equations will depend on the elementary length $\d x$. For example, the intensiveness of the diffusion process will be proportional to $(\d x)^{-3}$. It makes impossible to launch $\d x$ to zero as in mathematical analysis applied to the processes of classical physics. The status of equations we are going to write is determined by the chosen value $\d x$ of the grain of spatial resolution. We must choose it so that the approximation of density within $(\d x)^3$ corresponds to the considered process. After the fixation of $\d x$ we can consider the dynamical diffusion swarm of the corresponding intensity, and differential equations approximating its dynamics, which turn equivalent to Shredinger equation. 

Therefore, the pair of functions determines a state of dynamical diffusion swarm 
\begin{equation}
\rho(t,\bar r),\ \bar p(t,\bar r),
\label{pair}
\end{equation}
where the scalar function $\rho$ is the density of samples, the vector function $\bar p$ is the summary impulse of sample in the current point, defined as $\lim\limits_{dx\ar 0}P(r,dx)/(dx)^3$, where $P$ is the summary impulse of samples occurring in the cube around $r$ with the side $dx$. We assume that we can do $dx$ much less than the chosen value of $\d x$, which determines coefficients of equations on $\rho$ and $\bar p$. 

The dependence of equations of the grain $\d x$ reveals as follows. The total impulse $\bar p(t,\bar r)$ will change slowly with the change of $\bar r$ on the values more than $\d x$. But its derivative $\frac{\partial \bar p}{\partial t}$ will be very big: of the order $1/(\d x)^3$, and will change quickly as well. E.g., the graph of the function $\bar p(t,\bar r)$ is smooth enough if we look at it with the big grain $\d x$, but if we increase the resolution, by the decreasing the grain $\d x$, we see that the graph becomes something like a saw with sharp teeth. The sharpness of these teeth will be more if we increase the resolution $1/\d x$, and will be limited only by the limit of speed of the movement: $c$ in the unit if time (compare with \cite{FH}). This assumption is important for the following, we call it the non-relativistic approximation, and write it as $v\ll c$. 

In view of the isotropy on the diffusion process, mentioned above, the change of density of the swarm $\rho(r,t)$ and its second derivative can be found through the integration on the surface $S(r)$ of the sphere of radius $\d x$ by the formulas
\begin{equation}
\begin{array}{lll}
&\frac{\partial\rho(r,t)}{\partial t} &=\frac{3}{4\pi (\d x)^3} \int\limits_{S(r)}\bar p(r,t)\bar n(\bar r_1)ds(r_1),\\
&\frac{\partial^2\rho(r,t)}{\partial t^2} &=\frac{3}{4\pi (\d x)^3} \int\limits_{S(r)}\frac{\partial\bar p(r,t)}{\partial t}  \bar n(\bar r_1)ds(r_1).
\end{array}
\label{density}
\end{equation}
 These formulas follow immediately from the definition on the density of swarm and are true for any mechanism of the change of speeds of its samples. 

We derive the law $\frac{\partial\bar p}{\partial t}\bar a$ of the change of total impulse of the swarm in the small ball with the center in the point  $\bar r$, resulted from the movement of samples along the vector normal to the surface of this ball $\bar a$ of the unit length.
Three magnitudes bring the deposits to the change of the total impulse of the small ball:
\begin{itemize}
\item The penetration of samples obtained its speeds from the impulse exchange (diffusion) through the small element of surface of sphere.
\item The impulses of samples acquired by the action of the external potential $V$.
\item The impulse of samples preserved their speeds, e.g., moving on inertia. 
\end{itemize}
It follows from the definition of the diffusion that these deposits equal, correspondingly,
$-I(\ grad\ \rho )\cdot \bar a$, $-\kappa\rho (\ grad\ V)\cdot\bar a$ and $g\frac{d\bar p}{d\bar a}$, where $I,\kappa,g$ are the intensities of the corresponding processes. 

Indeed, let us consider the flow of samples through three points disposed along one axes $x$, where $a_1,a_2,a_3$ are total numbers of samples in these points, $p_1,p_2,p_3$ total impulses in these points. The first deposit is then proportional to $a_1-a_3$, the third is proportional to $p_1-p_3$. 

By the choice of system of units we can reach that $g=1$. The dependence of coefficients on the grain of space, which is needed to the approximation of Shredinger equation, has the form:
\begin{equation}
I=\frac{h^2}{2m^2(\d x)^3},\ \kappa = \frac{h}{m\d x}.
\label{intens}
\end{equation}

Here we use the supposition about the quick transmission of the impulse in swarm in comparison with the shift of samples. It means that the deposit of the last summand is small in comparison with the deposits of the first and the second for the small $\d x$, and we can thus omit it. It follows from that the intensity of the diffusion $I$ is proportional to $(\d x)^{-3}$ and if we take very small $(\d x)^3$, we can reach that $\d x\ g\ d\bar p/d\bar a$ is much less than $I\bar a\cdot \ grad\ \rho$ that is the deposit of the third term is negligible. We obtain the following approximate formula
\begin{equation}
\frac{\partial \bar p}{\partial t}\approx -I\ grad\ \rho-\kappa\rho\ grad\ V.
\end{equation}

We can thus obtain the following equation on the density of the diffusion swarm:
\begin{equation}
\frac{\partial^2\rho (r)}{\partial t^2}=\int\limits_{S(r)}(-I\ grad\ \rho -\kappa\rho\ grad\ V)\bar n(r')\ dS(r'),
\label{swar}
\end{equation}
where coefficients $I,\kappa$ are computed by formulas \ref{intens}. 

We will show that the quantum swarm satisfies the equality \ref{swar}, which means the approximation of quantum dynamics by the evolution of swarm. 

The method of collective behavior permits to give the simple algorithm for the computation of energy, impulse and momentum of impulse for quantum particle represented as the swarm of samples. Accordingly the rules of quantum mechanics, to obtain the value of any magnitude we must fulfill the averaging over the dual magnitude. For the computation of the impulse we must fix some sample and time segment $\Delta t$ and fulfill the averaging of impulses on any lengths of its run. This recipe gives in our notations the vector
\begin{equation}
(\sum\limits_r cm\frac{n_x(r)^+-n_x(r)^-}{n(r)}, cm\frac{n_y(r)^+-n_y(r)^-}{n(r)}, cm\frac{n_z(r)^+-n_z(r)^-}{n(r)})
\end{equation}
equal to the total impulse of all samples in the swarm computed over the run in the fixed time frame. The analogous computation of the momentum of impulse gives the average momentum of impulse of samples, of the potential energy - the average potential energy. 

To find the average kinetic energy we must fix a sample and the length $\d x$ and fulfill the averaging of the kinetic energy over all time moments $t$. It means that we must sum the energies of only those samples, which are moving in the current time instant, e.g., for any cube if the speed is $v=cn_x/n$ along the axes $x$ the average number of such samples is $nv/c$, that is their ration in the all set of samples is $n_x/n$. Here we use the non relativistic supposition that the ration of moving samples is small. The total kinetic energy computed according to this rule is 
$$
mc^2n_x n_x/n+mc^2n_y n_y/n+mc^2n_z n_z/n
$$
that coincides the kinetic energy found by the formula $Mv_{mean}^2/2=((cn_x/n)^2+(cn_y/n)^2+(cn_z/n)^2)nm/2$. The laws of conservation for the impulse, energy and momentum of impulse then follow from the classical laws and the non-relativistic supposition. In the next section we prove that the dynamics of the diffusion swarm well approximates the quantum dynamics. Using Erenfest theorems and the law of conservation for quantum mechanics we conclude that the proposed method of the computation of these magnitudes $A$ gives its quantum average values found by the formula $\langle A\rangle = \int\Psi^*(r)A\Psi(r)dr$. 

\subsection{About the diffusion swarm with the non uniform intensity}

The coefficient at the Laplace operator in the diffusion equation we call the diffusion intensity. The diffusion intensity determines the number of samples passing through the unit square in the unit time. In the simulation of quantum dynamics we need the diffusion process with the non uniform intensity. It means that the diffusion intensity depends on the chosen grain of space $\d x$. 

At first we consider the case when the external potential is uniform $grad\ V=0$.
The diffusion process with the non-uniform intensity can be obtained by the special mechanism, which we call the threads. We illustrate it on the following example. We suppose that all samples move not in all the space but along some closed trajectory (thread) determined by the homeomorphic inclusion of the circle to the space: $\g :\ S^1\ar R^3$. We also suppose that the speed exchange process goes along this trajectory only so that all samples remain on it, only its linear density changes. It is equivalent to the imposition of holonomic bounds to samples. Let us suppose that the linear densities as well as the module of the speed of samples are almost the same in each point of the space. We take a cube containing one point on this trajectory. The flow of samples through its border will not depend on the length of its side $\d x$, because there is only one trajectory. This type of the diffusion process will thus have the intensity proportional to $1/(\d x)^3$, because the quantity of samples penetrating in the unit of time into the cube with the side $\d x$, is independent of $\d x$, and the density is the ration of the total quantity to the volume. This example is not very good because many areas turn to be free of particles at all. 

 We consider the other example. Let the space be divided to cubes which are grouped on the layers $1,2,\ldots$. For each layer $j=1,2,\ldots$ the cubes of the layer $j+1$ consist of 8 cubes of the layer $j$ each, and their side is correspondingly, twice bigger. For any $j$ the exchange of samples between the neighbor cubes of the layer $j$, included in one cube of the layer $j+1$, goes only through the narrow passage which wide does not depend on $j$. The quantity of samples moving between the cubes of every fixed layer is thus independent of the number of this layer. We can guarantee it by the appropriate choice of pairs for the impulse exchange operation. This mechanism gives us the required intensity of diffusion, proportional to $1/(\d x)^3$, due to the definition of the density (\ref{density}). 

Now we consider the case of non-uniform external potential. At each step of the evolution for samples which lost their speed in the exchange we will use the rule from above, ensuring the diffusion intensity proportional to $1/(\d x)^3$. The samples obtained the speed from the external potential will move as usual, independently of the layers. It gives the formula (\ref{intens}).
This organization of space shows how in principle could we obtain the required non uniform intensity of the diffusion in the same swarm. 

However, the described method can be difficult for programming because we have to trace on the positions of all samples in the swarm, which means the refusal from the uniform space, or the passage to fractal space with the fraction dimensionality resulted from the non uniform lattice. 
We will fix the grain of special resolution $\d x$ from the very beginning such that the corresponding approximation of the wave function is sufficient for us, and will consider the dynamical diffusion swarm with this intensity found by the formula \ref{intens}. If we do not satisfied with the obtained dynamical picture, we have to choose the other value of $\d x$ and repeat all the work again.

\subsection{Equivalence of quantum and diffusion swarms}

The aim of this section is to prove that the sequence of diffusion swarms is the appropriate approximation of quantum unitary dynamics. We have defined the quantum swarm as satisfying the equation \ref{swarm} and which evolution is reduced to the local shifts of samples.  

At first we prove that the quantum swarm exists, e.g., it is possible to obtain the fulfillment of \ref{swarm} by the local shifts of samples. We then prove that the mechanism of the movement of the quantum swarm samples coincides with the diffusion, which gives the main result. 

Let us take up the quantum swarm. We start from Shredinger equation
\begin{equation}
ih\frac{\partial\Psi(r,t)}{\partial t}=-\frac{h^2}{2M}\Delta\Psi(r,t)+V_{pot}(r,t)\Psi(r,t),
\label{Sh}
\end{equation}
which we can rewrite as 
\begin{equation}
\begin{array}{lll}
&\Psi^r_t(r)&=-\frac{h}{2M}\Delta\Psi^i_t(r)+\frac{V_{pot}}{h}\Psi^i(r),\\
&\Psi^i_t(r)&=\frac{h}{2M}\Delta\Psi^r_t(r)-\frac{V_{pot}}{h}\Psi^r(r)
\end{array}
\label{Sh2}
\end{equation}
for the real and imaginary parts $\Psi^r,\ \Psi^i$ of the wave function $\Psi$. We need only the evolution of density of quantum swarm that is the function
$$
\rho(r,t)=(\Psi^r(r,t))^2+(\Psi^i(r,t))^2.
$$

We fix the value of $\d x$ and apply for the approximation of the second derivatives the difference scheme
$$
\frac{\partial^2 \Psi(x)}{\partial x^2}\approx\frac{\Psi(x+\d x)+\Psi(x-\d x)-2\Psi(x)  }{(\d x)^2}
$$
fro each time instant, assuming that the wave function satisfies all conditions of this approximation. The addition of any constant to the potential energy $V_{pot}$ does not influence to the evolution of the quantum swarm density. We then can treat instead of $V_{pot}$ the other potential $V=V_{pot}+\a$, where $\a=-\frac{3h^2}{m(\d x)^2}$, which leads to the disappearance of the summand $2\Psi(x)$ in the difference schemes for the second derivatives on $x,y,z$ (it gives the coefficient $3$) after their substitution to Shredinger equation. We introduce for the simplicity of notations the coefficient 
$$
\g = \frac{h}{2M}\frac{1}{(\d x)^2}.
$$
Since we yet do not know the mechanism of the movement of samples in the quantum swarm we simply suppose that we take some of them from one cell or place them to the cell from some storage. We divide the evolution of the quantum swarm to so short periods of duration $\d t$, that on each of these frame samples move only in between two neighbor cells. If we prove that the evolution of quantum swarm at this frame can be obtained by the diffusion mechanism, it would be true for all evolution as well, because our supposition about the exchange between the neighbor cells does not decrease the generality. We also assume that these cells differ one from another by the shift to $\d x$ along the axes $x$, also without the loss in generality. We denote the centers of these cells by $x$ and $x_1=x+\d x$. Then in our suppositions about exchange the summand $\Psi(x-\d x)$ in the difference scheme also disappears and at the short time frame the quantum swarm evolution results from the following systems of equations: 
\begin{equation}
\begin{array}{lll}
&\Psi^r_t(x)&=-\g\Psi^i(x_1)+V(x)\Psi^i(x),\\
&\Psi^i_t(x)&=\g\Psi^r(x_1)-V(x)\Psi^r(x),
\end{array}
\label{quant}
\end{equation}
and the analogous system of equations obtained by the replacement of $x$ to $x_1$ and vice versa. This system arises in the supposition that samples move from $x$ to $x_1$. For the opposite direction of the movement: from $x_1$ to $x$ we obtain the analogous system with the replacement of $x$ by $x_1$ and vice versa. Without loss in generality we can take periods $dt$ so short that on each of them we have only one from these possible direction of movement. Therefore, we need, due to the complete symmetry, to regard only the case when samples move from $x$ to $x_1$; the second case is completely analogous. Shredinger equation thus reflects the result of the common process of evolution consisting of these two cases, where $x$ and $x_1$ can be dispose by 6 different ways along 3 coordinate axes. Now by the period $dt$ we mean just so small frame in which the exchange goes only from $x$ to $x_1$.

For this period we have 
\begin{equation}
\begin{array}{lll}
\frac{\partial\rho(x)}{\partial t}&=2\Psi^i(x)(\g\Psi^r(x_1)-V(x)\Psi^r(x))&+2\Psi^r(x)(-\g\Psi^i(x_1)+V(x)\Psi^i(x))=\\
&=2\g (\Psi^i(x)\Psi^r(x_1)-\Psi^r(x)\Psi^i(x_1))&=-\frac{\partial\rho(x_1)}{\partial t}.
\end{array}
\label{1der}
\end{equation}
It means that the decrease of samples in one cell equals their increase in the other, e.g., the evolution of quantum swarm satisfies the condition of locality. Now for the comparison of this evolution with the diffusion dynamical swarm evolution we find the second derivative of the quantum density on the time:
\begin{equation}
\begin{array}{ll}
&\frac{\partial^2\rho(x)}{\partial t^2}=2\g [ (\g\Psi^r(x_1)-V(x)\Psi^r(x))\Psi^r(x_1)+\Psi^i(x)(-\g\Psi^i(x)+V(x_1)\Psi^i(x_1))-\\
&(-\g\Psi^i(x_1)+V(x)\Psi^i(x))\Psi^i(x_1)-\Psi^r(x)(\g\Psi^r(x)-V(x_1)\Psi^r(x_1))]=\\
&2\g^2(\Psi^r(x_1))^2-2\g V(x)\Psi^r(x)\Psi^r(x_1)-2\g^2(\Psi^i(x))^2+\\
&2\g V(x_1)\Psi^i(x)\Psi^i(x_1)+2\g^2(\Psi^i(x_1))^2-2\g V(x)\Psi^i(x)\Psi^i(x_1)-\\
&2\g^2(\Psi^r(x))^2+2\g V(x_1)\Psi^r(x)\Psi^r(x_1)= \\
&2\g^2((\Psi^r(x_1))^2+(\Psi^i(x_1))^2-((\Psi^r(x))^2+(\Psi^i(x))^2))+\\
&2\g [(V(x_1)-V(x))((\Psi^r(x))^2+(\Psi^i(x))^2)+o(\d x)],
\end{array}
\end{equation}
where $o(\d x) = (\Psi^r(x)\Psi^r(x_1)+\Psi^i(x)\Psi^i(x_1)-((\Psi^r(x))^2+(\Psi^i(x))^2))(V(x_1)-V(x))$. Now we compare it with the expression for the second derivative of the diffusion swarm density, found in the previous section, taking into account our agreement that the exchange goes between the neighboring cells along the axes $x$. Comparing with \ref{swar} in view of \ref{intens}, we conclude that the second derivative of the quantum swarm density asymptotically converges to the density of the diffusion swarm.

Now we choose as the initial state such a state of the diffusion swarm in which its density has Gaussian form and coincides the main state of harmonic oscillator. Then for the corresponding value of the energy $V=a(x^2+y^2+z^2)$ we have $\partial\rho /\partial t=0$ in the initial instant in any points in the space. As we saw, the second derivatives of the density of the quantum swarm and the density of the dynamical diffusion swarm are approximately the same, the diffusion swarm is the good approximation of the quantum density on some interval $\Delta T$. If we switch slowly some potential, it gives the approximation of any quantum unitary evolution in the limit of swarms for $n\ar\infty$.

The swarm approximating quantum dynamics of one particle depends on the choice of $\d x$. After the choice of the value for $\d x$ we obtain for the unlimited decreasing of $\d t$ the approximation of the exact wave function within $\d x$. Here the intensity of diffusion will be determines by the chosen value of $\d x$, namely, it will be $\frac{h^3}{m^3c(\d x)^3}$. E.g., if we want to decrease the step of spatial resolution we must admit the bigger ration of the moving samples in the unit of volume. It reflects the uncertainty relation "space-impulse": the dispersion of speeds of samples will increase with the decreasing of the grain $\d x$. In any case for the obtaining of the dynamical picture it is required to fix the grain $\d x$. 

If the total number $n$ of samples is limited we obtain the model of quantum dynamics with the inbuilt decoherence. We can generalized such a model to the case of many particles, and it will then serve as the approximation of quantum dynamics in the standard Hilbert formalism (see (\cite{Oz6}). The appropriateness of this scheme for the numerical simulation follows from the fact that it gives Born rule for quantum probability, which thus turns inbuilt in the algorithmic formalism, in contrast to Copenhagen formalism where this rule is merely the axiom. 

\subsection{Restoration of wave function from the dynamical diffusion swarm}

We have solved the problem of approximation of the quantum dynamics by the special dynamical diffusion process with non-uniform intensity. The dynamical diffusion swarm is determined by the pair \ref{pair}. This pair does not contain the notion of complex number which in the standard formalism induces the quantum interference and does not give the beautiful equations of Shredinger type for $\rho$ and $\bar p$. Moreover, the mechanism of dynamical diffusion we introduced for the simulation of quantum dynamics differs radically from the classical processes (for example, from the heat transfer of oscillations) because its intensity depends on the chosen grain of spatial resolution. We admit this for the sake of the main: the economy of the computational resources required for the simulation of quantum dynamics. 

Now to complete the picture we must solve the inverse problem: to show how we can restore the usual complex wave function $\Psi$ from a given state of the dynamical diffusion swarm. \ref{pair}. 
For this we turn to the equation \ref{1der}, and substitute to it the expression for the wave function by the density: $\Psi (r)=\sqrt{\rho (r)}\exp(i\phi (r))$. The problem is to find the phase $\phi (r)$ of the wave function. We mention that only the relative phase between the different points has the physical sense, and we thus can fix some point $r$ and regard the phase in the other point $r_1$ relative to $r$. If $r_1$ is close to $r$, the equation \ref{1der} givfes us 
$$
\phi(r)-\phi(r_1)=arcsin\ k(\d x)^2\frac{\bar p(\bar r-\bar r_1)}{\sqrt{\rho(r)\rho(r_1)}}
$$
that leadds to the following formula for the relative phase:
\begin{equation}
\phi(r_1)=\int\limits_\g k(\d x)^2\bar v\ d\bar\g
\label{phase}
\end{equation}
where the contour $\g$ goes from $r$ to $r_1$. This definition explicitly depends on the choice of the contour $\g$ and we thus have to prove its correctness, e.g., its independency of the contour $\g$. Since the phase is determined within the integer multiple of $2\pi$, the different contours can lead at most to the change of phase to such number that takes place, for example, for the excited states of electron in the hydrogen atom with the nonzero magnet number $m$ (say, $3d$). We show that the integration of the speed $\bar v$ of the swarm along the closed contour preserves its value in time the more accurate, the less is the grain of spatial resolution $\d x$. It will give that if in the initial moment the definition (\ref{phase}) was correct, it will be correct for the future moments as well.  

For this we consider the derivative of the integral of the speed of swarm along the closed contour $\g_c$. Using the formula (\ref{swar}) and taking into account $\partial\bar p /\partial t=\rho\ \partial\bar v/\partial t$, we obtain
\begin{equation}
\frac{\partial}{\partial t}\int\limits_{\g_c} (\d x)^2\bar v\ d\g=-\int\limits_{\g_c}I(\d x)^2\frac{grad\ \rho}{\rho}+\kappa (\d x)^2\ grad\ V.
\end{equation}
The first summand gives after the integration along the closed contour zero, because it is $grad\ \ln\rho$, the second summand gives zero after the integration by analogous reason.  

Now it is sufficient to convince that the definition (\ref{phase}) is correct in the initial instant that can be done explicitly for each concrete problem. In case when the wave function of the initial state can be obtained from the ground state in the Coulomb field where $\bar v=0$, the correctness follows from the proved because here there is no phase shift on $2\pi k$. If to obtain the initial condition we have to start from some excited state with the nonzero phase, we must check the correctness for this state.

Now, applying these computations we can write the formulas connecting the swarm parameters with the wave function:
\begin{equation}
\begin{array}{ll}
&|\Psi(r)|=\sqrt{\rho(r)};\\
&\phi(r)=\int\limits_{\g :\ r_0\ar r}k(dx)^2\bar v\cdot d\g,\\
&\bar v = a(dx)^{-2}\grad\ \phi(r),
\end{array}
\label{connection}
\end{equation}
Using these formulas, we can pass from the wave function to the swarm and vice versa with any required accuracy. The single peculiarity of the dynamical swarm description of quantum dynamics will be that the swarm parameters substantially depends on the grain of spatial resolution $dx$.

\subsection{The method of collective behavior for many particles}

We generalize the method of collective behavior to the case of quantum systems of many particles.  

We now regard the problem of quantum many body dynamics in the full generality. The main guiding line for us will be Hilbert many particle formalism. We will apply it not in the algebraic, but in the constructive form. For this, we introduce the important notion of bounds between samples. We will group samples of the different real particles in the corteges, so that each cortege represents one version of the dynamics of the whole ensemble. This formal model is convenient for the description of complex dynamical scenarios that we will see in the next section.

Let we are given a set of $n$ quantum particles which we will enumerate by the natural numbers: $1,2,\ldots,n$. We assume that the main act of the evolution is the reaction of scattering, when these particles fly at each other and can associate in the stable systems called molecules, or dissociate to compounds. Initial particles can consist of more elementary particles, which can regroup in course of the reaction. This regrouping forms the products consisting of the same elementary particles but in the different configuration. 

We divide the evolution to the scattering acts
\begin{equation}
S_1\ar S_2\ar\ldots\ar S_k
\label{ev}
\end{equation}
so that the main (physical) time is represented by the lower index of a state $S_j$ of our system, and the internal time inside each scattering act is negligible in comparison with the physical time, it will make sense only inside the scattering act itself. It means that each scattering act is so short that particles have no time to change their spatial positions but the state of the system changes: $S_j$ passes to $S_{J+1}$. The evolution goes not by the dynamics of particles, but by the change of its collective state, which we describe in this section applying Hilbert formalism. 

The simple example of this reaction is the scattering of the proton on the hydrogen atom. Here the flying proton (the proton number 1) flies to the standing hydrogen atom, which in turns consists of the proton number 2 and the electron. The possible products of this reaction are: 

\begin{itemize}
\item a) the separate proton number one and the hydrogen atom (nothing happened),
\item b) the separate proton number two and the hydrogen atom formed by the proton number one and the electron, 
\item c) the molecular ion of hydrogen,
\item d) two separate protons and the electron.
\end{itemize}

The most interesting to us are the cases of recombination of compounds (b) and the association to the molecule (c). 

We assume that the observable chemical dynamics does not depend on the sub nuclear states, e.g., electrons and nuclei as the integral objects determine the chemistry. This is in principle under question for very complex objects but may be true for them as well provided we treat not too large periods (the more general structure we discuss in the last chapter). We also assume that our consideration is based on quantum mechanics and the single essentially unconventional procedure is the model of decoherence. 

The most right model of decoherence, which we called absolute claims that decoherence represents the reduction of the quantum state

\begin{equation}
|\Psi\rangle =\sum\limits_j\la_j|j\rangle
\label{psi}
\end{equation}
in the instant where the memory of the simulating computer cannot hold the full notation of this state. We have shown above that this model gives Born rule for quantum probability that proves the correctness of this model. However, this form of the absolute model of decoherence cannot serve as the core of the simulating algorithm because we yet have no method of the simulation of unitary quantum dynamics of many particles but the computations by the matrix algebra, that are too recourse-intensive. 

To build the robust scheme of the simulating algorithm we need to apply the method of collective behavior sequentially, where the algorithmic reduction of quantum states is the inbuilt property. 
We consider the swarm representations of our $n$ particles $1,2,\ldots,n$, where  $S_1,S_2,\ldots,S_n$ are the swarms of samples corresponding to their states $|\Psi_1\rangle, |\Psi_2\rangle, \ldots, |\Psi_n\rangle$. The ensemble consisting of all these samples will represent the not entangled state of the form $|\Psi_1\rangle\bigotimes |\Psi_2\rangle\bigotimes \ldots\bigotimes |\Psi_n\rangle$. To represent the not entangled state of the form 
\begin{equation}
\Phi\rangle = \sum\limits_{j_1,j_2,\ldots,j_n} \la_{j_1,j_2,\ldots,j_n}|j_1,j_2,\ldots,j_n\rangle
\label{many}
\end{equation}
we need to introduce the new substantial element into the method of collective behavior. This is the notion of bounds between the samples from the different swarms. The basic state $j_i$ can be treated as the coordinates of the particle $i$ in the corresponding configuration space. The representation of the wave function in the form \ref{many} means that there exist the bounds connecting the points $j_1, j_2,\ldots,j_n$ in one cortege. The relative quantity of such bounds (their total number divided to the general quantity of all bounds) is $|\la_{j_1,j_2,\ldots,j_n }|^2$. 

In the method of collective behavior, we accept that the bounds connect not spatial points but samples of the different real particles. We can write these bounds as the corteges 
\begin{equation}
\bar s =(s_1,s_2,\ldots,s_n)
\label{cortege}
\end{equation}

where for any $j=1,2,\ldots,n\ s_j\in S_j$. The wave function $|\Phi\rangle$ is then represented as the set $\bar S$ of corteges $\bar s$ so that for any $j=1,2,\ldots,\ s_j\in S_j$ there exists exactly one cortege of the form \ref{cortege}. Any cortege plays the role of the so-called world in the many world interpretation of quantum mechanics.\footnote{Strictly speaking, for this we need to treat the numbers $t$ as the part of corteges. In the non-relativistic theory there is the common time $t$.} We consider this cortege \ref{cortege} as the probe representation of the system of $n$ particles and agree that all interactions go inside of one cortege, whereas the state of the real system resulted from the "interference of amplitudes" corresponding to all corteges occurring in the current spatial cell, or by some other process replacing the interference of amplitudes.  In the case of one particle, we saw that the impulse exchange in the dynamical diffusion swarm could be such a process. We will also generalize this process on the case of $n$ particles. We call $\bar S$ the swarm representation of the system of $n$ particles.

The density of the swarm $\bar S$ we determine as  
\begin{equation}
\rho_{\bar S}( r_1,r_2,\dots,r_n)=\lim\limits_{dx\ar\infty}\frac{N_{ r_1,r_2,\dots,r_n,\ dx}}{(dx)^{3n}},
\label{density}
\end{equation}
where $ N_{ r_1,r_2,\dots,r_n,\ dx}$ is the total number of corteges occurred in the $3n$ dimensional cube with the side $dx$ and center $ r_1,r_2,\dots,r_n$. 

If the wave function $|\Phi\rangle$ is the tensor product of one particle wave functions:
$$
\Phi\rangle = \bigotimes\limits_{i=1}^{n}|\phi_i\rangle 
$$
then the corresponding bounds can be obtained by the random choice of samples from the uniform probability distribution $s_j\in S_j$ for each $j=1,2,\ldots,n$, which thus form each cortege $s_1,s_2,\ldots,s_n$.
With this choice of corteges we find that the density of the corresponding swarm satisfies Born condition, which has the following form for swarms:
\begin{equation}
\sum\limits_{\bar r\in D}|\langle\bar r|\Phi\rangle |^2 = \frac{N_{\bar r, \bar S}}{N}
\label{born}
\end{equation}
where $D\subset R^{3n}$, $ N_{\bar r, \bar S}$ is the total quantity of corteges occurred in the area $D$. For the entangled state $|\Phi\rangle$ this choice of corteges for the list of swarms $\bar S$ does not give us the condition \ref{born}. We thus have to take \ref{born} as the definition of the choice of corteges in $\bar S$. To determine the swarm we must also determine the speeds of all samples. Namely, we must generalize the equality \ref{connection} to the case of $n$ real quantum particles.   

Let $\Psi(r_1,r_2,\ldots,r_n)$ be a wave function of the system of $n$ particles, 

$\Psi=|\Psi|exp(i\phi(r_1,r_2,\ldots,r_n))$ be its Euler decomposition. We denote by 

$\grad_j\phi(r_1,r_2,\ldots,r_n)$ the gradient of $\Psi$, taken by the coordinates of the particle $j$, where $j\in\{ 1,2,\ldots,n\}$ is the fixed number. The generalization of the formulas \ref{connection} to $n$ particles has the form 
\begin{equation}
\begin{array}{ll}
&|\Psi(\bar r)|=\sqrt{\rho(\bar r)};\\
&\phi(r)=\int\limits_{\bar \g :\ \bar r_0\ar \bar r}k(dx)^2\bar v\cdot d\g,\\
&\bar v = a(dx)^{-2}\bar \grad\ \phi(\bar r),
\end{array}
\label{connection_n}
\end{equation}
where $\bar r$ denotes $r_1,r_2,\ldots,r_n$, $\bar\grad$ denotes $(\grad_1,\grad_2,\ldots,\grad_n)$, and $\bar \g$ is the path in $3n$ dimension space. The rules \ref{connection_n} is sufficient for the determining of the swarm for a given wave function if we agree to join samples into corteges independently of their speeds. The microscopic mechanism of the swarm dynamics then acquires the following form. The exchange of impulses between two corteges of samples: $\bar s=(s_1,s_2,\ldots,s_j,\ldots,s_n)$ and $\bar s'=(s'_1,s'_2,\ldots,s'_j,\ldots,s'_n)$ is the exchange of impulses between two samples $s_j$ and $s'_j$ provided $\bar s$ and $\bar s'$ belongs to the same spatial cube in the configuration space $R^{3n}$ for $n$ particles. Here we choose $j$ arbitrarily from the uniform probability distribution. With this definition the reasoning represented above fro one quantum particle we repeat word by word, and we obtain that this microscopic mechanism of impulses exchange for $n$ particles ensures the approximation of $n$ particle quantum dynamics within $O(dx^{3n})$ for the determining of the wave function in the fixed time frame. 

This method of collective behavior gives us the base for the simulation of quantum evolution.

\subsection{Problems of simulation of the dynamical diffusion}

In the defined method of collective behavior, the influence of the external potential is transparent for the simulation but the diffusion model is not so clear. Its definition given in the previous section is such that in the practical realization of it the problem of the right distribution of the computational resources must be clarified. The exchange of impulse between the samples of one quantum particle goes on the basement of their proximity. It means that we have to search all the pairs of samples to determine is this exchange possible or not, or to store in the memory the contents of all spatial cells. It is desirable to find the more appropriate way of the realization of the collective behavior. 

Here we show one trick, which can be useful for this aim. It rests on the representation of the swarm as the vector of the form 
\begin{equation}
s_1(t),s_2(t),\ldots, s_l(t)
\label{ve}
\end{equation}
where the dependence of each samples of the time: $s_j(t)$ means that its coordinates depends on the time. Here the exchange of impulses in the dynamical swarm must look like the exchange between the neighbor samples: $s_j,\ s_{j+1}$. In the dynamical swarm when we proved the approximation of unitary dynamics, we used the model of impulse exchange as the simultaneous acquirement of the maximal mutually reverse speeds, or the reverse transformation. In the practical simulation, this trick is not good. If we ascribe the speed $v_j$ to every sample $s_j$ it means that we store this speed in the memory. Then it is more convenient to permit it to have all real values, than only two: $c$ and $-c$ along each axes, where $c$ is the maximal speed.  

\begin{figure}
\centering
\caption{Recharge in the scattering of proton on atom of hydrogen. The picture from the paper \cite{OO}.}
\vspace{150mm}
\makebox[180mm][l]{\includegraphics{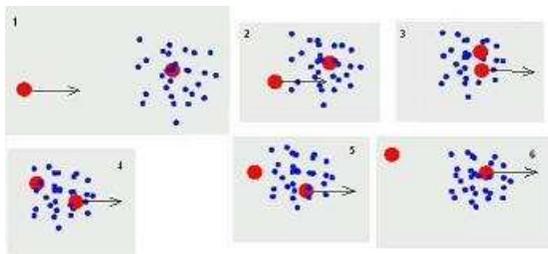}}%
\end{figure}

We can introduce the exchange of speeds through the averaging of speed on the neighboring samples in the vector \ref{ve}. We then have 
\begin{equation}
v_j(t+\delta t)=\a_j v_{j-1}(t)+\b_jv_j(t)+\g_jv_{j+1}(t)
\end{equation}
for some numbers $\a_j,\ \b_j,\ \g_j$. For this problem without photon emission it would be necessary to ensure the impulse conservation, that is $\a_j+\b_j+\g_j=1$, and to compute the energy as was shown above. The problem with one particle and photon emission we consider below. If we accept that all samples are identical, we have to require that $\a_j,\ \b_j,\ \g_j$ do not depend on $j$. It brings the little difficulty with $j=1$ and $j=l$. The simplest way is to asuume that $j$ takes values from the residue ring modulo $l$, e.g., to identify $l+1$ with $1$. This way is more appropriate for the states close to stationary. Fro the description of states depending on the time we can put $\a_1=\g_l=0$, that permit to avoid the usage of non existing values $v_0$ and $v_{l+1}$. 

This way of the simulation is more economical than the model with the impulse exchange between close samples. The point is that we avoid the scanning of all configuration space, which would be unavoidable if we use the notion of geometrical closeness. We must compare the swarm approach applied to the solution of differential Shredinger equation with the classical methods: analytical and numerical. These methods always are the schemes of integration. Classical methods (including the finite differences scheme) realize Riemann scheme of integration. Riemann scheme rests on the division of configuration space into small segments and the representation of the solution as the sum over all segments of the form 
\begin{equation}
\sum\limits_{j\in J_K} \Delta xf(x_j).
\label{ri}
\end{equation}
Riemann scheme \ref{ri} has drawbacks. The main is the non-effective expenses of the computational resources inbuilt to this scheme. This is the summing on the division $J_K$ of configuration space $K$. We can put up with it provided the configuration space dimensionality is not too much, as in the case of one particle. When the total number $n$ of particles grows the dimensionality $dim(K)$ grows linearly, as $O(n)$, but the size of the division $J_K$ (the accuracy of representation of any separate particles is conserved) grows as $exp(n)$ that makes Riemann scheme not scalable, e.g., inappropriate. This fundamental defect of Riemann scheme has the known corollaries. For example, it is unstable in the case of essentially discontinuous functions, which would be natural to treat as integrable, but not in Riemann scheme.\footnote{The easiest example: the function taking the value zero in the rational points and unit in irrational. Just such functions appear in quantum mechanics, when we apply path integrals. There the computation of speed of a quantum particle concentrated in the small area we obtain the unlimited dispersion of values. The graph of speed with the big magnification looks like a saw with the sharp teeth rather than as the smooth curve.}
The single plus of Riemann scheme is that it allows the analytical computations, again, for small dimensionalities $K$. 

The swarm method we proposed realizes the alternative, Lebesgue scheme of integration. In this scheme instead of the division of the configuration space we disperse samples randomly in the all space. After that we divide to the segments not the configuration space but the set of values of the function itself, and build the sum 
\begin{equation}
\sum\limits_i|G_i|f_i
\label{le}
\end{equation}
where $G_i$ is the quantity of samples for which the values of the function $f$ occur in the $i$-th segment corresponding to the typical value $f_i$. This scheme is not good for the analytical solution. However, it has the decisive advantage: it is extremely economical in the algorithmic sense. Here the summing does not depend on the complexity of the configuration space, e.g., on the total number of real particles. All collective effects contain in the samples themselves, which are completely at our disposal. 

The economy of such scheme makes it sensitive to the dynamical influence of the emitted photons. We consider by the swarm method the problem of the association of a free electron and the positive nucleus. The swarm representation of an electron in comparison with the classical representation of it as the point wise object leads to two features:
\begin{itemize}
\item the fast growth of the distance between samples when one end of the vector \ref{ve} approaches the nucleus,
\item the fast stabilization of the electron flow along the circular orbit after the addition of some impulse to the far end on the vector.
\end{itemize}
The first effect arises due to the fast growth of the Coulomb potential with the decreasing of the distance to the nucleus. The second effect arises because the impulse exchange along the vector \ref{ve} leads to the redistribution of samples to the more stable orbits corresponding to speeds they acquired in the rotation in Coulomb field.

In terms of swarms, we can even interpret the polarization of photons orthogonal to the impulse that arises from the character of the movement of the charged electron as the swarm in Coulomb field, namely, from the stabilization of circular orbits for the samples. This scheme does not give the exact expressions fro wave functions of stationary states, e.g., it yields to the standard way of the solution of Shredinger equation, where such states appear straightforwardly. The advantage of the swarm approach is the different: it gives the effective description of the dynamical picture, which can be extended to the case of many particles, e.g., to the area where Shredinger equation is not applicable. 

Here we at the first time meet the necessity to account the emitted photon for the obtaining right picture of the dynamics. Electro dynamical effects appear in the constructivism as the unavoidable effect of the drawing of right dynamical pictures.\footnote{Strictly speaking, the appearance of such methods of drawing is the reaction of the simulating algorithm to the expert conclusion of a user, which has the form of oracle in the final form of algorithm. We will touch the possibilities of the practical usage of such oracles in the computational devices.} If we have no QED, we could try to reproduce it by means of such algorithmic "drawing". Fortunately, this step is already passed\footnote{Not without the elements of "drawing". Maxwell himself used pinions as the vivid image of representation of photons in the deduction of electro dynamical equations. In constructivism, we must do something opposite.}
, and we then can use the advantages of electrodynamics as the exact formulas for the amplitudes of photon emission (Feynman's formulas, see below). Such usage of standard quantum theory (SQT) as the debugger for the simulating programs is typical for the constructive physics. It is applicable in the sphere of SQT, e.g., for the problems reducible to one particle. In the complex problems with many particles, SQT cannot help us, and there constructive methods of "drawing" will be our single support until the moment when we can use the heuristic of the other sciences, for example, biology.

\section{Simulation of quantum dynamics via quantum state selection}

 We now formulate the principle of quantum superposition of the many particle states in the form of genetic algorithm of building of the many particle scenario. We call this algorithm the selection of quantum states. It is easy and has a wide area of applicability; for example, it can describe the various chemical reactions and phase transitions. 

\subsection{Scattering of a proton on a hydrogen atom}

Qauntum state selection is the form of the method of collective behavior. We illustrate it on one example: the scattering of a proton on a hydrogen atom. 
We have the motionless in the given reference frame hydrogen atom consisting of the proton and the electron. The other proton flies to this atom from the large distance. The main channels of this scattering are the followings. 
1) The complete decay of the system to two separate protons and the electron. 
2) Conservation of the form of products (one proton and a hydrogen atom). 
3) The molecular ion of hydrogen, consisting of two protons and one electron. 

At first, we consider the channel giving the molecular ion. We make the following simplifying assumptions (later we refuse from them):

1) Protons, the hydrogen atom and the molecular ion we treat as classical particles, which are characterized by their coordinates and impulse. 

2) In course of the reaction, the electron is in the ground state in the field of two protons.

3) The reaction is the non-elastic scattering when the energy somehow disappears (it is necessary for the association).

This rough model makes possible to build the satisfactory video film of the reaction of association of the molecular ion. Instead of this rough scheme we can use the more refined considerations accounting passages between the stationary states of electron corresponding to each of protons (Landau Zener scheme), but it is not now significant. The main is that the protons we treat as the classical point wise particles, which in the reaction join to the new particle. Here the initial state will determine the final state. 

We now work out this scheme in detail, introducing the entangled states of GHZ type. Let for the distinctness we be given W and GHZ states. 
These states have the form \ref{GHZ_W}.
We choose as the measure of entanglement of a state $\Psi$ of bipartite system $S=S_1\cup S_2,\ S_1\cap S_2=\emptyset$ von Neumann entropy  $E_{\Psi,S_1,S_2} = Tr(\rho_{S_2}\ln \rho_{S_2})=Tr(\rho_{S_1}\ln \rho_{S_1})$ for the partial density matrix $\rho_{S_2}$. For a system $S$, consisting of many parts, we define the measure of entanglement as 
$$
E_{\Psi}=min_{S_1,S_2:\ S=S_1\cup S_2,\ S_1\cap S_2=\emptyset} E_{\Psi,S_1,S_2}.
$$
The other measures of entanglement are known, (see, for example, \cite{Ak}), but it is not important now.   

These states have the maximal measure of entanglement in some their vicinities in Hilbert space. Hence, the practical fabrication of such states is important for the perspectives of quantum computing and for the quantum information processing in general. Such states are obtained in the experiments with ions in Paul traps. The selection of such states follows from the simulation of molecular association. 

We determined the process of quantum state selection in the numerical experiments. We start from the semi classical description of the scattering process of one abstract particles called a bullet, on the other motionless particle, called a target. We write the parameters of them in this order.

We can represent one separate particle as the point wise object in the space possessing the set of attributes (mass, spin, charge, etc.). 
This representation we called the sample of this quantum particle. Each sample has its own trajectory as the classical particle. If we consider the particle as the quantum, it means that we represent it as the swarm of its classical samples. The density of the swarm is given by the formula
$$
\rho (\bar x)=\lim\limits_{\d x\ar\infty}\frac{N(\bar x, \d x)}{\d x^3}.
$$
The main property of the density connecting it with quantum mechanics is Born rule for the quantum probability:
$$
|\Psi (\bar x)|^2=\rho (\bar x).
$$
In principle, the last equation we can take as the definition of the swarm representation of the wave function.
We can use either collective behavior method, or Feynman path integrals. In the last case, the complex amplitude of samples is among its attributes. It is convenient to assume that all modules of amplitudes are equal, and they differ only in the phase. 
Each sample then moves along its own trajectory and accumulates the amplitude $\la$ depending on the action of this sample by the formula
$$
\la (t_1)=e^{-\frac{iS(\g,t_0,t_1)}{h}},\ \ S(\g,t_0,t_1)=\int\limits_{t_0}^{t_1}(E_{kin}-E_{pot})dt,
$$
the wave function appears after the summing of amplitudes of all samples occurred in the cube, as we defined earlier.

Alternatively, we can use the dynamical diffusion mechanism in the swarm of samples. This mechanism consists in the impulse exchange between close samples, which otherwise preserve their speeds. This mechanism ensures the admissible approximation of the unitary quantum dynamics determined by Shredinger equation, provided we fixed the grain of resolution. 

At last we can use the different algorithm for the determining of individual trajectories, for example, the semi classical representation of particles in the form of wave packages of Gaussian from (see \cite{FH}). We choose somehow the method of determining of the individual trajectories.

We consider the joint state of two real particles: the bullet and the target in the framework of the quantum formalism for many particles. 
This state has the following form:
$$
\sum\limits_{j,l}\la_{j,l}|\bar x_j,\bar x_l\rangle
$$
where $|\bar x_j\rangle$ and $|\bar x_l\rangle$ are the coordinates of bullet and target correspondingly. The pair $|\bar x_j,\bar x_l\rangle$ is the basic state of the joint system. The unitary evolution of such system induced by Shredinger equation contains two operators acting sequentially in each time instant: the operator of the kinetic energy $(-\frac{h^2}{2m}\Delta_{bul})\bigotimes(-\frac{h^2}{2m}\Delta_{tar})$, where Laplace operators $\Delta$ act independently to each coordinate of each particle, and the operator of the potential energy $V(\bar x_{bul},\bar x_{tar})$ which acts on two particles simultaneously. 

For the representation of quantum evolution by the method of collective behavior, it means that the action of the kinetic energy operator comes from the internal processes in the swarm of samples corresponding to the bullet and the target, whereas the potential energy operator action comes from from the interactions inside the separate pair of samples of the bullet and the target. 

We assume that the chosen model of evolution in the separate swarm of samples is so that we can trace the trajectory of every sample (Feynman path integrals do not satisfy this condition, but we can modify them by some artificial trick for that). We then can choose one sample from one swarm, the other sample from the other swarm and consider the interaction between these samples only, and the interaction of each of these two samples with the other samples of the same swarm. We must choose the partner-target for each sample of the bullet (all samples will be then joint in pairs) and gather the statistics. 

E.g., all interactions between two real quantum particles are reduced to the interactions inside of the selected pairs of samples from the different swarms. If we have the non-entangled state of the form $|\psi_1\rangle\bigotimes |\psi_2\rangle$ then we can choose pairs arbitrarily; if the state is entangled this is not the case. If $L$ is the general quantity of samples in each swarm we have $L^2$ possible pairs of samples. For $n$ real quantum particles we have corteges of the form $a_1,a_2,\ldots,a_n$ of their samples instead of pairs, as in the case of two particles. If the quantity of real particles grow the number of all corteges grows as $L^n$ that reflect the main computational problem of quantum theory for many particles. 

The computer simulation of many particle quantum dynamics thus meets the problem: how to choose the "essential" corteges $a_1,a_2,\ldots,a_n$ of samples, which gather the main part of amplitude. Let ${\cal B}$ denote the set of all corteges $a_1,a_2,\ldots,a_n$. 
The problem of selection is to find a small (not exponentially large) subset $B\subset{\cal B}$, such that $\int\limits_{B}|\Psi(\bar x)|^2\ d\bar x>1-\epsilon$ for the chosen error probability $\epsilon$. We call the problem of finding of such a subset $B$ the problem of state selection. If the bulk of $B$ consists of entangled states of some type $Z$, the selection problem can be called the problem of selection of quantum states of the type $Z$. If for some class of many particle quantum evolutions for their states $Z$ the set $B$ can be built by some effective algorithm, we say that for this class $Z$ the selection problem is constructively solvable. Here we assume that the set of initial states for the evolutions from $Z$ is simple (e.g., we can find it by the effective classical algorithm). 

We mention that the selection problem for quantum states not mandatory has the solution for any reasonable quantum evolution in the full Hilbert space. For example, it is unsolvable for the fast quantum algorithms characterized by that the amplitude is distributed almost uniformly on the exponentially large number of basic states. Fast quantum algorithms are not the single example. For Walsh- Hadamard operator $H^{\bigotimes n}$, where $H$ is Hadamard matrix the distribution of amplitude has this character, but it is not entangling. 

Nevertheless, we have reasons to expect that for some important types of quantum evolution the selection problem can be solved constructively. 
One of examples is the evolution of quantum systems which Lagrangian has quadratic form $L=\bar a\bar x^2 +\bar b\bar x\bar x_t +\bar c \bar x_t^2$ of the coordinates and impulses (for example, a system of harmonic oscillators). In the book \cite{FH} it is proved that for such systems all trajectories giving the significant deposit to the amplitude are classical that immediately gives us the effective method of the state selection, since we have only to trace trajectories generated by the simple algorithm describing the classical movement. 

The problem of state selection is actual for quantum computing as well. Its solution establishes the peculiar upper bound for the complexity of quantum evolutions, which we can simulate efficiently on a classical computer. The evolutions, which lie out of this area, require for its simulation a quantum computer. \footnote{Up to nowadays, fast quantum algorithms represented artificial constructions. Solving the state selection problem, we could try to find the more vivid examples of the fast quantum evolutions which themselves can serve as subroutines in the simulation of quantum systems. We can find such quantum subroutines among evolutions, which do not admit the constructive state selection.}

We describe the algorithm for our model system consisting of the bullet and the target. Factually, this restriction is not substantial, and this algorithm is applicable to systems with any number of particles. We start from some subset $B_0\subset {\cal B}$ which we can obtain by the random choice of a partner $a$ (a sample of target) for each sample of bullet $b$. This subset $B_0$ represents non-entangled state which we have in the initial instant. We have $B_0=\{ (a_1,b_1),(a_2,b_2),\ldots,(a_n,b_n)\}$. 

We consider the evolutionary operator acting on $B_0$ by the natural way: 
 $$
B_0\ar B_1=\tilde B_0=\{ (\tilde a_1,\tilde b_1),(\tilde a_2,\tilde b_2),\ldots,(\tilde a_n,\tilde b_n)\} .
$$
We now group all pairs from $B_1$ in some quantity of groups: $\Gamma_1,\Gamma_2,\ldots,$ so that for any $j=1,2,\ldots,s$
$\| a_j-\tilde a_j\|+\| b_j-\tilde b_j\|<\e_0$ and $\| v(a_j)-v(\tilde a_j)\|+\|v( b_j)-v(\tilde b_j)\|<\e_1$ for some small $\e_0,\e_1>0$, where $v(a)$ denotes the speed of sample $a$. In the other words we join in the same group pairs of samples having close spatial positions and close speeds of the corresponding elements of pairs. (Closeness of spatial positions implies the closeness of speeds, if we average the speeds on the position). We then select from all groups $\Gamma_j$ $m$ some groups: $\Gamma_{j_1},\Gamma_{j_2},\ldots,\Gamma_{j_m}$ such that there are more than $n_0$ elements in each of them fro some $n_0<n$. We call pairs from the selected $m$ groups the selected pairs. We move off the non-selected pairs from the consideration. The general quantity $\tilde n$ of all pairs $a_j,b_j$ now becomes smaller than the initial quantity $n$ in the set $B_0$. This step of the selection we call the rejection. 

To restore the initial quantity of pairs we add $n-\tilde n$ new pairs $(\tilde a,\tilde b)$ to the selected pairs. We can do that effectively; the canonical way to do it is the cross over. It means that we generate new pairs from two existing by the exchange of samples of the corresponding swarms. E.g., we form pairs $\tilde a_2,\tilde b_1$ and $\tilde a_1,\tilde b_2$ from pairs $\tilde a_1,\tilde b_1$ and $\tilde a_2,\tilde b_2$. Here the previous (parent) pairs we choose randomly from the uniform distribution. Such generation of pairs goes until we reach the initial quantity of pairs $n$ within one pair at most. This sequence of new states generation we call replication. The new set of pairs will now reflect the probability distribution fro the obtaining of the real pair of particles more effectively than the initial because we fulfilled the much concentration of amplitudes on states, which probabilities are larger. 

The step of selection this consists of two steps: rejection and replication.

We then repeat this procedure many times and obtain the sequence   
$$
B_0,B_1,\ldots
$$
of sets consisting of pairs of samples corresponding to our real particles: the bullet and the target. At the end of this sequence, which is determined by the period of the process, we have the resulted set of pairs, which gives the approximation of the final quantum state. 

Why this procedure of the quantum state selection corresponds to standard quantum description of the bipartite evolution? We show that the selected pairs at each step carry the prevailing part of quantum probability for the bipartite system in comparison with the rejected pairs. 

Indeed, for the selected $k$ pairs the amplitudes which they carry in sense of Feynman path integrals are close in the phase and equal in the module. The closeness in the phase follows from our method of grouping on the spatial closeness and formulas \ref{connection_n}. The close on spatial positions samples will be also close in the speeds, because the exchange of speeds goes fast. In the swarm, all interference is constructive in contrast to the computation of the wave function by the analytical method, for example, as the path integral. Since the resulted probability to find the pair (bullet, target) in the corresponding position will be approximately 
$$
|\sum\limits_{j=1}^k\frac{1}{n}| = \frac{k}{n},
$$
due to the almost constructive interference of amplitudes, whereas for the same total number of pairs which are not close in the sense of positions and impulses this probability will be about
$$
|\sum\limits_{j=1}^k\frac{\a_k}{n}|\approx\frac{\sqrt{k}}{n},
$$
that is much less than $\frac{k}{n}$ for large $k$. It follows from that the phases associated with the samples are distributed randomly for the pairs from the different groups. The selection process thus gives the good approximation to the standard quantum evolution if the total number of samples is large.

Practically, the quantity of samples in each swarm must be sufficiently large that we can recognize the features of quantum behavior of a real system following from all Feynman paths giving the substantial deposit to the final state. For example, for the experiment with the joint particle (bullet plus target) penetrating through two slits the total number of samples must be sufficient for the obtaining of the standard interference picture, etc.

The main advantage of the selection algorithm is that we need only the initial number of samples in each swarm, which depends on the quantity of particles linearly, whereas the standard method requires the memory depending on the quantity of particle as exponential.

The cost of such economy is that fast quantum algorithms lay out of this method if we use the standard computers. Beyond the area of its applicability lie also the processes where the amplitude is distributed on the large quantity of states without the visible concentration. Quantum entangled states obtained in experiments, like and W satisfy the conditions of applicability of the selection procedure. Therefore, the practical selection of these states in the numerical experiments must be our first aim.

\subsection{Effective selection algorithm for states of $n$ particles}

We show how to solve the selection problem fro the wide class of quantum evolutions of systems of the arbitrary number of particles. This class is determined by the requirement that the computational resource is expended by the most economical way corresponding to the absolute model of decoherence. We generalize the construction from the previous section to the case of $n$ quantum particles and show how to combine the description of unitary dynamics and decoherence. 

In the standard description of quantum dynamics of the wave function, the determining of the form of initial wave function represents the serious problem, because it essentially influence to the form of the resulted state. In our case we must take care that the choice of initial corteges \ref{connection_n} ensures the best approximation of the wave function of the initial state. However, in the evolution our model can substantially diverge from the exact solution of Shredinger equation or with the exact unitary solution of the scattering problem because our model includes decoherence. 
We can assume that one sample or one cortege (in the many particle case) carries the amplitude grain in the sense determined above, and we automatically will use the absolute model of decoherence. The choice of corteges is thus the main problem in the determining of the many particle swarm dynamics. 

The numerical experiments on the simulation show that we could hardly represent the choice of corteges as a transparent procedure. I propose for this aim the simple genetic algorithm for finding corteges by the sequential repetitions of the dynamical scenarios where the choice of the initial state for each scenario uses the result of the previous scenario. 

We will describe the genetic entanglement on the example of the scattering of $n$ quantum particles. We start from the non-entangled state of these particles in which the particle $j$ has the wave function $\Psi_j$, or, in the swarm representation, the swarm $S_j$. In the first scenario we determine the initial state of $n$ particle swarm $\bar S_{ini}$ so that corteges $\bar s$ are chosen randomly for the uniform distribution of samples inside each one particle swarm. 
After some small time $\Delta t$ of the swarm evolution we find its final state $\bar S_{fin}$. If the initial quantity of samples $N$ in swarms $S_j$ is sufficiently large, we repeat the reasoning from above and obtain the approximation of the wave function through $\bar S_{fin}$, where the final quantity of all samples serves as the factor of decoherence. The problem is how to manage with the strictly limited number $N$ of samples for the simulation of quantum dynamics with the admissible accuracy. Here the admissible accuracy means the right recognition of the products of reaction: for chemical reactions it will be the list of the possible resulted molecules with the corresponding probabilities of the outcomes, which depend on the initial state of all particles. 

With this limitation on $N$ we factually charge samples with two roles: at first, the simulation of quantum unitary dynamics of the wave function, at second, the (absolute) model of decoherence following from the amplitude grain. We mention that these two roles do not completely agree with each other. мы 
The approximation of the wave function requires the small distance $\|\Psi_{Shredinger}-\Psi_{swarm}\|$, whereas decoherence follows from the assumption that the grain of amplitude results in the nulling of small amplitudes (less than $\e$) of the summands, that can give the big disagreement with the unitary dynamics of the wave function in the dispersion of amplitude on basic states is large. It happens for the large value $n$ of real particles, but the cases where the amplitude somehow concentrates on small number of summands in the superposition. In these exclusive cases we observe quantum effects in the macroscopic pieces of matter. Such effects we discuss below; their description is one of the most probable areas of application of our approach. 

We have to choose corteges $\bar s$ so that the distribution of samples in them represents the most accurate form of unitary dynamics. We claim that the absolute decoherence, which appears in the course of this simulation will correspond to real experiments. 
It must be true for elementary scattering in the small period $\Delta t$. We call $R^{6n}$ the double configuration space for $n$ particles. Its sense is that we intend to consider the pair of states - the initial and the final. For each cortege $\bar s_{ini}$ in the initial swarm we suppose that the cortege $\bar s_{fin}$ is determined uniquely, if it results from $\bar s_{ini}$ in the evolution of this swarm. 

We choose the division of the double configuration space for $n$ particles to cells of the cube form, and join the resulting pairs $(\bar s_{ini},\bar s_{fin})$ of corteges in the groups ${\cal G}_1, {\cal G}_2,\ldots, {\cal G}_k$ so that each group consists of all pairs containing in the same cube of this division.
 
We enumerate these groups so that the quantities of pairs in them does not increase: ${\cal N}_1\geq {\cal N}_2\geq\ldots\geq {\cal N}_k$. We choose the first $k_1<k$ wrong pairs. Now we can give the definition of the initial state for the next scenario of scattering. We exclude all wrong pairs from $\bar S$, and regroup all their samples in a new fashion, accordingly to the rules of genetic algorithms. Here it would be appropriate to use such genetic methods as "cross-over" when the former members of wrong corteges is regrouped in order to make them close to right corteges. I will not stop on this item. Let us suppose that we somehow regroup wrong corteges so that the new set contains corteges closer to right than the former wrong corteges. For so defined initial state $\bar S_2$ of the second version of scenario we launch again the process of the dynamic diffusion, etc. It results in the sequence of pairs 
\begin{equation}
(\bar S^1_{ini},\bar S^1_{fin}),(\bar S^2_{ini},\bar S^2_{fin}),\ldots
\label{chain}
\end{equation}
where each pair $(\bar S^j_{ini},\bar S^j_{fin})$ represent the result of the repetition number $j$. 

The process of the passage from one pair to the next pair thus consists of three steps: 
\begin{itemize}
\item the dynamical diffusion of the swarm based on the impulse exchange, 
\item the selection of right pairs, and 
\item the replication of right pairs using samples from the selected pairs by the procedure of the "cross-over" type.
\end{itemize}

The impulse exchange between samples belonging to the spatially close but the different corteges, contained in the operator of the dynamic diffusion is the analogue of the mutation in the evolutionary programming. In terms of the many world interpretation of quantum mechanics a cortege is a world and the exchange of impulses between two samples from the different corteges represents the interaction between the different worlds. The chain \ref{chain} breaks when the number of elements in the selected groups becomes stabilize.   

The described method of state selection corresponds to the unitary dynamics for many particles. We turn to the representation of this dynamics in the form of Feynman path integrals (see \cite{FH}), where the wave function $\Psi$ in each time instant $t$ is determined by the equality:
\begin{equation}
\Psi(t,\bar r)=\int\limits_{R^{3n}}K(t,\bar r, t_1,\bar r_1)\Psi(t_1,\bar r_1)d\bar r_1,
\end{equation}
where $K$ is the kernel of our system, which is (within the coefficient) the amplitude carried by one cortege, if we accept that corteges carry complex amplitudes instead of their speeds as in the method of collective behavior. Amplitudes $K$, carried by corteges thus depend on the initial and final positions $\bar r_1$ иand $\bar r$ of corteges, and must be thus close for corteges with close initial and final positions. This is true if $\| r-r_1\|$ is small and there is no large variety of trajectories going from the point 1: $(t_1,\bar r_1)$ to the point 2: $(t,\bar r)$. We estimate the deposit to the probability $|\Psi(t,\bar r)|^2$ of two sets of corteges with $l$ elements each: the first set is a part of one group ${\cal G}_j$, and the first set consists of corteges which final positions ar eclose but the initial positions are not close, for simplicity let the initial positions be chosen randomly from the uniform distribution. The deposits of these two sets in the probability will be approximately 
$$
d_1=|\sum\limits_{s=1}^l\a |^2=k^2|\a |^2,\ \ d_2=|\sum\limits_{s=1}^l\a e^{i\phi_s}\approx k|\a |^2
$$
where the phases $\phi_s$ are distributed uniformly. The last approximate equality follows from the fact that the uniform distribution of phases 
$\phi_s$ gives the average divergence of the sum from zero of order the square root of the total number of summands. Consequently, the deposit of the first set will substantially prevail. Returning to the samples from the collective behavior we note that the information carried by the complex amplitudes are transmitted by the speed of samples, and the deposit of minor groups ${\cal G}_j$ for $j> k_1$ to the probability will be much less than the deposit of right corteges taken in the same quantity. It justifies the selection procedure we introduced.

We illustrate the action of this algorithm of state selection on our example of the association of two protons in the molecular ion of hydrogen. We suppose that the initial spatial position of the first proton and the atom of hydrogen are close enough to form the molecule. For the simplicity, we treat the electron not as the particle but as some factor inducing some attracting potential between two protons, which otherwise merely repulse from each other by Coulomb force. The exclusion of the electron from the consideration is not of principal, we do it for the simplicity. 

We suppose that the first choice of pairs (a sample of the first proton, a sample of the second proton) was made so that in many pairs protons are located a distance strongly different from the distance $r_0$ between protons in the stable state of molecular ion. Protons then will fly away to large distance in the short time and these pairs turn out to be in the different groups after our procedure of grouping. After some iterations of our process of the choice of scenarios it results in the growth of the quantity of pairs in which the distance between protons is close to $r_0$, and we obtain the prevailing quantity of pairs which form the molecular ion of hydrogen. 

\begin{figure}
\centering
\caption{Density of the proton pairs distribution depending on the distance between protons in the association of molecular ion.\newline Drawing from the work \cite{AO}.}
\vspace{120mm}
\makebox[500mm][l]{\includegraphics{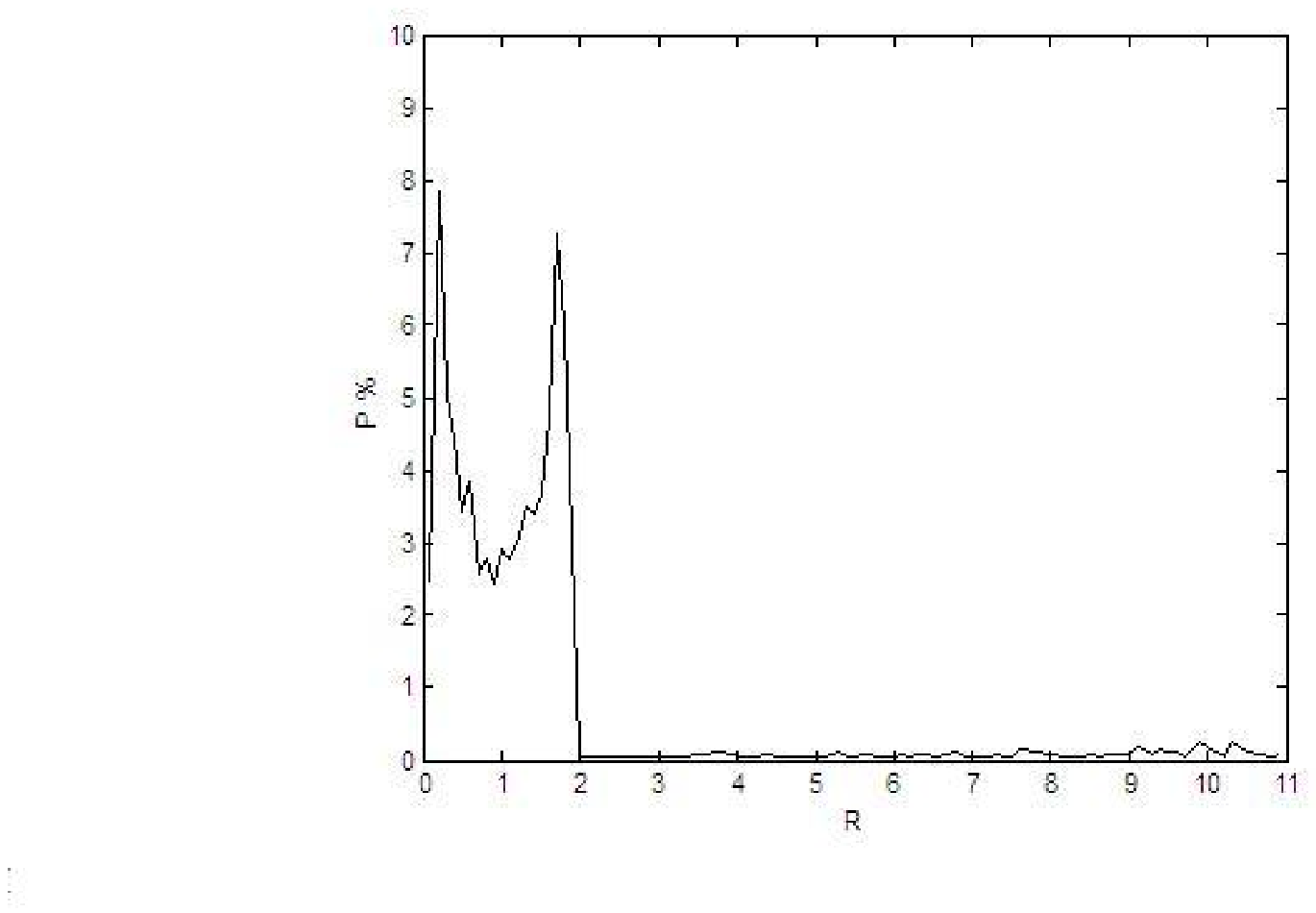}}%
\end{figure}

In the case when the initial positions of the first proton and atom are far this scheme can be extended. Here we have to account also photons emitted by the system, which decrease its energy and make possible the association of molecule. We can include photons in this selection scheme without problems but it requires some extension of its formalism, namely we need to add the time to the basic states. It will be done in the next paragraph. We mention that for the rough simulation it is not necessary to account emitted photons exactly. We can replace them by the kind of friction and divide the period $\Delta t$ to smaller segments so that at the last from them the spatial positions of reagents are close enough to apply the selection procedure. Of course, it gives only the rough picture of the association. Its more exact treatment requires the explicit inclusion of photons to the model.

The model of chemical reactions touches the basic notions of quantum theory. It uses Everett many world picture of quantum evolutions. Each cortege consisting of samples of real particles represents the separate quantum world. These worlds interact with each other, and the selection process plays the role of arbiter in this interaction. The interaction itself through the dynamical diffusion mechanism represent the exact form of pre-quantum fields (see \cite{Kh}), generating the visible quantum dynamics. 

\section{Identity of electrons from the viewpoint of collective behavior}

Since we suppose that the method of collective behavior can replace the traditional algebraic apparatus of quantum theory we need to interpret the identity of quantum particles of the same type. It shows as the symmetry or anti symmetry of wave functions of ensembles of these particles. We must interpret these properties of symmetry in terms of swarms. We consider the system of $n$ electrons with the same spin states. We start from the case $n=2$. Let al electrons have the same spin state, e.g., the wave function of this system in standard formalism, have the form $\Psi (r_1,r_2)$ where $r_1,\ r_2$ are the coordinates of the first and the second electrons, and satisfy the anti symmetry property $\Psi (r_1,r_2) =-\Psi (r_2,r_1)$. In the case of $n$ particles, which coordinates we denote by $\bar r=(r_1,r_2,\ldots,r_n)$ this condition looks as 
$$
\Psi(\bar r)=(-1)^{\sigma (\pi )}\Psi(\pi\bar r)
$$
where $\sigma (\pi )$ denotes the evenness of the transposition $\pi$. If we treat the coordinates of particles as the set of spatial and spin coordinates of them, the anti symmetry property concerns the full set of coordinates. 

The simplest way to build anti symmetric function is to consider a determinant of the form
\begin{equation}
D(f_1,f_2,\ldots,f_n;r_1,r_2,\ldots, r_n)=\frac{1}{\sqrt{n!}}det\left(
\begin{array}{lllll}
&f_1(r_1) &f_1(r_2) &\ldots &f_1(r_n)\\
&f_2(r_1) &f_2(r_2) &\ldots &f_2(r_n)\\
&\ &\ &\ldots\ &\ \\
&f_n(r_1) &f_n(r_2) &\ldots &f_n(r_n)
\end{array}
\right)
\label{det}
\end{equation}
where wave functions $f_j$ form the orthonormal system.  
Every anti symmetric wave function of $n$ variables can be represented as the row 
\begin{equation}
\Psi = \sum\limits_j^l\la_jD_j(\bar f_j;\bar r)
\label{sum_det}
\end{equation}
of such determinants for the different choices of one particle functions $\bar f^j$. 

The requirement of anti symmetry for wave functions relative to the full set of coordinates (including spin) reflects the Pauli principle. If we require the anti symmetry in the initial time instant $t=0$, then the solution of Shredinger equation corresponding to this initial condition will be anti symmetric for all further time instants, which follows from the general formula for the solution $|\Psi (t)\rangle =exp(-\frac{iHt}{h})|\Psi(0)\rangle$. 

In the standard formalism of quantum theory the anti symmetry of wave functions for systems of identical fermions is the axiom. 

We show how fermionic wave functions can be represented in the formalism of collective behavior. At first, we make one remark. Let us suppose that the supports of one-particle wave functions do not overlap. The swarm representation of the wave function of $n$ electrons will be equivalent to the representation of this function in the form of one determinant \ref{det}. Indeed, let us imagine that we compute the energy of the system with this wave function by the formula 
\begin{equation}
E_\Psi = \int\limits_{\cal R}\Psi^*(\bar r)H\Psi (\bar r)d\bar r.
\label{ene}
\end{equation}
Then for any cortege of the values of variables $\bar r =(r_1,r_2,\ldots,r_n)$ there exists no more than one set of indices $j_1,j_2,\ldots,j_n$, such that  
$f_{j_1}(r_1)f_{j_2}(r_2)\ldots f_{j_n}(r_n)\neq 0$, because in the opposite case the supports of these wave functions are overlapping. The sign of wave function in the computation of energy by the formula \ref{ene} does not play role because it cancels. Hence, the computation by the formula \ref{ene} in the case of non-overlapping supports can be fulfilled as if each particle is represented by its own swarm. 

We now show that the condition of non-overlapping supports of one-particle wave functions always can be guaranteed by the appropriate choice of the grain of spatial resolution. Really, let the supports of one particle wave functions $f_j$ overlap. We divide each cube $c^s$ in the division of configuration space for one particle to $n$ smaller parts (if $n$ is a degree of 8 there are cubes, in the other case it may be several cubes), which we denote by $c^s_1,c^s_2,\ldots, c^s_n$. Without loss of generality we can require that the support of the function $f_j$ in the cube $c^s$ is contained in $c^s_j$ for each $j=1,2,\ldots,n$. In the swarm terms it means that all samples of the swarm corresponding to a particle $j$, occurring in the cube $c^s$, belong to $c^s_j$. If this condition is not satisfied we always can shift samples of each particle to some distance, less than the side of initial cubes to satisfy this condition. Such shift causes no change in the wave function provided we put the grain of spatial resolution equal to the side $\Delta x$ of the cubes of initial division. After this small shift in the spatial position of samples we make the supports of one particle wave functions non overlapping. 

We mention that the condition of non-overlapping supports is always satisfied for the case of particles of the same type interacting by Coulomb law. Really, Coulomb potential induces the repulsion and is unlimitedly large in the point of disposition of each sample. Therefore, the situation when two samples of the same cortege are in the same spatial point is practically excluded. 

Therefore, the consideration of one determinant $D(\bar f;\bar r)$ equals to the consideration of our system of quantum particles of the same type in terms of collective behavior without entanglement. In this case, as we saw above, it is possible to consider merely the set of corteges in which samples from the different swarms are joint randomly. Hovewer, in the interaction states quickly become entangled. Here the wave function of the system with $n$ particles of the same type has the form \ref{sum_det}. In swarm terms it means the following. If $L$ is the number of particles in the swarm corresponding to one particle $j$, we divide this quantity to $l$ sets $S^1_j,S^2_j,\ldots, S^l_j$, so that their total numbers of elements are in the same proportion as the numbers $|\la_j|^2$ from the equation \ref{sum_det}. We now composw corteges so that they will be divided to the types $1,2,\ldots,l$. Here each cortege of the type $k$ will consists of samples belonging to the sets $S^k_1,S^k_2,\ldots,S^k_n$, where these samples from sets $S^k_j$ we choose randomly. It exactly corresponds to the rule of cortege forming for one determinant that we showed above. We see that in the method of collective behavior is not necessary to trace especially on the fulfillment of Pauli principle. It is satisfied for all charged particles of the same type due to the definition of the dynamical diffusion. Here we can build corteges from the determinant notation by the simple rule which generalizes the case of one determinant state \ref{det}. 

States expressed by one determinant of the form \ref{det}, we consider as non entangles in the fock space of occupation numbers. The correspondence between the method of collective behavior and Hilbert formalism we have defined, maps the states corresponding to \ref{det} in the swarm representation to non-entangled states. E.g., non-entangled states in Hilbert and Fock formalisms are transformed to the same class of states in the collective behavior formalism. 

We conclude that in the method of collective behavior particles are represented by their swarms of samples, which are joint into corteges corresponding to quantum "worlds". The identity of the particles of the same type is then the subject of the agreement for the simplicity of storage of large missives’ of such particles in the computer memory, and no more. The identity of particles of the same type is not thus the basic principle of the constructive version of quantum theory. It means the possibility to treat differently some particles belonging to the same type. We will return to the discussion of this question in the next chapter.

\section{Method of collective behavior for quantum electrodynamics}

The universality of the collective behavior formalism requires the reformulation of quantum electrodynamics on this language. It is not yet done. To do it we have to define how to represent the fundamental process of QED. This section is devoted to this question. Here we show arguments for that the required translation of QED is possible with the full conservation of the scheme of collective behavior we determined earlier. The translation of QED to the language of collective behavior must preserve the main intuitive representation of QED processes:

- the treatment of photons as particles moving with the maximal speed,

- the consideration of photons in the impulse-energy basis in the space of states,

- the rules for the computation of the probabilities of the main electrodynamics process: the interaction between the charged particle and photon, which gives Dirac equation for the electron in the electromagnetic field.

\subsection{Swarm representation of charged particles in electromagnetic field}

The swarm representation of charged particles given in the previous section is fully preserved. Photons must be represented in the form of special set ${\cal S}=\{ {\cal s}_1,{\cal s}_2,\ldots,{\cal s}_l\}$ of their samples. We thus do not separated photons as the particles, but include them in the description of systems as their samples. We call such photons virtual; they form the object called in QED the electromagnetic field. Each photon sample  ${\cal s}_j$ has the coordinates $\bar r_j$, the time $t_j$, the vector of impulse $\bar p_j$, and the vector of polarization $\bar \e_j$. 

A real photon is not a part of the signature of the system, but a set of corteges of the charged particles representing some quantum state of the type W, selected by the rules especially formulated by the given large system. These rules formalize the notion of the "click of photo detector" and are not universal for all systems. In the other words, the notion of a real photon in to universal, in contrast to a sample, but depends on the concrete system, which peculiarities determine what real photons will be in it. 

Electromagnetic field consisting of photon samples thus belongs to the signature of the swarm version of QED. Just this field and photon samples are fundamental objects, but not real photons. Real photons, e.g., detected as the separate particles appears from the division of the field to quanta, which in turn depends on the considered system including the detector. Therefore, the probability to detect a photon in a concrete point in the space makes no sense. The sense has only the question: what is the probability of the event, which is characterized as the "click of photo detector" relatively to this system of charged particles and field. This event does not contain directly photon samples. It represents the appearance of a state of W type of the charged particles, which is interpreted in this system as the "click", e.g., resulting in the oscillation of the air that a human can hear, if we include to the model the molecules of air and a human. The general definition of what should we treat as the "click" is the interesting question, which needs the separate investigations. It is possible only after the realization of the models we describe here. Now we can only fix that the fact of "click" essentially depends on the chosen system. Hence, "photons" detected by some type of devices must not coincide with "photons' detected by the other type of devices. 

As photon samples, they reveal themselves not only as the "clicks" of photo detector but also by their direct influence to the dynamics of charged particles through the fundamental interaction, which swarm description we give below. This influence gives Dirac equation for the charged particle in electromagnetic field, which is the ensemble of photon samples, and in the classical limits Maxwell equations and the laws of electrodynamics that follow from them.

Corteges in the swarm representation of QED must have the form:
\begin{equation}
s^1_{j_1}, s^2_{j_2},\ldots, s^n_{j_n}, {\cal s}_{k_1},{\cal s}_{k_2},\ldots, {\cal s}_{k_d},
\label{cor}
\end{equation}
where $s^m_{j_m}$ is $j_m$-th sample of the charged particle $m$, ${\cal s}_{k_d}$ is $d$-th photon sample. E.g., samples of particles in corteges preserve their belonging to the certain particle, whereas photon samples do not. The individuality of all samples is preserved. This permits in principle, to introduce photons as the special quantum particles to the signature of the swarm representation of QED. However, we meet the difficulty on this way, because a photon moves with the maximal possible speed. Hence, for a photon it would be difficult to describe the interference picture by the diffusion mechanism, as we did it for a particle with the nonzero mass.\footnote{Of course, if we do not introduce the notion of ether. We could do it, assuming that the bath with the ether samples depends on the considered system of charged particles. For example, we could accept that this bath is the ensemble of bounds connected the samples of particles. In the algorithmic approach it cause no problem because such ether makes no physical sense, for example, it does not break the light because the time of impulse exchange between photon samples and ether samples is the administrative time, but not the user time. All this will be legal provided these tricks give the effective simulation algorithm. Such representation of photons gives the model of interference of the light on two slits as for a massive particle.}.
The interference of photons can be described only through their interaction with charged particles. E.g., a photon can hardly be treated as a separate quantum particle; it appears in the detector as the special oscillation of charged particles. However, photon samples are fundamental objects, which influence immediately on the dynamics of the considered system and participate in the selection of many particle quantum states. It exactly corresponds to the method of the representation of W states we gave earlier.

We must introduce one addition to the method of quantum state selection concerned photon samples. We accept that two photon samples ${\cal s}_1$ and ${\cal s}_2$, are close if their impulses and energies are closed. It corresponds to the consideration of photons in impulse-energy basis. All other parts of the state selection scheme of many particles and the field, based on the grouping of corteges of the form \ref{cor}, remains unchanged. This addition to our scheme for QED is not substantial. The photon speed is large in comparison with the average speed of the movement of charged particles samples. If we use our former scheme where the closeness is treated in the coordinate-time sense, for photons we would choose the division by spatial cubes much exceeding the cubes for massive particles. Massive particles almost stay in their places in the period $\Delta t$ of the elementary scattering act, but photon samples shift, and we obtain that the proposed method of grouping corresponds the grouping on the values of impulses. 

\subsection{Swarm description of fundamental processes of QED}

We consider now the swarm description of the main process of QED: emission - absorption of a photon by the charged particle without spin. The amplitude of this process, accordingly to \cite{Fe} equals $\la = \sqrt{4\pi}ie(p_1+p_2)\cdot \e$ where $e$ is the charge of particle, $p_1,\ p_2$ the impulse of particle before and after the emission (absorption) of a photon, $\e$ is the photon vector of polarization. 

\begin{figure}
\centering
\caption{Emission (absorption) of photon by charged particle}
\vspace{120mm}
\makebox[80mm][l]{\includegraphics{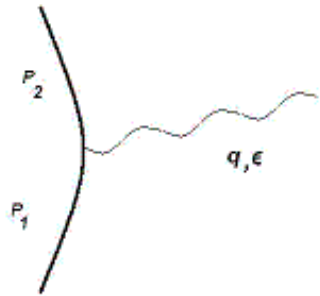}}%
\end{figure}

This process when the photon is emitted, and in opposite direction when it is absorbed. It is equivalent to the equation $p_1=p_2+q$ for the emission, and  $p_1=p_2-q$ for the absorption. 

In the swarm representation, we treat this rule as follows. Each sample of the charged particle emits a photon sample when the law of energy -impulse conservation is satisfied, with the polarization directed along the vector $p_1+p_2$ with the probability $|\la|^2$. 

We consider process that is a little bit more complicated. It is the interaction of two moving charged particles by the exchange of a virtual photon. One particle emits a virtual photon, which then is absorbed by the other particle. 

\begin{figure}
\centering
\caption{Interaction through the photon exchange}
\vspace{120mm}
\makebox[80mm][l]{\includegraphics{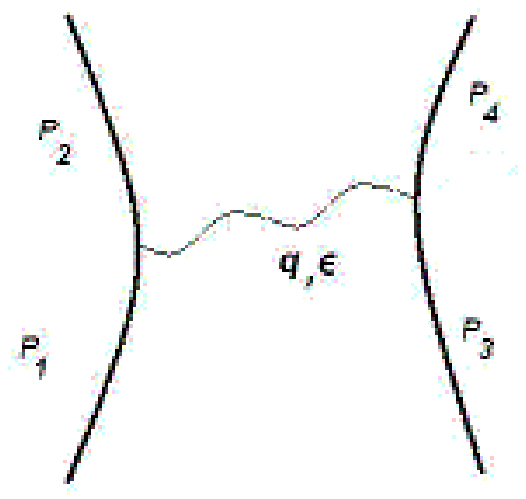}}%
\end{figure}

The amplitude of this process is $\la_2=(e\sqrt{4\pi})^2\frac{(p_1+p_2)(p_3+p_4)}{q^2}$ and it happens only with the conservation of energy-impulse. This is because the amplitudes in the parts of diagram must multiply, and the amplitude of a free photon movement is $1/q^2$. 

In the swarm representation this process means literally that two samples of charged particles exchange by the photon sample and the probability of such event is $|\la_2|^2$. We must take into account not only transversal photons, which polarization $\e$ is orthogonal to the impulse, but also longitudinal photons, which polarization is parallel to the impulse (it is the same thing that the deposit of photons of time photon, which polarization is parallel to the time axes, because for a photon the coordinate directed along the vector of impulse is the time multiplied to the coefficient $c$). 
The deposit of photons of the second type gives Coulomb field with the instantaneous interaction, and the deposit of transversal photons is the the so-called retarded potential)/ If Coulomb field for the samples of charged particles is accounted as the potential $1/r$, we must consider only samples of transversal photons with the probability of the fundamental process written above. If there is the huge total number of photon samples, this model gives Dirac equation, and in the classical limit Maxwell equations.

We now regard the question about the joining of samples of the particles and photons in corteges of the swarm formalism in more details. Such corteges have the form \ref{cor}. We saw above that in the reference frame of the type space-time the condition of the fundamental interaction is the being of the interacting particles in the same point of the space-time, e.g., their spatial and time coordinates must coincide. It means that the administrative time of these particles will be the same in course of all scattering reaction whereas the coordinates change accordingly to their speeds. In the other words, we can formulate the requirement to corteges of the swarm formalism of QED:

{\bf In corteges of the swarm formalism of QED the individual times of all samples must differ no more than in the time $\Delta t=\Delta x/c$,  where $\Delta x$ is the distance between the samples in the space.}

The parameter of the individual time in samples \newline $ s^1_{j_1}, s^2_{j_2},\ldots, s^n_{j_n}, {\cal s}_{k_1},{\cal s}_{k_2},\ldots, {\cal s}_{k_d}$ must then be the same. It means that the interacting particles exist only in the same time; if the individual times of samples are different they cannot interact at all. E.g., corteges of QED have the form $\bar C_1(t_1), \bar C_2 (t_2),\ldots$, where each of them corresponds to some time instant of the administrative time.  

We have to show only, how can we choose impulses $p_1,p_2$ in the case where samples either stay in places, or have the maximal possible speed. For this we must average their impulse on the time, e.g., to store all the history of each sample in all instants of the administrative time: $t_1,t_2,\ldots$. Of course, the sample, taken in the different time instant is not formally the same. We introduce the individual mark for each sample so that the coincidence of marks means that there is the same sample, but taken in the different time instant. The swarm formalism allows this. If there are photon samples, their coordinates must correspond to their movement with the speed of the light along its impulse, if there are samples of a charged particles the correspondence must be the analogous. 

We impose the last condition to the corteges of QED: the conservation of the individuality of the corteges themselves. It means that if in two corteges  $\bar C_1$ and $\bar C_2$ there are samples with the same individual mark, then all their samples have the pair wise equal marks, e.g., the marks of $s^1_1$ in them must coincide, the marks of $s^2_2$ must coincide, etc. E.g., such corteges represent the same cortege of samples, but taken in the different time instants. 

Therefore, we have the followng picture. Corteges of QED can be joined in the groups of the form 
\begin{equation}
\begin{array}{lll}
&\bar C_1(t_1), \bar C_2(t_1),\ldots, &\bar C_d(t_1),\\
&\bar C_1(t_2), \bar C_2(t_2),\ldots, &\bar C_d(t_2),\\
&\ldots, &\ldots,\\
&\bar C_1(t_l), \bar C_2(t_l),\ldots, &\bar C_d(t_l),
\end{array}
\label{qed_}
\end{equation}
where $C_j$ is the cortege with the constant members taken in the different time instants, and the elements of the different corteges $C_j$ do not overlap (have the different individual marks).  

We now can average the impulse on the time for each sample, and the expression for the probability of fundamental processes obtains the sense. In course of all administrative period in which we consider the reaction we should choose only one instant for the emission or absorption of a photon. We also can regard the processes of higher orders, when two or more photons are emitted or absorbed simultaneously. The corresponding probabilities follow from the amplitudes of these processes. 

\subsection{Procedure of quantum state selection with photons}

We now can formulate the procedure of quantum state selection with photon samples. The swarm of QED has the form \ref{qed_}. We will join corteges of this swarm into groups ${\cal G}_1,{\cal G}_1,\ldots,{\cal G}_z$. Here exactly the corteges located in the same cell of the division of the extended configuration space fall in the same group. Here the extended configuration space also includes photons. There is the dimension in this space corresponding to the polarization of photons. If we intend to account the spins of charged particles, the configuration space includes also the dimensions for all participating spins. Now we can apply the selection procedure shown above to states of QED systems. The administrative time will behave as the spatial coordinate. 

Summarizing the above statements, we conclude that each step of the evolution of the state of charged particles and the field consists in the transformation of of our swarm corresponding to the unitary operator of the current fundamental interaction and the following selection procedure. 
The sequence
\begin{equation}
Q_1\ar Q_2\ar\ldots\ar Q_k
\label{sequ}
\end{equation}
of such steps is the model of the evolution of the systems of charged particles and the field in the method of collective behavior for the particles without spins. 

In the case of particles with the spin tehir wave function will also depend on the spin. It means that we must add to the configuration space the spin dimension an dto reformulate our recipe for this case literally. 

\subsection{Description of chemical association with the implicit account of photons}

There are many types of reactions, for which the laser impulses or the visual light is necessary. In these cases, it would be natural to include photons in the description of these reactions. At the same time, many reactions go without the external source of photons. The formalism needed for the description of chemical reactions must not include photon samples. In the standard description in the framework of quantum chemistry, such reactions go either with the decreasing of the potential energy of reagents, or with the external source of the heat, e.g., after the kinetic hits. We consider how we can simplify our formalism of the collective behavior for the reactions going without external photon source. 

Here we will regard the reduced version of QED swarms, which we can obtain from the full swarm \ref{qed_} by the exclusion of all photon samples from corteges. We call such a swarm a reduced swarm. It differs from the ordinary swarm for the ensemble of particles only in that the time exists between the attributes of all samples, and corteges are joint with the conservation of the history of each cortege as we defined earlier. We show what happens if we apply to this swarm the selection of quantum states. 

We consider the reaction of the association of two protons in the molecular ion of hydrogen, and let $r_0$ be the radius of the stationary position of protons in this ion. We join in one-cell corteges with the close coordinates and time because now, in contrast to the case of simple quantum mechanics without photons there is the time in the attributes of samples. Now in the same cell of the configuration space we have corteges with the different times  $t$. It means the finer distinguishing of corteges than in the previous case, without the time, where corteges were different only in their spatial positions. 

\begin{figure}
\centering
\caption{Different devergence of samples in the time}
\vspace{150mm}
\makebox[230mm][l]{\includegraphics{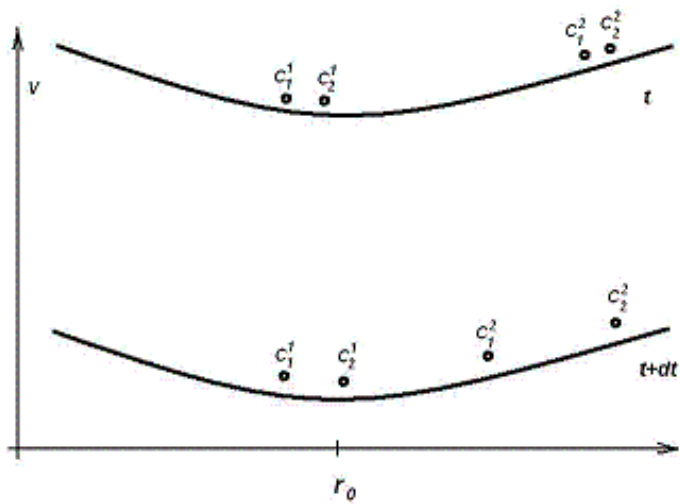}}%
\end{figure}

Indeed, let us assume that corteges of two samples $s_1,s_2$ of two protons have such coordinates of samples that the distance between the samples differs from $r_0$ substantially and they have the different values of the time. We call such corteges non-canonical, whereas the other corteges, where the distance between samples is close to $r_0$, we call the canonical. Then during the small time frame $\Delta t$, in which we regard the redistribution of corteges in the selection, two samples of the form 
\begin{equation}
(c^1_1,s^2_1),\ (c^1_2,c^2_2)
\end{equation}
with the different time $t(c_1)\neq t(c_2)$ will diverge strongly in space, much stronger than such pairs in which the distance between the samples is close to $r_0$. It happens because $r_0$ is the distance ensuring the sable state of the molecule. Therefore, in the selection process the non-canonic corteges more probably break up than the canonic corteges. It corresponds to the full version of the selection with the photon samples. Indeed, we represent the movement of a cortege as the movement of a harmonic oscillator. For the non-canonic corteges the amplitude of oscillations, and correspondingly, the acceleration of protons will be more than for the canonic corteges. Since the probability to emit a photon will grow with the growth of acceleration of protons, we obtain that the probability to emit a photon will be more for the non-canonic corteges. Since the photon emission adds the filled new dimension to the configuration space (not emitted photon by the definition has the same coordinates as the corresponding samples of particle) the groups containing non-canonic corteges will have the less quantity of samples than the groups with the canonic corteges. It gives the increasing of the ratio of corteges corresponding to the molecule.

We mention that this model of the appearance of the molecule with the explicit account of the time is rougher than the full model with photons. For example, we absolutely do not account the polarization of emitted photons, and effects induced by it. Nevertheless, this reduced model is much simpler in the work because here we must not consider the emission of photons as in the full version. The reduced version thus operates with samples of chemical objects only: atomic nuclei (and, perhaps, electrons, if we consider the electron transfer between atoms by the proposed method). 

\subsection{About scalability of QED}

The question about the scalability fo QED belongs to the most interesting questions of quantum theory. Computational difficulties coming from the summing of Feynman diagrams for a long time keep this question in the background and nowadays it is not studied in details. Even the scheme of the application of the hypothetical quantum computer for the analysis of this problem is not yet elaborated, in contrast with the analogous scheme for the ordinary quantum mechanics. (see the chapter 5).

However, the problem of scalability of QED is principal. Its solution is necessary for the concordance between quantum theory and general relativity. It is known that the application of general relativity in the cosmology requires its advancement to the traditional area of quantum mechanics. It concerns, for example, massive objects like black holes. Here we meet the necessity to apply the methods of quantum theory to large ensembles, e.g., the scalability of QED. Even more, it is important because in view of that the theory of nuclear interactions is built in the fashion of quantum theory. The scheme of concordance between QED and the general relativity would have the universal character for the cosmology.

The scheme we proposed in this chapter could help in the solution of this serious open problem. By means of representations of the collective behavior, we can learn more about the role which fundamental interactions play in systems with the large number of particles. It is well known that this role is valuable, for example, in the massive bodies, or in the high-energy processes like cosmic explosions where the renormalization of mass is substantial. One could speculate about the application of this method to the definition of the notion of the physical time in the system with the large quantity of particles. 

We see that the simulation of the elementary scattering act by the selection procedure takes some time, which, generally speaking, is the administrative time of our model. If we consider the sequence of scattering acts \ref{sequ}, then the total number of these acts is the real physical time. One could try to evaluate the portion of the administrative time in the simulation of sufficiently long process. The administrative time factually must be excluded from the physical longitude of the process. For the different processes, the portion of administrative time may be different. In the other words, the value of the administrative time segment $\Delta t$ is different for the different types of movement of particles. For example, it must be relatively more for the movements with the large speeds and accelerations. Let us imagine that the some large ensemble the local administrative time for some particles becomes greater. It is natural to admit that it results in the decreasing of the physical time for these particles comparatively to the others. If it happens with some small probability with all particles in the large ensemble, it leads to the decreasing of the real physical time for the whole ensemble. Of course, it is the pure speculation.

\section{Bounds between samples}

We saw that the method fo collective behavior gives us some new possibilities for the investigation of complex systems comparatively to the tradition approach based on wave functions. For the advancement of this method, we need to formulate the heuristic, which makes possible to apply this method in the simulation of real problems. Now we formulate the important notion of this heuristic: bonds between samples. We have already met the particular case of bonds: corteges consisting of samples of the different elementary particles. We applied corteges for the description of the dynamics in the framework of Hilbert formalism, but they do not possess the full scalability, because, at first real particles can be non-elementary, at second, all real particles can turn to be parts of some big particle which we must treat as quantum particle as well. 

We consider again the hydrogen atom consisting of the heavy proton and the light electron. This atom behaves as the integral particle, which has its own samples and swarm. For example, this atom can interfere when passing though slits. How then can we tie electron samples corresponding to the different samples of the atom? Cortege technique requires hiding the electron when we regard the atom as the integral particle. It works only if the internal state of the atom (the electron state in it) does not depend on the spatial position of the atom itself! However, this dependence can take place, for example, if we give a laser impulse to this atom, or if our atom approaches to the other atom and the association begin between them that change the electron state. In these cases, the cortege technique works badly, and we must generalize it by the notion of bonds. This notion is free from such drawbacks, and it makes the collective behavior method completely scalable. 

\subsection{Bonds for one real particle}

In the collective behavior method each real particle is represented by the swarm of its samples $\bar s=(s_1,s_2,\ldots,s_L)$s. 
Earlier we saw that the intensity of the diffusion exchange for such swarm required for the approximation fo Shredinger equation is proportional to $\delta x^{-3}$. Such intensity will be if we place samples along some curve $\g$ in the sequence $s_1,s_2,\ldots, s_L$ and require that the impulse exchange goes only between the neighboring samples: $s_j$ and $s_{j+1}$. Then choosing the value of $\delta x$ we obtain that the change of the swarm density resulted from the passage of one sample along this curve $\g$ will be inversely proportional to the value of the volume. We call this curve $\g$ the thread, corresponding to one real particle. We can treat it as a rope of the negligible weight rolled into a ball but not entangled. Samples of this particle we can regard as beads beaded on the rope with the approximately equal intervals. If the rope can roll into a ball, it permits to make the density of real particle as big as needed in the small spatial volume. 

We can introduce the averaging of impulse along the length of thread, and consider the average impulse in the segment of the form $s_j,s_{j+1},\ldots,s_{j+l}$ for some $l$, or some more complicated rule like impulse exchange. The thread itself influences to the dynamics of samples: the impulse exchange must go such that samples do not depart widely from the thread. We can reach this if we require that the average impulses are directed along the tangent vector to the thread $\g$ in the corresponding point. It factually imposes some limitation to the appearance of pairs with the opposite speeds in the dynamical diffusion. Just this is the influence of the thread to the dynamics of the swarm. Here the thread behaves as the rope, which can roll into a boll and unroll from this ball stretching itself along the direction of the average impulse. Such representation can be productive if we simulate the decay of the main state of an electron in the field of two protons, provided these protons fly away to the opposite sides. The thread will then roll around one of them and leave the other with high probability. The more exact representation of this process requires the introduction of the photon samples emission that we take up further. 

However, the representation of a swarm in the form of only one immutable thread is not always productive. For example, we consider one of the basic experiments on quantum mechanics: the interference of a particle when passing from two slits. If this particle is represented by one thread which simple changes its spatial position, we could not obtain the interference clot in the middle of the slits. This clot appears only is we accept that the thread representing one particle can break to the different parts and immediately join otherwise. Let us trace for the thread flying to the screen with two slits (see the picture). To obtain the clot between the slits besides the screen we suppose that the thread breaks in two points so that its medium part becomes isolated, and immediately joins so that this medium part links to two free ends whereas its former edges link together that gives the circle. The exchange of impulses between neighboring samples along the thread gives the real interference picture. \footnote{For the comparison, we show the result of the simulation of interference on two slits by the method of eigen functions.}

\begin{figure}
\centering
\caption{Interference of a particle on two slits. Method of wave functions. \newline (Visualization of Semenihin, the picture from the site http://qi/cs/msu/su )}
\vspace{190mm}
\makebox[800mm][l]{\includegraphics{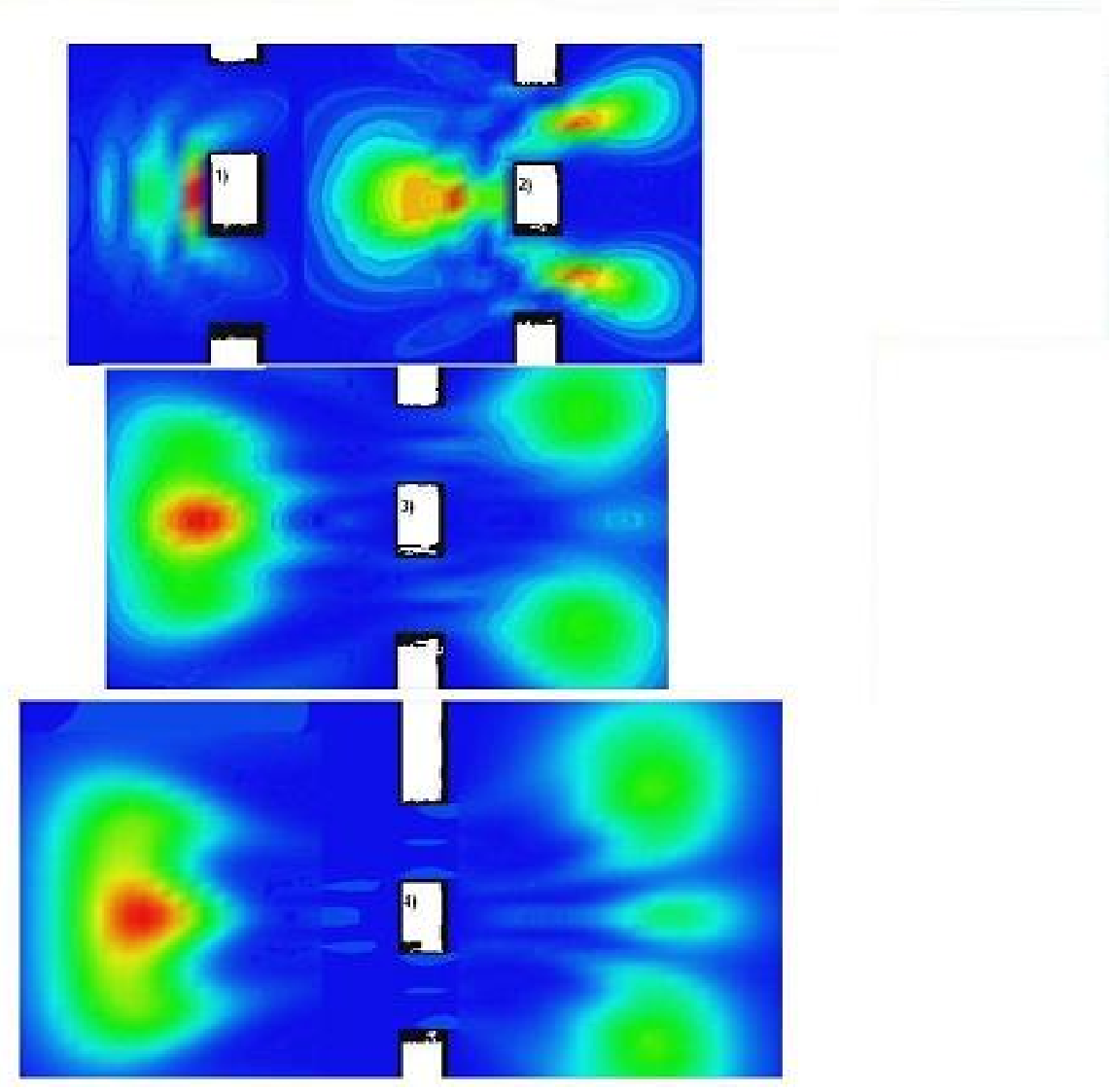}}%
\end{figure}

\begin{figure}
\centering
\caption{Interference of a particle on two slits. The method of collective behavior}
\vspace{140mm}
\makebox[800mm][l]{\includegraphics{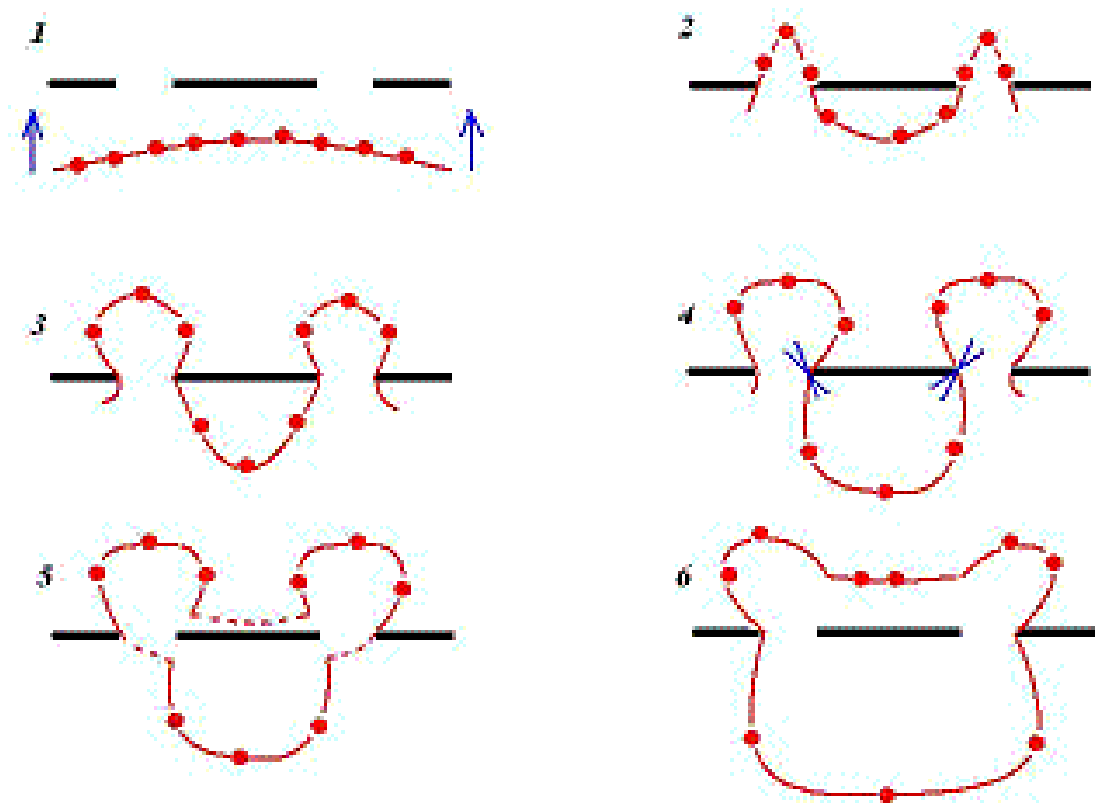}}%
\end{figure}

This reasoning with the thread is the part of the heuristic of collective behavior, which we will consider. Such heuristic is needed for the building of fast algorithms simulating quantum dynamics. Any representation of a swarm in the computer is the kind of vector consisting of samples of this swarm. This is why the rebuilding of the thread we described is easy to program. This is the advantage of the thread over the scanning of the three dimensional space that we would have to do in the methods of matrix algebra (e.g., in the already discussed Riemann scheme of integration).

\subsection{Bonds for many particles. Quantum "worlds" ordering}

We now consider what changes we must do in our scheme if there are many particles instead of one. Let we are given $n$ real particles enumerated by numbers: $1,2,\ldots,n$, and we represent each particle $j$ as the swarm $s^j_1,s^j_2,\ldots, s^j_l$. If we identify each real particle $j$ with one of its samples $s^j_{k_j}$, we obtain the cortege $\bar s$ of the form $s^1_{k_1},s^2_{k_2},\ldots,s^n_{k_n}$. Such cortege $\bar s$ we call a quantum "world". The choice of numbers $k_j$ for $j=1,2,\ldots,n$ is the separate question; it is important because the total number of such choices grows as the exponential of $n$, hence this choice must be limited from the very beginning. The simplest (but not the single possible) restriction is to require that there exist the numbers $d_2,d_3,\ldots,d_n$, such that for any $q=2,3,\ldots,n$ and any $k_1=1,2,\ldots, l$ $k_q=k_1+d_q$. It means that the possible "worlds" are created by the sequential choice of samples one by one (see the picture). This rule we call the order of bonds.

However, this scheme is not yet general, because we strictly fix real particle themselves, which samples participate in the proper bonds. This scheme cannot give us the dynamics of bonds necessary for the simulation of the measurement process: the sequential appearance of the new bonds and the break of the old bonds. In this process the situation when the different samples of the same real particle form bonds with the samples of the other real particles is unavoidable. 

We thus come to the more general definition of the many particle swarms in the form of a net.

{\bf A bond} between samples $s$ and $s'$ we call an object which is put into correspondence to the pair of these samples in the model. We call this bond the bond, connecting these samples. If $S$ is the set of all samples the set of bonds is some subset of the form $S_c\subset S\times S$. A bond connecting two samples of one real particle we call a {\bf thread}, or non-proper bond. All other bonds are called the proper bonds.  

{\rm \bf definition 1} The net is the graph ${\cal N}$, which vertices are marked by the different samples of real particles of the system at hand, and its verges are marked by bonds. We require that the maximal connected subgraph $G\subseteq {\cal N}$, which all verges are threads, embraces all samples of one real particle. Such sub graph we call the swarm for this particle. A way on proper bonds we call a sequence of vertices $v_1,v_2,\ldots v_k$ of the graph ${\cal N}$, in which for each $j=1,2,\ldots,k-1$ $v_j$ and $v_{j+1}$ are connected by some proper bond.

{\rm \bf Definition 2} Quantum world of the net ${\cal N}$ is the maximal sub graph $M\subseteq {\cal N}$, such that it doe not contain two vertices connected with the thread, and for each of its vertices $v$ any path along the proper bonds containing $v$, entirely contained in $M$. 

The example of the net is shown at the picture.  
The dynamics of the net is determined by the rules of rebuilding. One of these rules says that the rebuilding happens in the moment when the distance between samples connecting by a bond (proper or thread) exceeds some critical limit $\D_{bound}$. Here $\D_{bound}$ depends on types of samples and on the type of the bond. The rule of rebuilding looks as follows.

1) If this bond is a thread and after its elimination, the connectivity of the swarm breaks for some particle, then the less (by the quantity of samples) component of the swarm is declared as the photon swarm, where the speeds of samples of these components are redistributed so that the broken connectivity becomes restored. 

2) If the distance between two samples of the different real particles becomes less than some limit $d_{bound}$, then the new proper bond appears between these samples, provided the order of bonds is satisfied.

Two these rules can serve as the basement for the drawing method of the dynamics visual picture, if $d_{bound}\ll D_{bound}$. This condition means the conservatism of proper bonds: the break of a bond requires some work, equal to the energy of the emitted photon in this process accordingly to the point 1). We can represent proper bonds as ties consisting of glue, which fastens onto a sample on a small distance but is able to stretch up to the point of the break. Threads we represent as the rolled peaces of a strong rope, which ends are fastened by glue, so that this rope quickly unrolls when stretched but it does not break. Of course, this description is not formal, and it is the subject of the specification for the concrete evolutions. Nevertheless, this analogy can serve as the details of heuristic based on the collective behavior for many particles.

\section{Heuristic of collective behavior}

The aim of quantum constructivism is the obtaining of the algorithmic tools fro the building of dynamical models for many particle systems. Arguments shown in the previous chapters and the experience in the practical simulation permits to state, that 

{\bf It is impossible to build dynamical models only by the laws of quantum theory. For this, we need special rules of the building of algorithms.}

We call these rules {\bf heuristics}. The method of constructivism via the collective behavior gives us such heuristic, which we describe here. 

Can we reach our aim applying one idea for classical movements and the other for quantum? This method presumes the existence of some exact barrier between two methods, for example, based on the action: if it is less than $h$, we apply Shredinger equation, if it is more than $h$, we apply the classical methods. This is how matters stand in the molecular dynamics modeling with standard applied packages: one package for the solution of Shredinger equation, and the other for the classical mechanics. From the chapter 5 we know that this way does not agree with constructive mathematical analysis, because it presumes the exact knowing of the action of the particle, whereas the action is determined within some accuracy only. We conclude that for the constructive quantum theory we need the uniform method of the representation of the dynamics with the accuracy available in the current moment. For example, for the collective behavior we need the deterministic algorithm determining the dynamics of any sample of each real particle, which state is given in the model within some accuracy. 

The flow block of this algorithm, we call the heuristic of the model. We give the description of such heuristic in this section\footnote{We understand this term more precisely than usually, in the sense that we require the distinctness from the heuristic. The status of this notion in constructive physics is higher than in the standard physics that follows from the especial role of algorithms.}. Our heuristic rests on the collective behavior, and we show in general what we hope to reach with it. Of course, I do not claim that this heuristic is the single possible. Somebody could try to build the other heuristic for the constructive quantum physics; this is the legal situation in the light of principal pluralism of the constructivism. However, one must remember that the comparison of two heuristic we can fulfill through scenarios they give, and not otherwise. E.g., it is senseless to discuss one or the other element of these heuristics separately, for example, to ascribe some "physical sense" or such characteristics as the mass, to bonds, etc.

The dynamical heuristic is the finite list of elements for building of the simulating algorithms. 

In the previous section we factually used the heuristic of selection, e.g., a step of the real time is the result of the selection with few steps in the administrative time, which is determines by the comparison of the total number of elements in groups $G_j$. The heuristic of selection is typical in quantum theory, and it is factually described in the canonical manuals on quantum theory. Its formal expression is the method of path integrals. Factually, the heuristic of selection is the single possible in the standard quantum theory. It well corresponds with the probabilistic spirit of this science and can work successfully, for example, for the description of chemical reactions with two or three atoms.

But this merit of the proximity to the standard language has the underside, because it gives to the selection heuristic the main defect of standard quantum theory: its insuperable complexity barrier. When the total number of particles grows the effective selection requires the exponentially bigger number of samples of each real particle and we again turn out to be thrown back to the starting point: the approach based on the selection only cannot be scalable to large systems. 

For the good work of the selection on the large quantum ensembles, we need to eliminate the waste of computational resources in the selection method. We need to limit substantially the range of cortege for the selection because just here is the main channel of the leakage of the time of the simulation. We saw that the conservation of the individuality of samples represents the main way of the computational economy in the case of Shredinger equation, where this individuality helps us to avoid the non-effective methods of the matrix algebra. Just this idea of the individuality of samples helps us in the general case, for the many particles quantum heuristic. We must determine the concrete rules of reach sample describing its states in each step. The selection procedure does not give us the conservation of individuality: samples containing in the cortege of the small groups will be redistributed on the new corteges. This reforming of corteges gives no guarantees that the new formed corteges will not turn again into small groups and this situation continues in future. E.g., the selection method allows the existence of the "unpromising" samples, which always lie in groups subject to culling. These unpromising samples must be replaced by the new samples, which states are close to the perspective samples containing in the corteges in large groups. 

In this procedure of replacement of the unpromising samples, the information about their history gets lost, whereas this information can be useful for all swarm. The preserving of this information is the main aim of the proposed heuristic. We explicitly point the mechanism of the return of unpromising samples to the swarm. It concerns the photon emission when the samples of charged particles accelerate. The replacement of unpromising samples represents the grouping of the swarm in the state close to the break to the several components of connectivity. The idea is the following. We represent the swarm of samples as the elastic thread to which these samples are beaded. In the moment when the thread stretches too much (that corresponds to the large distance between its samples) its break happens. In this case, the return of photon samples emitted by the samples of particle close to the point of the break smoothes the strain of the thread. The points of strain will arise in the other points of the thread where photon samples will be emitted as well, etc.
The photon emission is thus the mechanism ensuring the integrity of the swarm and the redistribution of the unpromising samples. 

This mechanism will work with the additional condition: photon samples must not belong to the fixed corteges. E.g., the sample of photons we consider must be common for the different quantum worlds. It concerns the so-called real photons, e.g., photons treated as the free particles. Free photon samples will be thus the objects ensuring the visual dynamics of the many particle swarms. The photon emissions smooth the trajectory and the form of this swarm. The community of real photon samples for the different worlds means the valuable correction towards the selection heuristic. In the constructive heuristic, where the unpromising from the selective view point samples do not disappear, but have the explicit program of the return to the swarm all the "worlds" are real equally. Contra intuitive elements of quantum mechanics thus transform to the small deviations of initial speeds and positions and in the states of real (none absorbed) photons. 

We thus refuse from the joining corteges into groups, and the swarm dynamics is determined by three elements:
\begin{itemize}
\item Threads connecting different samples of the same particle.
\item Proper bonds, connecting the samples of the different particles.
\item Impulse exchange inside of one cortege and photon samples emission. 
\end{itemize}
 
We see that real photon samples do not belong to the concrete corteges but nothing forbids them to keep the memory about the sample of charged particle, which emitted them. Since now, all samples have the certain history we can include real photon samples in all corteges. The situation is possible when a real photon sample emitted by one sample is absorbed by the sample from the other cortege. For example, one electron sample can emit a photon sample, which will then absorbed by the other sample of the same electron. This self-influence of a particle is well known in QED and leads to the renormalizing of a mass and a charge. We can agree that the impulse exchange between the neighboring samples of the same particle connecting by the thread is the exchange of such a sample of a virtual photon. We can accept the possibility of photon samples exchange through several samples in the same thread. A sample of virtual (absorbed) photon can travel between samples of the different particles connected by the bonds in the cortege. However, it cannot move between the different corteges. If a photon sample emitted by some sample of particle $p_1$ in some cortege $K_1$ is absorbed by a sample of particle $p_2$ belonging to the cortege $K_2$, it means that either $p_1$ equals $p_2$, e.g., there are samples connected by the thread of one real particle, or $K_1$ equals $K_2$, that is there are samples connecting by the thread of the same cortege, or the corteges $K_1$ and $K_2$ contains only samples of non overlapping sets of quantum particles and in this exchange the joining of $K_1$ and $K_2$ to one cortege $K$ happens. 

We conclude that the real photon samples carry bonds, which tie the different corteges, e.g., the different quantum worlds. 
The click of photo detector signaling of its work happens not after the absorption of one sample of photon by a sample of particle belonging to the corteges inside the photo detector. A click is the oscillation of real atom samples, and it happens after the absorption of many photon samples. Even if there was no click we can speak about the joining of samples of a particle emitted this photon and such sample of a particle in photo detector, which absorbed it to the same cortege. In the other words, the forming of the bond between the object of emission and the detector not mandatory entails with the click of the detector. However, this bond is the real object of the administrative part of the model, and it can influence to the detecting of the following photon samples. 

\begin{figure}
\centering
\caption{Fall of electron to proton.
The red thread represents the electron. The dotted line points the probable movement of the thread without the photon emission. On step 1) the emission brakes the electron, on step 3) - accelerates preventing the fall. The vector of polarization of photon sample is parallel to the tangent to the thread in the point where the emission changes the trajectory of thread.}
\vspace{150mm}
\makebox[230mm][l]{\includegraphics{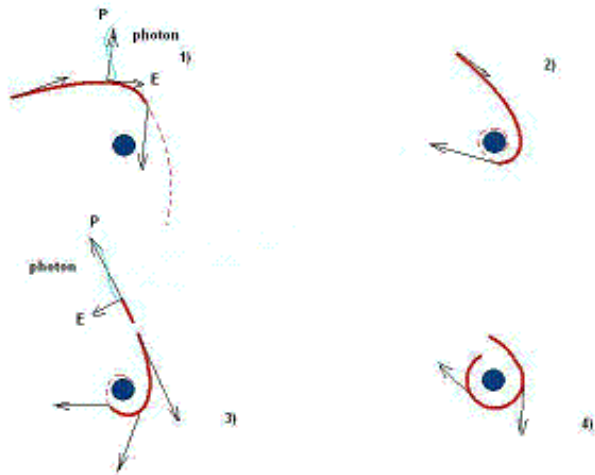}}%
\end{figure}

The procedure of the emission of photon sample in the strained thread of the electron swarm is valid for a free electron only. If electron samples are joint in the cortege with the samples of the other particles, all remains true, but we need to replace the notion of "strain of the thread" by the "strain of the net", in which besides the thread there are proper bonds, connecting samples in corteges. These cortege bonds are much weaker than threads connecting the samples of one particle. The critical strain of proper bonds leading to its break (decay of the cortege) is many orders less than for a thread. It guarantees the stability of the elementary particles in QED. 

Now we are ready to formulate the heuristic of the collective behavior ready for the creating of simulating algorithms. It must be fully scalable in the following sense. Its elements must be sufficient for the building of model of any complexity. It must permit the addition of new particles to the system, structuring of particles that were considered as elementary. 

There are three elements:
\begin{itemize}
\item elastic collisions of particles,
\item rules of transformations, and
\item conservatism of bonds.
\end{itemize}

The elasticity of collisions, e.g., the energy and impulse conservation expresses the main law of classical physics on the level of few particles, including fundamental interactions of QED on the level of separate samples of charged particles and photons.

The rules for transformations say what must result from the collisions, for example, of a photon with a charged particle. 

The conservatism of bonds is the central moment of heuristic. It means that a bond preserves up to the moment when the distance between samples is not more than some threshold $\d_0$, depending on the type of the particle. If this threshold is exceeded the transformation happens, described in the point 2. The state of swarm after the transformation must restore the former bonds. The further evolution can lead to the exceeding of this threshold again, and this procedure then repeats, etc. 

We consider the process of association of a free electron with a proton in the framework of the described heuristic. We do not distinguish the stationary states in discrete as well as in the continuous spectrum, and will treat the association, when the hydrogen atom is formed, or the electron flies past. Correspondingly, we suppose that only one photon is emitted. In the reality there are many photons and the electron turns to be associated in the different stationary states in the vicinity of the proton. All these details are determined by the initial state of electron and proton, which we cannot know even in principle, and we must replace it by the generating of randomness. Here we only describe the heuristic, which we can further specify, and suppose that exactly one photon is emitted and the association results in the ground state. 

Let the bonds between samples be determined by their neighborhood in the vector $\bar s$. The set of such bonds geometrically represents the thread to which electron samples are beaded. Each electron sample is joint to one cortege with some proton sample accordingly to the order of the swarm vector. The step of evolution represents three actions:

\begin{itemize}
\item exchange of impulses between neighbor samples,
\item change of impulse of each sample by the Coulomb field of the corresponding proton sample,
\item free movement of all samples.
\end{itemize}

In the movement of such thread in Coulomb field of proton, the end of thread closer to the proton will fly faster. This speed by the impulse exchange quickly comes to the opposite end, which is farer from the proton. The addition to the speed obtained by the farer end of the electron thread gives to the samples in it the speed sufficient for the fly past the proton provided the middle part of the thread does not change the direction of movement. From the other hand, the impulse exchange along the electron thread causes the brake of samples closer to proton. It changes their trajectories closer to proton which results in the increasing of the speed. We see, that in the general case the thread of electron samples is the system with the negative feedback, that is the divergence of speeds increases, that leads to the exceeding of the threshold $\d_0$ between its samples. In this moment, two possibilities arise:
\begin{itemize}
\item the portion of electron samples located farer from the protons claimed the samples of photon emitted by this electron, or 
\item the portion of electron samples located closer to the proton is claimed the samples of such photon.
\end{itemize}
In the first case the closer to the proton part of the thread turns on the stationary orbit and we have the association of hydrogen atom. In the second case the photon will be emitted to the proton and the electron flies away from it. 

The emitted photon we treat as the real if its samples are emitted by all samples of the electron. In the both cases, the photon is emitted, and the only difference is in its impulse. Our heuristic must give the method of finding the photon impulse by the initial states of electron and proton. Again, we consider the rough situation and assume that the impulse of the emitted photon is directed either to the proton, or in the opposite direction, and has the same absolute value. In the last case, the association of the atom happens, in the first the electron flies away from the proton. Our model must thus choose the prevailing type of corteges:
\begin{itemize}
\item $s_1^{electron},s_2^{proton},s_{assoc}^{photon}$ or 
\item $s_1^{electron},s_2^{proton},s_{dissoc}^{photon}$
\end{itemize}
The choice of initial cortege determines this choice in the higher degree than for the situation when there is no photon. Indeed, we always have some tolerance in the value of coordinates and speeds of all particles. We suppose that the accuracy of the determination of the coordinates and speeds for each sample is $\delta$. It means that we fulfill the round up of all magnitudes to the integer multiple of $\delta$. Let samples determining the fact of association $s_{assoc}^{photon}$ or $s_{dissoc}^{photon}$ differ in no more than $\delta$. The choice of the first cortege means the choice of such approximation for one of photon samples. The configuration space in the case of one proton and one electron will contain $ L^{2\cdot 3}$ elements, where  $L$ is the total number of divisions on one dimension axes. If there is the photon, this number of elements will be $ L^{3\cdot 3}$. If we consider the situation with the absorption and re-emission of the photon, the dimensionality increases so that even very roughly we can deal with small number of samples in the cortege only. In the other words, adding to our system a few new particles and correspondingly lengthening corteges, by the changing of photon coordinates in the framework of the accuracy, we can obtain the same result as in the selection of entangled states. 

\begin{figure}
\centering
\caption{Net for the hydrogen atom. Bonds between electron and proton samples}
\vspace{150mm}
\makebox[180mm][l]{\includegraphics{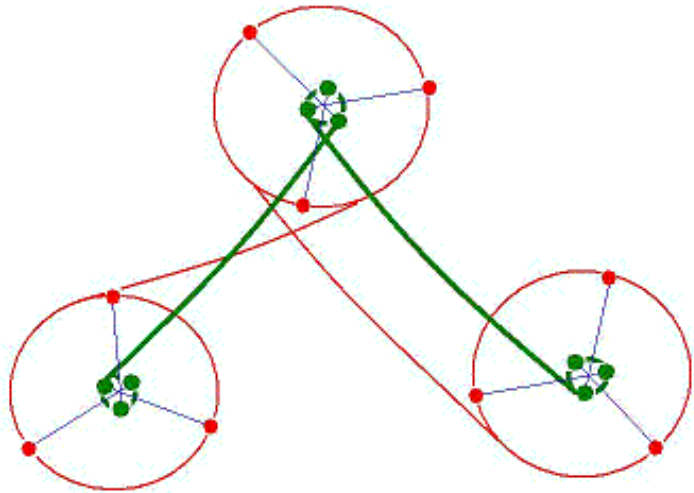}}%
\end{figure}

\section{Evolution of net}

The model of many particle system based on the collective behavior represents the net consisting of samples of real particles and bonds between samples. We represent the net as the graph, which vertices are samples and edges are bonds. The bonds connecting the different samples of the same real particle we call threads, the bonds between the samples of the different real particles we call proper bonds. 

Let us consider the sub graph, which vertices do threads connect, and which samples correspond to one real particle. We call this sub graph the swarm corresponding to this real particle. If we then consider the other sub graph, which vertices do proper bonds connect, this sub graph we call "quantum world". 

The fixation of the notion of real particles means the choice of interaction type, which we are going to consider. For example, if we choose electrons and atomic nuclei, we are going to consider electromagnetic interactions. An evolution of the net in the time goes step-by-step, and we can represent it as a chain $S$ of the net states called a scenario:
\begin{equation}
S:\ N_1\ar N_2\ar\ldots\ar N_l
\end{equation}
The dynamic model of evolution $D$ is the set of such scenarios: $D=\{ S_1,S_2,\ldots S_h\}$. The model $D$ is the subject of the selection heuristic. It means that we can apply to the different scenarios the methods of grouping, mutations and rejection, e.g., the genetic computational methods. We divide the steps $N_j\ar N_{j+1}$ to two types: the simple evolution and the rebuilding. In the first case the set of real particles remains unchanged. The net then evolves by the rules given in the previous section: all samples of particles fly, emit and absorb photon samples. Here the threads transform to themselves, e.g., particles cannot transform to other particles (emissions and absorptions we do not account). The evolution concerns proper bonds only. This is the decay of entangled states and the forming of the new entangled states. 

The rebuilding means the change of real particles. For example, joining of two protons to the nucleus of helium, accompanied with the photon emission. The rebuilding changes threads. For example, the joining of protons means that instead of two threads we obtain only one, corresponding to the nucleus of helium. There is the exact quantitative criterion for the rebuilding, for example, we can accept that it happens if and only if the distance between the neighboring samples in one thread exceeds some threshold $\Delta_{rebuild}$. One could think that the establishment of the exact threshold contradicts to the principle of constructive mathematical analysis. This is the imaginary contradiction. The real evolution resulted from the genetic procedure of the selection of scenarios belonging to the model. In these scenarios the initial parameters (initial states of the particles) are randomly distributed and this averaging eliminate the influence of the accuracy of the choice of $\Delta_{rebuild}$ to the result of the simulation. We always can specify the resulting scenario, taking the less grain of spatial resolution, choosing the new swarms as we define it above, that give us the new, more accurate model of the dynamics. E.g., we always can choose the more accurate approximation of the real scenario that agrees with the constructivism. Some values of parameters make possible the rebuilding with the more probability. It happens when the acceleration is high. 

\begin{figure}
\centering
\caption{The net dynamics in the reaction $H_2+O\ar H_2O$. One electron thread is shown}
\vspace{150mm}
\makebox[680mm][l]{\includegraphics{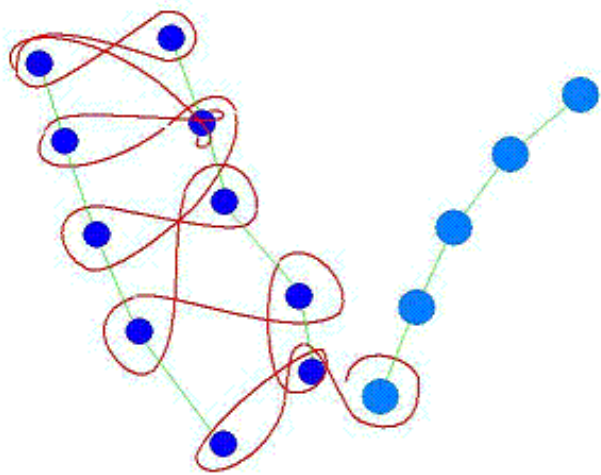}}%
\end{figure}

We show how our heuristic for the change of the net looks in the simple decay of quantum states caused by decoherence. We consider for determinacy, the electron in the ground state of the field of two Coulomb centers (for example, molecular ion of hydrogen), when the distance between the protons grows. We represent the thread corresponding to the electron, in the form of the circle. It is not important in the sense of concordance with Shredinger equation but is convenient if we study stationary states. It is also convenient to assume that the impulse exchange goes through the photon samples that are emitted by one sample of the charged particle and is immediately absorbed by the neighboring sample. We then obtain that if the emitting sample is the extreme in the thread, the appeared photon sample flies away. Let a photon appear when the threshold $d$ for the distance between neighboring samples is exceeded. The evolution of the net including the samples of electron and photon looks then as follows. The distance between samples in the swarm will grow when Coulomb centers diverge up to the level when this distance exceeds $d$, then the photon sample will be emitted that equals to the break of the thread. After the break we can suppose that the thread preserves the distance less than $d$ between samples for some time, so that the new photon emission does not happen. The movement of the thread then is described by the return from the emitted photon sample and the thread probably comes off the Coulomb center in which vicinity it broke. Because the beginning of the thread we define randomly, we obtain that the probability that the electron swarm turns out to be about one or the other Coulomb center is roughly proportional to the number of its samples that were about this center from the very beginning. We remember, that the quantity of samples in the unit volume is the density of swarm, which approximately equals $|\Psi |^2$. It gives us the agreement with Born rule for quantum probability of the transfer to one of stationary states: about one or the other Coulomb centers after the decay of the ground state in the field of two Coulomb centers. 

We consider the process of the association of hydrogen atom and proton into the molecular ion of hydrogen. At first we take up Born - Oppenghaimer model e.g., will treat protons as the classical particles, whereas the quantum electron in the filed of two Coulomb centers - as the source of the attracting potential between them. The graph of the potential energy of two protons with the attracting of electron and Coulomb repulsion is represented at the picture 4.3. If we consider merely the movement of nuclei in this potential, we will never obtain the association due to the law of energy conservation. The simplest model of the association appears if we agree that the potential acts only if the protons are in the so-called tied state. The tied state of protons appears from the independent states if the distance between protons becomes less than some bound $d$, and it transforms to the independent states back when this distance becomes more than some greater value $D>d$. In this case, we have the association of two protons in one molecular ion.

From the viewpoint of the net, it looks as follows. At the beginning, the electron thread is rolled as a ball around one of protons. The rapprochement of protons caused the jump of the electron thread, which will move in the vicinities of two protons simultaneously. This jump happens when the bound $d$ is overcome. The thread continues to envelop two protons if the distance $R$ between them less than $D$, and comes off one of them after the exceeding of this bound: $R>D$. This scenario looks likely for that we choose it as the element of heuristic.\footnote{There is the simplest chemical reaction: the electron transfer, which some chemists compare with the harpoon-throwing. The visual model of this reaction confirms this analogy. Perhaps, it is more right to compare it with the lasso throwing, because here the situation is more equitable.}

\subsection{Pointers}

For the description of the net dynamics, the useful tool can be the explicit separation for each sample of a real particle the other samples, which this sample is connected by bonds. The net dynamics can be then described as the change of these pointers. We will not give the general algorithm for such change but instead consider one example, in which the dynamics of pointers helps (as we hope) to build the qualitative dynamical picture. 

This shows in the dependence of the atomic nucleus stability of the state of its electrons. This dependence is surprising from the view point of QED, since the energy of electrostatic interaction between nucleus and electrons to many orders lower than the energy of interaction between nuclons inside of nucleus. The experiment concludes in the pumping of the nucleus by neutrons that result in its transformation to the boundary state, which is stable in case of completely filled electronic envelopes only. The ionization of such atom immediately results in the decay of its nucleus, e.g., the taking off neutrons. It certainly witnesses about the connection between electrons and protons, which influences to the nuclon-nuclon interactions. We mention that it is hardly possible to explain this effect by the "hydro dynamical" pressure of electrons to the nucleus, which decreases after the ionization and thus makes the nucleus unstable, like the removal of the atmosphere leads to the evaporation of ponds on a planet. The point is that the size of electron is much bigger than of the nucleus (its magnetic momentum is much more whereas the mass is much less). Hence, the "hydrodynamic" way of the building of the model of this process meets the serious difficulty. 

We consider the approach to the simulation of this process in terms of the net. Let the charge of nucleus equal $k$ and there is $q$ neutrons there. Here one quantum world of the model is the cortege of samples of all nuclons and electrons contained in the atom:
\begin{equation}
s^p_{1,i_1},s^p_{2,i_2},\ldots, s^p_{k,i_k}, s^n_{1,j_1},s^n_{2,j_2},\ldots, s^n_{q,j_k}, s^e_{1,l_1},s^e_{2,l_2},\ldots, s^e_{k',l_{k'}}.
\label{nuclo}
\end{equation}
The upper index here denotes the type of the sample (proton, neutron or electron), in the neutral state of atom $k=k'$, in the ionized $k'<k$. We introduce the additional supposition that the pointers of any sample in this cortege are designed so that they point to the samples of the small number of other particles (2-3), which remain unchanged independently of the choice of this sample inside one swarm (that is the proper bounds are somehow regulated). For the neutral atom, the pointers must remain in the stable state because nothing happens there. The ionization (taking away) leads to the unavoidable change of pointers because in the ionized atom they will not stable. The originated shortage of electrons must lead to the competition for them between protons that results in the permanent varying of pointers. The inevitability of the permanent change of pointers follows from the dynamical diffusion model because the absence of electron (hole) must behave as an electron, e.g. must show quantum properties; in particular, its swarm must spread in the time. 

For the bonds between samples, we can introduce some analogue of elastic forces, which must quantize so that the bonds break if the burden exceeds some threshold, which can be called the strength of this bond. The absence of the real electron must lead, due to the described mechanism, to small growth of the tension of bonds inside of the corteges \ref{nuclo}. If the strength of bonds between the samples of protons and neutrons is smaller than this burden, it leads to their break. 

This model is purely heuristic and it cannot give any numerical results until we define the numerical parameters of the bonds. However, such description of the process may help to build its visual picture in contrast to the other ways of explanation that is the argument foro the collective behavior method. We note that this method is not limited by QED only, because the dynamics of the net depends on the nuclear forces as well.

We showed in general how the collective behavior, in terms of net, describes the decay of the ground state of electron in the molecular ion of hydrogen, with the photon emission, and the absorption of the free electron by Coulomb field of the proton. This heuristic is applicable in the more complex cases, and it requires using computers. We note that this picture does not go out of the framework of standard quantum theory. In quantum electrodynamics we can consider this and analogous simple problems in terms of the wave functions. This standard consideration cannot be extended to the case of many particles (even for 3) by the fundamental reasons. The way of the collective behavior is free from this drawback. This constructive way contains such notions as samples and threads that do not belong to standard formalism, and which makes possible to "draw" such processes accordingly to the constructivism. This method is scalable, because the more complex systems can be regarded by the same manner. 

\subsection{Separation of spatial and spin variables}

We consider from the viewpoint of collective behavior the example with the mechanical movement of the magnetic piece of metal, when its magnetic field turns following the mechanical movement of the piece. Standard Hilbert formalism makes impossible to treat this situation uniformly with the state of separate electrons, since the mechanical movement does not touch spin variables. This is the serious difficulty of the standard formalism. The single influence to spin variable makes spin-orbit interaction, which is very weak, and cannot explain the effect of the turn of the magnetic field accordingly to the mechanical movement of the magnet. 

The method of collective behavior represents the piece of metal in the form of the net. Each quantum world of this net embraces the single sample of each particle participating in the process. Let this world have the form $s^1_{i_1},s^2_{i_2},\ldots, s^n_{i_n}$ where $n$ is the number of all particles. Such a world participates in the dynamical diffusion with the other worlds and exchanges impulses with them, as we defined earlier. Let the total number of real particles $n$ be small. The center of masses of this world will then shift to the distance of the same order $\d x$, as the separate samples. If $n$ is large, this shift will be much less.  

The orbital moment of the separate electron is the real momentum of impulse of the swarm of samples representing this electron. The electronic spin we can represent as the momentum of impulse, but only if we suppose that the electron consists of several parts so that each part has its own swarm of some size $\d_s x\ll \d x$. In this case, there are no separate spin and coordinate spaces of states. The configuration space is the unique and spin turns connected with the spatial coordinates. Let the considered system contain the small number of particles, for example, this is the atom in vacuum. There are no long proper bonds then, which connect its samples with the samples of particles forming the reference frame. Spin then remains stable in the reference frame. If there are such bonds, as in the piece of metal, we obtain the strict connection of spins with the mechanical movement. There are no separate coordinate and spin configuration space in the piece of metal. 

Of course, this reasoning about the “pieces” of electron is the conditional description of the administrative parts of the model. This method does not mean the possibility to divide the electron practically on real parts. It is guaranteed by our agreement about threads connecting the different samples of the same swarm, namely that they must be very firm. 

The scalable representation of the large ensemble is its representation in the form of the net. Only this representation gives the agreement with Galileo principle about the relativity of movement. The movement of the coordinate axes must have the description in the same terms as the movement of the s=considered system. It requires the uniform approach to the small and large ensembles, what we have in the method of collective behavior. The standard quantum formalism considers the large ensemble (reference frame) and the small ensemble (quantum system) as two principally different systems. 

\subsection{Constructive treatment of uncertainty relations}

Uncertainty relations $\Delta x \Delta p=h,\ \Delta t \Delta E=h$ of the type “coordinate – impulse” and “ time – energy” are contained in the basement of quantum theory. We have to give the constructive treatment for them. For the first relation the constructive treatment follows from the proven correspondence between the dynamical diffusion swarm and Shredinger equation. It concludes in that for the exact determining of the impulse of samples in the swarm we have to gather them from the large spatial area along the corresponding coordinate. 

We consider the second uncertainty relation “energy – time”. We could say that the exact determining of the energy requires the large period, and it would be true. However, the time represents the specific magnitude. In the constructivism, it is the number of steps of the simulating algorithm.  We know that in the constructivism there are two types of the time: physical and administrative. The first one represents the number of cadres of the video film for the user. The second one is the internal time which algorithm spends to the creation of this film. Let the longitude of the film be $T$. At each step we allow the error $\e$ in the scenario. The resulting error will be at most $T\e$ because due to the linearity of quantum theory in the worst case errors sum. Hence, to obtain the right scenario for $T$ steps we must determine each step within the accuracy $1/T$. It means that for the creation of the long video film we have to spend the big time for the processing of each cadre. The processing of the cadre means that we find what is happening in it, which particles and where move. The building of this detailed scenario requires the exact distribution of events in the time. This is the administrative time that algorithm spends to the preparation of cadres, and the user does not see this work. 

Now we look what happen if the user wants to know what energies have the participating particles in each step. It means that the user is interested the physical process going in each cadre of the film, which he (she) observes. In this case, the administrative time of the model turns to the real physical time. If we have the exact distribution of events in the time, the uncertainty relation “energy – time” gives the large uncertainty of the energies. We obtain the serious conclusion. If we want to observe the substantial film of the large longitude, we have to refuse from the knowledge of the exact energies of particles participating in this film. The more valuable film we observe, the lesser we know about the energies, and vice versa.   

The main for us is the building of the video film. The computation of energies thus bears the auxiliary character and serves for the finding of the right scenario for each cadre. For example, the law of energy conservation in QED reactions is ensured by the photon absorption and emission. These photons are valuable for us only is they are somewhere absorbed or emitted again. Their bonds create the properties of the net, classical electromagnetic field, e.g., they influence to the next dynamical scenarios. The role of conservation laws thus in that they help to create the dynamical scenarios for the computer realization. The following history of the participating particles often is not important for us. For example, for the molecular association we need to know that the energy is taken away by photons, but their further way is not interesting for us. However, if we need to account these particles for the more refined scenario we always can include them into consideration.

\section{Features of the description of QED by nets}

\subsection{Photons and entanglement}

All that we formulated for the simulation with nets remain valid for the case where one of the considered types of real particles is photons. Photons differ from the particles with nonzero mass in that they carry electromagnetic interactions, e.g., they create the potential $V$ in Shredinger equation in the usual quantum mechanics. The right net description of photons rests on the following reasoning which we reproduce here by the book \cite{Fe}. 

We consider two moving charged particles without spins (conditionally we can treat them as electrons without spins), and denote the corresponding currents $j=e\ v$ by $j^a$ and $j^b$ ($e$ is the charge, $v$ is the speed), which components we denote by the lower indices. Their electromagnetic interaction concludes in that they exchange by photons with the amplitude computable via Feynman rule we use here not vectors in $R^3$, but vectors in $R^4$, for which pseudo scalar product is defined as 
\begin{equation}
a\cdot b=-a_1b_1-a_2b_2-a_3b_3+a_4b_4.
\label{sca}
\end{equation}
We use the fact that the photon flight brings the deposit to the amplitude equal to $1/q^2$ where $q$ is its impulse. This agreement does not follow from everywhere; it is taken only with the aim to obtain Dirac equation, e.g., in the final, Maxwell equations. For this in \cite{Fe} factually the heuristic of collective behavior works, without the explicit mentioning, for example, in the analogy between the charged particle and the source of real photons flying to the all directions. Coulomb law follows from these assumptions and the determining of the relativistic corrections to it as well. 

We then use Feynman rule for the finding of amplitude of the process of emission - absorption of a photon. If our particles exchange by one photon with the impulse $q$ and the polarization $\e$, then the deposit of this exchange to the amplitude will be $j_\mu^a(q)\e_\mu\frac{1}{q^2}j_\mu^b(q)\e_\mu$ where the silent summing on $\mu$ goes with the account of signs, as in \ref{sca}. Let the photon impulse be directed along the third coordinate axes (the fourth is the time). The fourth coordinate in $q$ is the charge, and we obtain the expression of the amplitude in the form 
\begin{equation}
\Delta \la=\frac{1}{\w^2-Q^2}(j_4^aj_4^b-j_3^aj_3^b-j_2^aj_2^b-j_1^aj_1^b),
\end{equation}
where $Q$ is the three dimensional photon impulse. 
The first two summands gives Coulomb field, the second two the deposit of transversal photons. Using the current conservation $q_\mu j_\mu (q)=0$, we obtain $\w j_4=Qj_3$, which gives us the general expression for the amplitude of interaction between two charges 
\begin{equation}
\Delta \la=-\frac{j_4^aj_4^b}{Q^2}-\sum\limits_{\w^2\approx Q^2}(\w^2-Q^2).
\label{sla}
\end{equation}
We note that here we can sum because we cannot distinguish the deposits of the different photons: the photons disappear in it!

If we pass to the spatial - time coordinates in Hilbert space of states by 
$$
j_4(Q,\w )=\int\rho(x,t)\exp (-i(Qx-\w t))d^3 xdt,
$$
and find the deposit to the amplitude from the first summand in \ref{sla}. The integration on $d\w$ gives $2\pi\delta (t_a-t_b)$, e.g., the interaction caused by the first deposit must be instantaneous. The integration on $d^3Q$ gives $-\frac{4\pi}{|x_a-x_b|}.$ We thus obtain the instantaneous Coulomb interaction. By the same manner we can find the deposit of the transversal photons coming from the last summand in \ref{sla}, which gives the lagging interaction, e.g., spreading with the speed of the light the action of transversal photons. It gives Ampere law for the interaction of currencies, if the charges move not in vacuum, but in the conductor, and experience collisions with its particles. 

Such reasoning is incorrect in the standard analytic view point, but we can make it correct in the constructivism, if we assume that $Q^2$ and $\w^2-Q^2$ are separated from zero by some threshold $\e>0$. It gives the specification of thread dynamics connecting photon samples. We must establish that the transmission of impulse along such thread goes practically instantaneously, though samples themselves move with the speed of the light, e.g., with the finite speed. The transmission of impulse transversely to the photon thread goes with the light speed , e.g., not instantaneously. We also must agree that Coulomb interaction spreads along photon threads, and what is called transversal photons - at right angle to them. If the photon thread we mean as the string stretched between two samples of charged particles, then the transversal photon for which $\w\approx |Q|$ looks like the transversal wave spreading along this string. 

Lengthwise photons cannot transmit information because they create the thread strain only, but not its oscillation (the oscillations are always transversal). The information can be carried only by oscillation of thread, which are the transversal real photons which vector of electric and magnetic fields together with the impulse form the orthonormal basis in $R^3$ (for each sample the separate basis). Due to this representation of photons we may not to consider them as real particles at all,, if it is not dictated by the technical necessities of the model (for example, by its program realization). The existence of photons as isolated particles we can treat as the property of the bond between samples of two charged particles. For example, the break of the electron thread in the fall to the proton represented in terms of the dynamics of this thread (the simplest: the break happens when the limit $\delta x$ of the distance between samples is exceeded), is the criterion of the photon emission. It gives the corresponding return to the electron samples and proton samples connected with them.

In these terms, we can represent the difference between the field and the separate photons. The reasoning from above shows that we obtain the field if its source is the entangled state of many particles; in terms of collective behavior, their samples must be connected with bonds. We saw that there are the bonds, which transmit the instantaneous Coulomb interaction, which means that all samples in it have the same time. 

If we have sources of the separate photons (for example, the million of lasers), there are no such bonds between particles in the photon source, that is the time of all these samples are slightly different. It then becomes evident why a field must act much stronger than the separate photons. We account the time as the attribute of samples, not only their spatial coordinates. In the case of field, the net has the large connectivity components on the proper bonds whereas in the case of separate photons there are no such components, because the times of source samples are different (see the picture). If we consider nets as the real mechanical it leads to that in the case of $n$ separate photons their action will be $n$ times less than in the case of filed of $n$ photons, as quantum mechanics predicts (see the reasoning from the book \cite{Fe}, that we showed above). 

Of course, it is impossible to use this reasoning for the immediate numerical estimations. Its aim is to show the appropriateness of the language of the collective behavior for the building of algorithms coordinated with the standard quantum theory.

\begin{figure}
\centering

\caption{Fields and separate photons}
\vspace{200mm}
\makebox[800mm][l]{\includegraphics{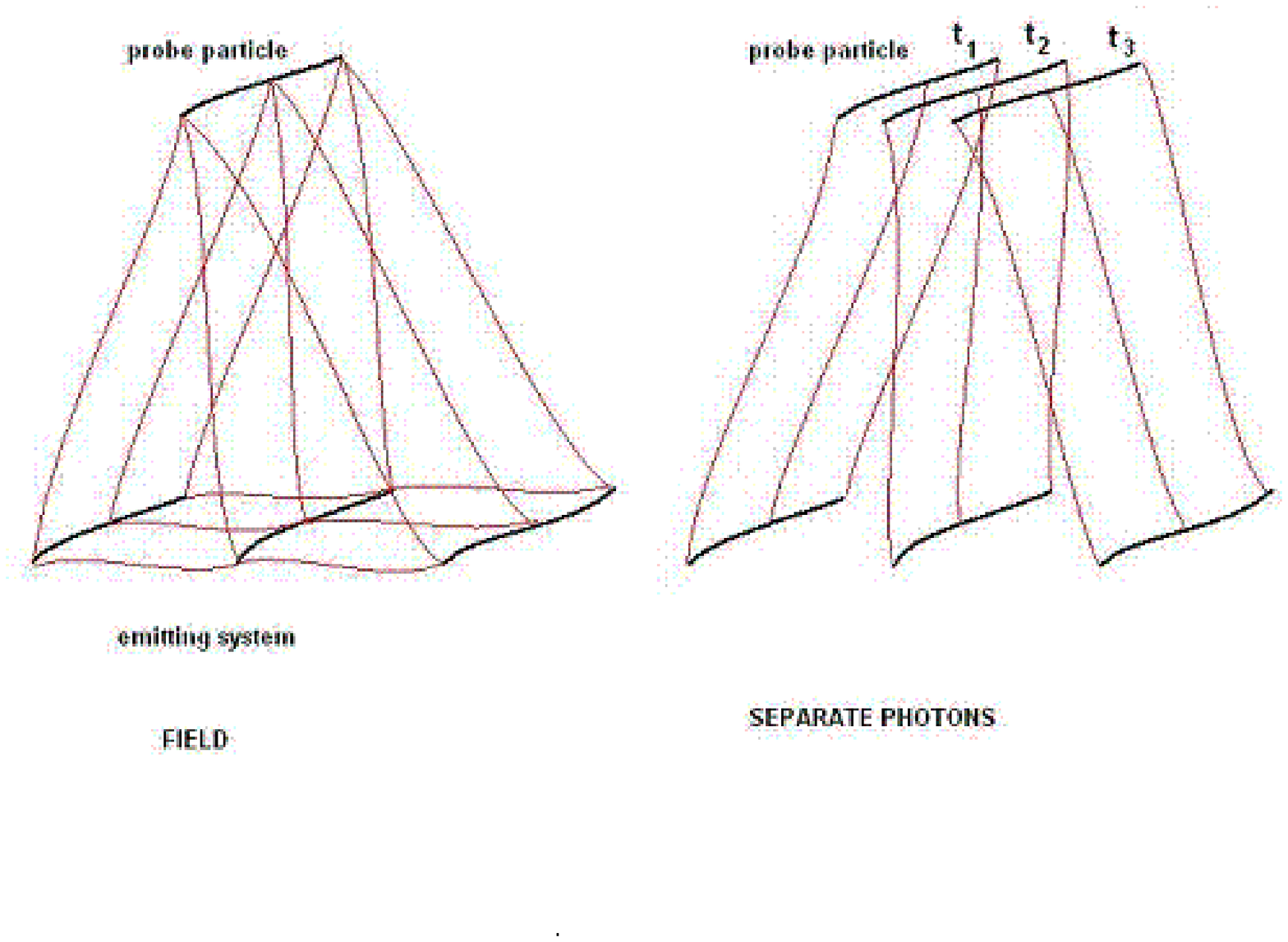}}%
\end{figure}

However, this model cannot give us Ampere law of the interaction between conductors. We assume that two electrons move in parallel with the same speed. Then only Coulomb force can act between them. Really, the exchange of transversal photons is impossible because they are not emitted in the uniform movement. If we pass to the coordinate system with the reference point in their center of masses, they will immovable in it and the single force between them will be Coulomb force resulted from the exchange of longitudinal photons. When the exchanges of transversal photons go? It takes place only if they move with the speedup. Such speedup in turns may follow only from one cause: their interaction with some other lengthwise photons, e.g., with some Coulomb field. For example, if one of electrons will be attracted by the positive charged nucleus and after the speedup emits the transversal photon accordingly to Feynman rule. 

Now we consider the other situation: the electron, moving in the conductor under the voltage. Such electron permanently collides with atoms of the lattice, where each collision means the action of the repulsion potential from the other electrons, which surround nuclei of atoms in this lattice. This collisions will go even if the speed of electron is zero due to the smearing of its wave function (see the kernel \ref{ker_free}). We can represent it as the movement of electron in the coreless tube when it collides with its walls. The net of samples of nuclei and electrons forming the lattice forms the wall. From the viewpoint of standard quantum theory, the wall is the system in the entangled quantum state of the form $GHZ$. Hence, the same thread will tie the photon samples, emitted in these collisions. It corresponds to that it is impossible to determine which collision of electron created the given photon (see the quantum system simulation). In this case, as we saw, the amplitude of the photon emission equals the sum of photon wave functions $\sum\limits_j\phi_j$, which gives us the vector potential of electromagnetic field, created by the conductor. In terms of the net it means that the existence of the thread connecting the samples of nuclei and electrons in the lattice of the conductor, all photon samples will be tied by the same thread, which gives us the possibility to sum amplitudes, not the probabilities. If we sum probabilities, we should divide the resulting field to the total number of photons that would decrease it in many times. 

Hence, the net description of electromagnetic field is substantially determined by the existence of the thread, connecting the samples of emitted photons. This thread cannot appear as is itself, it can only inherit from the thread connected samples emitted these photon samples. Moreover, the division of this integral thread to the separate photons is conditional because only the mechanical oscillation induced by the contact with this thread in the big ensemble can be detected immediately. If we take the big number of lasers instead of the conductor we could not create such a filed, because in this case we obtain the separate photons which field is much weaker than the field created by the integral thread. 

\subsection{Photon threads}

We now consider the corollaries, which the notion of thread gives for the case of photon thread. We hope that it will help us to get rid of the difficulties appeared in QED due to the application of the classical mathematics. One of such difficulties is the impossibility of the correct representation of a field as a stream of point wise particles. One could formally go round it, refusing to consider point wise particles at all, which is traditional way. Nevertheless, this representation of a field as the stream of particles, serves as the single source of heuristic for QED, and in the many particle problems, the role of heuristic is very important because the traditional way meets the serious difficulties. Their essence is as follows. In the previous section we saw that Coulomb (scalar) field must spread instantaneously that would make possible the instantaneous far communications. Just this makes impossible the point wise representation of a field in the standard approach.

When we use photon threads, this difficulty disappears. The impulse along the thread transmitting Coulomb field (longitudinal photons), really, can spread instantaneously. However, it cannot serve as the carrier of the user information, because the movement of the thread itself can transmit this information. The thread moves through the transversal shifts of its samples (that is by transversal photons). The transversal wave along the thread means the unreeling of the new thread which direction approximately equals the direction of the initial thread, and this unreeling goes with the speed of the light. Therefore, if in some place some neutral body dissociates to two charged particles which fly away with the huge speed, in the remote point located at the distance $s$ from the point of explosion the effect from the change of Coulomb field will be observable in time $s/c$, but not instantaneously, as in the absence of threads.  

The second difficulty concerns the divergence of the row of amplitudes of the showed in the picture process of the emission of two photons. We look at this situation keeping in mind that the photon samples are emitted by the samples of he charged particle connected by the thread. The emission and the following absorption of the same photon by the same particle we must treat as the transmission of the impulse along the thread. There is no real photon in this process, and only the segment of thread works between the neighboring samples. As for a real photon that can we detect, represents the thread of its samples, moved away from the thread of particle. The diagram leading to the divergence of the row in QED is forbidden in the language of collective behavior. No other events like photon emission-absorption can happen between the emission and absorption of the same photon by the same particle. 

\begin{figure}
\centering
\caption{Allowed and forbidden diagrams in the method of collective behavior}
\vspace{100mm}
\makebox[100mm][l]{\includegraphics{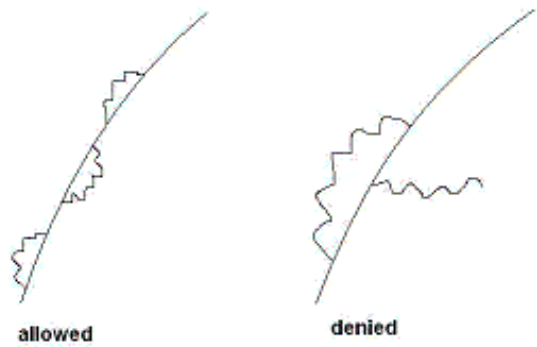}}%
\end{figure}

\section{What do we hope to reach with the heuristic of collective behavior}

Method of the collective behavior is designed for finding of effects of the concrete type in the complex systems that we can call effects of the global synchronization. We consider two atoms, each of which possess more than one valent electron and is able to establish the valent bond. We assume that these atoms are in the big molecule $A$, and their positions are fixed. There is the other molecule $B$ which in turn has two atoms with the analogous properties which are located at the same distance. Let these molecules be in the mixture and experience intense hits from all sides. Molecules of the types $A$ and $B$ can form the association if the two pairs of atoms associate by their valent electrons. We represent the attracting potential induced by the valent bond as the pseudo potential of quadratic form. This is the hypothetical model, which we use for the illustration.  
We suppose that the real potential is alternating and consists of the interchangeable zero potential and quadratic attractive potentials. We compare two situations:

1) The both pairs of atoms, which must associate, have synchronized alternation of potentials (for example, the attractive segment appears simultaneously for the both pairs).

2) The both pairs of these atoms have non-synchronized potentials.

These two cases we can distinguish in the experiment. When potentials are non-synchronized the external hits are possible to molecules in instants when one of atomic pairs attracts, and the other not. This can lead to the effect reversal to association because the pair of molecules appears which are linked in one point only. If the potentials are synchronized we always have twp point of fastening and the dynamics of molecules $A$ and $B$ will be the other that we can detect.  

We have regarded the artificial example, illustrating the global synchronism. In the method of collective behavior this synchronization are reached through the bonds between valent electrons that is it has the explicit description. Such effects we can hope to understand by the proposed heuristic. We underline again: the matter concerns the dynamic models of many particle systems represented as a video film.

Consider the other example about the detecting of entangled EPR photon pairs (the so-called biphotons). We will denote by $A^2_1,A^2_2,\ldots$ the all possible areas in the configuration space of two photons for which we can detect a hit of a photon pair to them (the lower index $1$ corresponds to the analogous areas in the one photon configuration space). The model of one photon detection can have a little dependence on the existence of its EPR- counterpart. This dependence cannot be found if we have access to one photon detector only, because the statistics will be the same for two cases: a separate photon or one photon from EPR pair. Nevertheless, it may turn evident when we compare the statistics of two photon detections. It can influence in a small degree to the statistics of biphotons, which reveals in some divergence of the experimental probabilities $P(A_j)$ from the predictions of quantum theory, supposedly observed in the experiments (see \cite{Kh2}). The physics of biphotons can differ from the physics of separate photons that we could observe in the probabilities $P(A^2_j)$, but not in $P(A^1_j)$. 
This is the possible effect of the global synchronization.  

The global synchronism, or quantum non-locality is the wide spread phenomenon, one of which manifestations is the instantaneous action of Coulomb field. Any form of constructivism must possess the special administrative element for the representation of this non-locality. In the method of collective behavior, bonds between samples play this role. Such administrative elements are inaccessible to a user, and therefore cannot transmit user information. The evidence of their reality may be only one: the necessity of the introduction of such elements for the creating of a dynamical model. 

\section{Back door in quantum informatics}

The main problem of quantum informatics is the application of quantum theory to the many particle systems. In the framework of standard approach to quantum theory the main contents of this discipline is the building and analysis of quantum computers, and quantum cryptography. Constructivism gives the possibility of some new turns into the traditional problems of this discipline. These new statements of problems follow not from the constructivism itself, but rather from its sense. Constructivism in physics serves not for the solution of the search problems; its aim is the simulation of many particle systems. Nevertheless, the success of constructivism can influence to the traditional problems of quantum informatics for which the more efficient solutions than classical were already established. We speak about quantum algorithms and about quantum cryptography. 

Constructivism imposes the new limitations to the computational capabilities of quantum systems factually submitting them to some classical supercomputer. We obtain the new object - bonds between samples of particles.\footnote{We speak about the method of collective behavior. In the other forms of constructivism, this object can have the different form, but it is necessary in all cases.} This new object does not belong to the standard formalism as the samples themselves. However, the presence of such objects in constructive formalism makes possible to speak about their objectivity. This can open new ways in the solution and statements of traditional problems of quantum informatics: fast quantum algorithms and protected quantum informational channels. We try to discuss these possibilities in general form keeping in mind that the representation about them is only the speculation based on the method of collective behavior, which we choose as the basement of quantum heuristic.

This new look at the quantum informatics problems would be more right to all "back door" because it presumes not the solution of these problems in their initial formulations but the change of these formulations. The typical statements of quantum informatics problems rest on the standard understanding of quantum theory, therefore we cannot treat them as correct when the mathematical foundations of quantum theory change. We will illustrate it on the example of two problems: quantum solution of search problems and quantum cryptographic protocol. 

\subsection{Constructive treatment of search problems}

We trace the influence of the quantum constructivism to the approaches to the search problem. In the standard formalism there is Grover search algorithm, which gives the most efficient solution of it, that is proved optimal. The search problem is the solution of equation 
\begin{equation}
f(x)=1
\label{ww}
\end{equation}

where $f:\ \{ 0,1\}^n\ar\{ 0,1\}$ is a Boolean function. There are two possibilities to define the function $f$: either in the form of gate array, or as the black box. In the first case we have the full information about the method of its generation, in the second we have no access to the device generating $f$ and can use as an oracle only. Earlier we did not distinguish these two cases. Our argument was: in the classic mathematics it is impossible to use the knowledge about the internal design of the device realizing $f$ for the obtaining of the roots of \ref{ww}. In the constructivism, we should reconsider this thesis and consider these two cases separately. I recall that the argument we used is not the exact theorem, it was something as the experimental fact in mathematics, e.g, it rests of the experience worked out in the classical mathematics, though it concerns algorithms. Now, looking at the situation from the position of constructivism we reformulate the conditions of the search problem for the case when the function $f$ is determined as the gate array.  

This situation is not typical and it needs the more detailed discussion. We speak not about the automatic application of the constructive mathematics, but about the interpretation of the physical internal design of the concrete algorithm determining $f$ from the viewpoint of constructive physics. We suppose that quantum physics already has the constructive form and try to trace what would it mean for the solution of the search problem by the methods of constructive quantum mechanics. 

It turns that the search problem from the viewpoint of the constructive physics\footnote{May be it would be formally more correct to call it the viewpoint of PCN (see below). However, we call it the search problem because we think that our speculations in this section concern every real embodiment of the search problem.} could admit the more fast solution than Grover algorithm gives for standard quantum theory. This solution in principle can have even the complexity $O(n)$, which is the logarithm of the classical solution of the search problem. This speedup can follow from the reconsideration of the formulation of the search problem only. Indeed, this is the result of the reconsideration of the statement of the search problem: it is the constructive understanding of the fact that there is the gate array realizing the function $f$. 

I stress again that we speak not about the established fact that the search problem can be solved in the time $O(n)$, but about the hypothesis. This hypothesis does not belong to the classical algorithm theory, where this question is equivalent to the fundamental problem ${\cal P}={\cal NP}?$, but it concerns the understanding of how a device can work from the constructive physics view point. The hypothesis is {\bf any physical device $Y_f$, computing the function $f$, must contain the information about sets of arguments on which this function takes each of its values.} In particular, a scheme of functional elements (gate array) determining, must contain the information about the solutions of \ref{ww}. This information must exist in the physical form and the question is only how can we extract it. This information becomes accessible in the moment when we launch $Y_f$ on some concrete value of the argument $x$. 

We suppose that the logical part of a quantum computer consists of $n$ particles, where the role of $j$-th qubit plays the state in which any sample of $j$-th particle can be. We design these states by $|0\rangle$ and $|1\rangle$. Quantum computations are built just on these states, and we can apply the constructive formalism to these situation. We then have to consider the net which contains threads, joining the samples of particles $1,2,\ldots,n$ in the corteges of the form 
\begin{equation}
s_1,s_2,\ldots,s_n.
\label{cort}
\end{equation}
 Since we have no other objects transmitting the information we assume that all information is material, then one can suppose that {\bf if there is only small number of solutions of \ref{ww}, then some cortege of the form \ref{cort} corresponds to some solution of \ref{ww}}. It means that thhe answer to our question is already in the net corresponding to the state of our $n$ particles after the first launch of $f$. 

This paradox situation can be clarified on the example of standard quantum computations. We consider Grover operator $G=-I_{\tilde 0}I_{j{tar}}$. Quantum algorithm of the fast search is the applications of $G^{t_q}$ to the initial state $|\tilde 0\rangle=\frac{1}{\sqrt N}\sum\limits_{j=0}^{N-1}|j\rangle$. If we slightly change $|\tilde 0\rangle$, by the nulling the coefficient corresponding to the target state $j_{tar}$, e.g., if we assume that $|\tilde 0\rangle=\sqrt{\frac{1}{N-1}}\sum\limits_{j\neq j_{tar}}|j\rangle$, this slightly changes Grover operator itself, whereas it is impossible to detect these both changes by the local measurements for huge $N$. However, now the application of the operator $G^{t_q}$ gives us nothing ! The algorithm will not work. E.g., to obtain $j_{tar}$, in the result of GSA we must be ensured that the coefficient of $|j_{tar}\rangle$ in the state $|\tilde 0\rangle$ is nonzero.  

It is easy to obtain $|\tilde 0\rangle$. We need to apply $H^{\bigotimes n}$ to $|0\rangle$. Let us imagine that there exists the amplitude grain so that factually in the state which we treat as $|\tilde 0\rangle$ does not contain not only $|j_{tar}\rangle$, but the bulk of possible states $|j\rangle$. Neither the direct measurement, nor the quantum tomography method can permit us to detect this divergence. To detect it we need to have not only the exponential time. We need to know exactly that all states we treat as the one particle components in $H^{\bigotimes n}|0\rangle$, are equal. The existence of the amplitude grain makes these states different in any generation and we cannot detect it ! There is the single way to resolve the question about the quality of $|\tilde 0\rangle$ to launch Grover algorithm. The correct work of the fast quantum algorithm is the single way to verify quantum theory in the many particle area. No round ways can replace a fast quantum algorithm. However, if the Launch of GSA does not give the certain answer, then the movement to quantum constructivism becomes unavoidable. It involves the corresponding reconsideration of the problems of quantum informatics. 

We could treat the problem of solution of \ref{ww} s follows. Let we be given a matrix of the form
\begin{equation}
l,\ f(l),\ l=0,1,2,\ldots,2^n-1
\label{tab}
\end{equation}

determining the function $f$. The application of $f$ to the value $x=l$ then means that we choose the concrete line in the matrix \ref{tab}. It does not correspond to the constructive form of quantum theory because the matrix \ref{tab} has exponential size. We must assume that the matrix \ref{tab} contains not all rows of the form $l,\ f(l)$, but only those which correspond to the rare values of the function $f$; in particular, if we are interesting the solution of \ref{ww}, and $1$ is not the prevailing value among the other values of the function $f$, then just the row with $f(l)=1$ must occur in the matrix \ref{tab} after the attempt to get any value of $f$. This statement radically differs from the traditional because we thus obtain that the target solution of of the problem is yet contained in its statement,\footnote{As $|j_{tar}\rangle$ is contained in $|\tilde 0\rangle$ and in $T_{\tilde 0}$.} and we only have to extract it from the initial state of the system or from the state close to the initial (obtained after the one address to the function $f$ on any of its arguments). This reformulation will be typical for the standard problems of quantum informatics in the constructive version of quantum theory! 

Factually, in the constructive physics only one problem remains: the building of the more and more detailed models of complex processes. All the other problems will be completely submitted to it. 

{\bf If we need (for something) to solve the equation \ref{ww}, in the constructivism we will analyze mainly not the function $f$, but rather this "something".} 

We can illustrate this surprising situation by the traditional example with Shredinger cat. It which sits in a closed box so that its life or death depends on the absorption or non absorption of one single photon that can be detected by some detector, which in turns launches the long sequence of chemical reactions resulting in the mechanical movement leading to the death of the animal. The life and death of the cat is thus in the EPR state with the photon state. The standard description of this situation leads to the possibilities to transfer the cat from the "dead" state to the "living" and vice versa, and to the gobbledygook of this kind. 

This situation in constructivism has the routine form. The net of samples of the charged particles represents the state of cat and this entire complex system to which supposedly influences the single photon. It makes the question like as if there are no photons at all. Most likely, that all will be decided, long before mechanical movements will begin. E.g., the cat's organism in all likelihood, has no chance to influence to the process\footnote{the opposite situation, in principle is not excluded completely, would be the sensation phenomenon and the demonstration of some super ordinary abilities, that must be studied thoroughly. It would mean the existence of very long bonds connecting samples of particles from the cat's organism with samples of particles in the laser emitting the photon.} and the experiment thus looses any scientific sense and ethical defense. 

All practical problems leading to the solution of the equations of type \ref{ww}, has purely physical formulation, which is reducible to the simulation of some real process that must be fulfilled by the methods of constructive physics. Then, if we meet some analogue of the equation \ref{ww}, we will have instead of this equation the equation with many limitations, plus many additional conditions to the function $f$, like in the problem of structure search (see Appendix). In any case we will have a cortege of the form \ref{cort}, pointing the target solution among the other corteges of this form. Here we could hope to find the target cortege \ref{cort} by the bruit force because the quantity of all corteges due to our requirements to the models must not exceed the total number of samples for one quantum particle. In the other words the requirements we imposed to the models must give the solution of the search problems arising in practice in the linear time $O(n)$. Here we speak about the complexity in the cellular automata model of computations in which we can directly simulate the nets of samples of particles; in the other models, the complexity will be polynomial. 

Hence, the matter concerns not a speedup of the search problems on a quantum computer. Such a speedup is impossible, because it would contradict to the known results \cite{BB}, \cite{BBHT}, \cite{Oz2}, \cite{Oz3}, \cite{Za2}, and some others. We speak about the different possibility: the formulations of the search problems arising in the practical simulation, which could be specified in the constructive quantum physics such that they will admit very fast solution, much faster than their standard formulations on a quantum computer. 

\subsection{About the application of limited quantum processors in supercomputers}

Constructivism establishes the new limitation on the scalability of a quantum computer, which follows from the absolute model of decoherence. This not only does not decrease the importance of experiments in quantum processors, but even makes them more valuable. We spoke about the theoretical significance of these experiments earlier. Here we look at their role in the perspectives of the development of classical computational devices. The main aim of quantum computing - simulation of the dynamical scenarios of the micro world - remains in the focus of attention in the creation of modern supercomputers. We yet have no a robust quantum computer, even limited; moreover, its perspectives and the modern vision of a constructivism are mutually exclusive. It points to the necessity to use classical computers and to adopt quantum theory of many particles to these type of computers. 

At the same time, the usage of limited quantum processors on the classical (super) computers looks promising. One could try to use in the computations the connectable devices, not based on the silicon technology, as the protein structures, for their application for the simulation of similar structures in the exclusive moments of computations. The big speed of work of protein structures and their closeness to the simulated objects gives the hope to use the queries to them as the insertions in the computational process. Whereas they cannot serve as the full alternative to the silicon technologies due to the quick abrasion.\footnote{Protein based processor is able to fulfill only several operations before the destruction; this unreliability follows from the peculiarity of a living tissue where proteins are permanently repaired by DNA molecules, which is not the case when we use them in computations.}
The using of the EPR photon pairs is possible as well, especially for the simulation of non-locality. In any case, we could hope to expect some effect from the tricks of such type only in the simulation of processes analogous to going in these devices. For standard problems, for example, from linear algebra, these tricks will slow down computations almost certainly. 

In the modern quantum electronics, the silicon technologies are completely prevailing. These technologies realize the classical computational model for which the range of problems solvable on computers is limited by standard computational tasks as the solution of finite difference schemes by the methods of linear algebra, fast Fourier transform, search problems and some others. The simulation of physics stands apart from this row because it has no commonly accepted algorithms. This problem factually represents the big direction that is now mainly hidden from us. 

I mention only one of numerous ways of application of alternative technologies in the creation of computational models of complex quantum systems. This trick concerns the generation of pseudo random numbers for complex systems. Standard models typically use program generators of pseudo random numbers for this aim. Here we completely loose the connection of these numbers with the simulated object. Accordingly to quantum theory, the randomness on the level of separate elementary particles bears the fundamental character, e.g., the corresponding event cannot be predicted more precise than Born rule says. Nevertheless, in complex systems as we saw, the notion of the wave function is not applicable, and Born rule must transform to the classical urn scheme for the choice from the limited number of possibilities. Here we go out of the area of applicability of quantum theory, and the prediction of randomness can no longer be impossible. It means not the divergence from Born rule. The point is that the mechanism of Born rule could be partially accessible in the case of many particles. Quantum probabilities in such complication of systems can transform (and transform) to the choice with the probability close to unit! However, the exact computation of these probabilities can be very complex, if we try to formulate it in the numerical language in a standard computational system. Just this part of the simulating algorithm we can try to simplify by hardware-based way. 

Practically it means that the stochastic methods for the systems of the different complexity levels could contain not only program generators of randomness’s but also the real measurements of such systems. This way of the generation of randomness gives the dependent random values for the different levels of complexity. The dependence can have the deep source. If such dependences play the role in the simulated dynamics, we can reach more by these models than by the standard digitalization. Due to its simplicity this idea, of course is not new. One can expect some return from it only for the complex models we regard. In standard situations (for one particle), this physical generation of randomness may be useful only because the quality of randomness can be higher than in the case of program generation, though it is open to question.

Constructivism in any case opens new ways for the usage of supercomputers with quantum elements. These ways are not limited by the search of maximally large processor realizing quantum algorithms. \footnote{The race for records in this area brings use only if we analyze the work of the quantum processor in details; in the opposite case we obtain the version of the roundabout way of standard formulation of problems, without explicit mentioning, like above.}

We see that the quantum constructivism encourages those who work at the problems requiring the application of quantum computer. Factually it concerns the most of those who work at one or the other aspect of quantum computations we can conclude that the prognosis of the constructivism is favourable for QC-community. It means that the community of quantum computer can expect the valuable increase of the practical output (application to concrete problems) of their activity from the successful constructive revision of quantum theory. 

\subsection{Influence of physical constructivism to quantum cryptography}

The surprising fact is that the influence of constructivism to quantum cryptography can turn in some sense opposite to its influence to quantum computing ! \footnote{That gives me the deep satisfaction because in that game my sympathy is always on the Eve side.}
The possibility to build the model of big systems can give us the new possibilities for the attacks to the information transmitted by quantum channels. Constructivism in quantum theory rather strengthens the side of cryptography attack than the side of defense. The new objects of these attacks can be not so much the channel itself, but rather the communicating subjects (Alice and Bob). It would be harder to defend against Eve, armed by models of complex processes with the predicting capabilities. It concerns all parts of quantum cryptography channel and the most vulnerable part: Alice and Bob themselves. 

This wonderful feature of quantum constructivism is general and independent of the details of cryptography schemes, for example, of quantum protocols. The possibility to build the robust models of complex processes always plays on the side of opening of the information and against its concealment. It does not mean that cryptography will loose its importance, on the contrary, its role only increases. The cause is not only the sharpening of the resistance of Alice and Bob against Eve, if the last obtain the more powerful arms. The main cause is the increase of the role of information exchange itself. If in standard quantum physics the area of the penetration of informational systems is limited by the standard formalism where we cannot say anything about the results of measurements, then in constructivism the situation is slightly different. Namely, there are the principal tools for the predictions of such a sort in constructivism. It is impossible to use these tools for the traditional devices resting on the statistical averaging of phenomena to which these tools are aimed. However, the fact of their existence as is gives the principal possibility to use these phenomena revealing in the complex systems. It means that in constructivism, the connection between the information and the real processes is deeper, that rises its importance. 

Doubtless, that in the practical sense the point of cryptographic attacks lies mainly on the participants of the connection, and only in the second order on the cryptography channel and the devices of interface (photo detector and laser). We consider the easy hypothetical version of this attack when Eve has the plausible model of the net of samples of electrons in the photo detector and laser participating in the photon transmission in quantum cryptography. 

For the determinacy, we consider protocol BB84. Here Alice and Bob must either generate a secret key $k_1,k_2,\ldots, k_l$ where $k_j=0,1;\ j=1,2,\ldots,l$, which nobody knows but they, or detect the presence of an eavesdropper Eve in the channel. The protocol looks as follows. Alice in each step $j$ chooses one of bases:
\begin{equation}
\begin{array}{lll}
&B1:\ &|0\rangle ,\ |1\rangle ,\\
&B2:\ &|\tilde 0\rangle =\frac{1}{\sqrt{2}}(|0\rangle+|1\rangle ),\ |\tilde 1\rangle =\frac{1}{\sqrt{2}}(|0\rangle-|1\rangle )\\
\end{array}
\label{basi}
\end{equation}
and if the basis $B1$ is chosen, she sends the state $|k_j\rangle$, if the basis $B2$ is chosen, she sends the state $|\tilde k_j\rangle$. When Bob receives all qubits, he measures them in the random basis. Approximately in the half of cases the basis chosen by Bob coincides with the basis which Alice has chosen for this $j$. Then Alice and Bob open to each other (by the open channel) their choices of the basis for each of qubits, and Alice send to Bob the values $k_j$ for a small quantity of numbers $j$, which Alice chooses randomly. Bob compares these values with its own values $k_j$ obtained by the measurements from the experiments in which their basis is the same. The complete coincidence of values from Alice and Bob for such $k_j$ says about the absence of Eve in the channel. After this Bob can use the values of $k_j$, for which his basis coincides with the basis of Alice - it will be the secret key, which Alice and Bob know and which is inaccessible to Eve. 

We suppose that Eve attacks the channel and catches the photons flying from Alice to Bob. To extract the information from these photons Eve has to measure the photon states. Since Eve knows the basis’s \ref{basi} (as all the information about the protocol but only the numbers $k_j$) themselves, she measures photon states in one of these basis’s. Eve does not know the choice of Alice's basis, and she will choose the basis randomly (she has no other way) and in the about half of values $j$, where the choice of Alice and Bob is the same, Eve chooses another basis. Eve's measurement unavoidably changes the state, which makes its presence evident for Alice and Bob that proves the security. 

The security of quantum cryptographic protocols thus rests on one thesis, which lies in the basement of standard quantum theory. It is the impossibility to predict the result of the quantum state measurement with the more reliability then follows from Born probability distribution. For two results $|0\rangle$ and $|1\rangle$ of quantum measurement in our case Born distribution gives the issues with the equal probabilities $1/2$.

This thesis of standard quantum theory completely preserves its truth in the constructivism due to the correspondence but only if a photon has exacly two states $|0\rangle$ and $|1\rangle$. The states $|0\rangle$ and $|1\rangle$ are the idealization. In reality there are no $|0\rangle$ or $|1\rangle$, but there are photon states emitted by Alice’s laser and transmitted by the optical fiber (or in space) and detected by Bob's detector. Such a state has the form 
\begin{equation}
|\psi\rangle=\sum\limits_{x,t}\la_{x,t}|x,t\rangle
\end{equation}
provided the photon does not interact with anything. In case of the unavoidable interactions we have the state of photon + all particles which it interacts with, of the form 
\begin{equation}
|\Psi\rangle=\sum\limits_{\bar x,\bar t}\la_{\bar x,\bar t}|\bar x,\bar t\rangle
\end{equation}
and the studying of this state by the methods of standard quantum theory is not possible due to already known to us quantum computing phenomenon. 
How quantum cryptography evades its difficulty? We suppose that the dispersion of the time is small and the cortege of times $\bar t$ is factually the scalar parameter $t$. We then can pass to the non-relativistic notations where the wave function has the form $|\Psi (t)\rangle$, and overlook its spatial part, e.g., we suppose that the photon polarization does not depend on the coordinate $x$. At last, we can accept the model of standard measurement, which we treat as the random variable with the inaccessible source of randomness. Now the formal apparatus of quantum cryptography works, including the theorems about the security of quantum protocols. 

The security of quantum cryptography thus rests on the supremacy of quantum theory in its standard form for all many particles world. Of course, the support of quantum theory makes quantum cryptography much more reliable than classical cryptography. Nevertheless, the success of constructivism could influence to this reliability.

What would we obtain if apply the constructive approach to cryptography scheme? We represent the measurement as the process in which the new and new particles are sequentially involved so that the state of photons, electrons and nuclei we describe by the net of samples of these particles. We accept the described type of evolutionary models with the criteria of transformations of threads and samples. We then obtain that all behavior of the system "laser + optical fiber + detector" \footnote{We can also include to the systems Alice and Bob themselves. These turn leads us out of quantum cryptography, but is is not impossible for the real scalable model. Of course, this variant belongs to the range of considered ways in the defense of secrets, because just the humans represent the most unreliable part of systems of information security. Constructive models would be useful here as well.} depends on the initial parameters only, which is subjects for varying; for fixed parameters we can certainly say what a state will the measurement give. The entire urn scheme will be in our hands, but the parameters we do not know. This supposition is now the hypothesis, because there are many particles in this scheme, but we can simplify the system as usual. The main that must remain is the deterministic dependence of the result of measurement of the initial parameters. The measurement in constructivism will be the usual evolution, which differs from the unitary dynamics only by the huge number of real particles involved in the system that exhausts the stock of samples and initiates the rebuilding of the net and photon emission. 

The uncertainty of the result of measurement in constructivism is reducible to the uncertainty of the initial parameters of the net. If we suppose that we can determine these parameters, knowing the details of the structure of all devices composing the atomic level of cryptosystem we could choice the values of the unknown parameters likely. Now we make the sequential attempts to break the same cryptosystem. Since we do not know the needed parameters of the net, we will sequentially fail. However, if we know sufficiently much about nets (it is not the case now) we can hope to find some algorithm specifying the choice of the unknown parameters, and consequently, finding the result of measurements with the more reliability than the simple generation of the key accordingly to the probability distribution in the standard quantum theory. Any possibility to learn more about the result of the measurement besides the standard distribution means the violation of the cryptographic channel security. There are the new possibilities that constructivism gives to Eve. Of course, these possibilities can turn something deserving the practical attention when only we obtain the robust constructive model of the interaction of a photon with an atom. 

Constructivism thus can give the new possibilities to the side of attack that would be stimulating for those who take up the cryptographic defense, and necessitates searching for new ways of the defense against these attacks. However, we must understand that this will be the situation on the high level of development of the cryptography in its quantum form. In the choice between classical and quantum cryptography the last has the unconditional priority, because the classical cryptography cannot resist against this kind of attacks even hypothetically. 

\section{Review of algorithmic modification of quantum theory}

I summarize here the method of collective behavior, which pretends to be the most appropriate formalism for constructive quantum theory. It looks as if the constructive approach to physics must rest on this formalism. However, it does not mean that the method of collective behavior is equivalent to the algorithmic approach itself. The application of traditional computational methods (for example, for the finding of stationary states) may be useful for many cases, for example, in the representation of chemical reactions.  

Why the formalism of collective behavior is more convenient than the standard formalism of wave functions? A wave function is not a directly observable magnitude. It arises not in the course of one experiment, but follows from the statistical processing of the big number of independent identical experiments.  Therefore, it is reasonable that the method of collective behavior rests not on the notion of wave function, but on the swarm of samples. The gathering of statistic required for the establishment of the wave function in any experiment corresponds to the decreasing of the grain of spatial resolution. It is interesting that the collection of statistic in the collective behavior method uses the dependent experiments; the initial condition of each experiment depends on the result of the previous experiment. The selection of quantum states reflects this dependence. 

Moreover, in standard quantum theory the averaging proceeds on experiments in each of which we choose the new system in the same quantum state. Here we accept that for the other choices of systems in the same state the averaging gives the same result. This supposition lies in the basement of standard quantum theory. It presumes the hypothetical possibility to carry out the unlimited quantity of experiments and then to fulfill the averaging. In the other words, it is assumed that the result comes after the potentially unlimited number of experiments. Of course, it radically disagrees with the principles of constructivism; in particular, the definition of the probability through the limit of infinite sequence of experiments is unacceptable in the constructive analysis. 

In the formalism of collective behavior, we deal not with the wave function but with the dynamical scenarios, which results are detected in experiments. It completely agrees with the constructive mathematical analysis. For any approximation of the resulting wave function initial value $|\Psi (0)\rangle$ through the swarm $S(0)$ we effectively find the corresponding approximation of the resulting wave function $|\Psi (T)\rangle$ through the swarm $S(T)$. It It concerns to the scattering problems with photons in full measure, that is to the swarm formulation of quantum electrodynamics. 

The method of collective behavior thus entirely agrees with the constructive analysis. Since this method can be treated as the possible constructive version of quantum theory for many particles. The heuristic of collective behavior including the notion of the bonds between samples and the nets as the forms of the complex system description is very natural. Applying it, we can hope to obtain the right dynamical scenarios for many particles in the area where the wave function method is not applicable in principle. The debugging of the details of this heuristic by its comparison with the standard quantum mechanics seems to me the surmountable difficulty. 

 \chapter{Program container of natural sciences}

The main advantage of the method of collective behavior in the constructive quantum mechanics concludes in its potential scalability to the large systems. It permits to speak about the creation of the more universal program designed to the simulation of systems which belongs not to quantum mechanics, but to chemistry and even biology. Such a hypothetic program we call the program container of natural sciences (PCN). This project in contrast to constructive quantum mechanics yet stands farer from the practical realization because it touches several disciplines. Though we trust in the doubtless priority of physical laws, there is the specific ideology in these disciplines that we cannot ignore. Hence, the reader should consider this chapter not as the purely scientific but also science fiction.  

Nevertheless, PCN is important because for the creation of algorithms for the constructive physics we need the heuristic and the basement of this heuristic must be common for the different areas of natural sciences. We discuss the perspective of the creation of PCN and some practical tricks that can we use for it. 

\section{Actuality of PCN}

The development of all natural disciplines entails the permanent appearance of the new divisions of fundamental knowledge. It leads to the geometric growth of scientific information that one can see analyzing the volume of articles in the electronic archives and its dynamics. This quantitative growth is impossible without the decreasing of the quality of scientific publications because the feedback becomes superficial. The other manifestation of this phenomenon is the permanently growing degree of specialization of areas of the sciences, making typical the misunderstanding between the specialists from the different areas, even if they study the same real object. This situation is objective. One cannot change it by the establishing of some filters or the increasing of the requirements to the selection of scientific texts. We do not intend here to analyze this problem, we only mention that it is deep and it is senseless to expect further breakthroughs from the science without its solution, at least, partial. This concerns the contents of science, not its form. No external administrative measures can solve this serious problem because the direction of development including the character of possible experimental breakthroughs is determined by the contents of science. 

I think that it is right to form on the basement of fundamental knowledge the program solutions valid for the technical applications as well as for the development of the knowledge itself. For the systems in any level of complexity, there are the special dynamic scenarios not reducible to the scenarios of the lower degrees directly. We measure the complexity of a system in the total number of its particles. There are the definite scenarios for one particle. For two particles, there are new effects not reducible simply to the one particle effects, for example, the existence of entangled states. For three particles, we have again the new effects not reducible to bipartite effects. For example, any state of two particles can be reduced to Schmidt type that is not the case for three particles. For 10 particles we observe the new effects again, this is true for 100 particles, for 1000, etc. An attempt to understand how a complex system functions without preliminary considerations of the more simple systems is an imprudence\footnote{For example, it is impossible to build the model of even simple living thing without attracting for this all the known physics, if we speak about the model, not the animation. The animation can have only advertising, not scientific value.}. For example, any progress even for very complex systems as living things presumes the preliminary consideration of its more simple parts\footnote{The discovery of bacteria, DNA spiral, etc., one can continue the list.}.

Is it possible to held sequentially this idea of hierarchy beginning with elementary particles to the complex systems? We abstract our mind from the technical difficulties unavoidable in any attempt to realize this idea practically. The matter concerns the principal realizability of this way. It is wittingly impossible with the traditional interaction of natural sciences, because their division just follows from the delimitation of their areas. I think that the modern programming methods make possible to overcome this obstacle. This is the aim of PCN project, which we consider in this chapter. 

\section{What is the data compression and why it does not satisfy us}

Since we meet the necessity of the severe economy of the memory, we consider at first the existing methods of the data compression in order to extract the useful tricks for our aim. What if we quantize all known scientific tricks and try to use them in this form. We suppose that we have to write in the most compact form a word $L$ in the alphabet $\{ a_1,a_2,\ldots,a_k$. We introduce the auxiliary alphabet $a_{k+1},a_{k+2},\ldots,a_l$ and will record the replacement of variables by means of these letters. Let we have a word $L=L_0$ at the starting step (basis). At the step $j$ we have the word $L_j$ in the extended alphabet and the system of equations in words of the form $a_g=a_da_m$. One step of our process is the simultaneous replacement all occurrences of words $a_da_m$ by $a_g$ in the word $L_j$ and the introduction of the equality $a_g=a_da_m$ in our system. We agree to make this operation only in the case if it gives the decreasing of the common length of the word and equalities. 

Applying this trick we can easily encode some regular grammars, for example, the results of replacements of the form $a\ar ab,\ b\ar ba$. The programs for the data compression use this scheme. Nevertheless, a word compressed by this way becomes non-functional. Indeed, the compressing method ignores the fact that letters $a_d$ and $a_m$ in this replacement can belong to the occurrences in the word with the different functions. We thus cannot effectively work with the word after data compression that makes this trick not appropriate for us. In our case, just the functionality is the aim, and the naive quantization methods here will be useless. 

\section{About the language of PCN}

Our nearest problem will be the creation of the general format for the description of big systems by means of the small memory, appropriate for the simulation of their dynamics. As we mentioned the scalability of our models requires that the memory for the record of a system with $n$ particles grows as $log(n)$. This requirement is mandatory because in the opposite case the simulation of big systems would require the cost of the same order to the cost of their creating that makes the simulation senseless. 

The condition of the logarithmic growth of the memory is much more severe than even the linear growth that we imposed earlier. We thus must simulate in the limited memory the behavior of exponentially large systems that presumes the special ideology.

As the basement of PCN, we take the spatial classification of systems and processes. This classification is not absolute: for example, the uncertainty relation in quantum physics says that the tendency to very small distances in the exact localization of a light object leads to the corresponding dispersion in its impulse, and consequently, to the large distance to which this object flies in the next time instant. However, this difficulty yields to the great advantage we obtain from the sequential division of the space because of the exponential economy of the memory. Factually, quantum phenomena should be treated as the kind of exception. We will see how we could fight against these exceptions in the order to preserve the logarithmic complexity of the model we announced. The uncertainty relation is not the single exception, which we have to process especially in the control system of PCN. The second important exception represents entangled states, which are able to influence to the processes of the macroscopic scope. These exceptions compose the main cause why we make a fuss (otherwise we could manage by only classical physics and mechanics, applying, for example, equations for the typical mechanical processes\footnote{We note that not all such processes can be described by the equations, for example, the turbulentness.}). They lie in the basement of complex evolutions we are going to describe, for which we intend to develop constructive physics. 

All the complex system we consider must be divided to layers so that inside each layer all its components must be described, for example, by means of finite automaton. This division is not necessary connected with the spatial nesting of parts but it is convenient to represent it just as the spatial nesting. 

Let us regard the following division of the space to the sequentially nesting cubes. We choose the grain of spatial resolution $\delta r$, and divide the space to cubes with the side $\delta r$. We ascribe the zero level to this division. Now we join the neighboring cubes to the bigger cubes with the side $2\delta r$. We have 8 cubes of zero level inside each of these cubes. We call it the division of the first level, etc. We associate with each point of three-dimension space its coordinates as follows. We take the smallest from cubes with the initial vertex in some fixed point O called the reference point, which contains this point. Inside this cube, we take the cube of the level smaller to unit, which contains this point. The position of this cube inside the first cube we can express by the three bits that we denote by $E_1$. These three bits will be the rough coordinates of our point inside the first cube. Continuing the process of the division of cubes we obtain three numbers $E_2$, such that six bits $E_1E_2$ gibe the approximation of the initial point in twice accuracy, etc., while we reach the maximal accuracy $\delta r$, and obtain the full coordinates of our point of the form $E_1E_2\ldots E_k$. We could divide the side of a cube not to two, but to more parts, it is not important. In any case $k=log|r|$, where $|r|$ is the distance from our point to the reference point. These reference frame represents the cubit configuration space for points with the accuracy $\delta r$, which we yet considered in the simulation of quantum systems. We now look at it in the different manner. 

We can divide the cortege $E_1E_2\ldots E_k$ to the several segments: $E_1\ldots E_{j_1}E_{i_1+1}\ldots E_{j_2}\ldots$ so that each segment $E_{j_l+1}\ldots E_{j_{l+1}}$ we associate with some natural discipline. For example, the levels from $1$ to $30$ correspond to the nuclear physics, the levels from  $30$ to $60$ with the atomic physics, from $60$ to $80$ with chemistry, etc. Of course, this division is approximate, because there are overlapping zones here. But it is good because of its absolute character, e.g., it does not depend on the chosen unit system; moreover, its dependence of the division of the side of cubes gives the coefficient only. 

Let us imagine that we are given a large number of point wise objects, each of which has its own spatial coordinates, and some type from the finite number of types. We then can ascribe to each such object its coordinates by this way. Analogously, we can ascribe speeds, keeping in mind that the speeds will decrease when the level increases: in the nuclear physics, they are large, in the atomic less, in chemistry yet less, etc. Such objects represent the smallest parts of the matter. Choosing $\delta r$ we can make the coordinates of the different objects different. The complex system will be then quantized in the form of coordinates $\bar E$ of its objects. Their total number may be of the order of $10^{40}$ in the considered system and it requires about  $120$ levels, e.g., $k=120$. Using logarithmic length of the notation of coordinates of each object we can by the economy way, applying only arithmetic operations, encode such processes as spreading of waves, heat transfer, diffusion process, etc. For this, we must regard finite difference scheme, representing this equation and transform it to the qubit form which is not a big problem for these processes. 

We thus can embed to PCN the area which conditionally can be called the mechanics of uniform matter (just embed, because it does not give the principally new ways for investigations). Here the structure of layers of some level that finite automata give will be enough for us. It means that any increasing of complexity here even with the growth of the number of elements of the division of cubes can be represented as the increasing of the degree of the occurrence of a word in the regular grammar. 

We consider how could we obtain the kernel for a free particle \ref{ker_free} by means of regular grammars. We take the sequential division of the space to cubes we describe earlier. We must obtain the more and more accurate approximations of the function $e^{imx^2/2ht}$ in terms of our hierarchy of cubes, e.g., the approximations of the function $\cos (mx^2/2ht)$ and $\sin (mx^2/2ht)$. We can do it having the algorithm of the sequential approximations for these functions, based, for example, on their Tailor expansions. At each level $j$ we have some functions $COS(\bar E^j),\ SIN(\bar E^j)$ where $\bar E^j$ are the approximation of the argument coordinates of the level $j$. We thus can represent any process described by the differential equations. 

We call this representation of the mechanical processes the qubit representation, because we identify qubits with the digits of coordinates $\bar E$ in the configuration space. The difficulty mentioned above arises here from the uncertainty relation leading to non-locality. We consider again the kernel for a free particle \ref{ker_free}. This kernel is the wave function of a free particle in the moment $t$ provided its state in the initial moment $0$ in the coordinate representation is delta function concentrated in the reference point. We see that the smaller is $t$, the faster oscillations go when the value $x$ is fixed. We thus have the process, which we cannot simulate in the real time mode in the sense of the sequential approximations from rough to more exact, as, for example, the association of molecules we considered above. Here the smaller the time $t$ is, the more exact approximation in all space we must take to draw the right picture. The application of the method of collective behavior permits to limit a speed of flight of the free particle samples only, but does not eliminate the difficulty itself, which concludes just in that we must take the most accurate approximation in all space, especially, in the far areas from the reference point for very small $t$. It means that we cannot use the method of sequential approximations in the real time mode, and must apply the model of dynamical scenarios and the selection of these scenarios as was described earlier. In the scenario selection we must make the passage along the qubit coordinate of the particle $\bar E$ in the limits determined by the considered interaction. For example, in the nuclear physics it is $30\leq j\leq 60$, and we then have to compare the scenarios and apply the selection specifying the needed values (for example, the behavior of the wave function in the far points as for the flight of free particle. As we agreed earlier, the time spent to the processing of scenarios is the administrative, not the real time, and just in PCN is releases in the full measure. 

The next difficulty touches the existence of entangled states and the clarifying of their principal role in the macroscopic bodies evolution. We can go round it as in the first case. The existence of entangled states is expressed in the method of collective behavior in the form of a net connecting the samples of real particles. These samples connected by threads and bonds can be on the different levels of the qubit hierarchy in the processing of the net. The time spent to these travels also cannot be treated as the real, it is the administrative time. 

Qubit spatial coordinates is not the single form for the storage of particle massive, which forms a complex system. We could apply this principle of the nesting hierarchy not to the spatial nesting, but to the particles themselves that are able to join into bigger particles by the hierarchical way. This method differs from the spatial hierarchy like Lebesgue integration scheme differs from Riemann scheme, analogously to the situation with the description of many particle ensembles by means of a net (see above). We consider the simple example of a system with 10000 identical atoms of ferromagnetic joint into the non-regular crystal lattice consisting of 100 clusters with 100 atoms each. Let this lattice be subject of the strong magnetic field. The initial orientation of spins of these atoms caused by the prevailing orientation of electron spins from the partially occupied orbits is unknown for us. If we study the process of establishing of spin orientation under the external magnetic field we can assume that the initial orientation of electron spins inside each cluster is the same. If it is not the case we can change their orientation and simulate from the very beginning in order to check that it does not influence to the result. 

Analogously, in the investigation of a complex biochemical reaction we can treat that all hydrogen atoms in its reagents are in the same state but a small number of atoms immediately colliding with each other in this instant, and exchanging by electrons with neighbors. Analogously we can identify the shapes of complex molecules forming the environment for some reaction but a few separate molecules, etc. This identification permits to economize the computer memory but it cannot give the logarithmic complexity that we want to have! 
If at each level we permit to have the different states at least to two particles we cannot avoid the memory expenses of the order $n^{\a}$, where $n$ - is the total number of real particles in the simulated system. 

To obtain the logarithmic complexity on the memory we have to account that for the bulk of hierarchy levels, the states of particles play no role, and their type determines all. The limit case is that the state plays no role at all, and all the identical particles are in the same state. All the difference between them is then in the spatial positions of zero level particles samples composing them (in QED there are electrons and atomic nuclei and photons) in the moments of their association (or photon emission). Since the result of this collision is determined by the generator of random number (if we do not use the different criterion like $d$ and $D$ for the association in a molecule - but this criterion factually, also requires the generation of randomness, but at the lower levels of hierarchy), all divergence which we allow in the system concludes in the 
\begin{itemize}
\item a) different types of particles, 
\item b) different spatial positions, 
\item c) the possible individuality of elementary particles. 
\end{itemize}
The most dangerous type of the divergence, which excludes the usage of logarithmic memory, is the point b).

We can suppose that all spatial positions are simply chaotic, e.g., the presence of random order in them (for example, randomly aroused periodicity) does not influence to the result. This thesis we can check by the random change of the initial conditions of the problem - all scenarios must preserve. For example, the building of a DNA molecule does not depend on the exact initial positions of its nucleotides - they are in the chaotic states. Just the chaotic positions ensure the assembling of this polymer, e.g., here the order arises from the chaos. If it is the general law (we have reasons to think so), we have only two sources of the divergence of scenarios: a) and b). The divergence of types of particles in all likelihood is certainty predictable, that is we can separate it explicitly for each level (though there is one dark place here connected with the dynamics, for example, some ions can be non stable in the mixture, or the stability can depend on the frequency of photons flying to them, etc.). Even if it is not the case, we can go to one level down and consider the "suspicious" particles as ensembles form the particles of the lower level. In this case all the divergence is reduced to the divergence of the identical particles of zero level. 

We can obtain he logarithmic on the memory model in the case only if we can reduce all the divergence of scenarios to the manifestation of the divergence of elementary particles of zero level, e.g., of the truly elementary particles accordingly to our representations. Nevertheless, the usage of them will be the most risky because it rests on the ignoring of the divergences of the points a) and b). When should we try to create such models? It would be right only if the subject of investigation much exceeds the computational system. For example, if we simulate the evolution of a star. 

Models with the logarithmic memory are similar to the models of the dynamics of uniform media that we describe by the systems of differential equations independently of the existence of non-stable solutions. The non-stability towards the initial conditions signals to the approach to the limit level of hierarchy for such particles. It means that the form of solutions is determined not by the initial conditions themselves, but the lower levels of the hierarchy, e.g., by the scenarios for the compounds of particles (as for a heavy stone standing at one point its trajectory is determined not by the equation of its dynamics but by the molecular processes in this point). The finding of these scenarios has the great value, for example, in the building of aero plans where they determine the limits of the admissible turns for wheels. The more interesting theme - the control on the system in such point that requires the coordination between the different levels of the hierarchy.\footnote{It slightly resembles the problem which solves a cheater in a risky game: usage of the more slight mechanisms for the generation of the desirable result.}

Of course, this model will no work if by a type of particle we mean only its contents, for example, if the components are electrons and nuclei which differ only in their charge and mass. We should distinguish in what state, $1s$ or $2p$ the electron in hydrogen atom is, etc., e.g., we must treat these states as the different types of particles. It works well for the stationary states but is not so good for the dynamics, for example, for the electron transfer we need to distinguish many excited states, especially if we need the steps of this process that may influence to the scenario. We then need the other language in which the type of particle contains the information about possible dynamical scenarios with it. The method of collective behavior gives just such a language when the system of particles is represented as the net of their samples connected by bonds and threads. In terms of a net we can store the information about the limited number of spatial configurations of the net fragments (for example, about its connectivity components), and call each such configuration the separate particle. 

We summarize the basic notions about PCN. It must be hierarchical structure, in each layer of which, there are particles of the same type of complexity. We treat them as the elementary components of the element standing immediately above them at the next layer. A particle of the level $j$ we call a set of particles of the level $j-1$ with the specification for each of them its exact coordinates in the reference frame associated with their center of masses, which we treat as the coordinates o of this particle of level $j$. Correspondingly, each particle $\a$ of zero level contained in the considered system has the coordinate $\bar E(\a )$. Here we require the existence of the fixed set $\bar j$ of numbers $j_1,j_2,\ldots,j_l$, such that for each level $j$, which is not contained in $\bar j$ it is possible to store in the memory only one particle of the level $j$. We agree that all the other particles of the level $j$ are defined arbitrarily, such that the dynamical scenario does not depend on their definition. We also agree that for the level contained in $\bar j$, the number of the different particles is limited by some absolute constant. The models then will have the logarithmic complexity in memory depending on the total number of all particles. 

All systems described by differential equations (as the homogeneous media) are the models with the logarithmic memory. It follows from that we spend the computer memory not to the storage of each particle separately but to store the total number of particles. PCN is the generalization of such models to the complex systems. Here we try to reach the same effect to the right scenarios as is given by differential equations but in the cases when differential equations do not work. These cases concern the forming and decay of entangled quantum states, e.g. the transformations of chemical type. 

\section{Why PCN is needed}

PCN is designed for the joining the methods of the natural sciences and the obtaining of the new possibilities for the development. We can expect the serious effect resulted from the sequential application of the known quantum mechanics (QED and many particles entangled states) to objects studies in chemistry and biology; this possibility becomes real with the application of modern program methods. However, there is the other possibility: to develop many particle quantum heuristic in the area of many particle systems, attracting the analogies from chemistry and biology. We already discussed the narrow character of quantum heuristic in the area of complex systems and that the facade of standard Hilbert formalism can hide the unknown phenomena. These phenomena can have the form of collective effects of fundamental nature, which we can meet in the building of PCN. It require from us the different attitude to the computer simulation, which will have the more high status than now, when it plays the role of technical service for the specification of details and engineering constructions. 

There is the traditional understanding of the word "fundamental", which is reduced to the "smallest element of the matter"; today this notion has the wider treatment. The behavior of an integral system is irreducible to the behavior of its components - the truth well known to biologists is now estimated in quantum theory by means of entangled states. We believe that the development of PCN will help us to discover the new depths in the area of collective effects. 

PCN is the particular program. Its debugging requires the coordination of modules corresponding to the different levels in the initial complex system $D_v$:

\begin{equation}
D_0, D_1, D_2,\ldots, D_v,
\label{order}
\end{equation}
so that the systems of each next level $D_{j+1}$ are obtained by the joining of some systems of the level $D_j$. We can assume that the process of assembling of the systems accordingly to the ordering \ref{order}, is the model of physical process of the appearance of the system at hand. We temporarily ignore the entangled states. If we establish, accordingly to the traditions, the exact parameters of processes determining the evolution of systems of the level $D_0$, taking into account that we have no entanglement, we obtain the exact parameters for $D_1$ and all the other levels up to $D_v$. The question is then reduced to the definition of the initial states for all elementary particles contained in the systems of the level $D_0$. If we want to make this choice in order to obtain the behavior of the system $D_v$, known from the experiment, we have to search all possible initial conditions for all elementary particles inside $D_0$, that is impossible if there are more than 3 such particles. It implies that the specification of the initial conditions for elementary particles of the lower level we must do taking into account the processes going at the levels differing in 3 units. 
This situation takes place for systems studied in the theory of oscillations, in the homogeneous media, aero and hydro dynamics, and in any case where to describe an ensemble of large number of particles, we need only two-three levels. 

For the more complex systems we need all hierarchy of systems \ref{order}, and this simple reasoning will not be true. The cause is the entangled quantum states, which we must now take into account. There are no separate states of the particles, only the common state of the entire ensemble. If such ensembles are not contained in any separate systems of levels less than $D_v$, the establishing of the initial conditions of elementary particles of the lower level will not play any role because there will be no such states at all. Factually, the characteristics of such particles will be determined by what is happening on the upper levels, up to $D_v$. This consideration means the unavoidable using of many particles Hilbert formalism that in the constructive physics directly leads to the introduction of the individuality of elementary particles depending on which entangled states they participate in. It is evident for the swarm approach. For any other approach to the quantum constructivism, it follows from the general reasoning, which I try to formulate.

There are no parameters but the individuality of elementary particles, which can we use for the coordination of the different levels. If we treat the elementary particles as the identical simplest "bricks", from which we can build an adequate model of a system of any complexity, we unavoidably have to use the direct search through all possible initial conditions of the system at hand that is not a real way. In the other words the traditional way of the consideration of evolution by Cauchy method: the equation plus the initial conditions is non-real for complex systems. We can expect the success only if we treat scenarios by the method of genetic selection. The limited case of this selection is that we treat all considered particles including elementary as the carriers of the genetic information. This information must correspond to the state of the whole system, not to the state of this particle only. It means that in each elementary particle must be the information about its place in the whole system. 

We consider by this way the method of collective behavior. In this method the model of the complex system of $n$ real quantum particles consists of the set of scenarios of the form 
\begin{equation}
\bar C_1,\bar C_2,\ldots,\bar C_M
\label{sc}
\end{equation}
where each scenario $\bar C_j$ has the form $C_j(t_0),C_j(t_1),\ldots,C_j(t_k)$, and each $C_j(t_i)$ is the list of corteges of the form $c_j^1(t_i),c_j^2(t_i),\ldots,c_j^L(t_i)$, $ML$ is the total number of samples in each swarm for any real particle participating in our system. The set of scenarios \ref{sc} is the object of the selecting procedure described in the previous chapter. 

Any cortege $c_j^s(t_i)$ consists of $n$ samples: one for each individual swarm: $c_j^s(t_i) = [c_{j,s}^1(t_i),c_{j,s}^2(t_i),\ldots,c_{j,s}^n(t_i)]$. We know that the corteges on each step of assembling must "perceive" the presence of each other, e.g., must interact by the impulse exchange on the level of pairs of concrete samples. In this interaction not only a pair of samples exchanging impulses plays the role, but also all samples occurring with them in two corteges, because these corteges must be close in the natural metrics. 

\begin{figure}
\centering
\caption{Corteges of samples}
\vspace{150mm}
\makebox[230mm][l]{\includegraphics{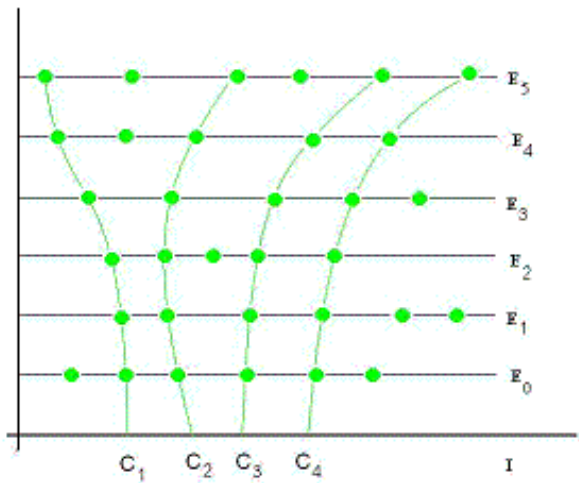}}%
\end{figure}

\section{Account of random factor in PCN}

In any concretization of the method of collective behavior (the version of diffusion, etc.) we have the random parameter determining the behavior of each sample form our corteges in any instant of administrative time. This random parameter influence to the result of selection, e.g., it determines the resulting scenario of our model. Since we agreed that all the models we consider require no more than the linear (for complex systems no more than the logarithmic) time depending on the size of the system, we can fulfill statistical experiments with these models. In particular, we can repeat the simulation many times, using the different sources of randomness. This work results in the understanding how strong is the dependence of the resulting scenario from the source of random numbers. The problem is to obtain the minimal dependence of this source. 

The minimization of the random factor corresponds to the commonly accepted logics accordingly to which the ideal model must be deterministic. However, in the standard quantum theory the randomness is the inbuilt factor, which has the absolute sense. We must introduce the random factor to PCN in order to preserve the integrity of the natural knowledge. The determinacy in quantum theory is the determinacy of Shredinger or Dirac equation, which includes wave functions of the considered systems. These functions are the results of statistical processing of the set of uniform experiments, but if we use the wave functions as the actual objects, our model will be deterministic in the sense that the random factor is completely removed from it. 

It must be treated as follows. If we are interested not what happens in the concrete experiment, but what happens in the result of averaging on the large set of uniform experiments hold with the different and independent sources of randomness, our model gives the certain result. In this sense, the random factor is completely removed from the model of simple systems. The same will be true for such processes as the standard reactions of scattering of two charged particles with spins, or simple chemical reactions, like the process of association of two protons in the molecule of hydrogen we considered earlier. For the processes of this type it makes sense to speak about the averaging on the large number of repetitions of experiments and the random factor must be removed. 

Factually, the removing of the random factor from the models of simple processes represents the other form of the known ban to the existence of the hidden variables in quantum theory, which was in the center of discussions yet in sixteenth by Bell, Bom and the others. We can treat this ban as the insurance in that the wave function is sufficient for the description of properties of any simple system. The ban on the hidden variables is thus the basic principle of the modern physics and must be accepted in quantum theory in the framework of its area of applicability. 

PCN, which we discuss, must embed not only quantum theory, but also (potentially) all natural sciences. This programming container must have the tools for the work with the phenomena, which do not belong to the area of applicability of quantum theory, even in its constructive version. Where are the borders of applicability of quantum theory? The first limitation is the total number of particles in the system, because it touches the problem of quantum computer we discussed. The second limitation is on the real physical time. 

We have already discussed that quantum effects reveal when the value of action $\Delta S=\int\limits_{t_0}^{t_1}L(x_t,x,t)dt$ becomes close to Plank constant $h$, that presumes the small time frames. Of course, it does not make impossible the application of quantum theory for larger periods but in this case, it gives trajectories close to the classical. Nevertheless, in the building of models of system with the large number of particles we cannot consider too small periods because of the following reason. Quantum mechanics operates with the wave function, which is the object of the statistical nature. Hence, its application gives the true result only for the large number of uniform scenarios, after the averaging\footnote{The measuring of spectra in not the exclusion because it requires the statistical sampling as well.}. When the quantity of particles in the considered system grows this large number of scenarios for the gathering of the statistics will grow as the exponential of the number of particles. However, the sense of the simulation is in that the result comes in the limited time. This is why in the complex cases we have to operate not with the reliable statistic sampling, but, factually, with the unique events, which give the visible scenario for the complex system. It means that the role of random factor, which we successfully factor out for simple systems (for example, in the scattering of one particle on the other), dramatically grows. 

We see, that the standard quantum mechanics contains the serious limitation in the time: {\bf the considered time frame must be sufficient for the revealing of the statistical character of all elementary processes in the complex system, that with the account of entangled states, makes the direct application of quantum theory to complex systems impossible.}

Do we thus owe to refuse from quantum physics in the studying of such objects as the complex chemistry or biology? Of course, the answer is not. Moreover, I am deeply convinced that such {\bf practices of the building of models not based on quantum theory is wrong,} and it cannot give us the systematic advances, despite of separate interesting guesses. 

The most important in quantum theory (besides the superposition principle) is the many particle Hilbert formalism of tensor products, which describes the states of many body systems. Just the constructive reduction of this formalism, which we discussed in the previous section gives the key for the building of PCN. The bonds forming corteges of samples for the scenario models hid the fundamental physical essence, which we yet do not completely recognize, because its manifestations becomes valuable at the level of complex systems only. 

The random factor is inbuilt in the standard quantum theory through the statistical nature of the wave function $\Psi$ itself, when its squared module is treated as the probability density of finding the particle in this point of configuration space. The constructive description of quantum dynamics of one particle gives to the random factor the form of the so called "pre-quantum fields fluctuation" (\cite{Kh}), if by pre-quantum fields we mean the cause forcing the particle to behave itself as the swarm. This constructivism gives not big because the source of randomness remains hidden from us. In the reality, the cause of this randomness can be so deep that we have no chance to understand it completely, but we do not owe to do it. If we observe the photo detector and wait for the click from a photon flied to it, practically, only the density of such clicks makes sense but not a concrete instant of a click. If only this instant is important (it can be in living things) the all system of emission and detecting of a photon is substantially involved in the process, and the problem becomes radically different in comparison with the waiting near the photo detector. 

If we turn to the notion of amplitude grain and apply the method of collective behavior, we can go farer along the way of understanding of the random factor in quantum mechanics: it turns to be connected with the forming of the long corteges of the samples of real particles. The fate of a separate cortege can depend on the behavior of one its member, e.g., potentially - of the individuality of each sample. It is impossible to analyze it by means of standard quantum theory: the probability density $|\Psi(r_1,r_2,\ldots,r_n)|^2$ is exponentially small, e.g., from the viewpoint of the constructivism is so small that simply does not exist: fro such ensemble the other algorithm must be applied. Here the border lies of the applicability of standard quantum theory: this border lies in the area of systems which total number of particles is bigger than 5 (for more numbers the direct method of wave functions is not applicable even on supercomputers). However, we know from the quantum computer theory that the little quantities are not the obstacle for the existence of the observable effects. This is the good injunction to the development of PCN from the traditional quantum mechanics. 
\section{About the individuality of elementary particles}

The natural question arises: is it possible to discover effects confirming the fruitfulness of this way, which we call constructive physics, and which leads to PCN idea on the modern experimental devices?\footnote{The negative results like the absolute model of decoherence does not belong to this row. Likewise, it does not touch the creating of the effective computer models. We take mean just the new effects, which (supposedly?) could extra support the idea of PCN among physicists.} One from such possibilities is the detecting of the individuality of elementary particles. It requires the sequence of the repeating experiments over the same elementary particle, for example, over one proton of atom. In the standard quantum theory, there is no individuality of elementary particles. At the same time there is no difference how many atoms do we use in experiments: one or many. The possibilities of experiments on the individual quantum particles appeared only recently, with the creation of tunnel microscope. We can compare statistics gathered for the different individual atoms, trying to detect something that may be characterized as the individual memory. Introduction of the individuality for elementary particle can be fruitful for the building of computer models, but we can try to detect this individuality directly, because the modern devices allow it. We must keep in mind that the manifestations of such individuality may be very rare, and it brings the question about the reliability of statistical conclusions. The rear disagreement with the typical statistics, which we will certainly obtain can fit into the supposition about the identity of atoms, if we apply to these statistics, for example, Person criterion $X ^2$. Hence, the more promising way must be the sequences of experiments on the more complex systems like molecules with the models of collective behavior that would explain the results by means of the individuality of atoms.

\subsection{Pierson criterion of agreement}

We give the short reminding about the main method of statistical processing of the results of sequential experiments, Pierson criterion. This criterion i also called $X^2$ criterion. 

Factually, the method of collective behavior includes not only the possibilities of elementary particle individuality, but the potential individuality of their samples. The random variable $X^2$ is defined as the sum of $n$ squared independent standard normal variables, divided to $n$, where $r=n-1$ is the number of degrees of freedom:
$$
X^2=\frac{\xi_1^2+\xi_2^2+\ldots+\xi_n^2}{n}.
$$
The distribution of this variable is well known. 

Let we be given a sampling $\bar x$ from the values of some random variable $\eta_0$. We want to determine whether the variable $\eta_0$ has some known probability density $p_{\eta}(x)$ or not. This hypothesis is called the hypothesis of agreement. We note that we know nothing about the variable $\eta_0$ but the sampling of its values $\bar x$. We must determine how this sampling agrees with the known distribution of the standard variable $\eta$. 

Pierson criterion permits to check the hypothesis of agreement. This criterion has numerous applications in the different areas. It makes possible not only to check the suppositions about the values of one or another variable (there are more simple criteria for it, like the confidence interval method, the most plausibility method), but about the agreement with the experiments and the theoretically found data.  

To apply Pierson criterion we should go the following. At first we divide the possible area of values of $x$ to $r$ intervals and enumerate them by the letter $k$. Then for each interval $\Delta_k$ we compute the probability to find the variable $\eta$ in this interval: $p_k=\int_{\Delta_k}p_{\eta}(x)dx$. Then we divide the elements of the sampling to the groups so that any group $k$ consists of elements occurring in the interval $k$, and let  
$$\tilde p_k=\frac{g_k}{G}
$$
be the relative number of the elements in the sampling belonging to the interval $k$. We choose the significance level $\a >0$, and find the corresponding threshold value $t_{r,\a }$ for $X^2$ distribution. At last we compose the check sum 
$$
S_{\bar x,\a,\bar\Delta}=\sum\limits_{k=1}^r\frac{(p_k-\tilde p_k)^2}{np_kq_k}.
$$
After that we compare the check sum $S$ with the level $t_{r,\a}$. If it turns to be less than the level, the hypothesis is accepted, if more - it is rejected. 

How can we apply Pierson criterion to our problem of the verification of the atomic individuality? We suppose that we try to find the dependence of the probability of excited state of atom from the longitude $t$ of laser impulse aimed to this atom. This dependence is given by the formula 
$P=\frac{|\eta |^2}{2\w^2}(1-cos\ 2\w )$, where $\eta=F_{mn}/h$, $F_{mn}$ is the matrix element of the excitation created by the laser, $\w$ is its frequency. (see \cite{LL}, page 176). The value $F_{mn}=\int \Psi_m^*(r)F\Psi_n(r)dr$ factually depends on the stationary wave functions which exact computation requires the application of the electro dynamical reaction of scattering of a photon on the atom. Rules of QED contain the renormalizations of a mass and a charge that depend on the environment in which the atom is placed, in particular, from the photon state that we do not know. If we suppose that the different atoms are in the different external conditions, the renormalizations will be slightly different for them, and it can be the source of individuality of atoms. This kind of individuality reveals in that the values of $p$ for the different atoms will be slightly different and we could try to detect this difference by Pierson method gathering the sampling of the measurements of longitudes of the excited states for each separate atom.  

In the bulk of experiments on quantum physics the individuality of the sampling was out of the focus, moreover, the experiments typically presume the common return from the large ensembles of atoms that certainly grades the individuality. The possibility to address directly to a separate atom many times arose relatively not far ago, with the invention of tunneling microscopes. It would be interesting to use this possibility for the gathering of the individual statistics along the lines we have described.

The principal objection could be looks as follows. We discussed the renormalization that are very small and lie besides the framework of accuracy of the initial experiment about the probability of the excited state. This objection is right only if we ignore the entangled states. However, these states arise in the simultaneous consideration of electrons of the atom, its nucleus, and all photons as the elements of the integral system. We saw that the entangled states arise in the fundamental interactions of QED, and thus we cannot get rid of them by means of the local tricks like the renormalization, or changing of the basis. The existence of entangled states can radically change the picture of standard approximations. It is possible due to the phenomenon of a quantum computer, which is described in details in the Appendix. The essence of this phenomenon is that there are conditions when very small amplitudes of extremely rare events are added constructively giving observable results, which would not be if all states are non entangled. The standard methods of approximations accordingly to which we omit high degree diagrams, etc. rest just on the ignoring of the entanglement. The ideology of a quantum computer says that the QC-like phenomena are rare but reliable. If we translate it to the language of real systems, we must conclude that the rare events are possible when the picture of standard approximations does not work, and it can be detected as the individuality of elementary particles, or the individuality of atoms. 

Of course, the way we propose does not guarantee any certain success. Statistical criteria for the simple experiments are not exhaustive. We consider the following abstract example. There is the team of monkeys, printing texts on keyboards, and their manager, who has no poetical gift but is able to distinguish the poetry from the abracadabra. This team works strongly and gives the result to the expert, who must determine who the author of the text is: Pushkin, who wrote the little known verse, or the team of monkeys. This is the situation when the expert can err. If the individuality of atoms is very rare phenomenon, we can turn in the position of such expert. Statistical methods cannot thus give us the exhaustive answer to the question about the individuality of atoms. 
атомов.

There is one more approach to the problem of individuality, which concerns the using of the time. We can investigate processes supposing that the accumulation of the information of the history of atoms takes place. Such experiments on molecules (\cite{M}) show the possible presence of the so called molecular memory, e.g., the memory about the reactions in which a given molecule participated. The molecular memory can be connected with the long living quantum states (for example, states of nuclear spins which can live hours), but not with the individuality of elementary particles, we must distinguish these two notions. 

\subsection{Individuality of samples of particles}

The method of collective behavior deals with the samples of the elementary particles, which can be the carriers of the individuality in the sense like the presence of the genes in each cell of a living thing that store the information about its individuality. The ascribing of the individuality to the samples is, in my opinion the right way in the method of collective behavior for the complex systems. \footnote{De Broil, who represented a quantum particle as a point located on the wave, used the analogous trick.}

We consider how it can looks in PCN. For this, we suppose that each sample of the same real particle contains the information about the form of dynamical scenario of its evolution in the time. Since the objects of our model are not the samples, but the corteges consisting of samples, we must assume that the information about the dynamical scenario is the attribute of the cortege of samples. Each cortege thus looks like the DNA molecule, which stores the information about the project of the whole organism. A cortege in the biological terms is a genome. Each sample in the cortege is similar to the certain gene, which is in charge of the behavior of the corresponding real particle. We can enumerate this behavior, corresponding to the gene $a$, by $S_a$. The list $S_a$ thus contains all parameters completely determining the concrete scatterings in which the sample $a$ participates. 
 
A dynamical scenario consists of the sequential acts of elementary scatterings, in each of which only the small number of real particles (2-3) participate. At each step of the scenario in the substantial interaction, only a few samples of the same cortege participate. It concerns also the diffusion mechanism of impulse exchange, only here in the elementary scattering participate samples lying in the different but spatially close corteges (we mean as usual, the closeness in the configuration space of many particles). If we want to determine the concrete gene, which is in charge of one or another turn in the evolutionary scenario, we could do it only by the comparison of this evolution with the different possibilities, using quantum state selection. These samples we call critical. On the picture, we show schematically the critical samples, which presence in the cortege determines the scenario of the evolution corresponding to this cortege. 

\begin{figure}
\centering
\caption{Critical samples of the model}
\vspace{150mm}
\makebox[230mm][l]{\includegraphics{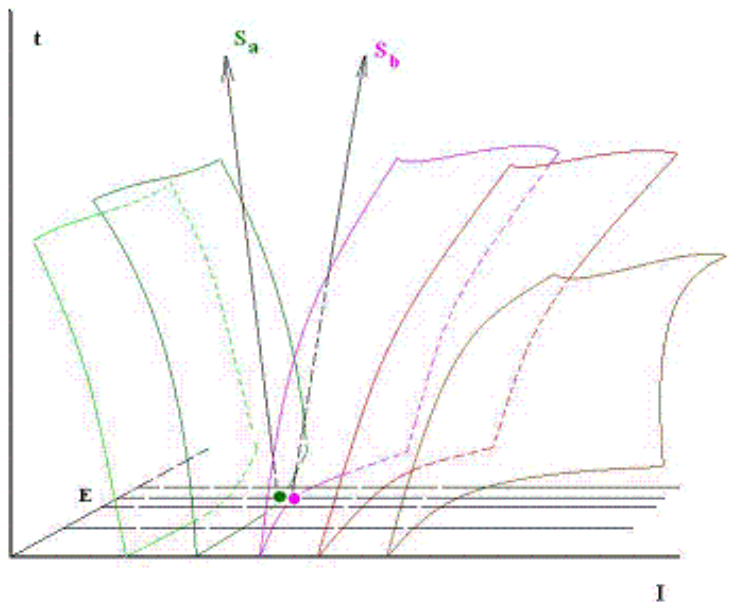}}%
\end{figure}

The destination of critical samples is, of course, the question of the convenience of the simulation, and belongs to the area of programming. Here the tools of the simulation is created, which depends on the PCN interface, this function can be placed to a user of the model. Nevertheless, this destination is not pure arbitrariness; moreover, the destination of critical samples is, probably, the serious question that touches the reliability of this model at all. For example, if we consider a system with the semi classical behavior, the critical will be the sample of the particle, which behaves as classical particle would behave. Each gene - sample represents the certain type of behavior of the real particle, and in the conditions of the severe economy of the memory of the simulating computer, each gene will represent the substantial type of behavior influencing to the whole scenario. In the other words, the majority of genes must be critical in one or the other instant of the administrative time. Using the analogy from the biology, we can suppose that the more right way to determine the scenario will be the sequence of the choices of the concrete genes active in the corresponding time instants $t_0, t_1,\ldots$.

The process of exchange of samples in the selection of quantum states corresponds to the biological process of the exchange of the genetic material, necessary for the evolution of living things. We have the analogy with the biological processes that was mentioned in the numerous works. PCN must permit us to use this analogy in the practical aims: for the debugging of the models of the processes on different levels, up to the fundamental interactions. It is doubtless that this width of the scope is necessary for the realistic models of biological objects. At the next picture, we represent the model of complex evolution consisting of many scattering acts. 

\begin{figure}
\centering
\caption{Evolution determined by the critical genes.
Blue color mark he change of cortege in the selection.}
\vspace{150mm}
\makebox[230mm][l]{\includegraphics{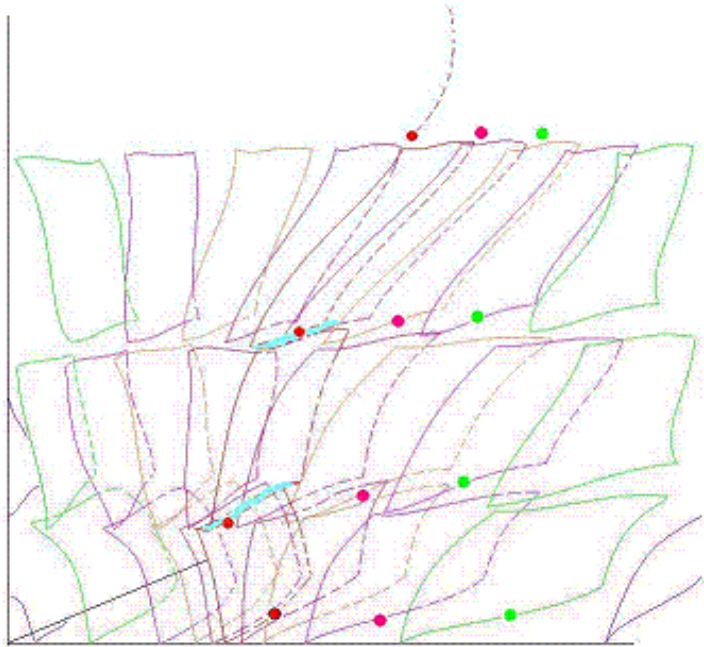}}%
\end{figure}

We consider the simple example, which kind of individuality of elementary particles we can treat. Let we be given a set of abstract atoms of some types $A,B,C,\ldots$, and we are interested in their association on the molecule of the form of the chain:
$$
C_1 -C_2-C_3-\ldots -C_k-C_{k+1}-\ldots
$$
where the type of each atom $C_j$ is determined in advance. Since atoms of the type $A$ can have the different spatial positions in this chain and their neighbors are atoms of the different types, we can for the determinacy consider the situation when the types of atoms will be distributed as:
\begin{equation}
B-A-C-\ldots -D-A-E-\ldots
\label{occur}
\end{equation}
We suppose that the equilibrium position of atoms in this chain, which is determined by the covalent electron bonds between the neighboring atoms is such that the distance $B-A$ equals $r_{B-A}$ and the distance $D-A$ equals $r_{D-A}$. We can also suppose that for the classical potential for the first and the second occurrences of atoms of the type $A$ in \ref{occur}, quadratic potential holes will differ not only by their centers coordinates $r_{B-A}\neq r_{D-A}$, but also by their steepness’s. We ignore the influence of the right neighbor. The selection of quantum states then makes us to accept the different properties of two atoms of the type $A$, because they will be associated with their neighbors at the different distances. 

We can yet complicate our example, if instead of $B$ and $D$ we consider not atoms but molecules with the complex structure. In the selection it will be then important in what point of the space the sample of atom $A$ is located when its left neighbor is in the fixed position. It means that the different samples of the same atom of the type $A$ will mainly associate in the corteges at the different distances from their left neighbors. The individuality must be then ascribed not to the real atoms but rather to their samples. The individuality of atoms shows through the individuality of their samples. The situation is analogous to the genetics where (we simplify it) only genes possessing the certain autonomy are inherited, whereas the genome consists of its set.

The autonomy of samples of elementary particles is convenient fro the swarm representation of QED. The emission of a photon sample by the sample of particle can depend on the individuality of the last sample. We could assume, for example, that each sample of the charged particle has its own stock of virtual photon samples, which this sample can emit up to the instant when this stock exhausts. The photon stock increases through the absorption of photon samples. The notion of the individuality of samples can turn fruitful just in QED. The reason is that QED is not the accomplished theory. The lack of the universal way to overcome the divergence of rows for amplitude can testify to the productivity of ascribing the individuality to the samples of charged particles and photons. In particular, this individuality can determine the renormalization of the charge and the mass of particles, because this renormalization depends on the environment of them.
 
The individuality of samples opens for us new possibilities in the using of PCN when it will be admissible to treat all evolutions as preprogrammed. This hides the serious danger. This supposition when we apply it ultimately makes PCN completely individualized, which means the loss of objectivity. Fortunately, there is the antidote for this danger, and it is in the basement of constructivism. I mean the idea of pluralism, inbuilt to in. Any user is free to ascribe objects from PCN by any individual properties admitted by its signature, in particular, by the individuality of the elementary particle samples. This possibility is important for the future development of the natural sciences, especially for physics. However, one must keep in mind the unavoidable selection that touches not only scenarios of quantum evolution we discussed, but any other scenarios realizable with PCN. Therefore, we should not fear that the physics looses its objectiveness. 

The new possibilities coming with PCN are much more important. It represents the new level of the scientific programming. These possibilities deserve that we take up this project despite of big uncertainty.

\section{Conclusion}

We familiarized ourselves with ideas joint by the common name of constructive physics. These ideas do not form the separate discipline, as the constructive mathematics is not some separate science. Constructivism is the direction, which aroused in mathematics, namely, in its foundations, and in its development absorbs physics. This process is unavoidable and wholesome. I will be glad if this book helps a reader to form more definite attitude to the constructivism, even more if it excites the desire to take up the development of this direction immediately. I permit myself to enumerate some problems which solution seems to me accessible in the framework of constructivism right now. 

The first of these problems is the creation of the effective heuristic for the traditional quantum mechanics, which would make possible to traslate it to the constructive language completely. The basement for this heuristic may be the method of collective behavior, we studied above, but the other ways are also possible. Here the simplest success criterion may be the right description of stationary states of electrons in the hydrogen atom. 

The second problem connected with the first is to create the effective heuristic for quantum electrodynamics, which will not rest to the tricks with vague status, like the summing of asymptotic rows, and would give for the atomic collisions the simple criterion of the forming of chemical bonds and their break. As the intermediate problems, we can pose - the creation of the qubit formalism for QED, common with the qubit formalism for the ordinary quantum mechanics. We saw that the direct method does not work here, and the heuristic of collective behavior we proposed in the chapter 5 was connected partially with this. Perhaps, there are some unknown elements, which make possible to build the qubit formalism for QED; the heuristic of nets could be then simplified, and its program realization concretized. These unknown elements can be connected with the more certain description of photons than we used.

The third problem is to create the prototype of PCN for the description of systems with $10^{12}$ atoms of one - two types with the probe rule of the forming of the net. This program we could compare with the existing simulators of many particles without entanglement. 

The solution of all these problems is interconnected. I stress that in any case the solution of each of these problems presupposes the essential coming out of the standard formalism of quantum theory. The success criterion is the heuristic giving effective algorithms in the standard cases (electronic states in atoms, the one particle dynamics, etc.). The solution of these problems will be the serious step on the way of quantum constructivism, and correspondingly, will facilitate the building of instruments for the real progress in future. 

At last, I underline again the importance of the new for physics feature of constructive mathematical apparatus - the existence of pluralism in the logical estimations. Following along the line of constructivism in physics, we must get into the way of this pluralism. It is inbuilt in the nature of constructive mathematical logic, and thus is unavoidable. This pluralism has the various manifestations not immediately concerning to the logical aspects. For example, it is well known that this feature arises everywhere in the applications of algorithms and programming, right up to the various program realizations of the same constructive heuristic. In the world of classical mathematics, where the main instrument is formulas, there is no such pluralism. 

We would never leave the cozy and deterministic world of formulas, if it is effective for our survival, but the facts show that its effectiveness is almost exhausted. We have to enter into the world of constructivism, to the way with many forks, where it is necessary to address to the expertise based on the visual images, and the thing of this sort, which traditional mathematics took out of the physical theories. Alas, this is the inevitable payment for the effectiveness, which we need. Taking this path, we must hope that the legality of constructivism in the science gained by our predecessors will save us from the washing out of the subject of our science itself. This legality gives us the clear understanding what an algorithm is and how human evaluations and arbitrariness are connected with the exact execution of commands by a computer. Nobody has been let down by this understanding of the legality, and we have no reason to doubt that constructivism will serve us in physics so reliably, as classical mathematics did. Only following this way we can reach the border, which separates our subject from the world of things called living.

\newpage 
\appendix{{\LARGE {\bf Appendix. Theory of quantum computations}}}

\section{Formal definition of quantum algorithms}

In this section, we give the more formal version of the definition of quantum computer. We define a quantum computer with oracle for some word function, preserving the length of the input word. We mention that the exactly analogous definition can be given for the oracle of the ordinary Boolean function. We describe sequentially the both parts of the quantum computer. 
\nn

{\bf Quantum part}
\n

It consists of two potentially infinite types: the working and the questionnaire, the finite list 
${\cal U}$ of unitary transformations, which we treat as easily realizable on standard physical devices, and the infinite list
$F=\bigcup\limits_{n=1}^\infty
F_n$ of unitary operators called the oracle of the preserving length function 
$f:\ \{ 0,1\}^* {\ar}\{ 0,1\}^*$, where each $F_n$ acts on $2^{2n}$ - dimensional Hilbert space generated by vectors $\{ 0,1\}^{2n}$ by the following way:
$$
F_n | \bar a,\bar b \rangle = | \bar a ,f(\bar a) \bigoplus \bar b \rangle,\ \ \
\bar a ,\bar b \in\{ 0,1\}^n,
$$
where $\bigoplus$ means the component-wise addition modulo 2.

Cell of the tape we call qubits. Each qubit takes values from the complex circle of radius 1: $\{ z_0 \mbox{{\bf 0}} +z_1 \mbox{{\bf 1}} \
| \ z_1 ,z_2 \in \mbox{{\tt C}}, |z_0 |^2+|z_1 |^2
 =1\}$. Here {\bf 0} and {\bf 1} is basic states of the qubit, forming the basis of $\mbox{{\tt C}}^2$.

In the course of the whole computation the both tapes are limited by two markers each, which occupy the constant positions so that on the working (query) tape only qubits $v_1 ,v_2 ,\cdots ,v_\tau$ ($v_{\tau+1} ,v_{\tau+2} ,\cdots ,v_{\tau+2n}$)
are accessible in the computation with the time complexity $\tau=\tau(n)$
on the input word of the length $n$. We put $Q=\{ v_1 ,v_2 ,\cdots ,v_{\tau+2n} \}$.
A basic state of quantum part is a function of the form $e:\ Q{\ar}\{ 0,1\}$.
Such state we can encode as $|e(v_1 ) ,e(v_2 ) ,\cdots ,e(v_{\tau+2n})
\rangle$ and naturally identify with the corresponding word in the alphabet $\{ 0,1\}$. Let $K=2^{\tau+2n}$; $\ e_0 ,e_1 ,\cdots ,e_{K-1}$ be all basic states, taken in some fixed order, ${\cal H}$ is $K$ dimension Hilbert space with the orthonormal basis $e_0 ,e_1 ,\cdots ,e_{K-1}$. We can treat ${\cal H}$ as the tensor product 
\newline ${{\cal H}}_1 \bigotimes {{\cal H}}_2 \bigotimes
\cdots \bigotimes{{\cal H}}_{\tau+2n}$ of two dimension spaces, where ${{\cal H}}_i$ is generated by the all values $v_i ,\ i=1,2,\cdots ,\tau+2n$.
A state (pure) of the quantum part is such an element $x\in{\cal H}$, that $\| x\| =1$.

Two types of unitary operators on its state determine the evolution of the considered quantum part in the time: working and query. Let a pair $G,U$ be chosen randomly, so that $G\subset\{ 1,2,\cdots ,\tau+2n\}$,
$U\in{\cal U}$ is the unitary operator in $2^{\card (G)}$ dimension Hilbert space.

{\it Working transformation} $W_{G,U}$ on ${\cal H}$ has the form $E\bigotimes U'$,
where $U'$ acts as $U$ on $\bigotimes\limits_{i\in G} {{\cal H}}_i$ in the considered basis, $E$ acts as the identical operator on $\bigotimes\limits_{i\notin G} {{\cal H}}_i$.

{\it Query transformation} $\Qu _f$ on ${\cal H}$ has the form $E\bigotimes F'_n$,
where $F'_n$ acts as $F_n$ on $\bigotimes\limits_{i=\tau+1}^{\tau+2n} {{\cal H}}_i$
and $E$ acts as the identical on $\bigotimes\limits_{i=1}^\tau {{\cal H}}_i$.

{\it Measurement} of the quantum part. If the quantum part is in the state $\chi =\sum\limits_{i=0}^{K-1} \la_i e_i ,$ the measurement is the random variable taking a value $e_i$ with the probability $|\la _i |^2$.

{\bf Classical part}

The classical part of the computer also consists of two tapes: working and query, and the cells of them is in one-to one correspondence with the qubits of quantum tapes of the computer, and have the limiting markers on the corresponding positions. 
Each cell of classical tapes contains a letter from some finite alphabet $\w$.
The classical Turing machine $M$ with a few heads on both tapes and the set of joint states of heads:
$\{ q_b ,q_w ,q_q ,q_o ,\cdots\}$ determines the evolution of the classical part. We denote by $h(C)$ the joint state of heads for the state $C$ of the classical part.

Let $D$ be the set of all states of the classical part.

{\it Rule of correspondence} between the quantum and classical parts has the form
$R:\ D{\ar} 2^{\{ 1,2,\cdots ,\tau+2n\} } \times\cal U$, where $\forall C\in D$
$R(C)=\langle G,U\rangle$, $U$ acts on $2^{\card (G)}$ dimension Hilbert space so that $U$ depends only on $h(C)$, and
elements of $G$ are exactly the numbers of cells in the classical part, containing the special letter $a_0 \in\w$.

A state of quantum computer is a pair $S=\langle Q(S),C(S)\rangle$
where $Q(S)$ and $C(S)$ are states of the quantum and classical parts respectively.

{\it Computation} on quantum computer. We call a computation a sequence of transformations of the following form:
\begin{equation}
S_0 {\ar} S_1 {\ar} \cdots {\ar} S_\tau ,
\label{1.1}
\end{equation}
where fro each $i=0,1,\cdots ,\tau-1$ $C(S_i ){\ar} C(S_{i+1} )$ is the passage determined by Turing machine M, where the following properties take place:

if $h(C(S_i ))=q_w$ then $Q(S_{i+1} )=W_{R(C(S_i ))} (Q(S_i ))$,

if $h(C(S_i ))=q_q$ then $Q(S_{i+1} )=\Qu_f (Q(S_i ))$,

if $h(C(S_i ))=q_b$ then $i=0$, $Q(S_0 )=e_0 ,\ C(S_0 )$ is the initial condition corresponding to the input word $a\in \{ 0,1\}^n$,

if $h(C(S_i ))=q_o$ then $i=\tau$,

in all other cases $Q(S_{i+1} )=Q(S_i )$.

We say that this quantum computer (QC) computes a function $F(a)$
with the probability $p\geq 2/3$, if for the computation (1.1) on any input word $a$ the measurement of $S_\tau$ with the following fixed routine procedure of the processing of the result gives 
 $F(a)$ with the probability $p$. We note that for $p<1$ we always can reach any more value of the probability $p_0 >p$ if we repeat the computations with the same input word and take as the final result the prevailing result in these sequence of computations. It leads to the at most linear delay of computations in comparison with the former level of probability $p$. 
These computations are the computations with the fixed error probability. In the case $p=1$ we have the exact computations.

If we use an oracle in computations, by the main complexity measure of the computation (1.1) we take not $\tau$, but the total number of the query transformations in this computation. The choice of Turing machine as the model of the classical part evolution thus becomes not principal. 

Instead of it, we can choose any other model of classical computations: cellular automata, normal algorithms, etc., the complexity of computations will not change.

\section{Why QC makes search surprisingly fast}

In this section we consider the famous quantum algorithm for the searching of a solution of the equation $f(x)=1$, proposed by L.Grover (see \cite{Gr}). This algorithm finds the solution in the time of the order of square root of the classical time, e.g., for the obtaining of the solution we need $O(\sqrt{N})$ queries to the oracle corresponding to $f$. This algorithm has the large number of potential applications to the computational problems, since all tasks with the bruit force are reducible to this problem. Indeed, let us suppose that we have to find some row with some easily verifiable property $P$. Despite of the simplicity to check this property for a given row, it is not easy to find the raw with such property. For this, we have to search the large number of rows sequentially checking each. It can be proved (I will not trouble the reader) that in the reasonable specification of formulations there is no other way of search but the bruit force. This property $P$ can be represented as the Boolean function $f$ such that exactly the solutions of the equation $f(x)=1$ will satisfy this property. Moreover, if $P$ is given in the form of some algorithm, we can represent $f$ in the form of scheme of functional elements (gate array), realizing the oracle corresponding to $f$. For the building of this algorithm GSA (Grover search algorithm), we need two important subroutines. 

\section{Grover algorithm}

We consider the problem of finding of a solution of equation 
\begin{equation}
f(x)=1
\end{equation}
for the function $f:\ \{ 0,1\}^n\ar\{ 0,1\}$. Let, for the simplicity, this solution de the single, we denote it by $x_t$. We consider Hilbert space of quantum states of $n$ qubits generated by the basis of elements $|j\rangle ,\ j\in\{ 0,1\}^n$, e.g., $j=0,1,\ldots,N-1,\ N=2^n$. We add to it one qubit, called ancilla (ancillary register) which is initialized by the state of the form $|\phi=\frac{|0\rangle -|1\rangle}{\sqrt{2}}$. We then have:
\begin{equation}
\begin{array}{lll}
&Qu_f|x_t\rangle\bigotimes |\phi\rangle &=-|x_t\rangle\bigotimes |\phi\rangle,\\
&Qu_f|j\rangle\bigotimes |\phi\rangle &=|j\rangle\bigotimes |\phi\rangle,\ j\neq x_t
\end{array}
\end{equation}
We note that the state of ancilla remains unchanged; it is in the beginning and in the end the tensor multiplier of the common state. 
We define for the arbitrary vector $|a\rangle$ of our Hilbert space the operation of reflection of all space along this vector, acting as:
\begin{equation}
I_a|b\rangle = \left\{\begin{array}{lll}
&|b\rangle,\ &if\ \ \langle a|b\rangle=0,\\
&-|a\rangle,\ &if\ \ a=b.
\end{array}
\right.
\label{inver}
\end{equation}
We then can write the equality $Qu_f=I_{x_t}$, where we use the limitation of the action of $Qu_f$ to vectors of the form $|a\rangle\bigotimes |\phi\rangle$. 

In addition, we can, using ancilla, consisting of $n+1$ qubits, realize the reflection along the vector of unit length $|0\rangle$.

\subsection{Walsh - Hadamard transform}

We consider the important example of unitary transformation - Walsh-Hadamard transform. Its one qubit variant looks as:
$$
W=\left(
\begin{array}{lll}
&\frac{1}{\sqrt{2}}\ &\frac{1}{\sqrt{2}}\\
&\frac{1}{\sqrt{2}}\  &-\frac{1}{\sqrt{2}} 
\end{array}
\right) .
$$
$n$ qubit variant we define as the independent action of $W$ on each qubit: $W^{\bigotimes n}$. It is evident from this definition that the aplitude of passage from one basic state to the other is the same and the sign changes so many times as many units stands on the same places in the both these states. We can write the general formula of this transform as $w_{i,j}=(-1)^{(i\cdot j)}$, where in the exponent stands the scalar product of binary notations of the numbers $i$ and $j$. It means that if numbers $i$ and $j$ have the form $\sum\limits_{s=0}^{n-1} i_s 2^s$ and $ \sum\limits_{s=0}^{n-1} j_s 2^s$ correspondingly, then $i\cdot j=\sum\limits_{s=0}^{n-1} i_s j_s$. It is easy to understand, that $(W^n)^{-1} =W^n$. 

This transform is remarkable because it does not lead to the entangling of qubits. If the initial state of qubits was not entangled, the same will be the result of the transform $W^n$. If we apply this transform to $n$ qubits in the zero state, we obtain the state of the form
\begin{equation}
\phi_0=\frac{1}{\sqrt{N}}\sum\limits_{i=0}^{N-1}|e_i\rangle .
\label{inistate}
\end{equation}
This state is uncommonly by that all its amplitudes are equal. By virtue of chapter 2, the last property uniquely determines the state, because the phase multiplier $e^{i\phi}$ has no physical sense, and we can conclude that it is the single state with this property.

\subsection{Operation of reflection and its realization on a quantum computer}

The unitary operator of reflection along the state $a$ is defined by the formula \ref{inver}.
Such definition is formally incomplete, but in view of linearity of unitary operators, it is defined by this equation uniquely. Representing the situation geometrically, we understand that this operator is the regular reflection of all the space relatively the subspace orthogonal to the vector $a$ (picture 3). How can we realize this operator on a quantum computer?

At first we take up the case when the vector $a$ is known. For the simplicity we set $a=0$ (cases of the other basic states differ from this not essentially). The simplest idea: to build the algorithm so that it looks sequentially at all cells of $a$ and searches units, and if it find one, it does nothing, but if it finds no units, e.g. $a=0$, it changes the sign of the state. The sequential search is easy to organize for a classical computer. In the quantum case there is the little difficulty, because there can be several units, and we must somehow fix it in the passage through the number $a$ in order to obtain the sequence of only unitary operators. We propose the following plan. We organize the ancillary qubit called $res$, which will signal about the presence of a unit in the number $a$. In addition for the ensuring of the unitarity, we set up for each valuable qubit its double. All these ancillary qubits will be initially in the state $0$. There will be two passages: at the first, we establish does $a$ coincide with the zero string, at the second we move in the opposite direction and fulfill the reverse operators that restores the contents of all ancillary qubits, e.g., we make the ancilla zero again. The last operation is necessary to make this procedure many times. This dustbin cleaning has the other sense in the quantum computing that has no analogues in the classical case. The point is that the addition qubits are in the entangled states with the valuable, which means that, for example, their measurement can cause change of valuable qubits and break the computation. 

We will do at each step of this passage one fixed operation $V$ on three qubits: the main, its double, and $res$. The special pointers will be point to these qubits. After that, we shift the pointers to the main qubit and the double to the step to right and all repeats (see the picture 4). After the first passage the contents of $res$ will indicate: has we met at least one unit or not. We then change a sign provided the state contains only zeroes, and at last fulfill all the reverse operators in the reverse order to clean the "dust". Now all is reduced to the definition of the operator $V$. It must change the contents of $res$ if it meets the unit first time, and it must be unitary. Here we need the qubits-doubles. It is easily to check that $V$ must act like this: 
$$
\begin{array}{lll}
&0\ 0\ 0\ \ &\ar\ 0\ 0\ 0\ \\
&1\ 0\ 0\ \ &\ar\ 1\ 0\ 1\ \\
&0\ 0\ 1\ \ &\ar\ 0\ 0\ 1\ \\
&1\ 0\ 1\ \ &\ar\ 1\ 1\ 1\ 
\end{array}
$$
On any other basic states, the action of this operation is extended so that it is one-to-one. The reflection operator relatively zero-state $|\bar 0\rangle$ has the form of the following sequence of the working operators:
$$
V_n V_{n-1}\ldots V_1\ Sign \ V_1^{-1} V_2^{-1}\ldots V_n^{-1},
$$
where $V_j$ denotes the operation $V$ fulfilled in the moment $j$, and $Sign$ has the matrix of the form 
$$
\left(
\begin{array}{lll}
&-1\ &0\\
&0\ &1
\end{array}
\right) .
$$
in the basis $|0\rangle , |1\rangle$.
It results in the transformation from the state $\psi\bigotimes\ \bar 0$ to the state $\psi'\bigotimes\ \bar 0$, where the first multiplier differs from the second only in the sign at the $|e_0\rangle$. 

Now it is easy to build the reflection along the state $\phi_0$, defined by the equality (\ref{inistate}). Using the known definition of Walsh-Hadamard, we obtain the equation: $I_{\phi_0}=W^n I_{\bar 0} W^n$. Consequently, this inversion can be realized on the quantum computer easily. The more general fact takes place, which can be checked straightforwardly. Le vectors $|\bar a\rangle$ and $|\bar b\rangle$ be connected by the relation $|\bar a\rangle=U|\bar b\rangle$. Then the reflections along them will be connected by the equation $I_{\bar a} =UI_{\bar b} U^{-1}$. We can thus fulfill the reflections along any vectors if we know how to find them from the zero-vector. For example, we can choose randomly the operator $U$, and we are then able to fulfill the reflection along any randomly chosen vector. We note that up to now all the state vectors along which we fulfilled the reflections were known to us in the sense that we had the methods of obtaining the corresponding state in the quantum memory. 

Is it possible to fulfill the reflection along the unknown vector? A classical computer could do the reflection only along the known vector. For q quantum computer it is not necessary to find this vector, it is sufficient to know that it is the solution of the equation $f(x)=1$, and to have the oracle for $f$. At first, we describe the not the best way to do it. For the simplicity at first we restrict ourselves by the case when there is only one solution of the equation $f(x)=1$. We denote this solution (and the corresponding basic state) by $tar$. Our aim is to build the subroutine realizing $I_{tar}$. We apply the quantum oracle corresponding to $f$, which has the form (\ref{oracle}). Then we change the sign depending on whether $b$ equals unit, or not, and apply the oracle once more. It is easy to understand that the last action cleans the contents of $b$, because th esing changes just as we want. What is the drawback of this method? The point is that the oracle we apply twice. Is it possible to gain the same result applying the oracle only once?

The answer is yes. The ancillary qubit we can initialize in any state, not only in the zero. What is important is that this state restores after the operations. We initialize the qubit corresponding to $b$, in the state $\psi_0 =\frac{1}{\sqrt{2}}(|0\rangle +|1\rangle )$. After the operation $Qu_f$ of the form (\ref{oracle}) we obtain that $1$ is added to the ancilla exactly when $a$ is the target root of the equation. Nevertheless, this change only the sign of ancilla, that follows directly from its form, because the unit and zero exchange places. This method requires only one query to oracle, and thus is the optimal. 

We have met the first miracle of quantum informatics: the possibility to reflect the space along the state vector, which we do not know!

\subsection{GSA}

We are ready now to describe the method of the fast quantum search. This method is surprisingly simple: is concludes in the sequential applications of the transformations $G=-I_{\phi_0}I_{tar}$\footnote{The sign minus in the definition of $G$ stands for the beauty only. }
to the vector $\phi_0$ $\ \ t_0 =\left[\frac{\pi\sqrt{N}}{4}\right]$ times. The resulting state practically equals to the target $tar$. To check it we consider how the operator $G$ acts on the arbitrary vector. Geometrical imagination will help us here. At first it is easy to understand that the operator $G$ transforms the plane generated by vectors $|tar\rangle$ and $\phi_0$ to this plane (it follows from that the both these reflections do not lead us out of this plane). In the second, one reflection changes the orientation; hence, two inversions preserve the orientation. Thus, $G$ is the orthogonal transformation of the plane (unitary but without complex coefficients). Therefore, $G$ is the turn of this plane on some angle. Looking at the picture 5 we concludes that $G$ represents the turn to the sharp angle $\a$, which is formed by vectors $\phi$ and $\tilde 0$, orthogonal to $|tar\rangle$. We have $\a\approx\sin\a=|\langle tar|\phi\rangle |=1/\sqrt{N}$, within the $1/N^{3/2}$. Hence, if we apply $G$ $t_0$ times, with the error of the order $1/N$ the result coincides the target vector $tar$. Now we can obtain the target state by the direct measurement of the state of quantum part of the computer.

\subsection{How to find solutions if there are many of them?}

Up to this moment we considered the case of the single solution of our equation $f(x)=1$. What changes is there are many solutions? At first, we note that if there are very many solutions, for example, if their number is of the order of the total number of all possible values of $x$, the application of a quantum computer fro the search of any of these solutions has no sense because we could simply take $x$ at random. We then suppose that there are not many solutions, for example, that their number is of the order $O(\sqrt{N})$. 

It is possible to apply GSA formally in this case as well. The both reflections in it we can fulfill independently of how many solutions are. However, now the geometrical sense of the reflection $I_{tar}$ will be the different. Let $L_f$ denote the subspace generated by all solutions of the equation $f(x)=1$. Then $I_{tar}$, which is defined as for the case of one solution, will be the regular reflection of the whole space relatively to the subspace orthogonal to $L_f$. Indeed, each solution will change the sign, any linear combination of the solutions will change the sign, and the vector, orthogonal to all solutions remains unchanged. 

We make one interesting observation. As we know, the reflection on a quantum computer we can fulfill not relatively to the state $\phi_0$, but also relatively to any state $\tilde\phi_0$, which we can obtain for the quantum part of the computer. We consider the projection of the initial state $\tilde\phi_0$ on the subspace $L_f$, and now let $tar$ denote the new target state directed along this projection (it differs from the projection itself only in that it has the unit length). We then again, as in the previous case, obtain that the plane spanned by vectors $tar$ and $\tilde\phi_0$ transforms to itself after the application of Grover operator $G$. In our geometrical reasoning nothing changes if we replace $\phi_0$ by $\tilde\phi_0$. The single that we have to correct is the time $t_0$. The analogous reasoning as in the previous case lead us to the conclusion that the time is now $\left[\frac{\pi}{4|\langle tar|\tilde\phi_0\rangle |}\right]$. 

It brings the conclusion. In Grover scheme we can use the reflection relatively to any initial state $\phi_0$, even if it is randomly chosen. Indeed, if we choose it at random, the module of the scalar product $|\langle tar|\tilde\phi_0\rangle |$ will have the order $1/\sqrt{N}$. The square of this value is the probability to obtain the state $tar$ after the measurement of the state $\tilde\phi_0$ in any basis containing our vector $tar$. For reasons of the symmetry after the random choice of the initial state $\tilde\phi_0$ this probability must be equal for any vectors of this basis, and must be then $1/N$. It does not directly imply that the probability to obtain the square root from this value is $1/\sqrt{N}$\ \ \ \ \footnote{The difference between the average square of the value and the squared average value of it equals the dispersion of this value.}. However, it is possible to prove that the average value of the module itself has the same order, namely, $1/\sqrt{N}$ within the multiplication to the constant not depending on $N$. We thus can apply Grover scheme for the fast obtaining of the states provided we have some apparatus realizing the reflection relatively to this state. Here the initial vector (determining the corresponding reflection) we can choose at random.  

It remains unclear when to fulfill the finite measurement? For this we must know $|\langle tar|\tilde\phi_0\rangle |$. Otherwise, we turn in the position of the passenger at the ring road, who does not know where to leave the train. If we know the total number of solutions $l$, we understand that this module equals $\sqrt{l/N}$. What to do is this number is unknown, or we simply want to obtain the target state $tar$ by Grover scheme using the corresponding reflection? Here we meet the case when it is convenient to measure in the course of computations. 

\subsection{When it is convenient to measure frequently}

 Let the total number of solutions be unknown for us. If we stop the process of applications of Grover operators in the random instant, less than $C\sqrt{N}$, where $C$ is sufficiently large, what is the probability to obtain one of target states after the measurement? Looking at the picture 5 we grasp that this probability must not be much smaller than 1/2, since our current state vector and vectors $tar$ and $-tar$ with the approximately equal probabilities form angles bigger and smaller than $\pi/4$. Therefore, if we do not pursue for the fastest method of obtaining the target states, we can simply launch the GSA process and in the approximately $O(\sqrt{N})$ steps obtain the result with the high probability, using the repetitions of attempts. However, here we certainly loss the possibility of the fast answer in the case of many solutions. For example, if $l=O(\sqrt{N})$ the lost profit in the time will be $O(N^{1/4})$. Here the following simple idea helps: to scan the time segment to $C\sqrt{N}$ inclusive, and to choose at each step the time segment twice bigger than on the previous step. Here we will not loose the possibility to find the result quickly for the case of many solutions, we obtain the answer in the time $O(\sqrt{N/L})$ that is proper. It follows from that the time spent almost in vain on the measurements for small time segments up to the moment when the current segment reaches the proper value, is approximately the same as in the large segment, because the sum of a geometrical progression is of the order of its maximal member. 
 
We thus meet the second miracle of quantum informatics: the possibility to find the target state without the direct searching through all possibilities. 
This surprising possibility completely rests on the fundamental property of quantum objects - the amplitude interference and the existence of entangled states. It is easy to show that any quantum computation, which does not use the entanglement, e.g., which operates with independent qubits only, can be straightforwardly realized on the classical computer. At the beginning and at the end of computations along GSA scheme we have non-entangled states, such that $tar$ or $\phi_0$. Nevertheless, in the middle of this iterative procedure the entanglement grows, reaches some limit, and then decreases up to zero. The entanglement in quantum informatics thus represents the peculiar and necessary resource, like the Sun light for living things. The physical nature of this resource is not completely clear, and its understanding is the most interesting open problem of natural sciences.

\section{Quantum Fourier transform}

\subsection{What is common for color vision and integer factoring?}

Each reader certainly knows that the difference in the color of things is connected with the difference in the length (or frequency) of the wave of reflected light. We will not touch the complex and not completely clear mechanism of the photon distinguishing in the human visual analyzer, but give the schematic formulation of this problem based on Fourier transform. 

Let us imagine that we have the method of generation of some harmonic oscillation of the form $e^{2i\pi\ wx}$, where its frequency $\w$ is unknown to us. The problem is to find this frequency. In this statement we can assume that $\w\in [0,1)$. This problem is very general and we will specify it in each case, so that the particular cases will be: the search of eigen values of operators, the structure recognition, and even the factorization of integers. In the general formulation, it can be solved by the single powerful trick invented in XIX century by Fourier: the special integral transformation acting on functions. This transformation and the reverse to it have the form:
\begin{equation}
\begin{array}{lll}
&F\ f=\phi(\la)=&\frac{1}{\sqrt{2\pi}}\int\limits_D e^{-i\la x}f(x)\ dx,\\
&F^{-1}\ \phi=f(x)=&\frac{1}{\sqrt{2\pi}}\int\limits_D e^{i\la x}\phi(\la)\ d\la .
\end{array}
\end{equation}

Let we give to the input of this transformation our harmonic oscillation $e^{2i\pi\ wx}$. There is one small obstacle here, because the transformation is formally defined on all number axes $D=R$, and it requires strictly speaking the application of the general functions. To avoid it we represent $D$ by the large interval on the axes, for example as $-B,B$, where $B$ is the large number. We then have: 
$$
F\ f=\phi(\la)=\frac{1}{\sqrt{2\pi}}\int\limits_D e^{i(2\pi\w-\la )x}f(x)\ dx.
$$
Now we see that if the argument of the resulting function $\la$ is close to $2\pi\w$, then the interval is very large because the integrand is close to unit, in the opposite case this integral is small because the different part of interval will subtract, or as physicists say, interfere destructively. The resulting function will thus have the large peak in the point $\la =2\pi\w$. We now imagine that we somehow realize Fourier transform on a quantum computer. Then in the output state all the mass of amplitude will be concentrated about the number $2\pi\w$ and the measurement of the output state with the high probability gives the value of the unknown parameter $\w$. There is the second idea of the application of a quantum computer. We have only to realize Fourier transform on a quantum computer. 

\subsection{Quantum Fourier transform and its main property}

At last, we try to write quantum version of Fourier transform. By virtue of its linearity, it is sufficient to define its action on basic elements. Just basic elements must play the role of functions $f$ and $\phi$ in the definition. We consider the following definition of quantum Fourier transform. 
\begin{equation}
\QFT :\ |a\rangle\ar\frac{1}{\sqrt{N}}\sum\limits_{b=0}^{N-1}e^{-\frac{2\pi i\ ab}{N}}|b\rangle ,
\ \ \ \ \ \ \ \ 
\QFT ^{-1} :\ |a\rangle\ar\frac{1}{\sqrt{N}}\sum\limits_{b=0}^{N-1}e^{\frac{2\pi i\ ab}{N}}|b\rangle .
\label{qft}
\end{equation}

This definition is built analogously to the standard. We can straightforwardly check (the reader can do it) that this transform maps basic vectors to the mutually orthogonal vectors of the unit length, which implies its unitarity. We then can convince that these formulas really determine the mutually opposite maps. 

We now can demonstrate how this transform can reveal hidden periods. We introduce one notion playing the important role in the quantum computing. This is the conditional application of an operator. The idea consists in the following. Let we are given a unitary operator $U$. For the obviousness, we can assume that it is given in the form of scheme of functional elements, though it is not necessary. Further, let we have some auxiliary quantum register consisting of several qubits, which we call controlling. The aim is to apply the operator $U$ sequentially so many times as is written in the controlling register. We write it formally in the form:
$$
U_{cond} |x,\a \rangle \ar\left\{
\begin{array}{cc}
|U \ x,\a\rangle,\ &\mbox{if} \ \a =1,\\
|x,\a\rangle\ &\mbox
{if} \ \a=0.
\end{array}
\right.
$$
We will not discuss the question of the realization of this transform now (it will be done below). We only note that if the operator $U$ is determined as the scheme of functional elements, we can easily build the scheme of the same type realizing $U_{cond}$. For this it is sufficient to make conditional each operator contained in this scheme. We leave the details for the reader.

We now take up the important problem of the finding of eigen frequency of the operator $U$. This frequency will result from the measurement of some special register from $n$ qubits denoted by $\a$, n which we store the sequential binary figures of this frequency, with the limited accuracy. We suppose that the real frequency can be written in this register with the absolute accuracy, it is not important for the general scheme we describe. The reader who is interested in the general case, we address to the paper \cite{Oz5}. Our computer thus works with two registers: the register of the argument of the operator $U$, and the register of the value of its eigen frequency. The initial state we choose $|\xi ,\bar 0\rangle$, where $\xi=\sum\limits_k x_k \psi_k$, and $\psi_k$ are eigen states of our operator $U$, corresponding to the eigen frequencies $w_k$. 

The central trick for the revealing of the eigen frequencies is the operator, which was introduced by Shor for the particular case where $U$ is the numerical multiplication, and was generalized to the case of arbitrary unitary operators by Abrams and Lloyd. Its definition is the following.
\begin{equation}
\QFT_2\ U_{cond}\ QFT_2.
\end{equation}
Here Fourier transform is applied to the second register. We find what gives this procedure as applied to our initial state. The first Fourier transform gives the uniform amplitude distribution in the second register: $=\frac{1}{\sqrt{N}}\sum\limits_k \sum\limits_{\a=0}^{N-1} x_k |\psi_k ,\a\rangle$. The operator of condition application of $U$, by virtue of that $\psi_k$ are eigenvectors of $U$ gives $U_{cond}|\psi_k,\a\rangle=|U^\a \psi_k,\a\rangle= e^{2i\pi w_k \a}|\psi_k, \a\rangle$, therefore, all the state after the application of the conditional operator transforms to $\frac{1}{\sqrt{N}} \sum\limits_k \sum\limits_\a e^{2i\pi w_k \a}|\psi_k, \a\rangle$. At last, the final application of Fourier transform gives the state:
\begin{equation}
\frac{1}{N} \sum\limits_k \sum\limits_c\sum\limits_{\a=0}^{N-1} e^{2i\pi \a (w_k -\frac{c}{N})}|\psi_k, c\rangle .
\end{equation}
If $c$ is just the list of binary figures of $w_k$, then the exponential degree is zero and we obtain after the summing on $\a$ the sun of units of the total number $N$ so that the coefficient at the state with this $c$ will be $x_k$. It follows from this, due to the normalizing - the sum of squared modules of all $x_k$ is 1, that the amplitude of the basic vectors with the others $c$ equals zero. We can check it straightforwardly: 
$\sum\limits_{\a=0}^{N-1} e^{2i\pi \a \b} =0$ when $\b\neq 0$. Indeed, this is the sum of the geometrical progression with the ratio not equal 1 for which the first summand equals the last summand. 

Our procedure thus results in the state 
$$
\sum\limits_k |\psi_k ,w_k\rangle ,
$$
where by $w_k$ we mean its binary notation. If we thus observe this resulting state in the basis consisting of the eigenvectors of the operator $U$, we obtain as the addition to the eigenvector the binary notation of the corresponding eigen frequency. In particular, if the initial state $\xi$ was eigenvector itself, we simply obtain its frequency. 

\subsection{Realization of QFT on quantum computer}

To apply the described method of the finding of eigen frequencies we need a little: to build the quantum algorithm for quantum Fourier transform (QFT) that we take up now.
We agree to represent an integer of the form $a=a_0 +a_0 2+\ldots +a_{l-1} 2^{l-1}$ as the basic state $|a_0 \ a_1 \ \ldots\ a_{l-1} \ \rangle$ and will place all $a_j$ from the up to down. The same agreement we accept for the output, only the binary figures $b_j$ of the number $b=b_0 +b_0 2+\ldots +b_{l-1} 2^{l-1}$ we then write in the opposite order - from the down to up. 
\begin{picture}(500,315)(0,-35)
\multiput(80,50)(0,50){5}{\line (30,0){270}}
\put(95,50){\circle{10}}
\put(125,100){\circle{10}}
\put(170,150){\circle{10}}
\put(230,200){\circle{10}}
\put(310,250){\circle{10}}
\put(110,50){\line(0,1){50}}
\put(140,50){\line(0,1){100}}
\put(155,100){\line(0,1){50}}
\put(185,50){\line(0,1){150}}
\put(200,100){\line(0,1){100}}
\put(215,150){\line(0,1){50}}
\put(245,50){\line(0,1){200}}
\put(260,100){\line(0,1){150}}
\put(275,150){\line(0,1){100}}
\put(290,200){\line(0,1){50}}
\put(110,50){\circle*{3}}
\put(140,50){\circle*{3}}
\put(185,50){\circle*{3}}
\put(245,50){\circle*{3}}
\put(155,100){\circle*{3}}
\put(200,100){\circle*{3}}
\put(260,100){\circle*{3}}
\put(215,150){\circle*{3}}
\put(275,150){\circle*{3}}
\put(290,200){\circle*{3}}
\put(110,100){\circle*{3}}
\put(140,150){\circle*{3}}
\put(155,150){\circle*{3}}
\put(185,200){\circle*{3}}
\put(200,200){\circle*{3}}
\put(215,200){\circle*{3}}
\put(245,250){\circle*{3}}
\put(260,250){\circle*{3}}
\put(275,250){\circle*{3}}
\put(290,250){\circle*{3}}
\put(60,50){$a_4$}
\put(60,100){$a_3$}
\put(60,150){$a_2$}
\put(60,200){$a_1$}
\put(60,250){$a_0$}
\put(360,50){$b_0$}
\put(360,100){$b_1$}
\put(360,150){$b_2$}
\put(360,200){$b_3$}
\put(360,250){$b_4$}
\put(60,10){Picture 6. Quantum scheme for $\QFT^{-1}$. }
\end{picture}

The circles denote the transformation $W^1$, two qubit operators have the form:
\begin{equation}
U_{k,j}=\left(
\begin{array}{ccccc}
&1 &0 &0 &0\\
&0 &1 &0 &0\\
&0 &0 &1 &0\\
&0 &0 &0 &e^{i\pi/2^{k-j}}
\end{array}
\right) ,
\ k>j.
\label{shift}
\end{equation}

To verify this we consider the amplitude of the passage from the basic state $a$ to the basic state $b$. This notion is legal because this is the matrix element of the corresponding operator. Here we must fulfill the short computation with the simple idea, though it requires some attention. At first we note that the modules of all amplitudes are the same and as in the reverse transform they equal $1/2^{l/2}$, since we have tor trace for the phase shift only, e.g., for the argument $\phi$ of the complex amplitude $e^{i\phi}$. We take into account the phase change summing the deposits from the Walsh transforms and the deposits of the two-qubit phase shifts. We introduce the following short notation for the simplification of computations: $b'_j =b_{l-1-j},\ j=0,1,\ldots , l-1$ it is needed to account the reverse order of the binary figures dispositions in the states $a$ and $b$ in the proper instant. We represent how states vary in course of the passage from the left to the right along the wires of our scheme. The passage from $a$ to $b$ itself happens in the Hadamard operation, whereas the two qubit operators are diagonal and do not change basic states. They only add some summands to the phase. The deposit from the Hadamard operator is: $\pi a_j b'_j$. This number is non-zero if and only if the both $j$-th registers of our input and output numbers equal 1, which exactly corresponds Hadamard operator. The deposit from the two qubit operator when $j<k$ will be $\pi a_j b'_k /2^{k-j}$, because the state $a$ transforms to $b$ only in the passage through the Hadamard device. As we see from the picture 6, this two-qubit operation is fulfilled in the instant when $j$-th qubit is yet in the state $a_j$, whereas $k$-th is already in the state $b'_k$. Summing all these deposits to the phase shift, and remembering that the integer multiplier of $\pi$ can be excluded from the sum, we obtain the following:
\begin{equation}
\begin{array}{ll}
&\pi\sum\limits_{l>k>j\geq 0}\frac{a_j b'_k}{2^{k-j}} +
\pi\sum\limits_{l>j\geq 0} a_j b'_j=\\
&2\pi\sum\limits_{l>j+k\geq 0}\frac{a_j b_{k}2^{j+k}}{2^l}=\\
&2\pi\sum\limits_{l>j,k\geq 0}\frac{a_j b_{k}2^{j+k}}{2^l}=\\
&\frac{2\pi}{2^l}\sum\limits_{l>j\geq 0}a_j 2^j \sum\limits_{l>k\geq 0}b_k 2^j =\frac{2\pi}{2^l}.
\end{array}
\end{equation}
This is just what we need for the definition of the reverse Fourier transform. If we need to fulfill the direct transform it would suffice to reverse the order of all functional elements in the considered scheme and put the sign minus in the front of the phase shift in the definition of two qubit operators. 

We now look at the result of our constructions. The scheme we have built realizes Fourier transform, it contains about $l^2$ functional elements. We note that if we do not intend to gain the maximal accuracy of this transformation we could omit all two qubit operators, which touch the far qubits. Really, the denominator in $\pi/2^{k-j}$ make the ratio negligible, the exponential will be almost unit, thus these transformations are approximately eqaul to identity and we can omit them. The scheme wll be then substantially simplified. Its size will grow linearly of the order $C\ l$, where the constant $C$ will depend on the chosen accuracy. 

Let us look to it from the other side. The matrix of Fourier transform has the size $N\times N$, e.g., it is inaccessible for us for the direct simulation. The classical memory cannot even store the miserable part of this matrix, let alone to do anything with it. However, we have just built the scheme of quantum computer of the accessible (and even small) size, which fulfills this enormous work: operates with the matrix of Fourier transform. It means that this small computer can cope with such tasks as the finding of eigenvalues of operator, molecular structures and spectra recognition, and fulfills all tasks, which rest on it, that is the large class of computational problems! This is the third miracle of quantum informatics. 

\section{Factoring, optimization, simulation and recognition on QC}

\subsection{Factoring of integers}

The general method of finding eigenvalues, which we represented in the previous section, was invented by P.Shor for the particular case arising in the problem of factoring integers. The factoring problem or the problem of decomposition of an integer number to the integer multipliers is the famous computational problem. The known classical algorithms for its solution require of the order $e^{a\ n^{1/3}}$ steps. This problem therefore belongs to the class of (supposedly) difficult problems. Here we show the quantum algorithm giving the solution of his problem. Shor algorithm was the first fast quantum algorithm solving a problem of the practical significance. The point is that the cryptography protocol RSA rests on the difficulty of the factoring integers, namely the security of this protocol depends on the complexity of factoring problem. This protocol is used in the numerous commercial applications, for example in the defense of the Windows operational system. There to overcome the highest level of defense one must be able to factorize integers with 200 decimal signs. This problem is out of the capacity even for the modern supercomputers. The quantum computer with only 1000 qubits with the frequency about 1 GHz is able to cope with this problem in a few minutes. The practical construction of the quantum computer would mean the final of the modern cryptography.

Shor algorithm has also the theoretical value. It illustrates the significance of the method how Fourier transform is applied, namely the significance of the auxiliary transformations. The point is that the main time is spent here not to Fourier transform, complex from the classical viewpoint, but to the multiplication of integers. 

We take up the factoring problem. Let we have to find the nontrivial decomposition $q=q_1 q_2$ of the known natural number $q$ to the multipliers. This task can be reduced to the problem of the finding of the minimal multiplicative period $r$ of the arbitrary natural number $y$ modulo $q$: $y^r \equiv 1\ (mod\ q)$. In a few words, this reduction looks as follows. Let we have the method of finding of $r$. We will choose $y$ randomly and find $r$. Then with the non-vanishing probability $r$ turns to be even. We then have $y^r-1=(y^{r/2} -1)(y^{r/2}+1)\equiv 1\ (mod\ q)$, and one of the multipliers is the divisor of multiple $q$ number, we thus obtain with the non-vanishing probability the factoring of the number $q$ itself. We thus have only to learn how to find $r$ quickly given $q$ and $y$. This is analogous to the finding of the unknown period when in place of the operator $U$ stands the operator of multiplication on the number $y$.

We take $n$ such that $2^{n-1}\leq q<2^n$ and will work with the quantum memory of $n$ qubits. 
We consider the following operator $U$: $U |x\rangle\ar |y x\ 
({\rm mod}\ q)\rangle$, where $y x$ is the numerical multiplication. To make this operator acting on all our basic vectors we agree that this equality defines it on the numbers less than $q$, whereas on the rest: $q,q+1,\ldots , 2^n -1$ it acts as the identical operator. It brings the little difficulty: this operator can be not unitary. If $y$ and $q$ have the common divisors, some elements will "stick" together. To exclude this trouble we assume that these numbers are mutually disjoint: $(y,q)=1$. Since we choose $y$ at random, then this is the case with the non-vanishing probability. All is ready now. We can apply the powerful technique of the quantum computing we developed earlier. Eigenvectors of $U$ have the form $\frac{1}{\sqrt{r}}\sum\limits_{j=0}^{r-1} \exp (-2\pi i k j/r) |y^j \
({\rm mod}\ q)\rangle$ and the corresponding eigenvalues are $\exp (2\pi i j/r)$. If we apply the procedure of the revealing of eigen frequencies from the previous section, the measurement results in the number $j/r$\footnote{in the reality we obtain the approximation of this number within $O(1/N)$ with the high probability. The details of the proof can be found (in some equivalent form) in the paper \cite{Sh}}, or in \cite{Oz4}, \cite{Oz5}. If we know with the high probability the binary approximation of the fraction, it is easy to find its denominator, if we assume that this fraction irreducible. The formal algorithms for this search are based on the method of continued fractions, it can be found, for example, in \cite{Kn}. This fraction will be irreducible with the non-vanishing probability, since the all possible $j$ appear uniformly if we guarantee that the initial state $\xi$ for the procedure of revealing (see the previous section) is chosen arbitrary. We then repeat this procedure many times, which results in the frequent appearance of values $j$ mutually disjoint with $r$ and thus can find $r$ itself. This is Shor algorithm. 

We now come to the main thing: we estimate how good this algorithm is. As for Fourier transform all is clear here, it is very fast, generally speaking in the linear time relatively to the length of the notation of the number $q$, which we have to factorize. However, there is the other routine operation, which threatens to eliminate all the advantages of quantum Fourier transform. This is the operation of multiplication on the number $y$ containing in the operator of the conditional application $U_{cond}$. To find $U^\a$ we have to multiply to $y$ $\a$ times, which is about $q$ actions. This difficulty in the general case of the quantum Fourier transform application bears the principal character. It is irremovable for the arbitrary operator $U$. However, in the case of factoring we are lucky: we can fulfill the conditional application in the time of the order $\log^2 q$. To multiply to the number $y^\a$ we will obtain the number $y^\a$ by the sequential involution to the second power, beginning with $y$: $y,y^2, y^4, \ldots$. Of course, at each step we take the remainder from the division to $q$. We thus reach the closest to $y$ degree of two: $2^{l_1}$. We then take the quotient $q/2^{l_1}$ and do the same with it, etc. We then reach $y$ in the time of the order logarithm of $q$, e.g., in the number of steps of the order of the length of the $q$ notation. At each step we use about $log^2 q$ actions for the computation of the multiplication of numbers by the direct method, which results in the realization of the operator of the conditional application $U$ in the time $O(\log^3 q)$.

It is the complexity of Shor algorithm. We see, that the most difficult part of this algorithm is the routine operation contained in the preparation of the input state for Fourier transform - the sequential multiplication of natural numbers. 

\subsection{Solution of the problem of discrete optimization}

We continue to consider the examples of problems for which fast quantum algorithms can be obtained by some successful modification or combination of th emain quantum tricks: GSA and QFT. At first we take up the natural generalization of the search problem: the search of the extreme point of an integer function. Let a function $f:\ \{ 0,1\}^n \ar\{ 0,1\}^n$ be defined by its oracle (or the scheme of the functional elements). We treat it, as usual, as the integer function. The problem is to find its extreme point: maximum or minimum. We note that in this most general formulation we cannot apply any trick essentially simplifying the search, like simplex method or the differentiation. The classical solution of this problem thus requires of the order of $N=2^n$ actions. 

The idea of its quantum solution rests on the GSA algorithm. We try to find the poin of maximum by the sequential approximations. Namely, we place all argument into the order of the growth of the function $f$ on them: $f(x_0 )\leq f(x_1)\leq\ldots \leq f(x_{N-1}$. On each step $j$ the input value will be some $x_{j_k}$. We apply G-BBHT algorithm with the oracle taking the value 1 exactly on the arguments $x_j$, for which $f(x_{j_k})<f(x_j)$, e.g., on $x_{j'},\ j'>j_k$. After the regular observation and the check of correctness we obtain the following value $x_{j_{k+1}}$ etc., up to the step when we reach $x_{N-1}$. The detailed analysis (see \cite{Ho}) shows that the complexity of such algorithm has the order $\sqrt{N}$, that gives us the same acceleration as GSA. 

\subsection{Recognition of structures and functions}

The list of problems in which the application of quantum computations gives the principal gain is very big. We finalize this subject by the consideration fo problems of recognition of the devices that can be briefly formulated as the direct and reverse problems:

1) given a scheme of device determine its function,

2) given a function of the device determine its scheme.

The problems of this type we meet always in the engineering practice. We use the notion of the eigen frequencies for the specification of this formulation. We assume that the device is the scheme of quantum elements represented as the quantum gate array, and the function of it is the unitary operator it generates. The function will be completely determined if we establish the eigen frequencies and for each frequency point the corresponding subspace of eigenvectors. At first, we narrow the problem and consider only eigen frequencies ignoring eihgenstates. It brings the following questions:

a)  Given a scheme generating the unitary operator $U$, and the number $w\in [0,1)$ define it this number an eigen frequency of the operator $U$.

b) Given a set of numbers and the scheme determine is the spectrum (set of eigen frequencies) of the operator generated by this scheme contained in the given set.

c) Given a set of frequencies build the scheme generating the corresponding operator.

The problem of the other types arise if we use the physical functionality of the device with the unknown scheme:

d) Given an oracle for the operator $U$ or $U_{cond}$ find the quantum scheme generating it. 

We put these problems in the order of increasing complexity. We consider briefly the scheme of the solution for the first of them. We denote the operator revealing eigen frequencies, we built in the previous section, by $\Rev$. Analogously we can build the reverse operator. Now, we have the state $\sum\limits_k x_k |\psi_k\rangle$, and we can reverse the sign of that $\psi_k$, which corresponds to the given frequency $w$. It can be done by the following way. At first by the help of $\Rev$ we reveal the value of the frequency $\w_k$ it appears in the additional register, then we change the sign provided $w=\w_k$ - it can be done as the reflection along the zero vector in GSA, then by means of $\Rev^{-1}$ we clear the additional register. Now we have the reflection along the subspace of the eigen frequency $w$, and we can apply GSA scheme in the order to find the vector form this subspace, after that, again by means of $\Rev$ we check is this vector really eigenvector with the eigen frequency $w$. If $w$ is the eigen frequency, all the reflections we used will be non trivial, if it is not the eigen frequency, these reflections will be identical, and we instead of the rotation of the current vector by GSA method obtain the absence of any movement at all - the initial vector of the algorithm will be resulting. Here the inaccuracy appears from the lack of knowledge of the instant of the termination of GSA. We can successfully fight against it using the parallel computations\footnote{The details can be found in the paper \cite{Oz7}.}

We pass to the second problem. Here we have to organize two included GSA - processes acting on the different registers. The external (main) GSA - process acts on the register for frequencies, its aim is to find the frequency occurred in the given list and which is not eigen frequency. For this, we need the reflection along such "bad" frequencies. To build the reflection we use the algorithm from the previous section containing the internal GSA - process acting on register for the argument of $U$ provided the register of frequencies is fixed.

At last, for the solution of the third type of problems we use  GSA - process of the threefold nesting. For the main external process, we establish the new register in which we store the binary codes of the possible schemes. The algorithm we considered earlier will be included to the internal process, which aim will be the reflection along the code of the schemes generating operators with the appropriate spectra. This reflection is used in the external GSA - process that obtains the code of the target system. 

In the considered problems, we used only codes of schemes. The problem of the type d) has the significance difference. Here we need the oracle for the transformation $U_{cond}$. We will not discuss here the question about the possibility to use the operator $U$ itself. To search the scheme corresponding to this oracle we use GSA again which gives the process of threefold nesting. The sense of registers will be the same as in the previous case. 

\section{Generalizations of Grover algorithm}
\nn

We consider the more general formulation of quantum search, when we have $m$ types of states: 
the set of $N_1$ main states, which will not change in the reflections, the set of $N_2$ states on $\pi$, and are the solutions of the equation $f(x)=1$, and $m-2$ of all others types of states with total numbers correspondingly: $N_3 , N_4 , \ldots , N_m$ which will rotate on the $m-2$ different angles: $d_1 , d_2 , \ldots , d_{m-2}$ correspondingly. We enumerate elements of the set $N_s$ so that $N_s =\{ x^j_s ;\ j=1,2, \ldots , l_s \}$, $s=1,2,\ldots , m$.
Such target states will be $x_2^1 , x_2^2 , \ldots , x_2^{l_2}$, and the number of all basic states $\sum\limits_{s=1}^{m} l_s =N$
is the dimensionality of the main Hilbert space $H$. We denote this orthonormal basis $H$ by $\bar N$.

Our supposition about the angles of rotation can be reformulated as the limitation imposed to the unitary operator $U$, which will be used in the definition of GSA instead of $I_{tar}$. 
 Now let $U$ be the unitary operator on $H$ with the eigenvectors $ x^s_j \ s=1,2,\ldots , m,\ j=1,2,\ldots , l_s$, and eigenvalues of all $x^2_j$, equal  $-1$, for all $x^1_j$ - equal $1$, and for all $x^s_j$ ($s\geq 3$) - equal $e^{i(d_k -\pi )} ,\  d_k \in [0,2\pi )$. Let $v_k =e^{id_k }$.

 We define $k$ as the index for which $d_k$ is maximal. Let $d=d_k$. We define $\g =\sqrt{\frac{l_2}{N}}$. We put $\g\ar 0 \ (n\ar\infty )$ 
because otherwise the problem can be trivially solved on the classical computer, and the application of quantum computer makes no sense. 
We then define $e_j =\frac{1}{\sqrt{l_j}} \sum\limits_{x\in N_j} |x\rangle$. Thus, for example, $e_2$ is the superposition of all target states with the equal amplitudes. We define $\tilde 0$ as $\frac{1}{\sqrt{N}}\sum\limits_{x\in \bar N} |x\rangle$. This state will be the initial for our algorithm. We denote by $H_0$ the subspace, generated by all vectors $e_j$. The evolution of the state vector will be the rotation in this $m$ dimensional space. It directly follows from our definition that the initial vector $\tilde 0$ belongs to this subspace. On the other hand the action of $I_{tar}$, restricted to $H_0$ coincides with the restriction of the transformation $I_{e_2}$ on $H_0$.
 
Our main result will be the following.

\newtheorem{Theorem}{Theorem}
\begin{Theorem}
Let $l_1 /N \ar 1$, $(N-l_1 -l_2 )/d\sqrt{Nl_2} \longrightarrow 0$, 
$\sqrt{\frac{l_2}{N}}=o (d)$ and $t=\left[ \frac{\pi}{4} \sqrt{\frac{l_2}{N}} \right] $, then the iterated transformations $I_{\tilde 0} U$, applied to  
 $\tilde 0$ $t$ times with the following observation give the state $e_2$ with the vanishing error probability ( $n\ar\infty$ ).
\end{Theorem}

We will work with the subspace $H_0$ of the main space.  $I_{\tilde 0} $ 
and $U$ preserve this subspace. The idea of the proof is the following. We compute the eigenvectors and eigenvalues of the matrix $G=-I_{\tilde 0} U$ 
and compare them with the corresponding eigenvectors and eigenvalues of the matrix of standard GSA - operator $I_{\tilde 0} I_{tar}$. It will turn that the two dimensional subspace of $H_0$ generated by states $e_2$ и $\tilde 0$, and the eigenvectors and eigenvalues of $G$ will be equal to the corresponding values for GSA with the high accuracy. It means that the behavior of iterations of $G$ will be the same as GSA. In the other words, the evolution of the state vector induced by our algorithm can be with the high accuracy described by the evolution in the two dimensional subspace generated by the vectors $e_1$ and $e_2$.

\subsection{Computation of matrices}

\subsubsection{Approximate computations}

The residuary part of this section is devoted, mainly, to the proof of the Theorem 1. 
We compute all matrices in the basis $e_1 , e_2 , \ldots , e_m$ of the subspace $H_0$. We define $\langle 
\tilde 0 |e_2 \rangle =\sqrt{\frac{l_2}{N}} =X$,
$\langle \tilde 0 |e_j \rangle =\sqrt{\frac{l_j}{N}} =
Y_{j-2}$, $j=3, 4, \ldots , n$, $x_j =2X_j$.
In tis section we fulfil all computations within $\sum\limits_{j=2}^{m} 
l_j /N$. The substantiation of the legality of this approach will be given in the next section . 

We set $\sum\limits_{j\geq 2} l_j /N =\e$. We have: 
$\tilde 0=(\sqrt{1-\e}, X,Y_1 ,\ldots ,Y_{m-2} )^T$,
$\sqrt{\frac{l_1}{N}} =\sqrt{1-\e} =1-\frac{\e}{2} +o(\e )$.
Then 
 $\tilde 0 \approx \tilde 0_{app} =
 (1, X, Y_1 ,\ldots , Y_{m-2} )^T$ within  $O(\e )$, where the distance between vectors is estimated in Hilbert space $H_0$.

Directly from the definition follows that $H_0$ is the subspace invariant relative to the operators 
$U$ and $I_{\tilde 0}$. In $m$ dimensional subspace $H_0$ the matrix of the operator $U$ has the following form.
$$
U
=\left(
  \begin{array}{ccccc}
    1 &0 &0 &\ldots &0\\
    0 & -1&0 &\ldots &0\\
    0 &0 &-v_1 &\ldots &0\\
    \ldots &\ldots &\ldots &\ldots &\ldots\\
    0&0&\ldots &0 &-v_{m-2}
  \end{array}
\right) .
$$
In order to find the matrix for $I_{\tilde 0}$ in this subspace, we represent this operator in the form $I_{\tilde 0} \approx I_{\tilde 0_{app}} =
V^{-1} I_{e_1} V$, where the matrix of $V$ is such that $V\tilde 0_{app} =e_1$. It is because 
$V^{-1} I_{e_1} V\tilde 0_{app} =-V^{-1} e_1 =-\tilde 0_{app}$. 
We can straightforwardly check that the matrix $V$ and its reverse have the form
$$
V=\left(
  \begin{array}{cccccc}
    1 &X &Y_1 &Y_2 &\ldots &Y_{m-2}\\
    -X & 1&0 &0 &\ldots &0\\
    -Y_1 &0 &1 &0 &\ldots &0\\
    \ldots &\ldots &\ldots &\ldots &\ldots &\ldots\\
    -Y_{m-2} &0 &0 &0  &\ldots &1
  \end{array}
\right) ,
$$
$$
V^{-1}=\left(
  \begin{array}{cccccc}
    1 &-X &-Y_1 &-Y_2 &\ldots &-Y_{m-2}\\
    X & 1&0 &0 &\ldots &0\\
    Y_1 &0 &1 &0 &\ldots &0\\
    \ldots &\ldots &\ldots &\ldots &\ldots &\ldots\\
    Y_{m-2}&0&0 &0  &\ldots &1
  \end{array}
\right) .
$$

We then have:
$$
I_{e_1} =
\left(
  \begin{array}{cccc}
    -1 &0 &\ldots &0\\
    0 & 1&\ldots &0\\
    0 &0 &\ldots &0\\
    \ldots &\ldots &\ldots &\ldots 
    \end{array}
\right)
$$
And now the direct multiplication of matrices gives

$$
I_{\tilde 0_{app}} =V^{-1} I_{e_1} V=
\left(
  \begin{array}{cccccc}
    -1 &-x &-y_1 &-y_2 &\ldots &-y_{m-2}\\
    -x & 1&0 &0 &\ldots &0\\
    -y_1 &0 &1 &0 &\ldots &0\\
    \ldots &\ldots &\ldots &\ldots &\ldots &\ldots\\
    -y_{m-2}&0&0 &0  &\ldots &1
  \end{array}
\right) ,
$$
$$
G=-I_{\tilde 0_{app}} U=
\left(
  \begin{array}{cccccc}
    1 &-x &-y_1 v_1 &-y_2 v_2 &\ldots &-y_{m-2} v_{m-2}\\
    x & 1&0 &0 &\ldots &0\\
    y_1 &0 &v_1 &0 &\ldots &0\\
    \ldots &\ldots &\ldots &\ldots &\ldots &\ldots\\
    y_{m-2}&0&0 &0 &\ldots &v_{m-2}
  \end{array}
\right) .
$$

\subsubsection{The accuracy of approximation}

We now show that for the approximation of GSA our accuracy to
 $\e$ is sufficient. 
We work with the usual matrix norm given by the equation 
$$
\| A\| = \max\limits_{\|\bar x\| =1} \| A\bar x\| .
$$
The accuracy of our approximation $G=-I_{\tilde 0_{app}} U$
of the exact matrix $-I_{\tilde 0} U$ 
which we will denote by 
$G_{exact}$ is 
$\| G-G_{exact} \| =\| U\| \| I_{\tilde 0} -I_{\tilde 0_{app}} \| =$
$\| I_{\tilde 0} -I_{\tilde 0_{app}} \| = \| \tilde 0 -\tilde 0_{app} \| =$
$O(\e )$.
Consequently, $G$ is the approximation of $G_{exact}$ within to $\e$, and we can represent it as $G=G_{exact} +\Delta$, where $\|\Delta\| =O(\e )$. Let 
$t=\sqrt{\frac{N}{l_2}}$ be the order of the total number of iterations in GSA. We evaluate the number 
$\nu =t\Delta =\frac{\sum\limits_{j\geq 2} l_j}{\sqrt{Nl_2}}$. If $l_2 
=O(\sum\limits_{j\geq 3} )$, we can omit $l_2$ and use the conditions of the Theorem, which gives $\nu =o(1)$.
 In the opposite case 
$\sum\limits_{j\geq 3} =o( l_2 )$ and $\nu$ will be of the order $\sqrt{\frac{l_2}{N}}$ 
that is again $o(1)$.
Hence, in all cases $\nu =o(1)$.
We then have: $G^t =(G_{exact}+\Delta )^t = G_{exact}^t +O(t\Delta 
G_{exact}^{t-1} )=$
 $G_{exact}^t +o(1)$. It means that for the investigation of the behavior of  
$G_{exact}^t$ it is sufficient to use the approximation $G$ to
 $G_{exact}$, that substantiates our computations. 

\subsection{Finding of eigenvalues}

To find eigenvalues of $G$, we at first find the characteristic polynomial 
$p_{m-2} (\la )= |G-\la I |$, where $I$ is the identical matrix. Then, solving recursively the equation
 $p_{m-2} (\la )=0$, we find the eigenvalues. We have
$$
p_{m-2} (\la )= 
\left|
  \begin{array}{cccccc}
    1-\la &-x &-y_1 v_1 &-y_2 v_2 &\ldots &-y_{m-2} v_{m-2}\\
    x & 1-\la &0 &0 &\ldots &0\\
    y_1 &0 &v_1 -\la &0 &\ldots &0\\
    \ldots &\ldots &\ldots &\ldots &\ldots &\ldots\\
    y_{m-2}&0&0 &0  &\ldots &v_{m-2} -\la
  \end{array}
\right| =
$$
$$
(-1)^{m+1} y_{m-2} v_{m-2} \left|
\begin{array}{ccccc}
    x & 1-\la &0 &\ldots &0\\
    y_1 &0 &v_1 -\la &\ldots &0\\
    y_2 &0 &0 &\ldots &0\\
    \ldots &\ldots &\ldots &\ldots &\ldots\\
    y_{m-2} &0 &0 &\ldots &0
  \end{array}
\right| +(v_{m-2} -\la )p_{m-3} (\la ) =
$$
$$
y_{m-2}^2 v_{m-2} (1-\la )(v_1 -\la ) \ldots (v_{m-3} -\la ) 
+(v_{m-2} -\la )p_{m-3} (\la ).
$$
that gives the following recurrent relation 
\begin{equation}
p_{m-2} (\la ) = (v_{m-2} -\la )p_{m-3} (\la ) +y_{m-2}^2 v_{m-2} (1-\la )
(v_1 -\la ) \ldots (v_{m-3} -\la ).
\label{1}
\end{equation}
Using the basis of the recursion: \newline
$-p_1 (\la ) = (\la -1+ix)(\la -1-ix)(v_1 -\la ) +v_1 y_1^2 (1-\la )$
we can deduce from the main equation (1) be means of the transparent transformations
the general formula for the characteristic polynomial:
\begin{equation}
\begin{array}{ll}
p_{m-2} (\la ) =&(\la -1+ix )(\la -1-ix)(v_1 -\la )(v_2 -\la )\ldots (v_{m-2} 
-\la )+\\
& v_1 y_1^2 (1-\la )(v_2 -\la )\ldots (v_{m-2} -\la )+\\
&v_2 y_2^2 (1-\la )(v_1 -\la )(v_3 -\la )\ldots (v_{m-2} -\la )+\\
& \ldots +v_{m-2} y_{m-2}^2 (1-\la )(v_1 -\la )\ldots (v_{m-3} -\la ).
\end{array}
\label{2}
\end{equation}
We denote the first summand in (2) by  
$p_0 (\la ) = (\la -1+ix )(\la -1-ix)(v_1 -\la )(v_2 -\la )\ldots
 (v_{m-2} -\la )$, 
so that $p_{m-2} (\la )=p^0 +\d$, where
$$
\begin{array}{ll}
&\d = v_1 y_1^2 (1-\la )(v_2 -\la )\ldots (v_{m-2} -\la )+\\
&v_2 y_2^2 (1-\la )(v_1 -\la )(v_3 -\la )\ldots (v_{m-2} -\la )+\\
&\ldots +v_{m-2} y_{m-2}^2 (1-\la )(v_1 -\la )\ldots (v_{m-3} -\la ).
\end{array}
$$ 
It means that $p_{m-2} $ и $p_0 (\la )$ differs in only the shift to $\d$.
The roots $p_0 (\la )$ will be $\la_1 =1-ix,\ \la_2 =1+ix,\ 
\la_3 =v_1 ,\ldots , \la_m =v_{m-2}$. We denote the roots $p_{m-2}$ by 
$\tilde\la_1 , \tilde\la_2 , \ldots , \tilde\la_m$. 

We now have to estimate the difference between $\la_j$ and $\tilde\la_j$, using 
the evaluation $\d$ in the vicinity of two roots: $\la_{1,2}$
which play the main role in the dynamics of the considered algorithm. 

 Wenote that $|\la_1 -\la_2 |=o(|v_j -\la_1 |)$. On the other hand
$|\la_1 -\la_2 | \gg \d^2$. Our polynomial $p$ can be approximated by the quadratic polynomial
 $q$ with the major coefficient $A=\la -v=\Omega (d)$ in the vicinity of
$\la_1$ or $\la_2$ of the radius $|\la_1 -\la_2 |$. The derivative 
 $q$ of this quadratic polynomial in this vicinity is 
$q =\g (v_1 -\la )\cdots (v_{n-2} -\la )$.  
We denote the difference between the roots by $\s =|\la_1
-\tilde\la_1 |+|\la_2 -\tilde\la_2 |$.  Then $\s =O(\d
/q' ) = \sum\limits_{j=1}^{m-2} \frac{v_j y_j^2 (1-\la )}{\g (v_j -\la )}
$. Using equalities $v_j -\la =O(d_j )$, $1-\la =O(\g )$, we conclude that
 $\s =O (\frac{1}{d} \sum\limits_{j\geq 3} \frac{l_j}{N} )$ that is
$o(\g )$, since by the condition of the Theorem $\sum\limits_{j\geq 3} l_j =o (
d\sqrt{Nl_2 } )$. Therefore, $\tilde\la_{1,2} =1+-ix+o(\g
)$, $\tilde\la_3 =v+o(\g )$.

\subsection{Finding of eigenvectors}

At first, we take up eigenvectors for two first roots:
$\tilde\la_{1,2} =1+-ix +o (\g )$.

1). $\la =1-ix+o(\g )$. Let the eigenvector is the column of the form
$\bar a=(a,b,w_1 ,\ldots , w_{m-2} )^T$. The system of linear equations, determining
  $\bar a$ has the following form:
$(G-\la E)\bar a =\bar 0$ and it can be written in the form
$$
\left(
\begin{array}{ccccc}
    ix & -x &-y_1 v_1 &\ldots &-y_{m-2} v_{m-2}\\
    x &ix &0 &\ldots &0\\
    y_1 &0 &v_1 -1+ix &\ldots &0\\
    \ldots &\ldots &\ldots &\ldots &\ldots\\
    y_{m-2} &0 &0 &\ldots &v_{m-2} -1+ix
  \end{array}
\right)
\left(
\begin{array}{cc}
     &a\\
     &b\\
     &w_1 \\
     &\ldots\\
      &w_{m-2}
\end{array}
\right)=
\left(
\begin{array}{cc}
     &o (\g )\\
     &o (\g )\\
     &o (\g ) \\
     &\ldots\\
      &o (\g )
\end{array}
\right) .
$$
We rewrite this in the form of the system of linear equations and find
$$
\left\{
\begin{array}{ccccccc}
&ixa &-xb &-y_1 v_1 w_1 &-\ldots &-y_{m-2} v_{m-2} w_{m-2} &=o (\g )\\
&xa &+ixb &\ &\ &\ &=o(\g )\\
&y_1 a &\ &+(v_1 -1+ix)w_1 &\ &\ &=o (\g )\\
&\ldots &\ldots &\ldots &\ldots &\ldots &\ldots\\
&y_{m-2} a &\ &\ &\ &+(v_{m-2} -1+ix)w_{m-2} &=o(\g )
\end{array}
\right.
$$
We suppose that the eigenvector has the limited norm, which means that all its components are limited. Solving the system, we find:
$w_j =\frac{y_j a}{v_j -1+ix}$. We apply the conditions of the Theorem: 
$\sqrt{\frac{l_2}{N}}
=o (d)$ and conclude that $w_j =o(1)$,
with the accuracy up to $o(1)$ $a=i, b=-1$.

2). $\la =1+ix+o (\g )$. The corresponding computation gives
$a=i, b=1$ within to $o (1)$, all $w_j =o(1)$.

3). For all the other eigenvalues we have $a=0, \ b=0$ within to
$o (\g )$, since the corresponding vectors must be orthogonal to the subspace generated by $e_1 
, e_2$.

We then have the following situation:

if $n\ar\infty$, then eigenvalues associated with the subspace generated by $ e_1 , e_2$
 differs to $o (\g )$ from the corresponding eigenvalues of standard GSA, and he corresponding eigenvectors differ to
 $o (1)$ from the corresponding eigenvectors of standard GSA. 
It means that if the number of $t$ of iterations is of the order
 $1/\g $ as in GSA, the difference between the resulting states of GSA and the iterated applications of $G$ will be or the order $o (1)$. 
The Theorem 1 is proved.

\nn

 \section{Realistic models of quantum computers}\nn

In this section, we study fermionic computations in the formalism of occupation numbers proposed in the paper \cite{BK}. 
We will show that to fulfill an arbitrary quantum computation with only minor slowdown it is sufficient to control the external field and the tunneling only. Here the interaction of qubits of the diagonal type remains continuous and non controlled. The substantiation of this approach will be done by means of the reduction to the standard model of computations in Hilbert space that uses the results of the paper \cite{OF} about one qubit control. 

\subsection{About the usage of fermionic identity}

Quantum computer is the unexampled testing of quantum physics because it requires such level of control over nano-sized objects, which has been never reached artificially. Whereas the mathematical theory of quantum computing in the framework of standard quantum formalism is well developed its physical realization represents the serious challenge to our understanding of the Nature. This is why it is important to look for its simplest possible realization, so that it rests on the basic principles of quantum theory and includes the minimal technological difficulties. Two requirements for such schemes we can formulate: the adequate description of states forming the computational Hilbert space, and the realistic method of control over computations. 

Typically one computational element - a qubit we represent as some characteristic, like a spin , charge or a position of some elementary particle. This approach works well for the isolated qubit. For a system of several qubits, this approach meets serious difficulties. These difficulties come from the fundamental principle of the non-distinguishability (or identity) of elementary particles of the same type. To control a computation we must be able to address to a separate qubit, whereas the different particles are identical. Of course, we can distinguish particles placing them on the big distance one from another, but in this case, it will be difficult to keep them in entangled states what is necessary for quantum computations. On of the solutions of this dilemma is to use Fock space of the occupation numbers for the description of quantum computations. Here the natural identification of qubits with the energetic levels in Fock space is used, so that the unit is treated as the occupied level, and zero - as the free level. This approach gives the universal quantum computations by the high cost. It requires to control not only the external field and the tunneling, but also the diagonal interactions between qubits, and the contact with the superconductor, e.g., we have to control the coefficients $\a ,\b , \g$ in (\ref{ham}) and to control on the additional summand $\d a_k^+ a_j^+ +\d^* a_k a_j. $

In this section, we will see how to decrease this cost by means of the idea of the continuous and non-controlled interaction. To this, we need two things: The supposition that the initial Hamiltonian of interaction contains only the external field, the diagonal and the tunneling summands, and the modified correspondence between states in the occupation numbers space and the computational Hilbert space. To control quantum computation we then need only to switch the external field and the tunneling. Lasers can fulfill this type of the control. The main scheme we give further. It rests on the idea of the continuous interaction, proposed in the works \cite{OF}, \cite{Oz8} and adapted to the language of Fock space of occupation numbers.

\subsection{One qubit control in quantum computations}

 The main difficulty in the practical realization of quantum computations is that it is technically difficult to fulfill two qubit operators playing the necessary role in such computations. To fulfill these operators we must control the degree of entanglement of particles that is determined by the overlapping of the spatial parts of their wave functions. However, to fulfill computations we need to distinguish particles with certainty, which is possible only if the overlapping is sufficiently small. This is the evident contradiction in requirements for the physical realization of quantum computations. We see that it is much more difficult to realize two qubit gates then one qubit. The following approach would be appropriate here. Since the interactions between particles with the varying degree of entanglement follows from the wave equation and is confirmed in experiments, the two-qubit transformations go permanently in course of natural time evolution of a quantum system. As for the control over such evolution, we can fulfill it by one-qubit impulses, whereas two qubit gate will go in the non controlled background regime. This is the essence of the proposed quantum computations with one qubit control. This model is much more realistic than the abstract model of quantum computer, which supposes the control on two-qubit interaction. We temporarily leave the question about the general possibilities of this approach and demonstrate how the concrete problem about the simulation of the behavior of many bodies system with quadratic interaction of the diagonal form can be solved in the frameworks of the proposed model. The main difficulty of the proposed model is that two-qubit interaction is out of the control, in particular, it goes with the outside qubits that seriously distorts the picture of quantum computations.
To perform computations in such model we have to create the method of the correction of "undesirable" transformations by means of one qubit impulses. 

For the demonstration of abilities of this approach, we first show how to realize quantum Fourier transform in the framework of this model. The main supposition will be that the Hamiltonian matrix of two qubit interaction has the diagonal form. For the convenience, we at first impose some limitations on the speed of decreasing of this interaction with the distance. Namely, we suppose that the potential falls with the distance as Yukawa potential.. This method then can be applied to the more wide class of the diagonal interaction. Moreover, this method can be generalized to the case when the different qubit pairs interact differently. At last, we apply this approach to the system of many particles with the potential of quadratic type. 
\nn

\subsubsection{Realization of quantum Fourier transform on one qubit control} 
\nn

Quantum Fourier transform is the key subroutine in quantum computing. It is used in the big number of other algorithms. The quantum gate array realizing this transformation is represented at the picture 6. It was proposed and used for the fast quantum factoring by P.Shor (see \cite{Sh}). We agree to represent an integer of the form $a=a_0 +a_
0 2+\ldots +a_{l-1} 2^{l-1}$ by th ebasic state $|a_0 \ a_1 \ \ldots\ a_{l-1} \ \rangle =|a\rangle$ . These states form orthonormal basis for the input states of the quantum gate array. We place them from the top to the bottom. The analogous agreement we accept for the output state but the binar signs $b_j$ of the number $b=b_0 +b_0 2+\ldots +b_
{l-1} 2^{l-1}$ we place in the opposite order. 

This scheme fulfills the reverse transformation $QFT^{-1}$ in $O(l^2 )$ steps, whereas its matrix $N=l^2$ - dimensional. However, in this scheme, the two qubit control is required, it cannot be directly applied in terms of our model. We show how to do that. We treat the interactions of the form
 \begin{equation} {\rm A)}\ H=\left( \begin{array}
{ccccc} &0 &0 &0 &0\\ &0 &0 &0 &0\\ &0 &0 &0 &0\\ &0 &0 &0 &\rho \end{array} \right) ,\ \ \rho >0, \ \ \ \ \ \ \ \ \ \ \ 
{\rm B)}\ H=\left( \begin{array}{ccccc} &\rho_1 &0 &0 &0\\ &0 &\rho_2 &0 &0\\ &0 &0 &\rho_3 &0\\ &0 &0 &0 &\rho_4 \end
{array} \right) , \label{Ham} \end{equation} where all $\rho =\rho_0 \frac{e^{-br}}{r}$; $b=const$; $r$ is the distance between qubits-particles, and $\rho_1 +\rho_4 \neq \rho_2 +\rho_3$. We place $l$ qubits on one line with the equal intervals. Let the interaction between $j$ - the and $k$ - th qubits have Hamiltonian $H_{j,k}$ of the form (\ref{Ham}). This type of Hamiltonian arises, for example, in Izing model for particles with the spin 1/2. The required decreasing on the interaction with the distance we can reach placing qubits to the appropriate potential hole. Choosing the proper unit of the length we can ensure that $b=1$. 
At first we study the case of interaction of the form (\ref{Ham}, A) and then extend the results to the case (\ref{Ham}, B). 
\nn

{\bf Realization of QFT within phase shift} 
\nn

We remind that QFT and its reverse have the form: \begin{equation} QFT :\ 
|a\rangle\ar\frac{1}{\sqrt{N}}\sum\limits_{b=0}^{N-1}e^{-\frac{2\pi i\ ab}{N}}|b\rangle , \ \ \ \ \ \ \ \ QFT ^{-1} :\ |a
\rangle\ar\frac{1}{\sqrt{N}}\sum\limits_{b=0}^{N-1}e^{\frac{2\pi i\ ab}{N}}|b\rangle . \label{qft} \end{equation} 
The reverse transform can be then fulfilled by the following scheme. 
\newpage
\begin{picture}(500,310)(0,-30) \multiput(80,50)(0,50){5}
{\line (30,0){30}} \multiput(140,50)(0,50){5}{\line (30,0){30}} \multiput(200,50)(0,50){5}{\line (30,0){30}} \multiput
(260,50)(0,50){5}{\line (30,0){30}} \multiput(320,50)(0,50){5}{\line (30,0){30}} \multiput(110,40)(60,0){4}{\framebox 
(30,220)} \multiput(95,50)(60,50){5}{\circle{10}} \put(30,200){j} \put(30,100){k} \put(60,50){$a_4$} \put(60,100){$a_3$} 
\put(60,150){$a_2$} \put(60,200){$a_1$} \put(60,250){$a_0$} \put(360,50){$b_0$} \put(360,100){$b_1$} \put(360,150){$b_2$} 
\put(360,200){$b_3$} \put(360,250){$b_4$} \put(60,10){Picture 2. \ Rectangles denote continuous interaction (\ref{Ham}, A),} 
\put(120,-5){circles - Hadamard operators} \end{picture} Here rectangles denote unitary transforms of the form $U=e^{-i\tilde H}$, where $\tilde H=\sum\limits_{l>j>k\geq 0}\tilde H_{j,k}$, and each from $\tilde H_{j,k}
$ has the form (\ref{Ham} , A) with $\rho_0 =\pi$, $r=j-k$. If we choose the unit of the length such that Planck constant multiplied to 
 $\rho_0$ equals to $\pi$ and the unit of the length such that $r=j-k$, then $U$ will be exactly the transformation of the state vector induced by the considered Hamiltonian in the unit time. Here we suppose that the time of one-qubit operations is negligible, and the interaction between qubits cannot substantially change the phase in this time. This scheme can be obtained from the previous one by the insertion of the missing elements corresponding to the interaction going in the system with this Hamiltonian. To prove that this scheme really fulfills $QFT^{-1}$, we apply the method of the amplitude counting proposed in the paper \cite{Sh}. Let the basic input state be given: $|a\rangle$, we consider the corresponding output state. 
This output state is the linear combination of basic states $|b\rangle$ with some amplitudes. All modules of these amplitudes are the same and equal 
$1/\sqrt{L}$, and we have to look after their phases only. For the simplicity we introduce the notation $a'_j =a_{l-1-j},\ j=0,1,\ldots , l-1$. 
In the course of application of our scheme the value of qubits number $j$ and $k\leq j$ pass through elements form the picture 2 from the left to the right. Following this direction, we separate four types of interactions: The interaction of $a'_j$ with itself and $a'_k$ with itself in Hadamard gate, the interaction of $a'_j$ with $a'_k$ ($j>k$), the interaction of $a'_j$ with $b_k$ for $j>k$, and the interaction of 
$b_j$ with $b_k$ ($j>k$). The times of these interactions will be the following: zero, $k$, $j-k$ and $l-1-j$ correspondingly. Summing the deposits of these interactions to the phase we obtain the resulting phase of the form 
\begin{equation} \pi\sum\limits_{l>j>k\geq 0}\frac{a'_j a_k 
k}{2^{j-k} (j-k)} + \pi\sum\limits_{l>j>k\geq 0}\frac{a'_j b_k (j-k)}{2^{j-k} (j-k)} + \pi\sum\limits_{l>j\geq 0} a'_j b_j 
+ \pi\sum\limits_{l>j>k\geq 0}\frac{b_j b_k (l-j-1)}{2^{j-k} (j-k)}. \end{equation} 
We denote the first and the second summands by $A$ and $B$ correspondingly. Their deposit corresponds to the action of the diagonal Hamiltonians to $|a\rangle$ and $|b\rangle$ correspondingly. We temporarily leave these deposits. We take up the second and the third summands of this sum. After the replacement of $j$ by $l-1-j$ this part acquires the form 
\begin{equation} \begin{array}{cc} \pi\sum\limits_{l-1>k+j\geq 0}\frac{a_j b_k 2^{j+k}}{2^{l-1}}+
\pi\sum\limits_{l- 1\geq j\geq 0} a_{l-1-j} b_k &=2\pi\sum\limits_{l>k+j\geq 0}\frac{a_j b_k 2^{j+k}}{2^l} =2\pi S+ 2\pi
\sum\limits_{l>k,j\geq 0}\frac{a_j b_k 2^{j+k}}{2^l}= \\ &2\pi S+ 2\pi\frac{ab}{2^l} \end{array} 
\end{equation} 
for some integer $S$. The first summand does not change the phase and we obtain that is required within the deposit of $A$ and
$B$. 

\nn
{\bf Correction of phase shift}
\nn

 To account the deposit of diagonal summands $A$ and $B$ to the phase we apply one trick. At first we consider the summand $A$. It consists of the members of the form $A_{j,k}=c_{j,k} a'_j a'_k$, where $c_{j,k}$ depends only on $j$ and $k$, but not on $a$. We call $j$-th and $k$-th qubits separated. 
We will apply one-qubit operator NOT several times to each qubit but separated in order to suppress all interactions but the interaction going between the separated qubits. At first we consider the pair of not separated qubits with the numbers $p,q,\ q>p$. Their continuous interaction in time $\D t$ gives the summand $d_{p,q}\D t\ a'_p a'_q$ to the phase, where the real number $d_{p,q}$ depends only on how fast the interaction falls with the distance, but not on  $a'_p , a'_q$. For example, for the decreasing of Yukawa type we have $d_{p,q}=e^{-|q-p|}/|q-p|$. Now we invert one of these two qubits, no difference which exactly, let it be $q$-th, by means of the NOT gate. It state will be $1-a'_q$. Now the second period of the longitude $\D t$ 
of the continuous interaction gives the summand $d_{p,q}\D t\ a'_p (1-a'_q)$ to the phase. At last, we restore the contents of $q$-th 
qubit by the second application of NOT. The resulting phase shift in these four actions will be $d_{p,q}\D t\ a'_p$ and it depends on the contents of $p$-th qubit only. Now we can compensate this phase shift by means of single one qubit transformation. If we consider the pair of qubits with the numbers
 $p,q$, where one, say, $p$-th is the separated, the other is not separated, we then can compensate its interaction using only one qubit operations: two NOTs for $q$-th and some phase shift for $p$-th. Now we have to modify this method so that to compensate all influences of not separated qubits simultaneously. For this, we will fulfill NOT operations on each such qubit with the sufficiently small time intervals so that the deposits to the phase of non-separated qubits will cancel each other. There are two ways to do it: to use the random process for the generation of the moments for one-qubit operations, or to realize them periodically with the different periods for the different qubits. At first, we study the first approach. 
\nn

 {\bf Method of random processes} 

\nn
 For each not separated qubit number $p$ we consider the Poisson process ${\cal A}_p$, generating time instances $0<t_1^p <t_2^p <\ldots <t_{m_p}^p <1$ with some fixed density $\la\gg 1$. Let all ${\cal A}_p$ be independent. We fulfill NOT operators on each qubit with number $p$ in time instances $t^p_m$ sequentially. In the moment 1 we fulfill NOT on $p$-th qubit if and only if $m_p$ is odd. Therefore, after this procedure each qubit restores its initial value. We count the phase shift generated by this procedure. Interactions between the separated qubits remain untouched. We fix some non-separated qubit and count its deposit to the phase. It consists of two summands: the first comes from the interaction with the separated, the second - from the interaction with non-separated qubits. We find them sequentially. In view of high density $\la$ of Poisson process ${\cal A}_p$ about the half of all time $p$-th qubit will be in the state $a'_p$, and the remaining half - in the state $1-a'_p$. Its interaction with the separated qubit, say, with $j$-th, gives the deposit $\frac{1}{2} d_{p,j}a'_p a'_j +\frac{1}{2} d_{p,j} (1-a'_p)a'_j$ e.g., $\frac{1}{2} d_{p,j}a'_j$. 2. We consider the different non-separated qubits with the numbers $q\neq p$. In view of independency of the time instants on the fulfillment of NOT- operators on $p$м and $q$-th qubits and the high density $\la$, these qubits will be in each state ($a'_p ,\ a'_q$), ($a'_p ,\ 1-a'_q$), ($1-a'_p , 
\ a'_q$), ($1-a'_p ,\ 1-a'_q$) approximately the quarter of the all time. The resulting deposit will be $\frac{1}{4}d_{p,q} [a'_p 
a'_q +a'_p (1-a'_q )+(1-a'_p )a'_q +(1-a'_p )(1-a'_q )]$ $=\frac{1}{4}d_{p,q}$. The common phase shift coming from the presence of non separated qubits is found by the summing of the values from the points 1 and 2 for all $p\notin\{j,k\}$. It will be $$\frac{1}{2}[\sum\limits_{p\notin\{j,k\}}d_{p,j}a'_j +\sum\limits_{p\notin\{j,k\}}d_{p,k}a'_k ]+\frac{1}{4}\sum
\limits_{p,q\notin\{j,k\}} d_{p,q}. $$ This shift can be compensated by only one qubit gates because the first two summands depend on the values of qubits only, and the other are constants. We thus obtain the scheme with the continuous two qubit interaction and one qubit operations which fulfills the appropriate phase shift to $d_{j,k} a'_j a'_k$. 
If we take the time segment $\D t$ instead of the unit time in this procedure, we obtain the phase shift to $\D t\ d_{j,k} 
a'_j a'_k$. If we want to obtain the phase shift to $-\D t\ d_{j,k}a'_j a'_k$, we must at first apply NOT to $j$-th qubit, then the previous procedure, then again NOT to $j$-th qubit, and at last, add $-\D t\ d_{j,k}a'_k$ by the one qubit operation. Therefore, we can do any addition to the phase of the form $c\cdot a'_j a'_k$ for a real $c$ independently of its sign. The appropriate combination of these schemes gives the phase shift 
\begin{equation} \sum_{j,k} c_{j,k} a'_j a'_k \label{Phase} 
\end{equation} 
for any $c_{j,k}$. Disposing these operations before and after $QFT^{-1}$ in the procedure of the previous point, we compensate the summands $A$ and $B$ in the phase and obtain the scheme realizing $QFT^{-1}$. The errors appearing in this scheme come from the possible low quality of Poisson processes generating the moments of the fulfillment of  NOT operations, and from the interaction in course of these operations. They can be minimized by the increasing of the density $\la$ and the decreasing of the time of NOT operations comparatively with the typical time of two qubit transformations determined by $d_{j,k}$. 

We evaluate the slowdown induced by the insertions of NOTs with the high density in comparison with the abstract realization of quantum computations of quantum gate arrays. We fix the unit of time such that the application of one operation in the scheme requires the unit time. 
Let the time axes be divided to the equal short intervals of the length $\d t$ units, NOT-th can be applied in the moments of the form $k \d t$ only, for any integer $k$ with the probability $p=1/\la$, where $\la$ in the density of process. Let the time of the whole computation equal $T$, and $M=T/\d t$. The error in the phase shift coming from the low quality of this model of random process, will be $\d t\ D$ where $D$ is the dispersion of the sum of random variables taking values 1 and 0 with the probabilities $p$ and $1-p$ that is $O(\sqrt{M})$. Consequently, the resulting error will be of the order $T/\sqrt{M}$ and must be negligible. For QFT we have $T=O(\log N)$ and we obtain that $M=O(\frac{\log^2 N}{\e} )$ is sufficient for the negligible $\e$. We see that the method of random processes gives a bit more than the quadratic slowdown comparatively to the standard abstract model that is sufficient for so fast quantum algorithms as 

\n 
We now prove the universality of the proposed model of quantum computations. We suppose that the interaction between qubits depends on their spatial positions only that we set fixed. The single condition we impose to the interaction is that it must be diagonal. Thus if $j$ and $k$ denote the number of two qubits, Hamiltonian of their interaction has one of the forms
 \begin{equation} {\rm A)}\ H_{j,k}=\left( \begin{array}
{ccccc} &E^{j,k}_1 &0 &0 &0\\ &0 &E^{j,k}_2 &0 &0\\ &0 &0 &E^{j,k}_3 &0\\ &0 &0 &0 &E^{j,k}_4 \end{array} \right) , \ \ \ 
\ \ \ \ \ \ \ \ \ \ {\rm B)}\ H_{j,k}=\left( \begin{array}{ccccc} &0 &0 &0 &0\\ &0 &0 &0 &0\\ &0 &0 &0 &0\\ &0 &0 &0 &E_
{j,k} \end{array} \right) ,\ \ E_{j,k} >0. \label{Hamiltonian} 
\end{equation} 
At first we note that any interaction on the general form (\ref{Ham}, A) can be reduced to the form (\ref{Ham}, B) by the addition of the proper one qubit Hamiltonians $H'_{j,k}$, which matrices have the forms
 $$ \left( \begin{array}{ccccc} &a &0 &0 &0\\ &0 &a &0 &0\\ &0 &0 &b &0\\ &0 &0 &0 &b
\\ \end{array} \right), \ \ \ \ \ \ \ \ \ \left( \begin{array}{ccccc} &\a &0 &0 &0\\ &0 &\b &0 &0\\ &0 &0 &\a &0\\ &0 &0 &0 
&\b\\ \end{array} \right) . 
$$ 
This addition reduces Hamiltonian of the form (\ref{Hamiltonian}, A) to  (\ref{Hamiltonian}, B) and it can be realized by the one qubit gates, since all diagonal matrices commutes. We note that the different pairs of qubits can interact differently, they can be placed on the different distances, not necessary on one line, etc. To prove the universality of the computational model with the continuous interaction we have to show how to fulfill an arbitrary two-qubit operation. Let we be given a unitary operation induced by Hamiltonian (\ref
{Hamiltonian}, B) in the unit time: $U_{j,k}=\exp (-iH_{j,k})$ (Plank constant we set equal unit, as usual). We show how to make this operation on two qubits:  $j$-th and $k$-th, preserving all the rest If we can do it, we will be able to realize any two qubit operation on any pair of qubits. Then for the far interaction, we have at most the linear slowdown comparatively to the standard model, and for the short interaction, we have to fulfill SWAP operation to bring the required pair of qubits together. We thus obtain the multiplier to the time of computation proportional to the size of memory. 
To make transformations $U_{j,k}$ it is needed to apply the method of NOT operations on non separated qubits described in the previous point, in moments of time generated by the independent Poisson processes of high density. However, now the advantage of this method is not as evident as in the case of QFT, because, for example, the fast quantum search requires more than logarithmic time: the square root of classical time. For such cases, one can apply the following modification of our trick.
\nn

{\bf Method of periodic NOTs} 
\nn

 We will make NOT operations on each of $j$ qubits in the time instants of the form $jk\d t$ for integer $k$, where $\d t$ is again the small period. We then can repeat the construction described above, and get rid of undesirable phase shifts by means of appropriate choice of $\d t$. This method gives the slowdown as the multiplier of the order $n^2$ comparatively to the complexity of the abstract model of quantum gate arrays. Now it is sufficient to show how by means of transformations $U_{j,k}$ we can make any two gubit gate. For example, we demonstrate how to realize CNOT operator on this pairs of qubits. Let $j,k$ be fixed and we omit indexes. We denote $\Delta E=E_1-E_2-E_3+E_4$. If $\frac{\Delta E}{\pi} \notin Q$ ($\frac{\Delta E}{\pi}$ not rational, then (because the physical parameters of our system affecting on phases, for example, cycle periods, can be slighly changed to avoid the rational parameter, we can treat it irrational without loss of generality) we can fulfil CNOT operation 
$$
\left[ CNOT=\begin{pmatrix}1&0&0&0\cr0&1&0&0\cr0&0&0&1\cr0&0&1&0 \end{pmatrix}\right]
$$ 
on the chosen pair of close qubit using only one qubit operators and the fixed diagonal operation $E$ 
$\left[ E=\begin{pmatrix}\exp{\left(i E_1\right)}&0&0&0\cr0&\exp{\left(i E_2\right)}&0&0\cr 0&0&\exp{\left(i E_3\right)}&0\cr0&0&0&
\exp{\left(i E_4\right)} \end{pmatrix}\right]$ by the following way.  

I. We denote the sequence of rotations of the phase of the first qubit by 

$ A=\left[
\begin{pmatrix}1&0\cr0&\exp{\left(i \left(E_1 -E_3\right)\right)}\end{pmatrix}\right]$, of the second qubit by $ B=\left[\begin{pmatrix}\exp\left(- i E_1
\right)&0\cr 0&\exp\left(- i E_2\right) ,\end{pmatrix} \right]$ and the operation $E$ by $U$ $ U=E \, (A \bigotimes B) = \left[\begin{pmatrix}1&0&0&0\cr0
&1&0&0\cr 0&0&1&0\cr0&0&0&\exp\left( i \Delta E\right) . \end{pmatrix}\right]$ 		

II. Using the irrationality of $\frac{\Delta E}{\pi}$ it is possible to show that $ \forall \varepsilon > 0 \exists m \in N \exists n \in N : |\Delta E n - \pi (2 m + 1)| < \varepsilon , $ 
e.g., for any chosen accuracy $\varepsilon$ there exists $n=n(\varepsilon)$ such that $U^n$ approximates the operator $\Pi
$ $ \Pi = \left[\begin{pmatrix}1&0&0&0\cr0&1&0&0\cr 0&0&1&0\cr0&0&0&-1 \end{pmatrix}\right]$ within the given accuracy. 

 III. Using the equality $ (I 
\bigotimes H) \Pi (I \bigotimes H) = CNOT, $ where $I$ is the identity matrix and $H$ - 
is Hadamard operation $ H=\frac{1}{\sqrt{2}}\left[\begin{pmatrix}1&1\cr1&-1 \end{pmatrix}\right]$ we see that CNOT is obtained as the sequence 
$ (I \bigotimes H) \left(E \, (A \bigotimes B)\right)^n (I \bigotimes H) $ of one qubit rotations and operation E. 

\subsection{Formalism of occupation numbers}

We consider the system consisting of $n$ identical particles. At first, we make the not physical supposition that they can be certainly distinguished. Then the state of such system belongs to Hilbert space with the basis  $\psi (r_1 , r_2,\ldots , r_n )$ $=\psi_{j_1} (r_1 )\psi_{j_2} (r_2 )\ldots\psi_{j_n} (r_n )$ where $\{\psi_j \}$ is some basis for the one particle states, $j_s$ belongs to some fixed set of indexes $1,2,\ldots ,J$, so that $r_j$ includes spatial and the so-called spin coordinates as well. \footnote{Spin is the internal momentum of inertia of a particle which has the relativity nature and can take the different values depending on the type of considered particles. For example, for electrons the projection of spin to teh chosen axes takes only two possible values.}
The choice of basis means simply that the system can be found only in some of basic states after the observation. 

However, in the real system of identical particles they cannot be distinguished. Therefore, each basic state must contain all summands of the form $\psi_{j_1} (r_1 )\psi_{j_2} (r_2) )\ldots\psi_{j_n} (r_n )$ with some coefficients. Now we need some information about the nature of the considered particles. They can be fermions, like electrons or protons, or bosons (as photons). The difference between these two types of particles is that the maximal value of the fermionic spin is half integer (1/2, 3/2, ...) and for bosons, it is integer (0,1,2,...). For us it is significant that the wave function of the system of fermions must change its sign in the permutation of two particles, for bosons, the sign is preserved. Algebraic correspondence we established between functional notations and qubit formalism dictates the representation of the wave function for the system of $n$ fermions in the form of the determinant:  
\begin{equation}
\Psi=\frac{1}{\sqrt{N!}}\left|
\begin{array}{ccccc}
&\psi_{j_1} (r_1 ) &\psi_{j_1} (r_2)&\ldots &\psi_{j_1} (r_n )\\
&\vdots &\vdots &\vdots &\vdots\\
&\psi_{j_n }(r_1 ) &\psi_{j_n} (r_2 ) &\ldots &\psi_{j_n} (r_n )\\
\end{array}
\right| ,
\label{state}
\end{equation}
and for the system of bosons in the form of the corresponding permanent\footnote{The permanent of matrix differs from its determinant only in that there are pluses instead of minuses in its computation, so that it remains unchanged in the permutations of rows of columns.}.
 Such state we can treat as the situation when only the states $\psi_{j_s}$ for $s=1,2,\ldots ,n$ are occupied by particles of our system, whereas the rest   $\psi_k$ for $k\in\{ 1,2,\ldots , J\}$, which have not the form $j_s$ are free. If $\psi$ with indexes denotes an eigenvector of the one particle Hamiltonian we speak about the occupied or free energetic levels, but in general, $\psi_k$ can form the arbitrary orthonormal basis in the space of one particle states. 

The state of the form (\ref{state}) can be represented as the symbol $|\bar n_\Psi\rangle =|n_1 ,n_2 ,\ldots ,n_J \rangle$ where 
$n_k$ is the unit, if $k$-th energetic level is occupied and zero, if it is free. It is the natural representation of the fermionic ensemble state in terms of occupation numbers. Such vectors $\bar n$ form the basis of Fock space and the general form of a state of our system in this basis is $\sum\limits_{\bar n}\la_{\bar n}|\bar n\rangle$ with amplitudes $\la$. 

The operator of annihilation $a_j$ of the particle in the state $j$ ($j$-th energy level) and its conjugate operator $a_j^+$ 
(creation of the particle in this state), is defined as  
$a_j |n_1 ,\ldots ,n_J \rangle =\d_{1,n_j}(-1)^{\s_j}| n_1 ,\ldots
 ,n_{j-1},n_j -1,n_{j+1},\ldots ,n_J \rangle$ 
where $\s_j =n_1 +\ldots +n_j$. They possess the known commutative relations: 
$a_j^+ a_k +a_k a_j^+ =\d_{j,k}$, $a_j a_k +a_k a_j=a_j^+a_k^++a_k^+a_j^+=0$.

Let us suppose that any interaction in Nature touches no more than two particles. Each interaction in many body ensemble then can be decomposed to the sum of one - two particle interactions of the form $H=H_{one}+H_{two}$ with the corresponding potentials $V_1 (r)$ и $V_2 (r,r')$. Each of them can be represented through the operators of creation and annihilation in the form  
$H_{one}=\sum\limits_{k,l}H_{k,l} a^+_k a_l$, $\ \ H_{two}=
\sum\limits_{k,l,m,n}H_{k,l,m,n}a^+_l a^+_k a_m a_n$ где 
$$
\begin{array}{ll}
&H_{k,l}=\langle\psi_k |\ H_{one}\ |\psi_l \rangle =\int \psi^*_k (r) V_1 (r)\psi_l (r)dr,
\\
&H_{k,l,m,n}=\langle\psi_l ,\psi_k\ |H_{two}\ |\ \psi_m \psi_n \rangle
 =\int\psi^*_k (r)\psi^*_l (r') V_2 (r,r')\psi_m (r)\psi_n (r') drdr'.
\end{array}
$$
Hence, given potentials of all interactions and all basic states $\psi_i$, we can in principle find their representation in terms of operators of creation and annihilation, that is in the language of occupation numbers. 

We consider the ensemble with Hamiltonian of the form $H=\sum_i H^i_{ext. f.}+\sum_{i,j}
(H^{i,j}_{diag.}+H^{i,j}_{tun.})$, where Hamiltonians of external fields, diagonal interaction and tunneling are represented in terms of creation and annihilation operators as 
\begin{equation}
\begin{array}{lll}
&H^i_{ext.f.} &= \a_i a^+_i a_i ,\ \ \ \ \a_i\in{\rm R},\\
&H^{i,j}_{diag.} &= \b_{i,j} a^+_i a_i a^+_j a_j ,\ \ \ \ \b_{i,j}\in{\rm R},\\
&H^{i,j}_{tun.} &= \g_{i,j} a^+_i a_j +\g_{i,j}^* a^+_j a_i .
\end{array}
\label{ham}
\end{equation}

We note that to realize the control on the diagonal Hamiltonian would be difficult, because this interaction touches two arbitrary particles in the considered ensemble, which are non-distinguishable by the identity principle. It is thus natural to treat this interaction as constant and independent from our control, whereas we can effectively control the tunneling interaction. This form of control makes possible to realize any quantum computation. This type of the control looks as more realistic because we can realize the tunneling by means of laser impulses. 

\subsection{Computations controlled by tunneling}

To prove the universality of the proposed simplified scheme of control on fermionic computations we must make one technical preparation, namely, to establish some different correspondence between Hilbert space of qubits and Fock space of occupation numbers. This correspondence will be different from the natural correspondence we spoke earlier. 

We fix some division of the set of energy levels to two parts and choose some one-to-one correspondence between these parts. For the determinacy we can take the $k$-th level down from Fermi level $\epsilon_F$ and agree that it corresponds to the $k$-th level up from $\epsilon_F$. We denote $j$-th level don from Fermi border by the standard letter, and the $j$-th level up from this level by this letter with the stroke $j'$. We call the first level the $j$-th the lower level and the second level the $j$-th upper level. Fock space $\cal F$ can be then represented as ${\cal F}={\cal F}_1 \bigotimes {\cal F}_2 
\bigotimes\ldots\bigotimes {\cal F}_k$ where each ${\cal F}_j$ corresponds to $j$-th pair of corresponding energy levels. We consider the subspace $F_j$ in ${\cal F}_ j$, which is generated by two following vectors. The first will be: "$j'$-th level is occupied, $j$-th free", the second will be: "$j$-th level is occupied, $j'$-th is free". 
We denote them by $|1\rangle_j$ and $|0\rangle_j$ correspondingly. We will work with the subspace $F=F_1 \bigotimes F_2 \bigotimes\ldots\bigotimes F_k$ in Fock space ${\cal F}$. We define the function $\theta$, which maps our Hilbert space $\cal H$ to $F$ by the following definition on basic states: $\theta (|\xi_1 ,\xi_2 \ldots\xi_n \rangle )=|\xi_1 \rangle_1 \bigotimes |\xi_2 \rangle_2 \bigotimes\ldots\bigotimes |\xi_n \rangle_n$ where all 
$\xi_j$ are zeroes and units. Then the function $\theta$ establishes the non-standard correspondence between Hilbert and Fock spaces (see the picture 1).

One qubit state in Hilbert space corresponds to two-qubit state in the usual identification with qubits (one level - one qubit).
We will see that this identification better fits to our aims than the natural. Now all is ready for the representation of unitary operators in Hilbert space in terms of operators acting in the space of occupation numbers. We consider an arbitrary Hermitian operator $H$ in two-dimensional Hilbert space of one qubit states $\cal H$. It has the form $H_0
+H_1$, where
$$
H_0 =\left(
\begin{array}{ccc}
&d_1 &0\\
&0 &d_2
\end{array}
\right) ,
H_1=\left(
\begin{array}{ccc}
&0 &d\\
&\bar d &0
\end{array}
\right) .
$$

\setlength{\unitlength}{0.6pt}
\begin{picture}(720,420)(0,90)
\put(100,500){$\xi\in{\cal H}$}
\put(247,500){$|0\rangle_2$}
\put(332,500){$|1\rangle_2$}
\put(520,500){$|010\rangle$}
\multiput(200,230)(0,36){7}{\multiput(0,0)(17,0){10}{\line(1,0){7}}\multiput(260,0)(15,0){14}{\line(1,0){5}}}
\put(255,410){\circle{10}}
\put(340,266){\circle{10}}
\put(500,374){\circle{10}}
\put(536,266){\circle{10}}
\put(572,446){\circle{10}}
\put(340,410){\circle*{10}}
\put(500,302){\circle*{10}}
\put(536,410){\circle*{10}}
\put(255,266){\circle*{10}}
\put(572,230){\circle*{10}}
\put(100,338){$\theta (\xi )$}
\put(200,338){\line(1,0){170}}
\put(460,338){\line(1,0){215}}
\put(700,338){$\epsilon_F$}
\put(150,230){$3$}
\put(150,266){$2$}
\put(150,302){$1$}
\put(150,338){$0$}
\put(150,374){$1'$}
\put(150,410){$2'$}
\put(150,446){$3'$}
\put(0,150){Picture 1. \ \ Correspondence between Fock and Hilbert spaces}
\end{picture}

It can be straightforwardly verified that for operators $\tilde H_0 = d_1 a^+_k 
a_k +d_2 a^+_{k'} a_{k'}$ and $\tilde H_1 = d a^+_k a_{k'} +\bar d a^+_{k'} a_k$ 
(external field and tunneling) the following equalities take place: $ \tilde H_i \theta 
=\theta H_i$ for $i=0,1$. Using the linearity $\theta$, we find $(\tilde H_0 
+\tilde H_1 )\theta =\theta H$.
Now we consider one qubit unitary operator $U$ in Hilbert space. It has the form $e^{-iH}$ for Hamiltonian $H$ (we have chosen the appropriate unit system to get rid of Plank constant and the time). Due to the linearity of $\theta$ and the equation $\theta^{-1} H^s \theta =(\theta ^{-1}H\theta )^s$ for integer $s$ we find that for any one qubit unitary operator $U$ we can effectively find the corresponding Hamiltonian in Fock space containing only the external field and the tunneling, which makes the diagram A from the picture 2 closed.

We take up two qubit transformations in Hilbert space. Since all diagonal matrices commute, for all diagonal transformations in the spaces ${\cal F}_k 
\bigotimes {\cal F}_j$ м can effectively find the corresponding diagonal operator in Hilbert space, which makes the diagram B from the picture 2 closed.

Now all is ready for the transfer of the trick from the work \cite{OF} with one qubit control to Fock space. The combination of diagrams from the picture 2 gives the diagram from the picture 3.

\begin{picture}(700,380)(0,150)
\put(64,330){\vector(1,0){256}}
\put(424,330){\vector(1,0){266}}
\put(64,220){\vector(1,0){256}}
\put(424,220){\vector(1,0){266}}
\put(50,234){\vector(0,1){86}}
\put(330,234){\vector(0,1){86}}
\put(410,234){\vector(0,1){86}}
\put(700,234){\vector(0,1){86}}
\put(37,275){$\theta$}
\put(317,275){$\theta$}
\put(397,275){$\theta$}
\put(687,275){$\theta$}
\put(43,213){$\cal H$}
\put(323,213){$\cal H$}
\put(403,213){$\cal H$}
\put(693,213){$\cal H$}
\put(43,323){$\tilde F$}
\put(323,323){$\tilde F$}
\put(403,323){$\tilde F$}
\put(693,323){$\tilde F$}
\put(80,340){ext. field + tunneling}
\put(480,340){$\tilde F
$ diagonal}
\put(80,200){one-qubit on $\cal H$}
\put(480,200){$\cal H$ diagonal}
\put(150,150){A}
\put(550,150){B}
\put(120,90){Picture 2.\ \ Correspondence of operators in }
\put(180, 65){Fock and Hilbert subspaces. $\tilde F=F_j \bigotimes F_k .$}
\end{picture}

\begin{picture}(700,490)
\put(90,320){$\cal F$ diag}
\put(250,320){f+t}
\put(580,320){$\cal F$ diag}
\put(90,130){diag}
\put(250,130){one qubit}
\put(580,130){diag}
\put(40,300){\vector(1,0){125}}
\put(220,300){\vector(1,0){125}}
\put(550,300){\vector(1,0){125}}
\put(40,150){\vector(1,0){125}}
\put(220,150){\vector(1,0){125}}
\put(550,150){\vector(1,0){125}}
\put(30,165){\vector(0,1){115}}
\put(197,165){\vector(0,1){115}}
\put(700,165){\vector(0,1){115}}
\put(390,300){.\ .\ .}
\put(390,150){.\ .\ .}
\put(20,297){F}
\put(190,297){F}
\put(693,297){F}
\put(20,142){$\cal H$}
\put(187,142){$\cal H$}
\put(685,142){$\cal H$}
\put(100,80){Picture 3.\ \ Correspondence of computations in}
\put(180,55){Fock and Hilbert spaces}
\end{picture}

Let the diagonal part of Hamiltonian of interaction in Fock space be fixed and act permanently in the non-controlled mode. We then can find the corresponding diagonal interaction in Hilbert space, making closed the "diagonal" parts of diagrams from the picture 3. By virtue of the result of theh work \cite{OF} we can find one-qubit transformations realizing the control on the arbitrary quantum algorithm in Hilbert space, in the form of the lower sequence of transformations in the diagram. Al last, we can find the control of the form "field + tunneling" on the state in Fock space making closed the whole diagram. We note that all operators of creation and annihilation considered in the whole Fock space are non-local due to the multiplier $(-1)^{\s_j}$, which depends on a given state.   

For the diagonal operator $a_j^+ a_j a_k^+ a_k$ and the external field these multipliers compensate each other. The tunneling operator $a_j^+ a_{j'}$ in the space $F$ brings the multiplier $(-1)^{\s'}$ where $\s'=\sum\limits_{s+j}^{j'-1}n_s =j'-j$, which does not depend on the given state $|\bar n\rangle\in F$, because for such state exactly the half of levels between $j$ and $j'$ are occupied by fermions. The sign we can factorize from all states, and ignore.

We thus obtain the universal quantum computer on states in the space of occupation numbers, controlled by the external field and the tunneling only. 

\section{Error correction in quantum computations} 

The peculiarity of quantum states reveals in the especial sensitiveness to the external influence. It follows from their extremely small size. For example, it is impossible to isolate a group of atoms in the crystal lattice from the heat influence from the other atoms. We saw that there are two principally different types of such influences: external field included to Shredinger equation, and measurements of the considered system. Here if the influence of an external field causes unitary, e.g., reversible disturbance of the state of system, the measurement always causes the irreversible change of this state. It is thus impossible to give the effective and uniform description of these two types of external influences in Hilbert space formalism\footnote{Of course, we could describe these influences formally by means of the same tools, by the so-called super operators - unitary operators acting in the extended Hilbert space. However, this description cannot be very fruitful because it does not possess even the limited determinacy, which presents in the standard Hilbert formalism for unitary operators, as the possibility to find the exact the wave function of the studied system in each time instant.}. There are thus two types of the errors in quantum states: reversible, that can take place in classical states as well, and irreversible, which is specifically quantum. However, this classification of errors does not exactly reflect the essence of the question. We take, for example, the state $\frac{1}{\sqrt 2} |000\ldots 0\rangle +\frac{1}{\sqrt 2} |111\ldots 1\rangle$. The error in the first qubit can have the form: $|0\rangle\ar |1\rangle,\ |1\rangle\ar |0\rangle$ or $|0\rangle\ar |0\rangle,\ |1\rangle\ar |0\rangle$. The first will be the unitary transformation, the second - not. Our state then acquires the form $\xi' =\frac{1}{\sqrt 2} |100\ldots 0\rangle +\frac{1}{\sqrt 2} |011\ldots 1\rangle$ or $\xi' =\frac{1}{\sqrt 2} |000\ldots 0\rangle +\frac{1}{\sqrt 2} |011\ldots 1\rangle$ correspondingly. We see that factually, the initial states and "corrupted" states are very close to each other, and the both errors can be easily corrected by the unitary operator (in the second case we should use the qubits, which are not touched by the external influence). It happens because these errors factually touch the contents of one qubit only. Now we regard the change of one qubit state, let it be the first qubit. Let the system be initially in the state $ \la |0,0,0\rangle +\mu |1,1,1\rangle$. The error then leads to one of two states: $|0,0,0\rangle$ or $|1,1,1\rangle$, and we loose all information about the initial amplitudes. We see that if the qubits are in the entangled state, then the measurement of one of them unavoidably causes the error in all the others ! Therefore, it would be more right to classify all possible errors to local, e.g., touching the contents of only one qubit, and global, touching the unlimited number of qubits. Here in the case of entangled state local influence (measurement of a single qubit) can lead to the global error. The both these types of errors naturally appear in course of computations, and we have to fight with them. It is much simpler to correct local errors than global. We can also suppose that for the correction of global errors induced by the local influence to the system it is necessary to use purely quantum trick because this type does not occur in classical systems. If this trick exists, we must use it many times in course of computations, because the influence of environment leading to the errors is permanent. We must thus launch the program of the error correction in the background regime and correct the errors permanently. We suppose that the environment influence is local and its intensity is low enough to allow the error correction. It turns that such a method of error correction exists. In this section, we give its simplified scheme. 

At first we consider the principle of the error correction in classical case. It rests on the redundant coding. We encode the state of one bit by $m$ bits in which we merely duplicate the contents of the initial bit. Now, if the error results in the change of $\left[\frac{m-1}{2}\right]$ coding bits, we can easily restore the initial contents, if we take the prevailing value in the coding bits. If the error touches the minor number of bits, the majority of them ensure the correct restoration of the initial state. Let us analyze the possibility of the application of this trick for the quantum state correction. For this we have to "replicate" quantum states so easily, as we replicate the states of classical bits. Here the first trap waits us. It turns that there is no a real physical process copying quantum states ! This fact is known as the "no cloning theorem" but it is quite simple, and we will now establish it. Really, we know that all physical processes can be described either by unitary operators, or have the probability character. We suppose that there exists the physical process $W$, cloning quantum states. To be useful, it must act to all states in Hilbert space, e.g., it must have the form of function $W:\ H\bigotimes H\ar H\bigotimes H$, such that $W(|\Psi , 0\rangle =\la |\Psi , \Psi\rangle$ for any state $|\Psi\rangle$ where the constant $C$ depends on$\Psi$. This function must not be probabilistic, because we consider only pure states. Hence, $W$ must be unitary operator. Due to the linearity of $W$ we would obtain $\la |\Psi ,\Psi\rangle +\mu |\Psi_1 ,\Psi_1\rangle = W|\Psi , 0\rangle + W|\Psi_1 ,0\rangle =W|\Psi +\Psi_1,0\rangle =\nu |\Psi +\Psi_1,\ \Psi+\Psi_1\rangle =\nu (|\Psi ,\Psi\rangle +|\Psi_1 ,\Psi_1\rangle +|\Psi ,\Psi_1\rangle +|\Psi_1 ,\Psi\rangle )$, for all pairs of states, that is obviously wrong. Therefore, the cloning of quantum states is impossible. 

One may suppose that we could use the natural "pseudo cloning" by means of CNOT and ancilla: 
$CNOT\ (\la |0\rangle +\mu |1\rangle )\bigotimes 0\rangle = \la |0,0\rangle +\mu |1,1\rangle$ ? Of course, we will not have the cloned state of the initial qubit in the result: $ (\la |0\rangle +\mu |1\rangle )\bigotimes(\la |0\rangle +\mu |1\rangle )$ (let you check it , using the distributive property of tensor product). However, in some sense our initial qubit will be somehow encoded into two new qubits. Further, we can repeat this procedure with the following qubit, that results in the state $ \la |0,0,0\rangle +\mu |1,1,1\rangle$. This procedure in case of classical state of qubit: 0 and 1 really leads to the cloning of state. However, there is no use from this coding in the case of global errors, it follows from the considered case of one qubit measurements. This example illustrates also the prohibition to the quantum state cloning, because in the attempt to "clone" the state its "copies" turn in the entangled state with the "originals" and the measurement of "originals" unavoidably destroys "copies". For the correction of this influence the more complex trick is required that we now describe for the simplest situation. 

Factually two chosen types of errors - local and measurements of separated qubits, exhausts all possible types of errors at all. We consider, for example, the phase error in the first qubit $|0\rangle \ar |0\rangle ,\ |1\rangle \ar -
|1\rangle$. Hadamard operator leads to the same result than the inversion of the contents of this qubit. If we correct the error in the contents of this qubit then the following Hadamard transform restores the previous state. We now consider such influence of environment that can be represented in the form $U:\ \a\bigotimes\xi\ar\chi$ where $\a$ is the state of environment and $U$ acts on $\a$ and on the first qubit in our system. At first glance this type of errors is more general. Nevertheless, we can reduce it to the combination of the one qubit measurement and the local error. 

Let us consider the correction of local errors again. We encode our quantum state by means of some number of ancilla, and form the special ancilla playing the role of "dustbin". As the error happens, we by means of standard unitary transformations transmit the contents of the qubits touched by the error to this "dustbin", so that the initial states of the corrupted qubits is restored, and only the "dustbin" stores the information about the error. 

This scheme will work in the case of the global errors as well. Let the intensity of local influences of environment to our quantum system be limited. We transform the initial state in the corresponding encoded state of the system with ancilla and then can use the correcting procedure again and again in the short time frame $\Delta$ because the probability of that in course of this time the environment will influence to more than one qubit will be negligibly small. So, varying $\Delta$, we can maintain the system in the initially encoded state as long as needed. 

Of course, the correcting procedure must depend on the encoding method, and we at first describe this method as applied to one qubit state $\xi$. In case of many qubits, we can simply encode each qubit separately and apply our procedure of correction on all coding ensembles simultaneously and independently. Our correcting procedure restores the encoded state and preserves the entanglement so that  its simultaneous application to each coding ensembles preserves the encoded form of the initial ensemble. 

We thus concentrate on the method of encoding of one qubit state $\xi =\a |0\rangle +\b |1\rangle$. 
This qubit we encode by three qubits so that if one of them is measured (we do not know what exactly), then our procedure held on the result of the measurement will restore the initial entangled state. The encoding will be the linear operation, and it is sufficient to define it for basic states $|0\rangle$ and $|1\rangle$. Their codes will be correspondingly: 
\begin{equation}
\begin{array}{cc}
\tilde 0=&\frac{1}{2\sqrt 2} (|000\rangle +|100\rangle +|010\rangle +|001\rangle 
+|110\rangle +|101\rangle +|011\rangle +|111\rangle )\\
\tilde 1=&\frac{1}{2\sqrt 2} (|000\rangle -|100\rangle -|010\rangle -|001\rangle 
+|110\rangle +|101\rangle +|011\rangle -|111\rangle ).
\end{array}
\end{equation}

The code of the state $\xi$ will be thus the three qubit state of the form $\tilde \xi =\a \tilde 
0+\b\tilde 1$. We now look what happens if we observe one of qubits. We have three qubits, two states encoding basic one qubit states: $\tilde 0$ and $\tilde 1$ and two different results of measurement: $0$ or $1$. There exist 12 different results of the measurement of our two encoding states, if one qubit is measured, plus 2 unchanged basic states if no measurement happened. We denote the results by $\tilde i^j_k ,\ 
i=0,1,\ j=\emptyset ,1,2,3,\ k=\emptyset ,0,1$ which means the "state obtained in the measurement of $j$-th qubit of the state $\tilde i$, if $k$ is the result of this measurement". If there is no measurement, we have $j=k=\emptyset$ and $\tilde i^j_k
=\tilde i$. In all cases the resulting state will have the form:
\begin{equation}
\begin{array}{cc}
\tilde 0_0^1 &=\frac{1}{2} (|000\rangle +|010\rangle +|001\rangle +|011\rangle 
,\\
\tilde 0_1^1 &=\frac{1}{2} (|100\rangle +|110\rangle +|101\rangle +|111\rangle 
,\\
\tilde 0_0^2 &=\frac{1}{2} (|000\rangle +|100\rangle +|001\rangle +|101\rangle ,\\
\tilde 0_1^2 &=\frac{1}{2} (|010\rangle +|110\rangle +|011\rangle +|111\rangle ,\\
\tilde 0_0^3 &=\frac{1}{2} (|000\rangle +|010\rangle +|100\rangle +|110\rangle ,\\
\tilde 0_1^3 &=\frac{1}{2} (|001\rangle +|011\rangle +|001\rangle +|101\rangle ,\\
\tilde 1_0^1 &=\frac{1}{2} (|000\rangle -|010\rangle -|001\rangle +|011\rangle ,\\
\tilde 1_1^1 &=\frac{1}{2} (-|100\rangle +|110\rangle +|101\rangle -|111\rangle ,\\
\tilde 1_0^2 &=\frac{1}{2} (|000\rangle -|100\rangle -|001\rangle +|101\rangle ,\\
\tilde 1_1^2 &=\frac{1}{2} (-|010\rangle +|110\rangle +|011\rangle -|111\rangle ,\\
\tilde 1_0^3 &=\frac{1}{2} (|000\rangle -|010\rangle -|100\rangle +|110\rangle ,\\
\tilde 1_1^3 &=\frac{1}{2} (-|001\rangle +|011\rangle -|001\rangle +|101\rangle 
\end{array}
\end{equation}

We now define the action of correcting operator on the given vectors as 
\begin{equation}
U_{rest} :\ \tilde i_k^j \bigotimes\bar 0\ar\tilde i\bigotimes\tilde 0_k^j .
\end{equation}
Let us check that thus defined operator preserves the angles between all vectors $i^j_k$. We denote the subspaces generated by all vectors of the forms 
$\tilde 0_k^j$ and $\tilde 1_k^j$ through $O$ and $I$ correspondingly. We have: 
$\tilde 0\bot\tilde 1$, $O\bot I$, all the other angles equal $\frac{\pi}{3}$, because 
$\langle \tilde 1_0^1 |\tilde 1_0^2 \rangle =\langle \tilde 1_0^1 |\tilde 1_1^2 
\rangle =\frac{1}{2}$ and the same takes place for the basis of the subspace $O$. We see that all angles are preserved.
Therefore the operator $U_{rest}$ can be continued to the unitary operator on all the space of six qubit states $C^8 \bigotimes C^8$, which we denote by the same letter. 
We note that its action we defined for the states with zero ancilla. It means that we have to take care about the restoration of ancilla after each application of this operator $U_{rest}$ that is necessary for the right action of the next correcting operator. 

We can do it measuring ancilla and then placing zero to them (that is not unitary operator). We denote this procedure by $A_{rest}$. The full circle "decoherence - restoration" then acquires the form 
\begin{equation}
\a\tilde 0+\b\tilde 1\stackrel{\mbox{observation}}{\ar}\a\tilde 0_k^j +\b\tilde 
1_k^j \stackrel{U_{rest}}{\ar}(\a\tilde 0+\b\tilde 1)\bigotimes\tilde 0_k^j 
\stackrel{A_{rest}}{\ar}\a\tilde 0+\b\tilde 1 ,
\end{equation}
where we omit zero ancilla in the notation of the tensor product.  
We note that our quantum correcting transformation does not depend of what error happened. Its role is to transfer the error, if it takes place at all, into the "dustbin" formed by ancilla and to clear the initial encoded state, which will not be changed in the measurement and restoration of ancilla because they are factorized as the tensor multiplier in the common state. 

Let $\delta_{rest}$ denote the time required for this sequence of operators, $\Delta$ be the interval on which we make the correction procedure.
We suppose that the intensity of external influence is not too large so that the probability of the measurement of more than one qubit from our three in the time $\Delta$ is negligibly small comparatively with the probability of the measurement of exactly one. We also hold ancilla in zero states by permanent measurements in any time but the periods $\delta_{res}$. In addition let the error probability in any qubit in the time $\delta_{rest}$ be also negligibly small.  
The described iteration of the correcting operator with high probability will maintain our encoded state of one qubit despite of decoherence. To defense many qubit states it would be sufficient to encode all qubits separately and apply this correcting procedure to each coding six-qubit ensemble separately. 

\subsection{Necessity of measurements in course of quantum error correction}

We obtain the elementary procedure of quantum error correction, consisting of the application of one unitary operator and the following measurements of ancilla and transfer them to zero state. The second part of this repeating procedure is evidently non-unitary. Of course, we can represent that the "corrupted" ancilla every time are eliminated, and we use the new ancilla instead. Can we correct errors by means of unitary operators only, e.g., not applying the measurements and changes of the contents of ancilla? Let us suppose that it is possible and let $U$ be the corresponding unitary correcting operator replacing all the chain $A_{rest}$. 
It then must fulfill its function with the arbitrary ancilla, because now ancilla is the subject of decoherence as well. We denote registers containing all coding qubits and all ancilla by $a$ and $b$ correspondingly. For the initial state of the form $a\bigotimes b$ with arbitrary $b$ and results $a_1$ or $a_2$ of the measurements with any $b$ we must have $U\ a_i \bigotimes b =a\bigotimes b_i$ for some states $b_i$, touched by decoherence. The results $a_1$ and $a_2$ of the measurement are orthogonal, since $b_1$ and $b_2$ must be orthogonal due to the unitarity of $U$. Let the dimensionality of the ancilla space be $k$. We choose some orthogonal basis $b^1 , b^2 , \ldots b^k$ in the space of ancilla and let $U\ a_i 
\bigotimes b^j =a\bigotimes b_i^j$ for some states $b_i^j$ of ancilla. Subspaces, generated by vectors $b_1^j$ and $b_2^j$ must be orthogonal. On the other hand, each pair $b_i^j ,\ b_i^{j'}$ must be orthogonal for $i=1,2;\ j,j' =1,2,\ldots 
k,\ j\neq j' $, since $U$ is unitary. We obtain $2k$ orthogonal nonzero vectors $b_i^j$ in $k$ dimensional space that gives the desired contradiction. 

Purely unitary quantum error correction is thus impossible. Hence, the elementary scheme of QEC we represented cannot be radically simplified. In any case the correcting procedure must contain irreversible elements, like measurements or the usage of new properly initialized ancilla.  

\end{document}